\documentclass[10pt,twocolumn,letterpaper]{article}

\usepackage[pagenumbers]{wacv} % To force page numbers, e.g. for an arXiv version

\usepackage{graphicx}
\usepackage{amsmath}
\usepackage{amssymb}
\usepackage{booktabs}
\usepackage{multirow}
\usepackage{gensymb}
\usepackage{placeins}
\usepackage[pagebackref,breaklinks,colorlinks]{hyperref}

\usepackage[capitalize]{cleveref}
\crefname{section}{Sec.}{Secs.}
\Crefname{section}{Section}{Sections}
\Crefname{table}{Table}{Tables}
\crefname{table}{Tab.}{Tabs.}

\begin{document}

\title{DeepCA: Deep Learning-based 3D Coronary Artery Tree Reconstruction from Two 2D Non-simultaneous X-ray Angiography Projections}
 
\author{Yiying Wang$^1$ \qquad  Abhirup Banerjee$^{1,2}$ \qquad Robin P. Choudhury$^{2,3}$ \qquad Vicente Grau$^1$\\
$^1$Institute of Biomedical Engineering, Department of Engineering Science, University of Oxford\\
$^2$Division of Cardiovascular Medicine, Radcliffe Department of Medicine, University of Oxford\\
$^3$Oxford Acute Vascular Imaging Centre, Oxford, U.K
}

\maketitle 

%%%%%%%%% ABSTRACT
\begin{abstract}
Cardiovascular diseases (CVDs) are the most common cause of death worldwide. Invasive x-ray coronary angiography (ICA) is one of the most important imaging modalities for the diagnosis of CVDs. ICA typically acquires only two 2D projections, which makes the 3D geometry of coronary vessels difficult to interpret, thus requiring 3D coronary artery tree reconstruction from two projections. State-of-the-art approaches require significant manual interactions and cannot correct the non-rigid cardiac and respiratory motions between non-simultaneous projections. In this study, we propose a novel deep learning pipeline named \emph{DeepCA}. We leverage the Wasserstein conditional generative adversarial network with gradient penalty, latent convolutional transformer layers, and a dynamic snake convolutional critic to implicitly compensate for the non-rigid motion and provide 3D coronary artery tree reconstruction. Through simulating projections from coronary computed tomography angiography (CCTA), we achieve the generalisation of 3D coronary tree reconstruction on real non-simultaneous ICA projections. We incorporate an application-specific evaluation metric to validate our proposed model on both a CCTA dataset and a real ICA dataset, together with Chamfer $\ell_2$ distance. The results demonstrate promising performance of our DeepCA model in vessel topology preservation, recovery of missing features, and generalisation ability to real ICA data. To the best of our knowledge, this is the first study that leverages deep learning to achieve 3D coronary tree reconstruction from two real non-simultaneous x-ray angiographic projections. The implementation of this work is available at: \url{https://github.com/WangStephen/DeepCA}.
\end{abstract}

%%%%%%%%% BODY TEXT
\section{Introduction}
\begin{figure*}[!t]
\centering
\includegraphics[width=0.88\linewidth]{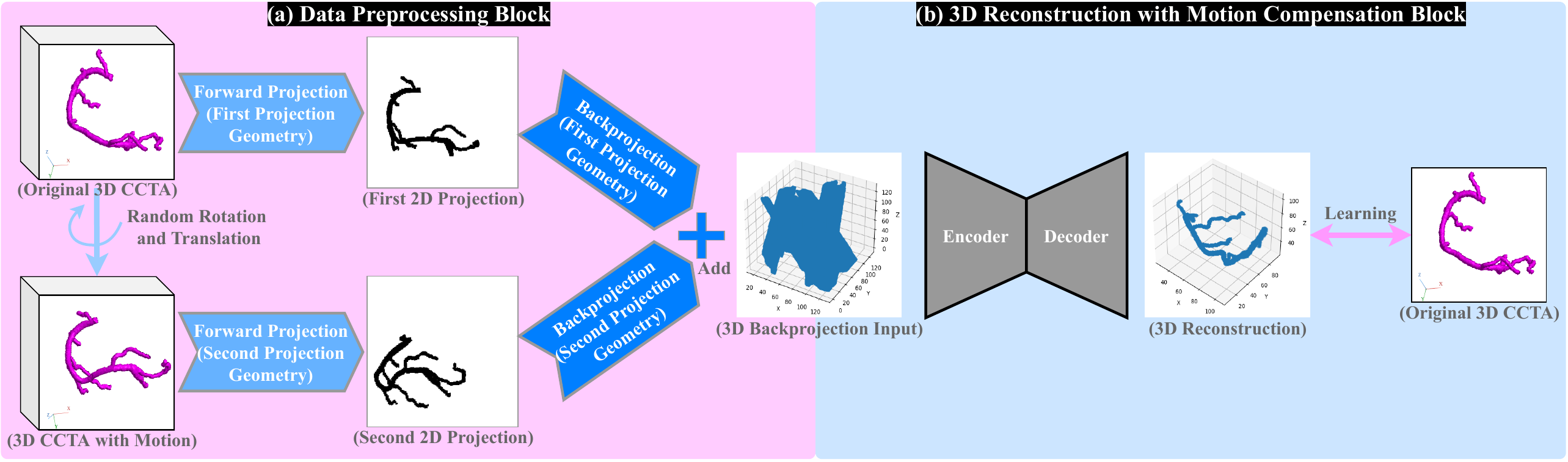}
\caption{The overall workflow of our proposed DeepCA pipeline consists of a data preprocessing block and a 3D reconstruction with motion compensation block. \textbf{(a)} The data preprocessing block generates two simulated ICA projections from 3D CCTA data, including simulated motion between projections, and then produces the 3D model input via performing backprojection on the two simulated projections. \textbf{(b)} The 3D reconstruction with motion compensation block receives the 3D backprojection input to train a deep neural network for 3D coronary tree reconstruction learned from the CCTA data, implicitly compensating for any motion.} \label{workflow}
\end{figure*}
Cardiovascular diseases (CVDs) represent a group of disorders of the heart and blood vessels. They are the most common cause of death worldwide, representing 32\% of all global deaths \cite{who2022} and posing a serious burden to healthcare systems. Invasive x-ray coronary angiography (ICA) is one of the most important imaging modalities for the diagnosis of CVDs and is most commonly utilised during real-time cardiac interventions \cite{lashgari2024fcvm}. However, ICA acquires only 2D projections of the coronary tree, which makes it difficult for cardiologists in clinical practice to understand the global 3D coronary vascular structure. This is complicated by the vessel overlap, foreshortening, complex vascular structure, and the artifacts caused by cardiac and respiratory motions and possible patient and device movements. These may all negatively affect the physicians' ability to locate the artery stenosis areas and navigate during clinical interventions \cite{green2005angiographic}. To alleviate this, cardiologists tend to use extra doses of contrast and x-rays to get more views for assessing the lesion. However, potential chemotoxic adverse reactions of radiographic contrast and x-ray radiation risk restrict the number of projections acquired to typically 2-5 projections. Therefore, the development of 3D coronary tree reconstruction based on limited (two) 2D coronary angiography projections has large clinical significance \cite{ccimen2016reconstruction}.

\par 3D coronary tree reconstruction from ICA images poses significant challenges: the complex vascular shape and limited projections provide limited information on 3D vessel structures. Most importantly, due to the non-simultaneous image acquisition, significant cardiac and respiratory non-rigid motions cause vessels to misalign between projections, which aggravates the difficulties of 3D reconstruction. Biplane x-ray angiography systems can simultaneously capture two projections and hence, unaffected by such non-rigid motions; however, they are expensive for clinical usage. Many traditional methods have been proposed for 3D coronary tree reconstruction from 2D non-simultaneous x-ray projections \cite{ccimen2016reconstruction}. However, they usually require substantial manual annotation and often cannot correct non-rigid cardiac motion. Recently, deep learning methods have achieved promising results in various tasks related to medical image analysis \cite{cciccek20163d,chen2022semi} including 3D reconstruction from 2D limited-angle projections. In terms of 3D coronary tree reconstruction using deep learning, previous methods have typically used synthetic data, coronary computed tomography angiography (CCTA), or ICA data from bi-planar scans, none of which suffer from non-rigid motion between projections \cite{bransby20233d,ibrahim20223d,iyer2023multi,maas2023nerf,uluhan20223d,wang2020weakly,wang2024neca3dcoronaryartery}. This limitation makes previous methods ill-suited for real non-simultaneous ICA acquisitions. Despite the improvement in deep neural networks, 3D coronary tree reconstruction from limited non-simultaneous angiographic projections has remained an open problem.

\par In this paper, we propose a novel deep learning pipeline named \emph{DeepCA}, leveraging the Wasserstein conditional generative adversarial network with gradient penalty, latent convolutional transformer layers, and a dynamic snake convolutional critic to implicitly compensate for the non-rigid motion to achieve 3D coronary artery tree reconstruction from two real non-simultaneous ICA projections. To resemble real non-simultaneous ICA projections, we simulate 2D projections in different planes from CCTA data containing real coronary tree geometries, with a rigid transformation applied to the CCTA data before forward projection on the second projection plane. We then use these simulated projections to learn from the CCTA ground truth to enable generalisation to real non-simultaneous ICA projections. In this way, we overcome the problems of both the limited number of real paired ICA data with projection geometry information and the unavailable 3D ground truth for real ICA data. We focus on the right coronary artery (RCA) in this study, because RCA undergoes more compressive strain and is affected more by motion artifacts than other coronary vessels. We provide an application-specific evaluation method to tackle the deformation in 3D reconstructions, the unavailability of 3D ground truth for real ICA scans, and the motion between projection planes, together with Chamfer $\ell_2$ distance. We validate our proposed model on a CCTA dataset and a real ICA dataset (the out-of-distribution domain), in comparison to four other models. The evaluation results demonstrate the promising performance of our proposed model in vessel topology preservation, recovery of missing features, and generalisation ability in 3D coronary tree reconstruction from real non-simultaneous ICA projections. The main contributions of this work are:
\begin{enumerate}
\item \textbf{3D coronary tree reconstruction using deep learning:} To the best of our knowledge, this is the first study that leverages deep learning to achieve 3D coronary tree reconstruction from two real non-simultaneous angiography projections.
\item \textbf{Generalisation:} Through simulating projections from CCTA data, we achieve generalisation on 3D coronary tree reconstruction from two real non-simultaneous ICA projections.
\item \textbf{Extensive evaluation:} We use a specific metric designated for this problem in the absence of motion-free 3D ground truth, which provides a baseline for future improvement in this area.
\end{enumerate}

%%%%%%%%%
\section{Related Works}
%%%%%%%%%
\subsection{3D Coronary Tree Reconstruction from Limited Views}
Much effort has been devoted to 3D coronary tree reconstruction from ICA projections by conventional mathematical methods \cite{ccimen2016reconstruction}. Several methods have been attempted for 3D reconstruction from limited views, including iterative reconstruction \cite{li2002accurate}, non-uniform rational basis splines \cite{galassi20183d,vukicevic2018three}, and point-cloud method \cite{8864089}. However, most of these methods are dependent on traditional stereo-vision algorithms, which usually require significant manual interactions such as label annotations, and they often cannot correct non-rigid cardiac motion \cite{Banerjee2018stacom,Banerjee2019premi}. 

\par Few studies have attempted deep learning-based 3D coronary tree reconstruction from several projections. In \cite{ibrahim20223d,uluhan20223d}, a supervised learning model was trained using 9 projections simulated from 3D synthetic coronary tree data. Iyer~\etal \cite{iyer2023multi} proposed a vessel generator to generate 3D synthetic coronary tree data, in order to train supervised multi-stage networks based on multi-layer perceptrons to predict both vessel centrelines and radii simultaneously, from 3 projections. Wang~\etal \cite{wang2020weakly} used CCTA data to simulate projections without motion and train a weakly-supervised adversarial learning model for 3D reconstruction from two projections. Maas~\etal \cite{maas2023nerf} proposed a neural radiance fields (NeRF)-based model to tackle the problem without involving 3D ground truth in training from 4 projections. Wang~\etal \cite{wang2024neca3dcoronaryartery} proposed an implicit neural representation-based method to reconstruct 3D coronary tree from two simulated CCTA projections. All of these aforementioned studies only used synthetic data or CCTA data for both training and testing without taking motion between projections into account, and were never applied to real non-simultaneous ICA projections. Bransby~\etal \cite{bransby20233d} used real ICA data to reconstruct a single coronary tree branch; but their acquisitions were based on bi-planar scans and so, no motion occurred between two projections.

\par To the best of our knowledge, our proposed method is the first to leverage deep learning to solve the problem of motion artifacts that exist between real non-simultaneous ICA projections acquired from single-plane x-ray angiography systems.

%%%%%%%%%
\subsection{3D Reconstruction from Limited Views in Medical Images}
Many deep learning algorithms have demonstrated promising results on the limited-angle computed tomography (CT) reconstruction problem recently, giving it potential for clinical diagnosis \cite{10253669}. In \cite{9176040}, a differentiable filtered backprojection-based neural network with an image domain U-Net was presented to achieve satisfactory reconstruction results from limited projections, while a fast self-supervised solution was proposed in \cite{zha2022naf} with promising results in terms of both speed and quality. However, the minimum number of projections required for the aforementioned methods were 145 and 50, respectively. Based on conditional generative adversarial networks, X2CT-GAN \cite{ying2019x2ct} was proposed to reconstruct CT from bi-planar x-rays. Further to this method, the CCX-RayNet was proposed with an addition of transformers \cite{ratul2021ccx}, achieving more accurate results than X2CT-GAN. An end-to-end encoder and decoder convolutional neural network was adopted in \cite{kasten2020end} for 3D knee bones reconstruction from bi-planar projections. Cafaro~\etal \cite{cafaro2023x2vision} proposed an unsupervised generative model with prior knowledge of anatomic structures to reconstruct 3D tomographic images of head and neck from bi-planar x-rays. However, these methods tackle the 3D reconstruction problem from simultaneous projections that do not suffer from non-rigid motion between projections. Park~\etal \cite{park20233d} proposed a framework for 3D teeth reconstruction from one panoramic radiograph using neural implicit functions, where the teeth are first segmented in the radiograph followed by 3D teeth reconstruction. However, their training involved the 3D ground truth, which is usually unavailable for our problem of 3D coronary tree reconstruction from two non-simultaneous ICA projections. 

%%%%%%%%%
\section{Proposed Pipeline}
Our proposed DeepCA method consists of two blocks: a data preprocessing block and a 3D reconstruction with motion compensation block, as illustrated in \cref{workflow}. In the data preprocessing block, we generate two simulated ICA projections based on 3D CCTA data, with simulated motion on the second projection plane, and then apply backprojection on them to produce the input to the model at the next block. In the 3D reconstruction with motion compensation block, we map the 3D backprojection input to the CCTA data for 3D coronary tree reconstruction via training a deep neural network, implicitly compensating for any motion.
\begin{figure*}[!t]
\centering
\includegraphics[width=0.9\linewidth]{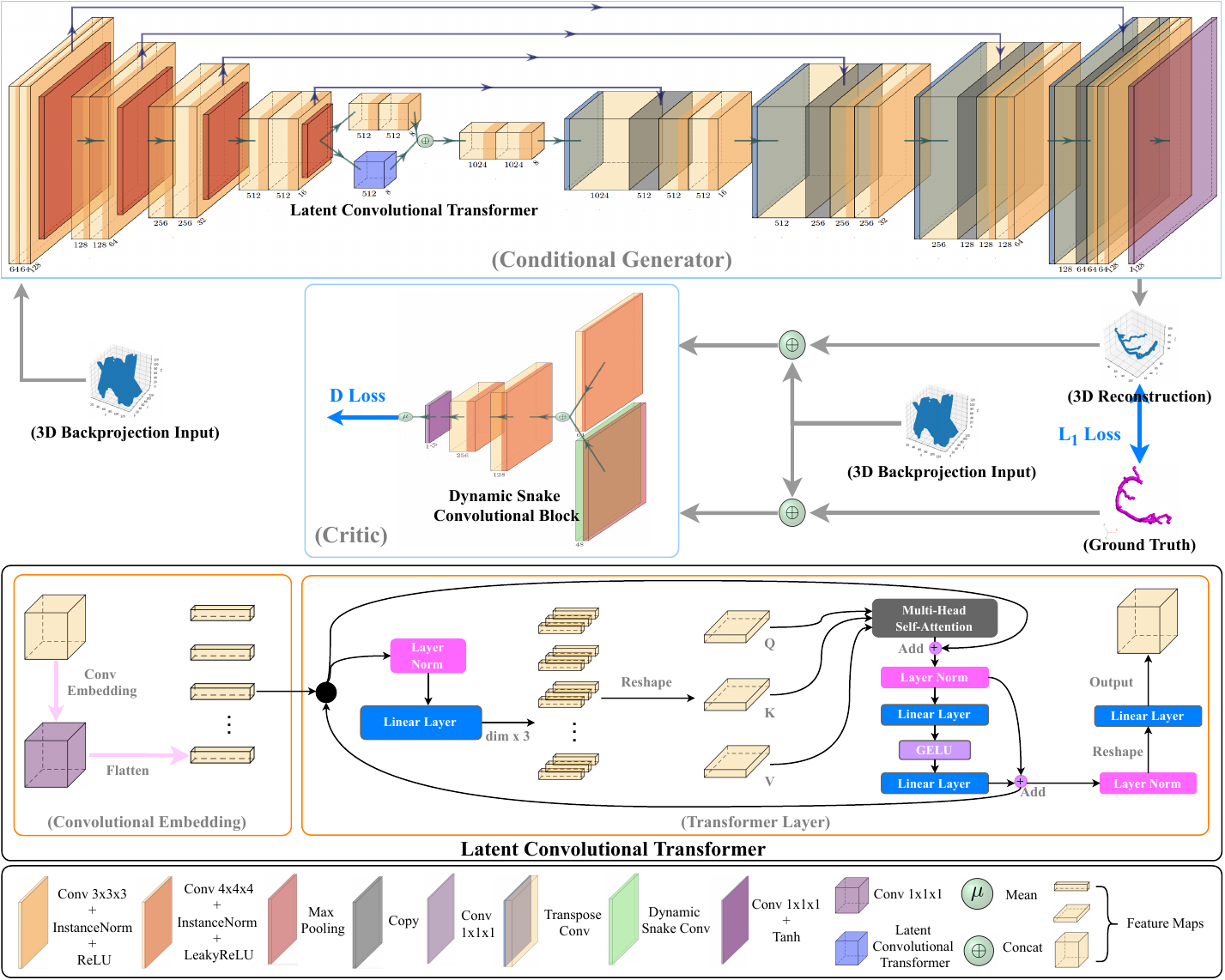}
\caption{The proposed DeepCA model architecture includes a conditional generator and a critic. The conditional generator is based on 3D U-Net with additional proposed convolutional transformer layers in the latent space. The generator produces corresponding reconstructed results according to the input condition. The latent convolutional transformers are built on convolutional embeddings following 8 transformer layers. The predicted results and the corresponding ground truth are concatenated with the input separately, which are then sent to the critic. The proposed critic uses both dynamic snake convolution and traditional convolution at the first layer to extract both global tubular and local features, and then applies several downsamplings to generate the critic loss.} \label{model}
\end{figure*}

%%%%%%%%%%%%%%
\subsection{Data Preprocessing Block}
Breathing and cardiac motions introduce deformations to the coronary tree between projections. We use deep learning to implicitly compensate for these motion artifacts. In the generation of simulated projections, we introduce rigid transformations to the CCTA data before performing forward projection on the second projection plane to simulate motion, as illustrated in \cref{workflow}. We use the projection geometry of real coronary angiography to simulate the two cone-beam forward projections. Details of the projection geometry parameters are provided in the supplementary material.

\par Since contrast injections used in real ICA projections change image intensity values between projections, we first segment vessels from the images to binarise them, where points on vessels are assigned as $1$ and background as $0$. In order to generalise to real ICA projections, we binarise the simulated projections with a threshold of $0$ as well, i.e. any points with values greater than $0$ are set to $1$ and $0$ otherwise. Using the known projection geometry, we perform backprojection on both binary simulated projections separately. We binarise the two 3D backprojections with a threshold of $0$ and add them together to generate a single 3D input to our model. 

%%%%%%%%%%%%%%
\subsection{3D Reconstruction with Motion Compensation Block}
We train a model to map the 3D backprojection result to its corresponding CCTA ground truth. Our DeepCA model architecture is based on the Wasserstein conditional generative adversarial network (WCGAN) with gradient penalty, latent convolutional transformer layers, and a dynamic snake convolutional critic, as illustrated in \cref{model}. Via mapping the input with non-aligned projections to 3D coronary tree data, most motion artifacts are corrected by our model. With the critic used, any residual uncorrected deformations are adjusted, while ensuring the connectedness of the coronary tree structures in the reconstructions and increasing the model's elastic generalisation capacity. So when generalising to real ICA projections, the non-rigid motion is compensated implicitly.
%%%%%%%%%%%%%%
\paragraph{WCGAN with gradient penalty (WCGAN-GP).}
The conditional structure of the GAN \cite{Isola_2017_CVPR} enables us to generate a desired output from a specific input, and the Wasserstein adversarial objective with an additional gradient penalty constraint \cite{NIPS2017_892c3b1c} improves training stability. The model consists of an encoder-decoder generator $G$ and a critic $D$. The 3D backprojection result $\mathbf x$ is the input to the generator $G$, which has the 3D U-Net \cite{cciccek20163d} as backbone, producing the predicted reconstruction $\mathbf{\hat y}$ as output. To ensure strict learning from the 3D backprojection input $\mathbf x$ to the corresponding ground truth $\mathbf y$, we keep the input $\mathbf x$ without any added noise, in contrast to previous style transfer applications. The predicted reconstruction $\mathbf{\hat y}$ and the corresponding ground truth $\mathbf y$ are then concatenated ($\oplus$) with the conditional input $\mathbf x$, respectively. 
\begin{equation}
\mathbf{\hat y} = G(\mathbf x), \quad
\mathbf{\hat y_x} = \mathbf{\hat y} \oplus \mathbf x, \quad
\mathbf{y_x} = \mathbf y \oplus \mathbf x.
\end{equation}

\par Next, $\mathbf{\hat y_x}$ and $\mathbf{y_x}$ are used in the critic to approximate the Wasserstein distance (or, Earth-Mover distance) $W((\mathbb P_r)_{\mathbf{y_x} \sim \mathbb P_r}, (\mathbb P_g)_{\mathbf{\hat y_x} \sim \mathbb P_g})$ and gradient penalty constraint $GP(\mathbb P_{\mathbf{\tilde y_x}})$ for each data batch, where $\mathbb P_r$ is the conditional ground truth data distribution, $\mathbb P_g$ is the conditional model generation distribution, and $\mathbb P_{\mathbf{\tilde y_x}}$ is the distribution sampling uniformly along straight lines between pairs of points sampled from the distributions $\mathbb P_r$ and $\mathbb P_g$.
\begin{multline}
W((\mathbb P_r)_{\mathbf{y_x} \sim \mathbb P_r}, (\mathbb P_g)_{\mathbf{\hat y_x} \sim \mathbb P_g}) = \\
\underset{\mathbf{y_x} \sim \mathbb P_r}{\mathbb E}[D(\mathbf{y_x})] - \underset{\mathbf{\hat y_x} \sim \mathbb P_g}{\mathbb E}[D(\mathbf{\hat y_x})].
\end{multline} 
\begin{multline}
GP(\mathbb P_{\mathbf{\tilde y_x}}) = \underset{\mathbf{\tilde y_x} \sim \mathbb P_{\mathbf{\tilde y_x}}}{\mathbb E}[(\left\Vert\nabla_{\mathbf{\tilde y_x}}D(\mathbf{\tilde y_x}) \right\Vert_2 - 1)^2],\\
\mbox{where}~~
\mathbf{\tilde y_x} = \epsilon \mathbf{y_x} + (1-\epsilon) \mathbf{\hat y_x}, \quad \epsilon \in U[0,1].
\end{multline}

\par During training, the generator $G$ tries to minimise $W(\mathbb P_r, \mathbb P_g)$ between distributions $\mathbb P_r$ and $\mathbb P_g$, while the critic $D$ tries to maximise this distance along with minimising the constraint $GP(\mathbb P_{\mathbf{\tilde y_x}})$. The objective function of the WCGAN-GP is presented in \cref{obj}, where $\lambda_1$ is the penalty coefficient. We use $\lambda_1 = 10$.
\begin{equation}
\label{obj}
\begin{split}
\mathcal L_{\text{WCGAN-GP}}(G,D)& = \\
&\arg \underset{G}{\min} \underset{D}{\max} ~W((\mathbb P_r)_{\mathbf{y_x} \sim \mathbb P_r}, (\mathbb P_g)_{\mathbf{\hat y_x} \sim \mathbb P_g}) \\
&+ \lambda_1 \underset{D}{\min} ~GP(\mathbb P_{\mathbf{\tilde y_x}}).
\end{split}
\end{equation}
\par We additionally impose an $\ell_1$ loss to enforce the reconstruction to align with the ground truth. Our final objective function $\mathcal{L}$ is presented in \cref{loss}.
\begin{equation}
\mathcal{L}_{\ell_1}(G) = \mathbb E_{\mathbf{\hat y}, \mathbf y} (\left\Vert\mathbf y - \mathbf{\hat y}\right\Vert_1),
\end{equation}
\begin{equation}\label{loss}
\mathcal{L} = \mathcal L_{\text{WCGAN-GP}}(G,D) + \lambda \mathcal{L}_{\ell_1}(G).
\end{equation}
The hyperparameter $\lambda$ is set to 100 after fine-tuning. The number of critic iterations per generator iteration is set to 2.
%%%%%%%%%%%%%%
\paragraph{Latent convolutional transformer layers (CTLs).}
Inspired by 2D compact transformers \cite{hassani2022escaping}, we implement our 3D latent CTLs that use convolutional embeddings following transformer layers in the latent space. Since our data contain a large empty background representing non-vessel regions, the usual patch embeddings are not suitable, while in the latent space, the transformer can help us extract the relations between feature maps and enforce their importance via attention modules. We do not require positional embeddings for feature maps as they are order invariant. The latent CTLs consider the max pooling results $f_{\text{input}} \in \mathbb{R}^{N \times C \times H \times W \times D}$ from the last layer as input, where $N$ is the batch size, $C$ denotes the number of channels, and $H, W, D$ stand for height, width, and depth, respectively. We next perform convolutional embeddings on the latent feature maps $f_{\text{embeddings}} \in \mathbb{R}^{N \times (H \times W \times D) \times C} = reshape(Conv_{1\times1\times1} (f_{\text{input}}))$. The feature map embeddings then go through the transformer layers as follows:
\begin{multline}
(Q, K, V) \in \mathbb{R}^{3 \times N \times (H \times W \times D) \times C} = \\
reshape( \text{LinearLayer} (\text{LayerNorm} (f_{\text{embeddings}}))),
\end{multline}
\begin{multline}
\mathcal{\mathbf z'} \in \mathbb{R}^{N \times (H \times W \times D) \times C} = \\
\text{LayerNorm}( \text{MHSA}(Q,K,V) + f_{\text{embeddings}}),
\end{multline}
\begin{multline}
\mathcal{\mathbf z} \in \mathbb{R}^{N \times (H \times W \times D) \times C} = \\
\text{LinearLayer}(\text{GELU}(\text{LinearLayer}(\mathcal{\mathbf z'}))) + \mathcal{\mathbf z'}, 
\end{multline}
where MHSA denotes a multi-head self-attention module, GELU the activation layer of the Gaussian error linear unit, $\mathcal{\mathbf z'}$ the intermediate result after MHSA, $\mathcal{\mathbf z}$ the output of one transformer layer, and $Q, K, V$ stand for query, keys, and values vectors, respectively. We use 8 attention heads and 8 transformer layers, where the output $\mathcal{\mathbf z}$ for one layer will replace the $f_{\text{embeddings}}$ for the next layer.

\par
\begin{figure}[!h]
  \centering
   \includegraphics[width=0.85\linewidth]{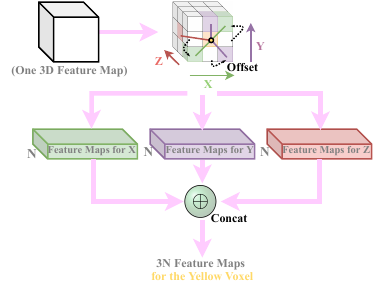}
   \caption{The dynamic snake convolution (DSConv). For each voxel (the yellow voxel in the figure as an example) in the feature map, the DSConv flattens the whole kernel along different axes with random offsets to extract different dynamic feature maps for X-, Y-, and Z-axes separately and then concatenates these feature maps together as the convolution output.}
   \label{fig:dscc}
\end{figure}

 After using the transformers to encode the relations between the latent feature maps, we use a linear layer to do final mappings on these embedded feature maps and reshape the results as the same size as input.
\begin{multline}
f_{\text{output}} \in \mathbb{R}^{N \times C \times H \times W \times D} = \\
reshape(\text{LinearLayer}(reshape(\text{LayerNorm}(\mathcal{\mathbf z})))).
\end{multline}
 %%%%%%%%%%%%%%
\paragraph{Dynamic snake convolutional (DSConv) critic.}
The coronary tree is formed of quasi-tubular topological structures, and the traditional convolutional kernel is not optimal for recognising thin local structures and variable global morphologies. DSConv kernels \cite{Qi_2023_ICCV} are designed specifically for such structures. Differing from the traditional 3D kernel with size $3 \times 3 \times 3$, DSConv flattens the whole kernel along different axes with random offsets to generate a dynamic snake-shaped kernel as illustrated in \cref{fig:dscc}. For 3D data, DSConv generates three dynamic feature maps for X-, Y-, and Z-axes respectively and then concatenates these three feature maps together as the output. Due to its nature of dynamic offsets, DSConv can extract tubular features more efficiently. We use the DSConv in the first layer of the critic to extract global tubular features and concurrently perform traditional convolution to extract essential local features as well. We then concatenate these global and local features together, and apply downsamplings to calculate the critic loss. This design of the critic can effectively distinguish vessel tubular structures.
 
%%%%%%%%%
\section{Experimental Settings}
\subsection{Datasets}
We use a public CCTA dataset \cite{ZENG2023102287} containing 3D binary segmented coronary trees for our study, and split the coronary trees into RCA and the left coronary artery. We use 879 segmented RCA data in total, dividing them into 75\% training, 15\% validation, and 10\% test datasets. We perform the cone-beam forward projection using the TIGRE toolbox \cite{biguri2016tigre}. The volume size is $128\times128\times128$, and the detector size is $512\times512$. Details of other projection geometry parameters are provided in the supplementary material.

\par We also collect a clinical ICA dataset of 8 patients for evaluation, who were admitted at the Oxford John Radcliffe Hospital with suspected coronary stenosis and provided informed consent. For two of the patients, three ICA projections were captured, while for the rest, there were two ICA projections. Our model takes 3D backprojection from binary projections as input, so these clinical ICA data are pre-segmented and then backprojected before evaluation.

\subsection{Baseline Models and Implementation Details} \label{baselines}
As there is no equivalent previous work on this problem using deep learning, we implement four models as baselines for comparative analyses. We replace the 3D U-Net in WCGAN-GP with Unet++ \cite{10.1007/978-3-030-00889-5_1} (termed as Un2+), with Unet+++ \cite{huang2020unet} (termed as Un3+), and with DSConv Net \cite{Qi_2023_ICCV} (termed as DSCN). We also implement the 3D convolutional vision transformer GAN \cite{10.1007/978-3-031-16446-0_49} (termed as CVTG). We use Adam optimiser for both generator and critic \cite{kingma2014adam}, with an initial learning rate of $10^{-4}$. The training was performed with a batch size of 3 on NVIDIA Quadro RTX 8000.

\subsection{Metrics}
We adopt the overlap using a sweeping distance threshold (\emph{Ot}($d$)) for evaluation metric, where $d$ is the distance threshold in $mm$~unit, as proposed in `A Coronary Artery Reconstruction Challenge' \cite{10.1007/978-3-319-67564-0_10}. $\emph{Ot}(d) \in [0,1]$ with $0$ representing no overlap and $1$ the perfect match; the metric is equivalent to the Dice score when $d=0$. The different $d$ values allow us to measure reconstructions under different degrees of deformation. Let us assume the set of all vessel points is $\mathcal P_{target}$ in the target data and $\mathcal P_{pred}$ in the prediction. Given a threshold $d$, every point $\mathbf p \in \mathcal P_{target}$ is marked as belonging to the set TPR($d$) of true positives of the reference if there is at least one point $\mathbf u \in \mathcal P_{pred}$ satisfying $distance(\mathbf p, \mathbf u) <= d$ and to the set FN($d$) of false negatives otherwise. Points $\mathbf u \in \mathcal P_{pred}$ are labelled as belonging to the set TPM($d$) of true positives of the tested method if there is at least one $\mathbf p \in \mathcal P_{target}$ satisfying $distance(\mathbf u, \mathbf p) <= d$ and to the set FP($d$) of false positives otherwise. The overlap \emph{Ot}($d$) for a certain distance threshold $d$ can then be calculated as: 
\begin{equation}
\emph{Ot}(d) = \frac{\lvert\text{TPM}(d)\rvert + \lvert\text{TPR}(d)\rvert}{\lvert\text{TPM}(d)\rvert + \lvert\text{TPR}(d)\rvert + \lvert\text{FN}(d)\rvert + \lvert\text{FP}(d)\rvert}.
\end{equation}
\par In addition, we use the Chamfer $\ell_2$ distance (CD($\ell_2$)) for measuring the corresponding voxel-wise or pixel-wise prediction errors ($mm$) in either 3D or 2D data according to their voxel or pixel spacing.

\par We evaluate the models on both the CCTA test dataset and the out-of-distribution real clinical ICA dataset. For the CCTA test dataset, we directly validate the results in 3D space after rigidly registering the ground truth to the predicted reconstruction using \emph{Ot}($d$) with $d=\{1, 2\}~mm$ and CD($\ell_2$). Since the training ground truth is the original CCTA data used to generate the first projection, there is no motion between the original ICA data and the reprojections of the predicted reconstructions on the first projection plane, while it exists on other projection planes. For this reason, we measure the Dice score between the ICA data and reprojections on the first projection plane (same as \emph{Ot}(0)). For the second and any additional projection planes, we first rigidly register the ICA data to the reprojections. We then compute the \emph{Ot}($d$) with $d=\{1, 2\}~mm$ and CD($\ell_2$) between them. All the 3D reconstruction results are binarised with a threshold of $0.5$ before evaluation, reprojection, and visualisation. All the 2D reprojections are binarised with a threshold of $0$ before evaluation and visualisation.

%%%%%%%%%
\section{Results and Discussion}
\begin{table*}[!t]
\caption{Quantitative performance of our proposed DeepCA model and 4 baseline models in terms of \emph{Ot}($d$) (\%) and CD($\ell_2$) ($mm$). The Dice score (\%) is equivalent to \emph{Ot}(0) (\%). All values represent mean ($\pm$ standard deviation), and the best results are annotated in \textbf{bold}.}\label{num_results}
\centering 
\begin{tabular}{|c|ccc|ccccccc|}
\hline
 \multirow{3}{*}{Model}              & \multicolumn{3}{c|}{3D CCTA Test Dataset}             & \multicolumn{7}{c|}{2D Real Clinical ICA Dataset (Unseen Domain)}                                                       \\ \cline{2-11}
 &                &                &                & $1^{st}$ & \multicolumn{3}{c}{$2^{nd}$ Projection}               & \multicolumn{3}{c|}{Additional Projection}            \\
                        & \emph{Ot}(1)          & \emph{Ot}(2)          & CD($\ell_2$)          & Dice         & \emph{Ot}(1)          & \emph{Ot}(2)          & CD($\ell_2$)          & \emph{Ot}(1)          & \emph{Ot}(2)          & CD($\ell_2$)          \\ \hline
Un2+ \cite{10.1007/978-3-030-00889-5_1}                    & \begin{tabular}[c]{@{}c@{}}51.99\\ $^{(\pm12.17)}$\end{tabular}          & \begin{tabular}[c]{@{}c@{}}64.75\\ $^{(\pm13.28)}$\end{tabular}         & \begin{tabular}[c]{@{}c@{}}4.64\\ $^{(\pm2.01)}$\end{tabular}         & \begin{tabular}[c]{@{}c@{}}65.42\\ $^{(\pm6.68)}$\end{tabular}          & \begin{tabular}[c]{@{}c@{}}22.97\\ $^{(\pm10.82)}$\end{tabular}         & \begin{tabular}[c]{@{}c@{}}31.05\\ $^{(\pm12.76)}$\end{tabular}         & \begin{tabular}[c]{@{}c@{}}12.79\\ $^{(\pm4.61)}$\end{tabular}         & \begin{tabular}[c]{@{}c@{}}25.25\\ $^{(\pm18.40)}$\end{tabular}          & \begin{tabular}[c]{@{}c@{}}39.37\\ $^{(\pm20.75)}$\end{tabular}        & \begin{tabular}[c]{@{}c@{}}8.49\\ $^{(\pm3.26)}$\end{tabular}         \\
Un3+ \cite{huang2020unet}                    & \begin{tabular}[c]{@{}c@{}}55.10\\ $^{(\pm10.72)}$\end{tabular}          & \begin{tabular}[c]{@{}c@{}}68.69\\ $^{(\pm10.96)}$\end{tabular}         & \begin{tabular}[c]{@{}c@{}}4.49\\ $^{(\pm1.67)}$\end{tabular}         & \begin{tabular}[c]{@{}c@{}}62.23\\ $^{(\pm9.49)}$\end{tabular}          & \begin{tabular}[c]{@{}c@{}}22.27\\ $^{(\pm8.30)}$\end{tabular}           & \begin{tabular}[c]{@{}c@{}}30.19\\ $^{(\pm10.51)}$\end{tabular}         & \begin{tabular}[c]{@{}c@{}}12.20\\ $^{(\pm3.77)}$\end{tabular}         & \begin{tabular}[c]{@{}c@{}}32.92\\ $^{(\pm23.99)}$\end{tabular}         & \begin{tabular}[c]{@{}c@{}}42.37\\ $^{(\pm26.30)}$\end{tabular}        & \begin{tabular}[c]{@{}c@{}}7.70\\ $^{(\pm1.59)}$\end{tabular}          \\
DSCN \cite{Qi_2023_ICCV}                    & \begin{tabular}[c]{@{}c@{}}61.74\\ $^{(\pm14.28)}$\end{tabular}          & \begin{tabular}[c]{@{}c@{}}72.21\\ $^{(\pm13.31)}$\end{tabular}         & \begin{tabular}[c]{@{}c@{}}3.49\\ $^{(\pm1.68)}$\end{tabular}         & \textbf{\begin{tabular}[c]{@{}c@{}}83.59\\ $^{(\pm4.01)}$\end{tabular}} & \begin{tabular}[c]{@{}c@{}}30.66\\ $^{(\pm11.30)}$\end{tabular}          & \begin{tabular}[c]{@{}c@{}}42.74\\ $^{(\pm11.68)}$\end{tabular}         & \begin{tabular}[c]{@{}c@{}}7.81\\ $^{(\pm2.19)}$\end{tabular}         & \begin{tabular}[c]{@{}c@{}}53.26\\ $^{(\pm6.74)}$\end{tabular}          & \begin{tabular}[c]{@{}c@{}}67.90\\ $^{(\pm4.77)}$\end{tabular}         & \begin{tabular}[c]{@{}c@{}}3.92\\ $^{(\pm0.48)}$\end{tabular}          \\
CVTG \cite{10.1007/978-3-031-16446-0_49}                    & \begin{tabular}[c]{@{}c@{}}61.53\\ $^{(\pm11.49)}$\end{tabular}          & \begin{tabular}[c]{@{}c@{}}73.71\\ $^{(\pm10.75)}$\end{tabular}         & \begin{tabular}[c]{@{}c@{}}3.51\\ $^{(\pm1.36)}$\end{tabular}          & \begin{tabular}[c]{@{}c@{}}76.84\\ $^{(\pm5.73)}$\end{tabular}          & \begin{tabular}[c]{@{}c@{}}32.98\\ $^{(\pm12.29)}$\end{tabular}         & \begin{tabular}[c]{@{}c@{}}44.85\\ $^{(\pm15.85)}$\end{tabular}         & \begin{tabular}[c]{@{}c@{}}8.23\\ $^{(\pm3.73)}$\end{tabular}         & \begin{tabular}[c]{@{}c@{}}49.50\\ $^{(\pm0.82)}$\end{tabular}          & \begin{tabular}[c]{@{}c@{}}64.63\\ $^{(\pm3.05)}$\end{tabular}         & \begin{tabular}[c]{@{}c@{}}4.22\\ $^{(\pm0.51)}$\end{tabular}          \\ \hline
\textbf{DeepCA}           & \textbf{\begin{tabular}[c]{@{}c@{}}64.21\\ $^{(\pm10.78)}$\end{tabular}} & \textbf{\begin{tabular}[c]{@{}c@{}}76.25\\ $^{(\pm9.72)}$\end{tabular}} & \textbf{\begin{tabular}[c]{@{}c@{}}3.22\\ $^{(\pm1.20)}$\end{tabular}} & \begin{tabular}[c]{@{}c@{}}83.31\\ $^{(\pm4.32)}$\end{tabular}          & \textbf{\begin{tabular}[c]{@{}c@{}}45.70\\ $^{(\pm6.79)}$\end{tabular}} & \textbf{\begin{tabular}[c]{@{}c@{}}58.39\\ $^{(\pm8.42)}$\end{tabular}} & \textbf{\begin{tabular}[c]{@{}c@{}}4.51\\ $^{(\pm1.29)}$\end{tabular}} & \textbf{\begin{tabular}[c]{@{}c@{}}58.58\\ $^{(\pm4.01)}$\end{tabular}} & \textbf{\begin{tabular}[c]{@{}c@{}}72.88\\ $^{(\pm0.90)}$\end{tabular}} & \textbf{\begin{tabular}[c]{@{}c@{}}2.81\\ $^{(\pm0.06)}$\end{tabular}} \\ \hline
\end{tabular}
\end{table*} 
%%%%%%%%%%%%%%
\subsection{Analysis on 3D CCTA Test Dataset}
\begin{figure}[!h]
     \centering
	\begin{subfigure}[h]{0.1\textwidth}
         \centering
         \includegraphics[width=\textwidth]{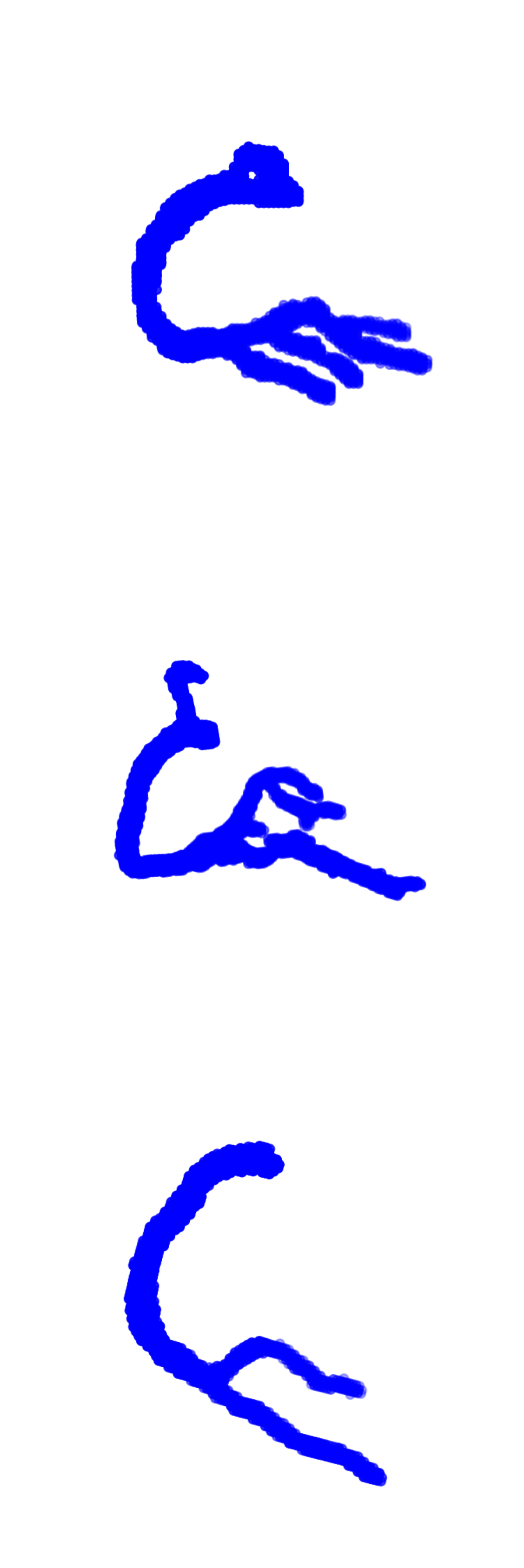}
         \caption*{Prediction}%{$D_1$: Prediction}
     \end{subfigure}
     \hfill
	\begin{subfigure}[h]{0.1\textwidth}
         \centering
         \includegraphics[width=\textwidth]{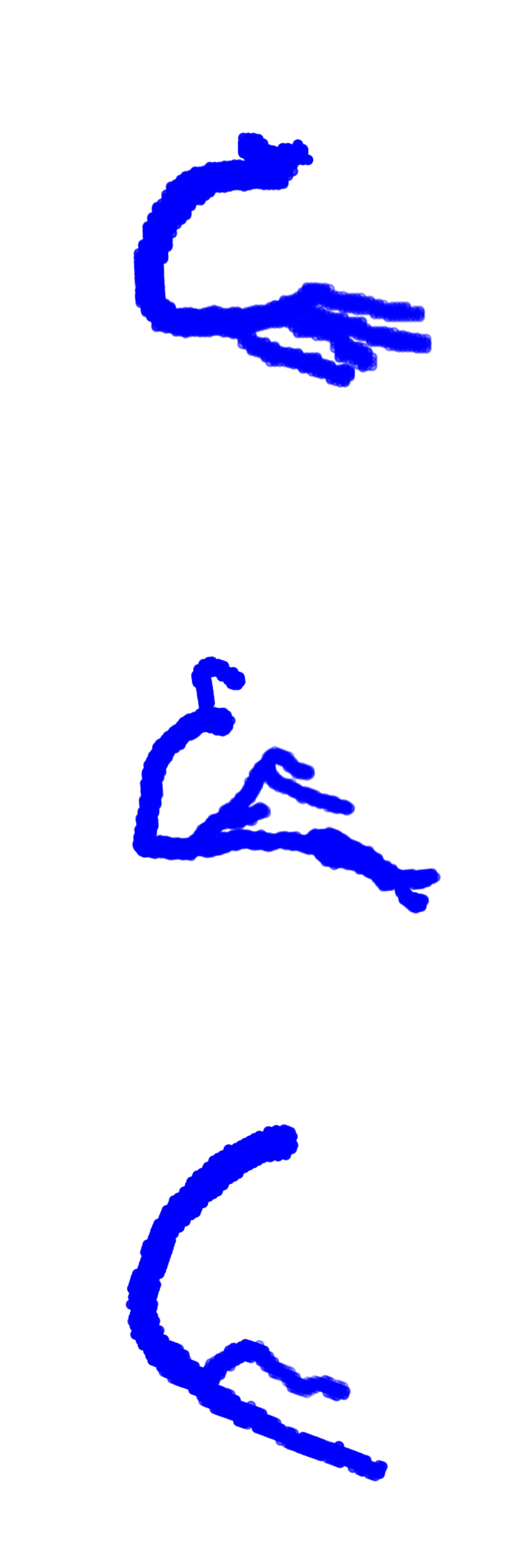}
         \caption*{Ground Truth}%{$D_1$: GT}
     \end{subfigure}
     \hfill
	\begin{subfigure}[h]{0.1\textwidth}
         \centering
         \includegraphics[width=\textwidth]{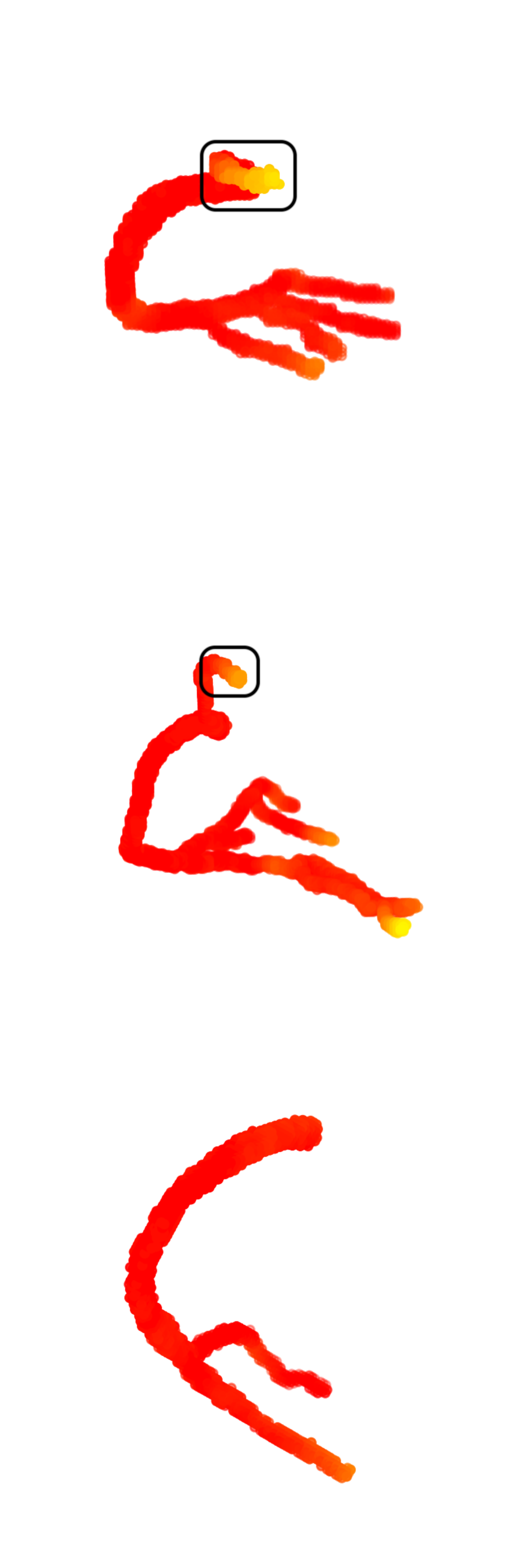}
         \caption*{Comparison}%{$D_1$: Comparison}
     \end{subfigure}
     \hfill
	\begin{subfigure}[h]{0.045\textwidth}
         \centering
         \includegraphics[width=\textwidth]{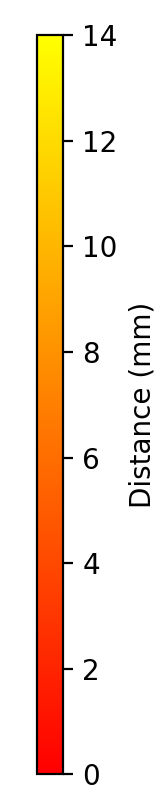}
         %\caption*{Comparison}%{$D_1$: Comparison}
     \end{subfigure}
        \caption{Three 3D reconstruction results on the CCTA test dataset by our DeepCA model. From top to bottom: three CCTA samples. From left to right: predicted reconstruction, ground truth, and the corresponding voxel-wise prediction errors in terms of CD($\ell_2$).}
        \label{ccta_visual}
\end{figure}

As demonstrated in \Cref{num_results}, our proposed DeepCA model achieves the best performance in all metrics on the CCTA test dataset compared to the 4 baseline models. This indicates our model can better capture the vessel topological structures in the source domain of the CCTA data. We also visualise the corresponding voxel-wise prediction errors in terms of CD($\ell_2$) between the ground truth and 3D predictions, as illustrated in \cref{ccta_visual}. We can see that our proposed model can reconstruct all the branches, though the reconstructed sinoatrial nodal arteries (marked by the black boxes) have larger offsets compared to the ground truth due to deformations. Additional qualitative results for all the models on the CCTA dataset are provided in the supplementary material.

\subsection{Analysis on 2D Clinical ICA Dataset}
From the quantitative results presented in \Cref{num_results}, we can observe that our proposed DeepCA model attains the best performance on real ICA data in all metrics on both the second and additional projection planes compared to the 4 baseline models. For the first projection plane, our model achieves the second best performance, with only 0.33\% behind the DSCN in terms of Dice score. However, in the second and additional projection planes, which are mostly affected by motion, our method performs the best. In particular, the results for the additional projection plane are the most significant, since the model is trained only based on two projections. Our proposed model presents large improvements of 38.57\% and 9.99\% in terms of \emph{Ot}(1), 42.25\% and 28.32\% in terms of CD($\ell_2$), for the second and additional projection planes respectively, compared to the best baseline models. This indicates our model has a better generalisation ability to the unseen domain.

\par
\begin{figure}[!h]
     \centering
	\begin{subfigure}[b]{0.23\textwidth}
         \centering
         \includegraphics[width=0.65\textwidth]{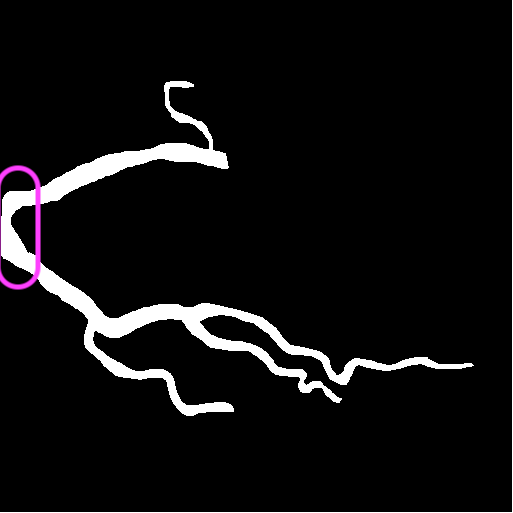}
         \caption*{Original ICA}
     \end{subfigure}
     \hfill
	\begin{subfigure}[b]{0.23\textwidth}
         \centering
         \includegraphics[width=0.65\textwidth]{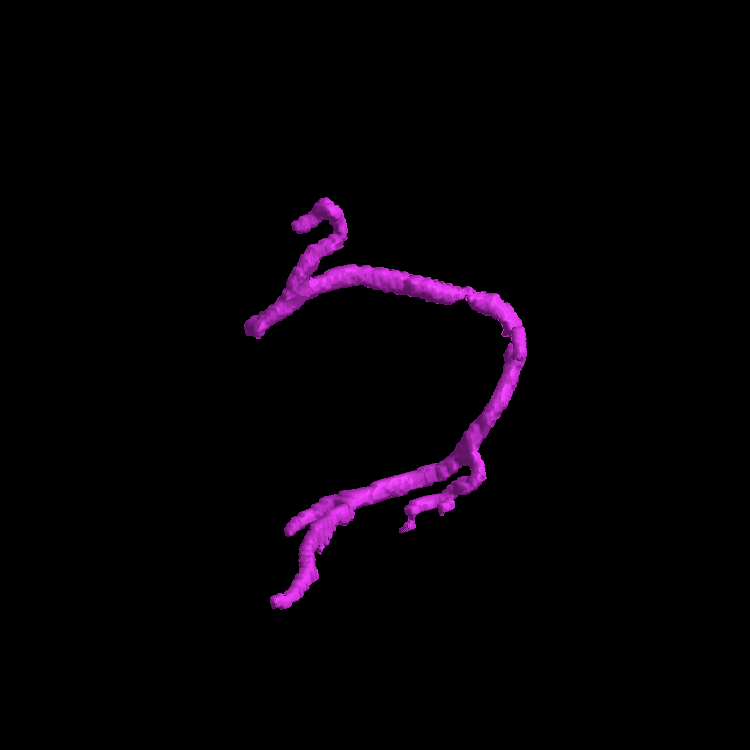}
         \caption*{3D Reconstruction}
     \end{subfigure}
            \caption{An example of 3D reconstruction for the RCA branch of a patient. Left: the original ICA data. Right: 3D reconstruction by our proposed DeepCA model.}
        \label{patient_1}
\end{figure}

 An interesting qualitative example is presented in \cref{patient_1}, where we observe a missing vascular section at the middle of the main RCA branch in the original ICA data, as marked by the pink box. This section is successfully recovered during 3D reconstruction, demonstrating our model's generative ability to recover missing vascular structures that may be missing due to acquisition or panning/zooming errors.

\par
\begin{figure}[!h]
     \centering
     $P_1$
	\begin{subfigure}[h]{0.145\textwidth}
         \centering
         \includegraphics[width=\textwidth]{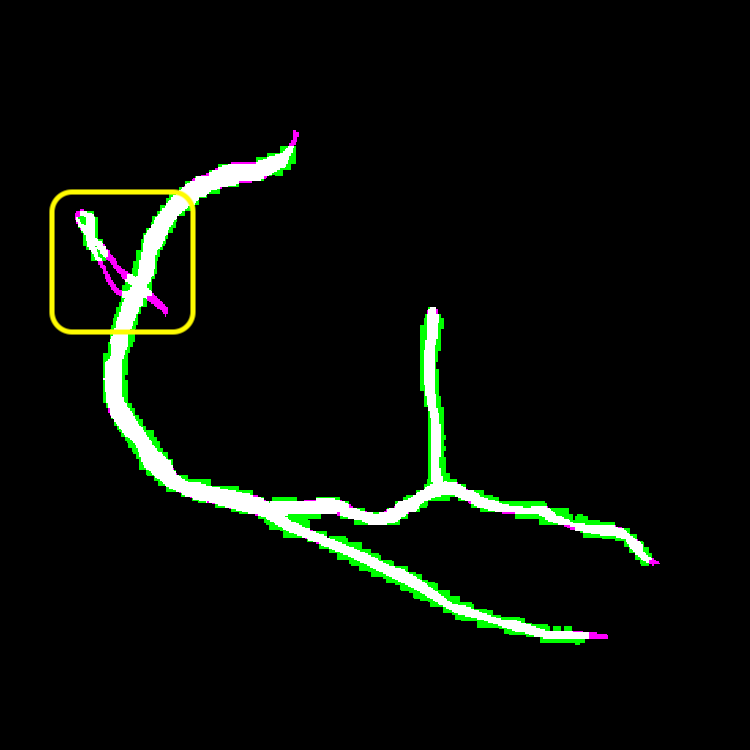}
         %\caption{$P_1: 1^{st}$ plane}\label{am_1}
     \end{subfigure}
     \hfill
	\begin{subfigure}[h]{0.145\textwidth}
         \centering
         \includegraphics[width=\textwidth]{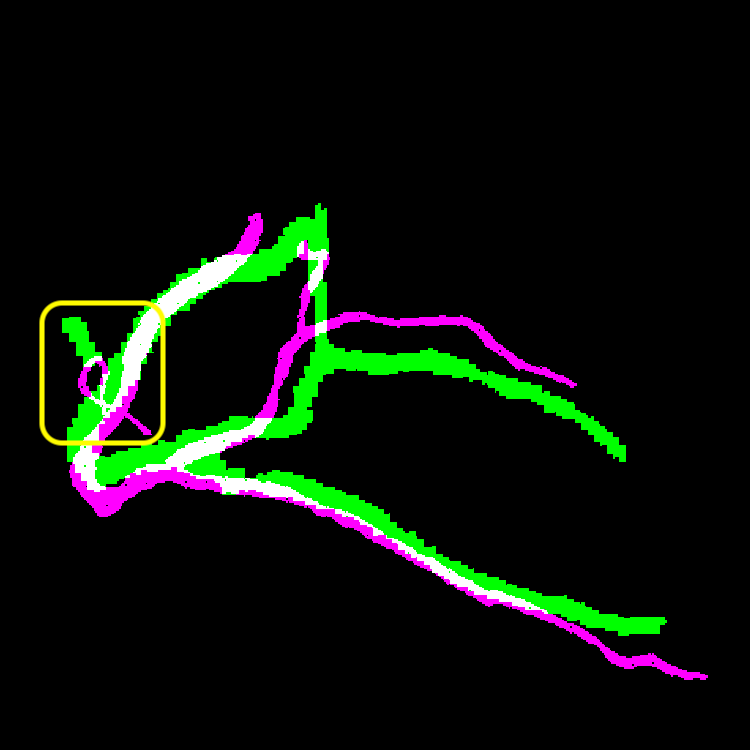}
         %\caption{$P_1: 2^{nd}$ plane}\label{am_2}
     \end{subfigure}
     \hfill
	\begin{subfigure}[h]{0.145\textwidth}
         \centering
         \includegraphics[width=\textwidth]{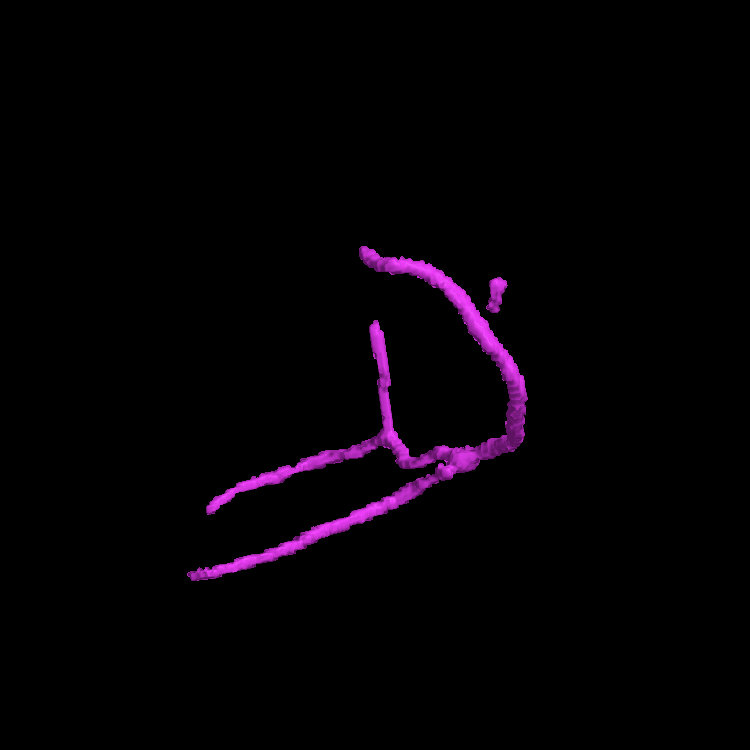}
         %\caption{$P_1$: 3D prediction}
     \end{subfigure}
     \vfill
     $P_2$
     \begin{subfigure}[h]{0.145\textwidth}
         \centering
         \includegraphics[width=\textwidth]{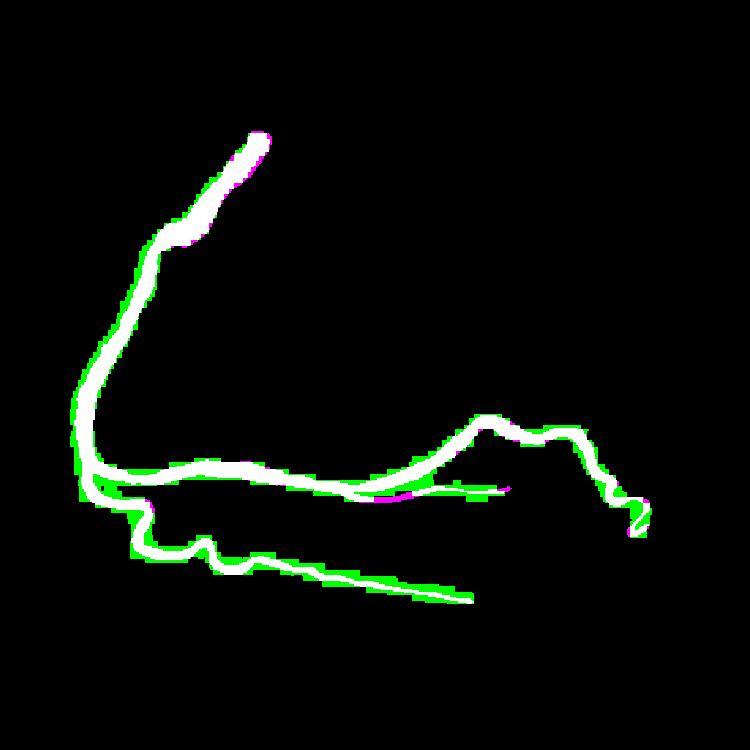}
         %\caption{$P_2: 1^{st}$ plane}
     \end{subfigure}
     \hfill
	\begin{subfigure}[h]{0.145\textwidth}
         \centering
         \includegraphics[width=\textwidth]{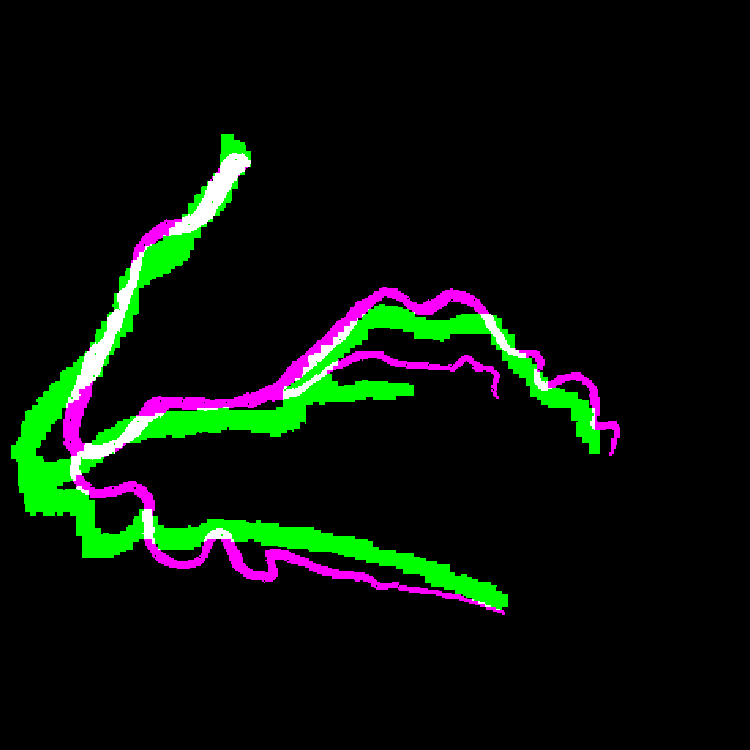}
         %\caption{$P_2: 2^{nd}$ plane}
     \end{subfigure}
     \hfill
	\begin{subfigure}[h]{0.145\textwidth}
         \centering
         \includegraphics[width=\textwidth]{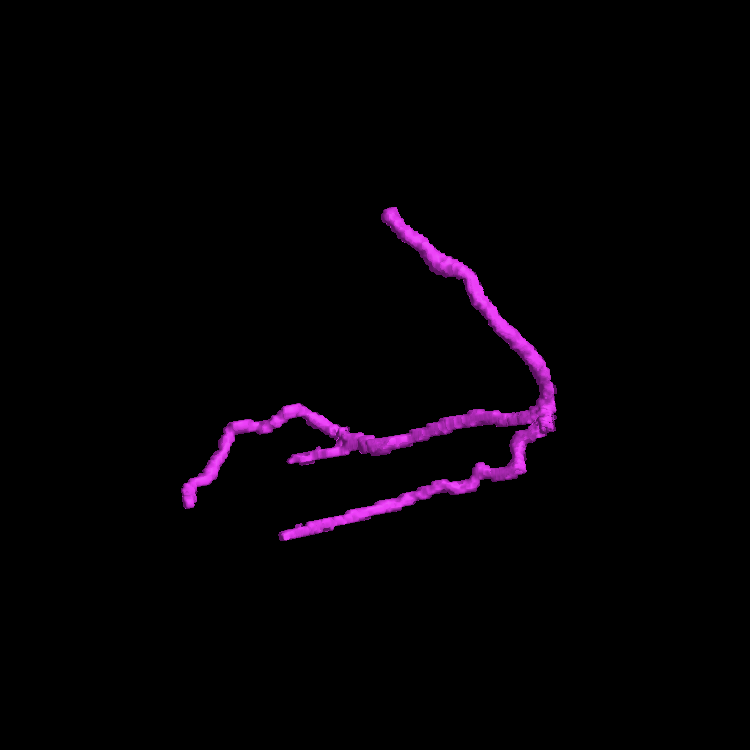}
         %\caption{$P_2$: 3D prediction}
     \end{subfigure}
     \vfill
     $P_3$
     \begin{subfigure}[h]{0.145\textwidth}
         \centering
         \includegraphics[width=\textwidth]{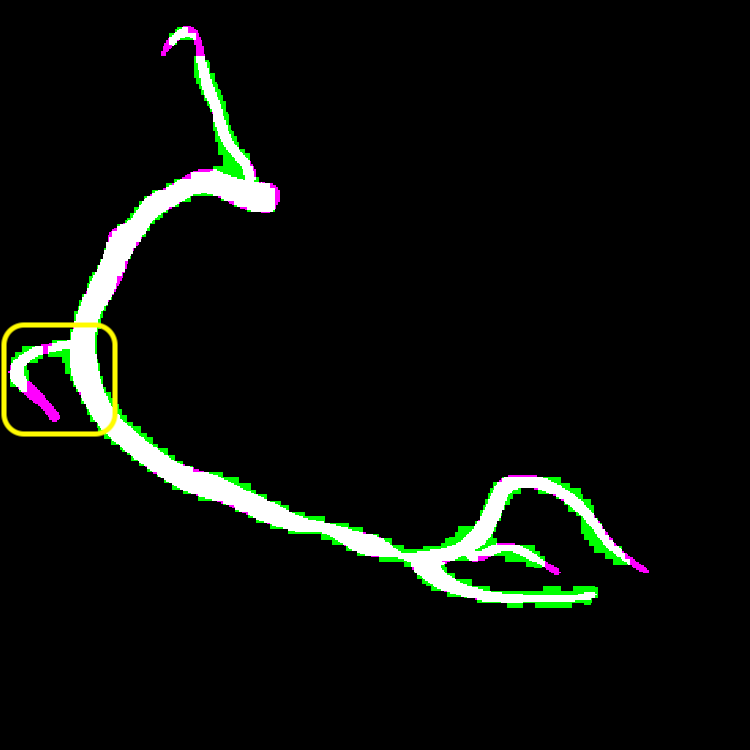}
         \caption*{$1^{st}$ Projection}%{$P_3: 1^{st}$ plane}\label{am_3}
     \end{subfigure}
     \hfill
	\begin{subfigure}[h]{0.145\textwidth}
         \centering
         \includegraphics[width=\textwidth]{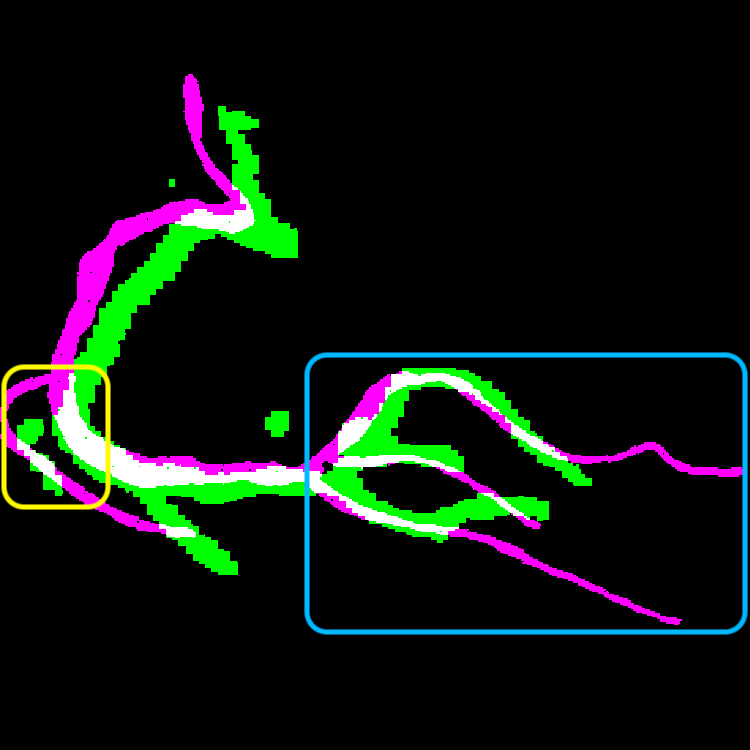}
         \caption*{$2^{nd}$ Projection}%{$P_3: 2^{nd}$ plane}%\label{short_branch}
     \end{subfigure}
     \hfill
	\begin{subfigure}[h]{0.145\textwidth}
         \centering
         \includegraphics[width=\textwidth]{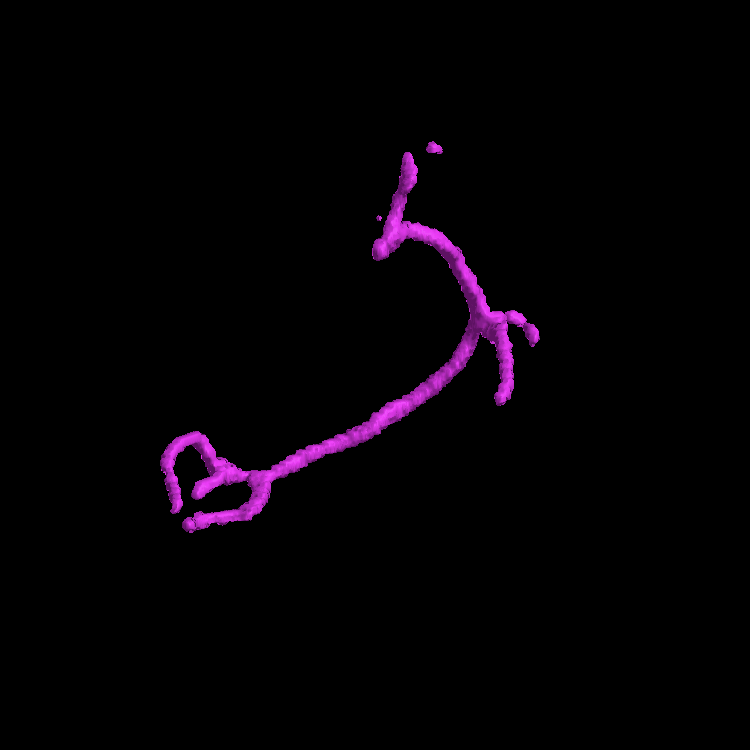}
         \caption*{3D Reconstruction}%{$P_3$: 3D prediction}
     \end{subfigure}
    \caption{Three qualitative examples. From top to bottom: three patients $P_{1,2,3}$. From left to right: the comparisons between the real ICA data and our reprojections on the first and second projection planes after rigid registration, and our DeepCA model's 3D reconstruction result. The colour purple represents ICA data, green represents reprojection, and white shows the overlap.}
        \label{visual_results}
\end{figure}

 \Cref{visual_results} shows three qualitative results by our proposed DeepCA model on real ICA data. The qualitative results for all the models on all the ICA data are provided in the supplementary material. We find that our model can reconstruct almost all the branches, especially the posterior descending arteries. Overall, the results illustrate good vessel connectivity, though there exist some broken acute marginal branches as marked by the yellow boxes in \cref{visual_results}. The reason behind this may be that those areas are affected heavily by motion, causing incomplete reconstruction. We also note that in the second projection of $P_3$ in \cref{visual_results}, the reconstructed posterior descending arteries are shorter as marked by the blue box. This may be the result of non-simultaneous ICA acquisitions where the contrast agent arrives at different distances between projections causing vessel differences in the different angiographic scans.
        
\par
\begin{figure}[!h]
     \centering
     $P_3$
	\begin{subfigure}[h]{0.145\textwidth}
         \centering
         \includegraphics[width=\textwidth]{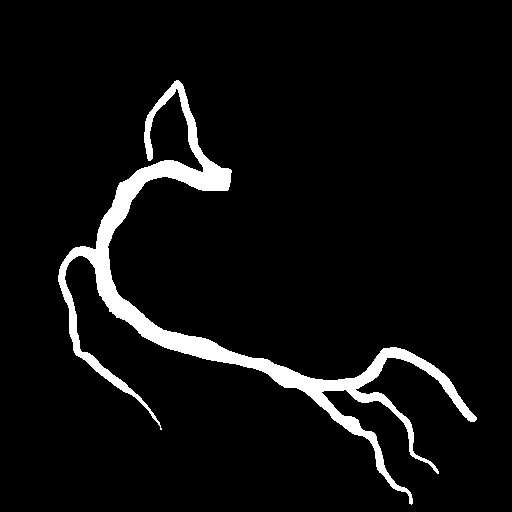}
         %\caption*{$P_3$: Original ICA}
     \end{subfigure}
     \hfill
	\begin{subfigure}[h]{0.145\textwidth}
         \centering
         \includegraphics[width=\textwidth]{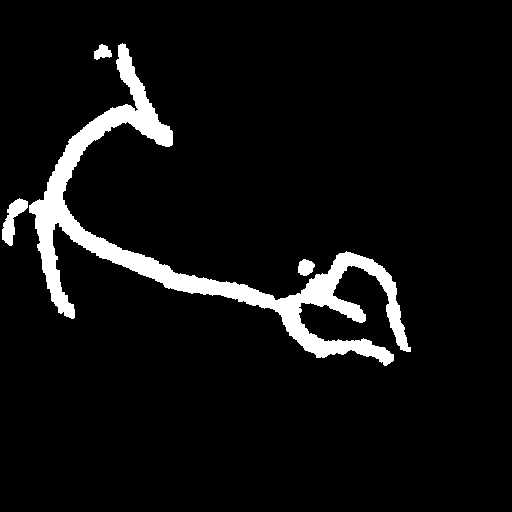}
         %\caption*{$P_3$: Reprojection}
     \end{subfigure}
     \hfill
	\begin{subfigure}[h]{0.145\textwidth}
         \centering
         \includegraphics[width=\textwidth]{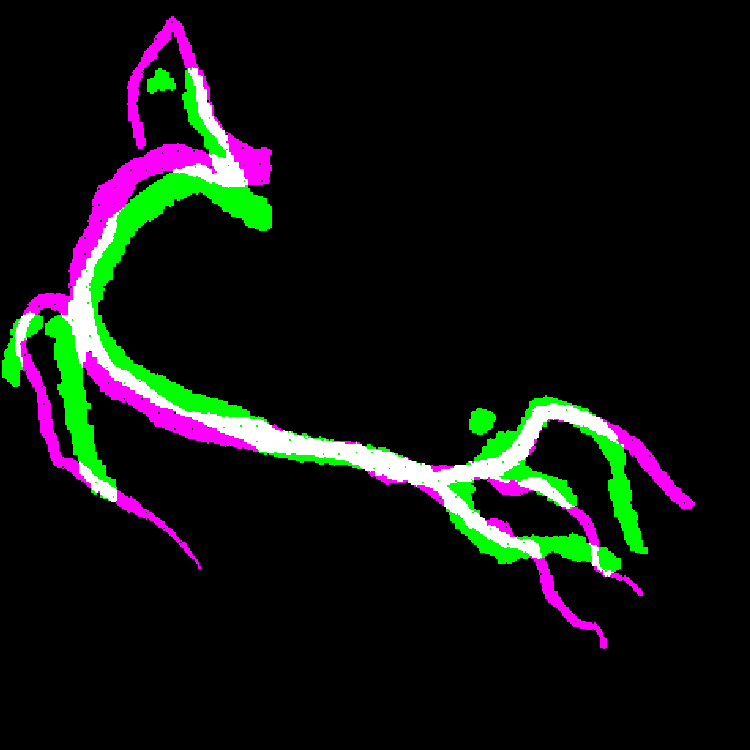}
         %\caption*{$P_3$: Comparison}
     \end{subfigure}
     \vfill
     $P_4$
     \begin{subfigure}[h]{0.145\textwidth}
         \centering
         \includegraphics[width=\textwidth]{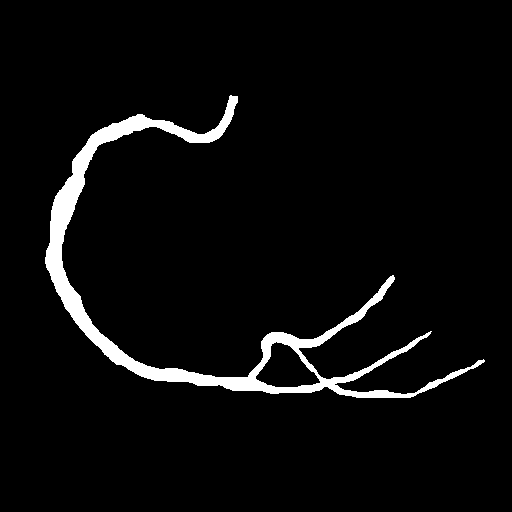}
         \caption*{Original ICA}%{$P_4$: Original ICA}
     \end{subfigure}
     \hfill
	\begin{subfigure}[h]{0.145\textwidth}
         \centering
         \includegraphics[width=\textwidth]{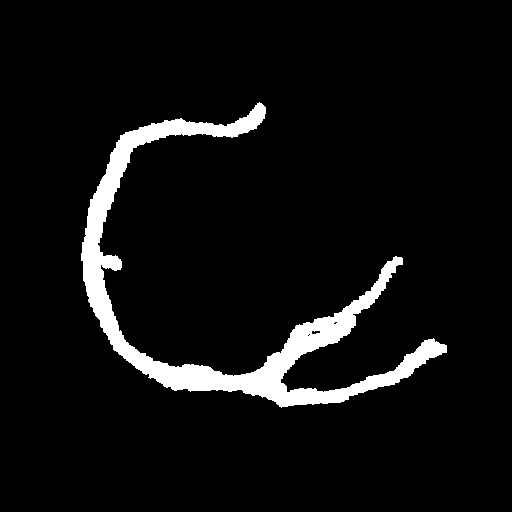}
         \caption*{Reprojection}%{$P_4$: Reprojection}
     \end{subfigure}
     \hfill
	\begin{subfigure}[h]{0.145\textwidth}
         \centering
         \includegraphics[width=\textwidth]{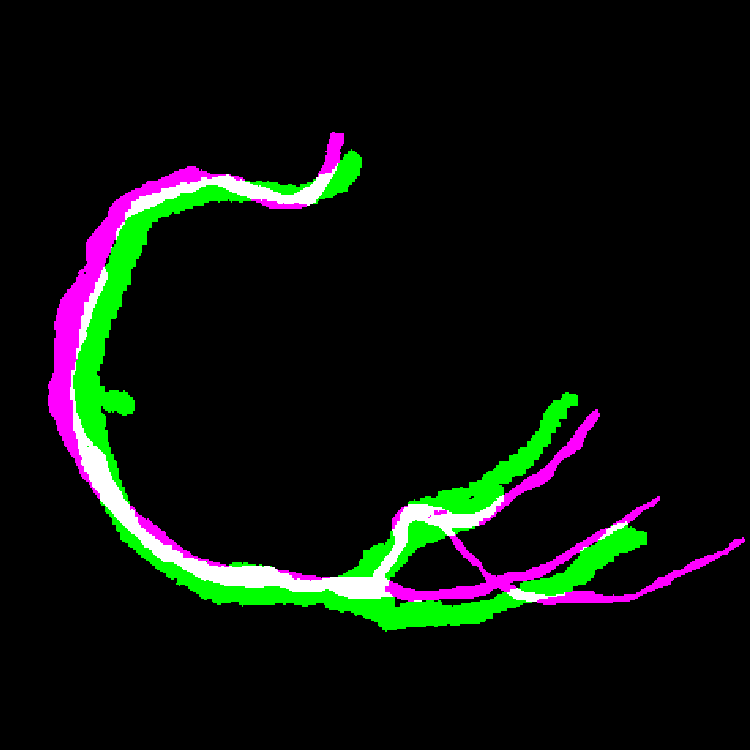}
         \caption*{Comparison}%{$P_4$: Comparison}
     \end{subfigure}
    \caption{Two example cases of our DeepCA model's reconstruction on the additional projection plane. From top to bottom: two patients $P_{3,4}$. From left to right: the original ICA data, the reprojection, and the comparison between them after rigid registration. The colour purple represents ICA data, green represents reprojection, and white shows the overlap.}
        \label{third_plane}
\end{figure}

 \Cref{third_plane} shows two example evaluations of our DeepCA model's 3D reconstruction on an additional projection plane. It demonstrates that even without involving the information of this projection plane in the input, our model can still reconstruct the accurate vascular structures.        

\subsection{Ablation Study} \label{ablations}
We evaluate the significance of different components of our proposed DeepCA model through an ablation study. We evaluate (1) WCGAN-GP (termed as WGP), (2) WCGAN-GP with latent CTLs (termed as +CTLs), and (3) WCGAN-GP with DSConv critic (termed as +DSCC). As demonstrated in \Cref{ablations_tab}, the combined models consistently provide improved performance. The qualitative results in the supplementary material on real ICA data also illustrate the advantages of each component of our model.
\begin{table}[!h]
\caption{Quantitative results of 3 ablation models in terms of \emph{Ot}($d$) (\%) and CD($\ell_2$) ($mm$). All values represent mean ($\pm$ standard deviation), and the best results are annotated in \textbf{bold}.}\label{ablations_tab}
\centering 
\begin{tabular}{|c|ccc|} 
\hline
\multirow{2}{*}{Model} & \multicolumn{3}{c|}{3D CCTA Test Dataset} \\ \cline{2-4}
   & \emph{Ot}(1)      & \emph{Ot}(2)      & CD($\ell_2$)      \\ \hline
WGP      & $62.06_{\pm10.61}$     & $74.38_{\pm10.01}$     & $3.43_{\pm1.29}$     \\
+CTLs    & $62.87_{\pm11.68}$     & $74.39_{\pm10.70}$     & $3.24_{\pm1.23}$     \\
+DSCC    & $63.46_{\pm10.85}$     & $75.14_{\pm10.00}$     & $3.24_{\pm1.23}$     \\ \hline
\textbf{DeepCA}     & $\textbf{64.21}_{\pm10.78}$     & $\textbf{76.25}_{\pm9.72}$     & $\textbf{3.22}_{\pm1.20}$     \\ \hline
\end{tabular}
\end{table}

%%%%%%%%%
\section{Conclusion}
In this paper, we propose DeepCA, leveraging the WCGAN with gradient penalty, latent convolutional transformer layers, and a dynamic snake convolutional critic for accurate 3D coronary tree reconstruction. Through simulating projections from CCTA data, we achieve generalisation on real non-simultaneously acquired ICA data. To the best of our knowledge, this is the first study that leverages deep learning in 3D coronary tree reconstruction from two real non-simultaneous ICA projections. The evaluations in this paper provide a baseline for future work in this area. Together with automated coronary vessels segmentation \cite{he2022semistudentteacher,He2023Stacom}, DeepCA can allow end-to-end automated real-time 3D coronary tree reconstruction during cardiac interventions.

%%%%%%%%% BODY TEXT
\section{Appendix}
We show additional qualitative results for our proposed DeepCA model, 4 baseline models, and 3 ablation models. All the 3D reconstructions on the CCTA test dataset and real clinical ICA dataset are binarised with a threshold of $0.5$ before evaluation, reprojection, and visualisation. All the 2D reprojections are binarised with a threshold of $0$ before evaluation and visualisation.
 
\par \textbf{Four baseline models:} We replace the 3D U-Net in WCGAN-GP with Unet++ \cite{10.1007/978-3-030-00889-5_1} (termed as Un2+), with Unet+++ \cite{huang2020unet} (termed as Un3+), and with DSConv Net \cite{Qi_2023_ICCV} (termed as DSCN). We also implement the 3D convolutional vision transformer GAN \cite{10.1007/978-3-031-16446-0_49} (termed as CVTG). 

\par \textbf{Three ablation models:} (1) WCGAN-GP (termed as WGP), (2) WCGAN-GP + latent CTLs (termed as +CTLs), and (3) WCGAN-GP + DSConv critic (termed as +DSCC). 

\subsection{Projection Geometry}
In the generation of simulated projections on the CCTA dataset, we use the projection geometry of real coronary angiography to simulate the two cone-beam forward projections, the parameters of which are illustrated in \Cref{pj_tab}. In order to simulate breathing and cardiac motions on the second projection plane, we rotate the CCTA data randomly for both primary and secondary angles ranging from -10$\degree$ to 10$\degree$ and add translations of -8~$mm$ to 8~$mm$ in both horizontal and vertical directions.

\subsection{Qualitative Results on 3D CCTA Test Dataset}
We present 5 CCTA test data samples for qualitative analysis. Before evaluation, we rigidly register the ground truth (original CCTA test data) to the 3D reconstruction. \Cref{ccta_recon} shows the original ground truth and the 3D reconstructions generated by all the models visualised from the front view. \Cref{ccta_cd} illustrates the corresponding voxel-wise prediction errors in terms of Chamfer $\ell_2$ distance (CD($\ell_2$)) between the ground truth and the 3D reconstruction. 

\subsection{Qualitative Results on 2D Clinical ICA Dataset}
We present the 3D reconstructions by all the models and the corresponding 2D reprojections when testing on the out-of-distribution 2D real clinical ICA dataset of 8 patients. \Cref{3d_ica} displays the 3D reconstructions by all the models. \Cref{ica_1} illustrates the comparisons on the first projection plane between the original ICA data and the reprojections. Before evaluation on the second and additional projection planes, we first rigidly register the original ICA data to the reprojections. \Cref{ica_2,ica_3} present the comparisons on the second and additional projection planes between the registered ICA data and the reprojections. 

\subsection{Discussion}
\begin{figure*}[!h]
     \centering
     %%%%%% 3d gt
     \begin{subfigure}[b]{0.13\linewidth}
         \centering
         \includegraphics[width=\linewidth]{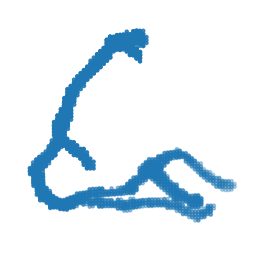}
     \end{subfigure}
     \hfill
	\begin{subfigure}[b]{0.13\linewidth}
         \centering
         \includegraphics[width=\linewidth]{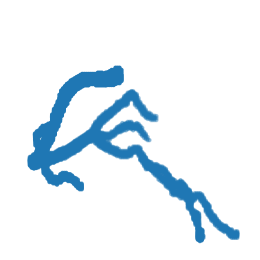}
     \end{subfigure}
     \hfill
	\begin{subfigure}[b]{0.13\linewidth}
         \centering
         \includegraphics[width=\linewidth]{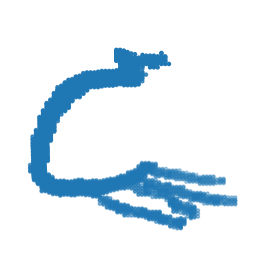}
     \end{subfigure}
     \hfill
	\begin{subfigure}[b]{0.13\linewidth}
         \centering
         \includegraphics[width=\linewidth]{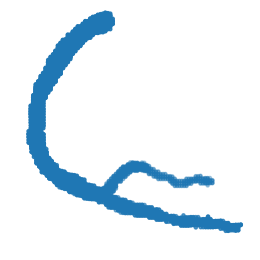}
     \end{subfigure}
     \hfill
	\begin{subfigure}[b]{0.13\linewidth}
         \centering
         \includegraphics[width=\linewidth]{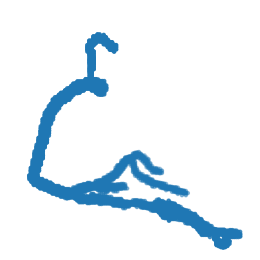}
     \end{subfigure}
     \vfill
     %%%%%% mine
     \begin{subfigure}[b]{0.13\linewidth}
         \centering
         \includegraphics[width=\linewidth]{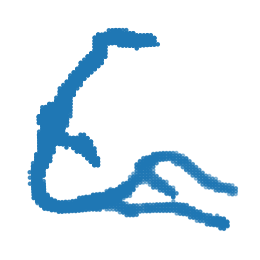}
     \end{subfigure}
     \hfill
	\begin{subfigure}[b]{0.13\linewidth}
         \centering
         \includegraphics[width=\linewidth]{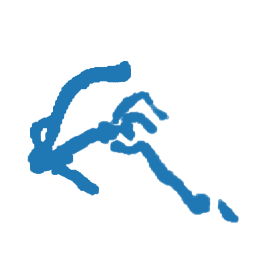}
     \end{subfigure}
     \hfill
	\begin{subfigure}[b]{0.13\linewidth}
         \centering
         \includegraphics[width=\linewidth]{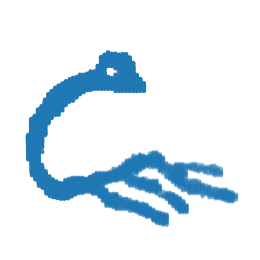}
     \end{subfigure}
     \hfill
	\begin{subfigure}[b]{0.13\linewidth}
         \centering
         \includegraphics[width=\linewidth]{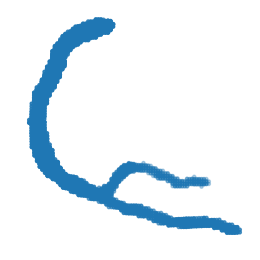}
     \end{subfigure}
     \hfill
	\begin{subfigure}[b]{0.13\linewidth}
         \centering
         \includegraphics[width=\linewidth]{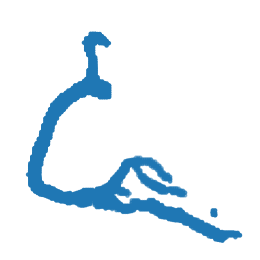}
     \end{subfigure}
     \vfill
     %%%%%% WGP
     \begin{subfigure}[b]{0.13\linewidth}
         \centering
         \includegraphics[width=\linewidth]{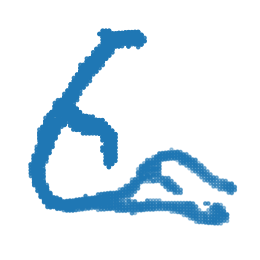}
     \end{subfigure}
     \hfill
	\begin{subfigure}[b]{0.13\linewidth}
         \centering
         \includegraphics[width=\linewidth]{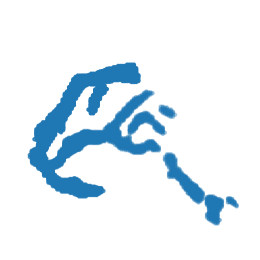}
     \end{subfigure}
     \hfill
	\begin{subfigure}[b]{0.13\linewidth}
         \centering
         \includegraphics[width=\linewidth]{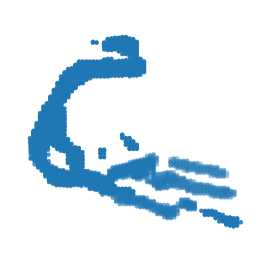}
     \end{subfigure}
     \hfill
	\begin{subfigure}[b]{0.13\linewidth}
         \centering
         \includegraphics[width=\linewidth]{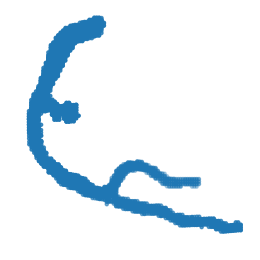}
     \end{subfigure}
     \hfill
	\begin{subfigure}[b]{0.13\linewidth}
         \centering
         \includegraphics[width=\linewidth]{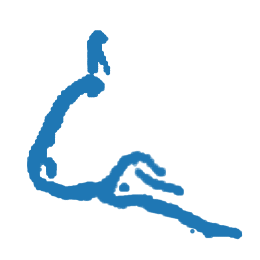}
     \end{subfigure}
     \vfill
     %%%%%% CTLs
     \begin{subfigure}[b]{0.13\linewidth}
         \centering
         \includegraphics[width=\linewidth]{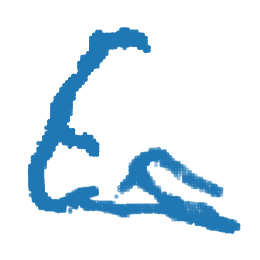}
     \end{subfigure}
     \hfill
	\begin{subfigure}[b]{0.13\linewidth}
         \centering
         \includegraphics[width=\linewidth]{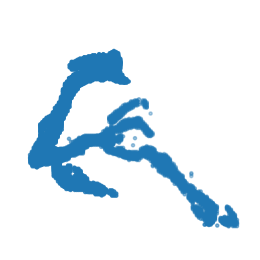}
     \end{subfigure}
     \hfill
	\begin{subfigure}[b]{0.13\linewidth}
         \centering
         \includegraphics[width=\linewidth]{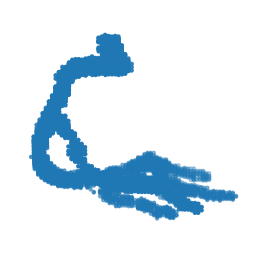}
     \end{subfigure}
     \hfill
	\begin{subfigure}[b]{0.13\linewidth}
         \centering
         \includegraphics[width=\linewidth]{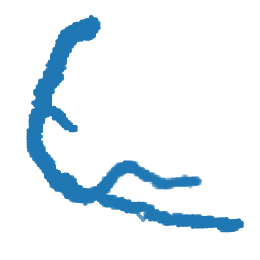}
     \end{subfigure}
     \hfill
	\begin{subfigure}[b]{0.13\linewidth}
         \centering
         \includegraphics[width=\linewidth]{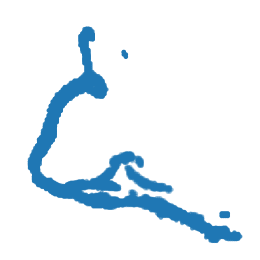}
     \end{subfigure}
     \vfill
     %%%%%% DSCC
     \begin{subfigure}[b]{0.13\linewidth}
         \centering
         \includegraphics[width=\linewidth]{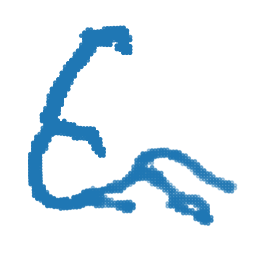}
     \end{subfigure}
     \hfill
	\begin{subfigure}[b]{0.13\linewidth}
         \centering
         \includegraphics[width=\linewidth]{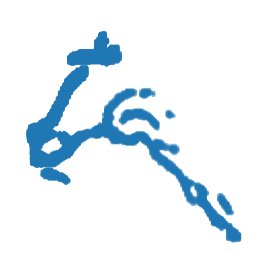}
     \end{subfigure}
     \hfill
	\begin{subfigure}[b]{0.13\linewidth}
         \centering
         \includegraphics[width=\linewidth]{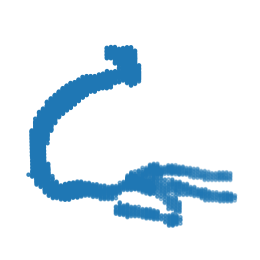}
     \end{subfigure}
     \hfill
	\begin{subfigure}[b]{0.13\linewidth}
         \centering
         \includegraphics[width=\linewidth]{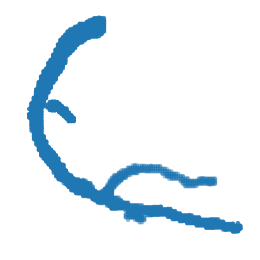}
     \end{subfigure}
     \hfill
	\begin{subfigure}[b]{0.13\linewidth}
         \centering
         \includegraphics[width=\linewidth]{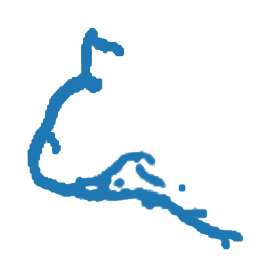}
     \end{subfigure}
     \vfill
     %%%%%% Un2+
     \begin{subfigure}[b]{0.13\linewidth}
         \centering
         \includegraphics[width=\linewidth]{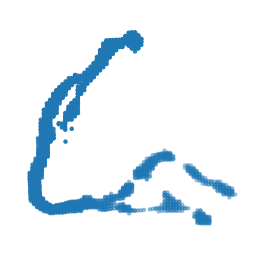}
     \end{subfigure}
     \hfill
	\begin{subfigure}[b]{0.13\linewidth}
         \centering
         \includegraphics[width=\linewidth]{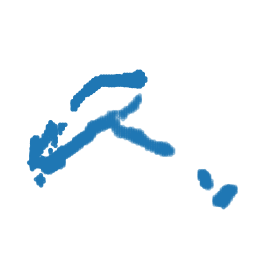}
     \end{subfigure}
     \hfill
	\begin{subfigure}[b]{0.13\linewidth}
         \centering
         \includegraphics[width=\linewidth]{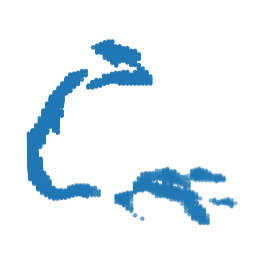}
     \end{subfigure}
     \hfill
	\begin{subfigure}[b]{0.13\linewidth}
         \centering
         \includegraphics[width=\linewidth]{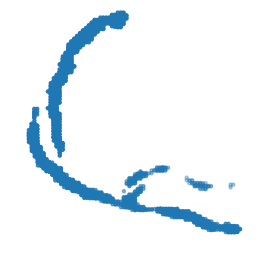}
     \end{subfigure}
     \hfill
	\begin{subfigure}[b]{0.13\linewidth}
         \centering
         \includegraphics[width=\linewidth]{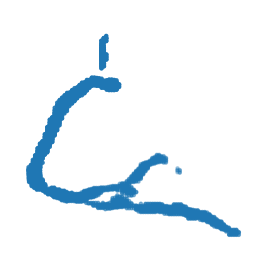}
     \end{subfigure}
     \vfill
     %%%%%% Un3+
     \begin{subfigure}[b]{0.13\linewidth}
         \centering
         \includegraphics[width=\linewidth]{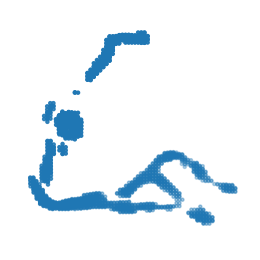}
     \end{subfigure}
     \hfill
	\begin{subfigure}[b]{0.13\linewidth}
         \centering
         \includegraphics[width=\linewidth]{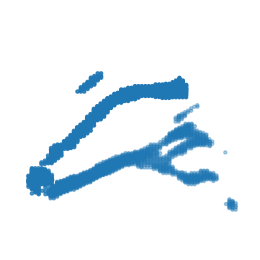}
     \end{subfigure}
     \hfill
	\begin{subfigure}[b]{0.13\linewidth}
         \centering
         \includegraphics[width=\linewidth]{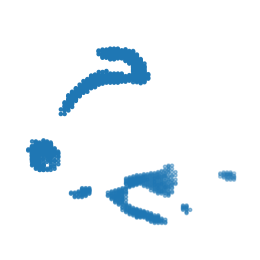}
     \end{subfigure}
     \hfill
	\begin{subfigure}[b]{0.13\linewidth}
         \centering
         \includegraphics[width=\linewidth]{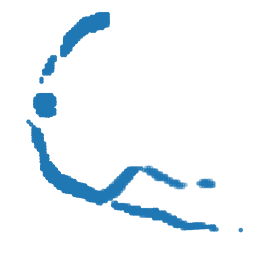}
     \end{subfigure}
     \hfill
	\begin{subfigure}[b]{0.13\linewidth}
         \centering
         \includegraphics[width=\linewidth]{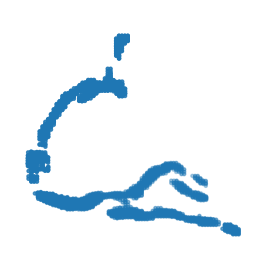}
     \end{subfigure}
     \vfill
     %%%%%% DSCN
     \begin{subfigure}[b]{0.13\linewidth}
         \centering
         \includegraphics[width=\linewidth]{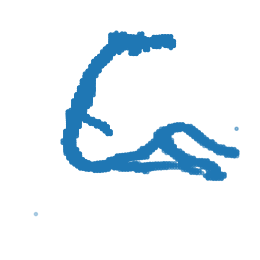}
     \end{subfigure}
     \hfill
	\begin{subfigure}[b]{0.13\linewidth}
         \centering
         \includegraphics[width=\linewidth]{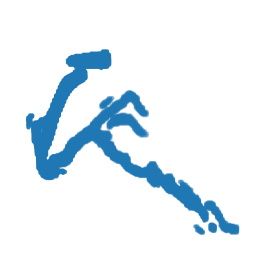}
     \end{subfigure}
     \hfill
	\begin{subfigure}[b]{0.13\linewidth}
         \centering
         \includegraphics[width=\linewidth]{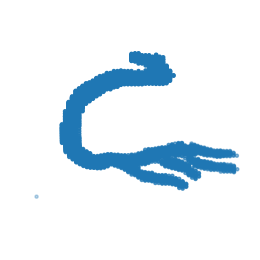}
     \end{subfigure}
     \hfill
	\begin{subfigure}[b]{0.13\linewidth}
         \centering
         \includegraphics[width=\linewidth]{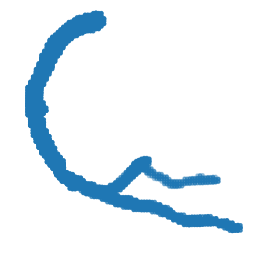}
     \end{subfigure}
     \hfill
	\begin{subfigure}[b]{0.13\linewidth}
         \centering
         \includegraphics[width=\linewidth]{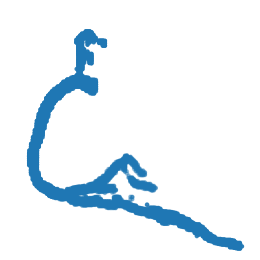}
     \end{subfigure}
     \vfill
     %%%%%% CVTG
     \begin{subfigure}[b]{0.13\linewidth}
         \centering
         \includegraphics[width=\linewidth]{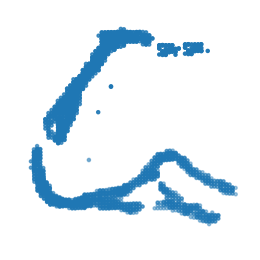}
     \end{subfigure}
     \hfill
	\begin{subfigure}[b]{0.13\linewidth}
         \centering
         \includegraphics[width=\linewidth]{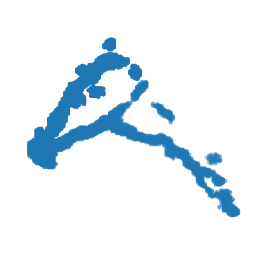}
     \end{subfigure}
     \hfill
	\begin{subfigure}[b]{0.13\linewidth}
         \centering
         \includegraphics[width=\linewidth]{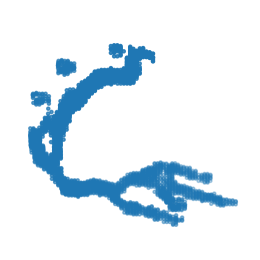}
     \end{subfigure}
     \hfill
	\begin{subfigure}[b]{0.13\linewidth}
         \centering
         \includegraphics[width=\linewidth]{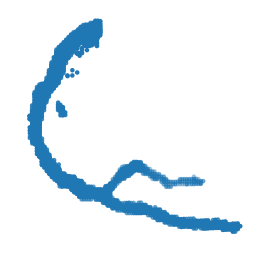}
     \end{subfigure}
     \hfill
	\begin{subfigure}[b]{0.13\linewidth}
         \centering
         \includegraphics[width=\linewidth]{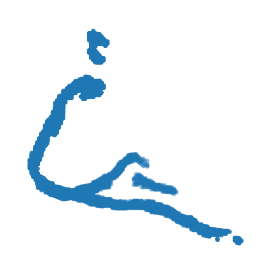}
     \end{subfigure}
     \caption{3D reconstruction results on 5 CCTA test data from all the models. From left to right: 5 CCTA test data samples. Row 1: The 3D ground truth. From row 2 to the end: The 3D reconstruction results by our proposed DeepCA model, WGP, +CTLs, +DSCC, Un2+, Un3+, DSCN, and CVTG.}\label{ccta_recon}
\end{figure*}

\begin{figure*}[!h]
     \centering
     %%%%%% mine
     \begin{subfigure}[b]{0.13\linewidth}
         \centering
         \includegraphics[width=\linewidth]{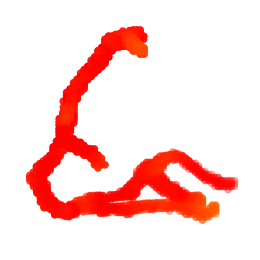}
     \end{subfigure}
     \hfill
	\begin{subfigure}[b]{0.13\linewidth}
         \centering
         \includegraphics[width=\linewidth]{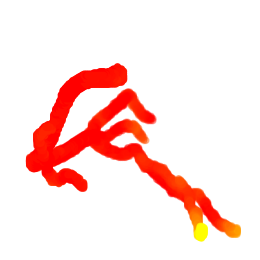}
     \end{subfigure}
     \hfill
	\begin{subfigure}[b]{0.13\linewidth}
         \centering
         \includegraphics[width=\linewidth]{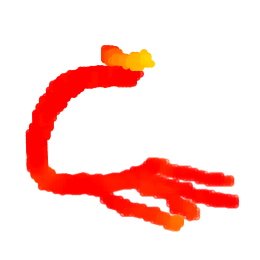}
     \end{subfigure}
     \hfill
	\begin{subfigure}[b]{0.13\linewidth}
         \centering
         \includegraphics[width=\linewidth]{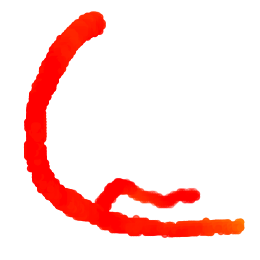}
     \end{subfigure}
     \hfill
	\begin{subfigure}[b]{0.13\linewidth}
         \centering
         \includegraphics[width=\linewidth]{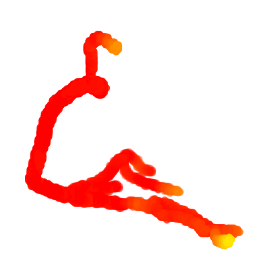}
     \end{subfigure}
     \vfill
     %%%%%% WGP
     \begin{subfigure}[b]{0.13\linewidth}
         \centering
         \includegraphics[width=\linewidth]{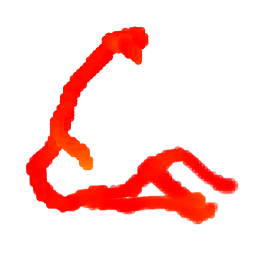}
     \end{subfigure}
     \hfill
	\begin{subfigure}[b]{0.13\linewidth}
         \centering
         \includegraphics[width=\linewidth]{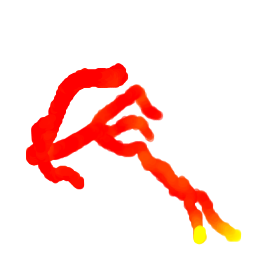}
     \end{subfigure}
     \hfill
	\begin{subfigure}[b]{0.13\linewidth}
         \centering
         \includegraphics[width=\linewidth]{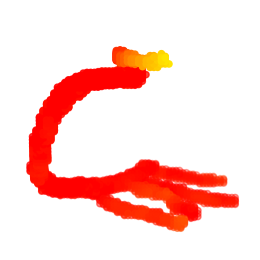}
     \end{subfigure}
     \hfill
	\begin{subfigure}[b]{0.13\linewidth}
         \centering
         \includegraphics[width=\linewidth]{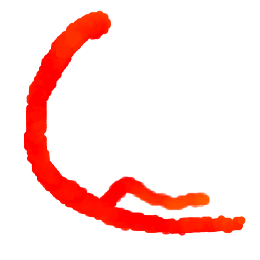}
     \end{subfigure}
     \hfill
	\begin{subfigure}[b]{0.13\linewidth}
         \centering
         \includegraphics[width=\linewidth]{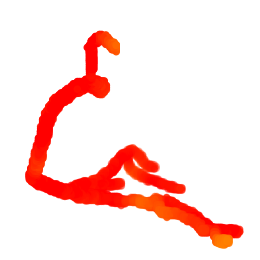}
     \end{subfigure}
     \vfill
     %%%%%% CTLs
     \begin{subfigure}[b]{0.13\linewidth}
         \centering
         \includegraphics[width=\linewidth]{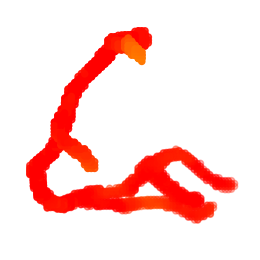}
     \end{subfigure}
     \hfill
	\begin{subfigure}[b]{0.13\linewidth}
         \centering
         \includegraphics[width=\linewidth]{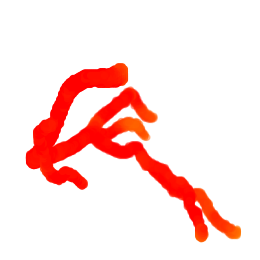}
     \end{subfigure}
     \hfill
	\begin{subfigure}[b]{0.13\linewidth}
         \centering
         \includegraphics[width=\linewidth]{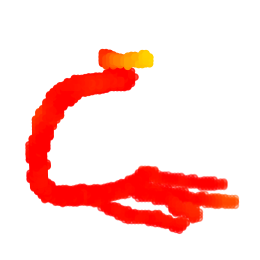}
     \end{subfigure}
     \hfill
	\begin{subfigure}[b]{0.13\linewidth}
         \centering
         \includegraphics[width=\linewidth]{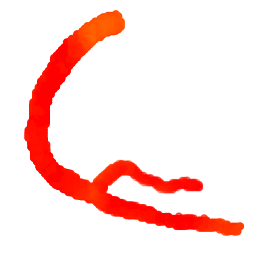}
     \end{subfigure}
     \hfill
	\begin{subfigure}[b]{0.13\linewidth}
         \centering
         \includegraphics[width=\linewidth]{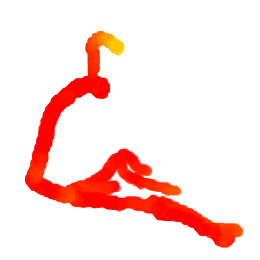}
     \end{subfigure}
     \vfill
     %%%%%% DSCC
     \begin{subfigure}[b]{0.13\linewidth}
         \centering
         \includegraphics[width=\linewidth]{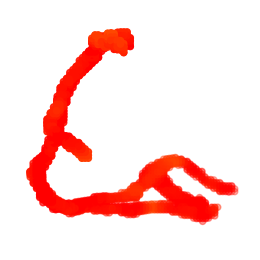}
     \end{subfigure}
     \hfill
	\begin{subfigure}[b]{0.13\linewidth}
         \centering
         \includegraphics[width=\linewidth]{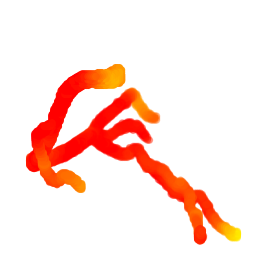}
     \end{subfigure}
     \hfill
	\begin{subfigure}[b]{0.13\linewidth}
         \centering
         \includegraphics[width=\linewidth]{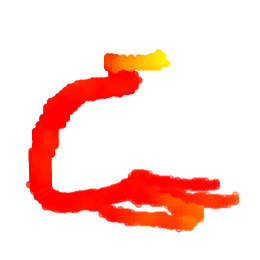}
     \end{subfigure}
     \hfill
	\begin{subfigure}[b]{0.13\linewidth}
         \centering
         \includegraphics[width=\linewidth]{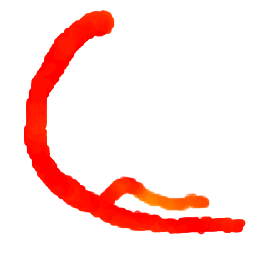}
     \end{subfigure}
     \hfill
	\begin{subfigure}[b]{0.13\linewidth}
         \centering
         \includegraphics[width=\linewidth]{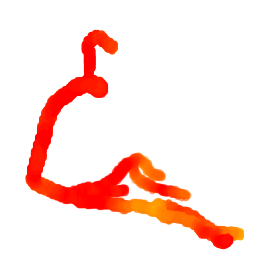}
     \end{subfigure}
     \vfill
     %%%%%% Un2+
     \begin{subfigure}[b]{0.13\linewidth}
         \centering
         \includegraphics[width=\linewidth]{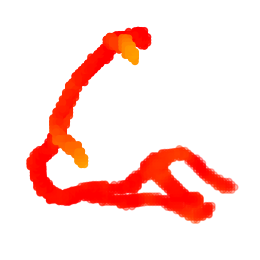}
     \end{subfigure}
     \hfill
	\begin{subfigure}[b]{0.13\linewidth}
         \centering
         \includegraphics[width=\linewidth]{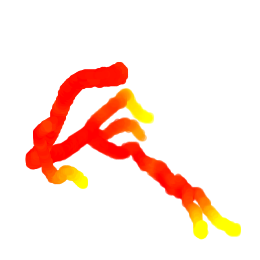}
     \end{subfigure}
     \hfill
	\begin{subfigure}[b]{0.13\linewidth}
         \centering
         \includegraphics[width=\linewidth]{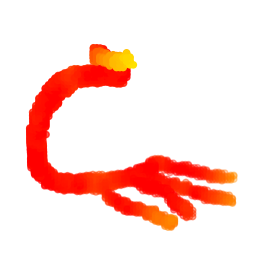}
     \end{subfigure}
     \hfill
	\begin{subfigure}[b]{0.13\linewidth}
         \centering
         \includegraphics[width=\linewidth]{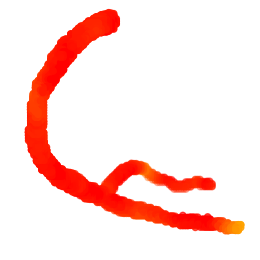}
     \end{subfigure}
     \hfill
	\begin{subfigure}[b]{0.13\linewidth}
         \centering
         \includegraphics[width=\linewidth]{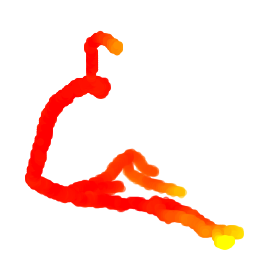}
     \end{subfigure}
     \vfill
     %%%%%% Un3+
     \begin{subfigure}[b]{0.13\linewidth}
         \centering
         \includegraphics[width=\linewidth]{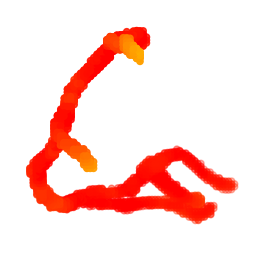}
     \end{subfigure}
     \hfill
	\begin{subfigure}[b]{0.13\linewidth}
         \centering
         \includegraphics[width=\linewidth]{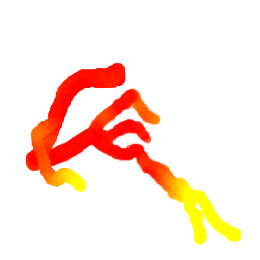}
     \end{subfigure}
     \hfill
	\begin{subfigure}[b]{0.13\linewidth}
         \centering
         \includegraphics[width=\linewidth]{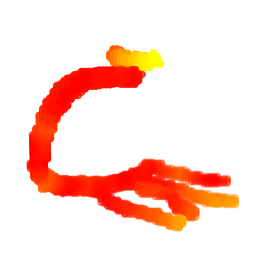}
     \end{subfigure}
     \hfill
	\begin{subfigure}[b]{0.13\linewidth}
         \centering
         \includegraphics[width=\linewidth]{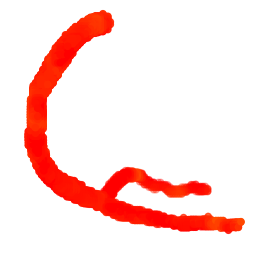}
     \end{subfigure}
     \hfill
	\begin{subfigure}[b]{0.13\linewidth}
         \centering
         \includegraphics[width=\linewidth]{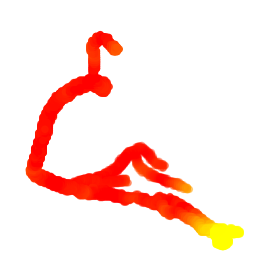}
     \end{subfigure}
     \vfill
     %%%%%% DSCN
     \begin{subfigure}[b]{0.13\linewidth}
         \centering
         \includegraphics[width=\linewidth]{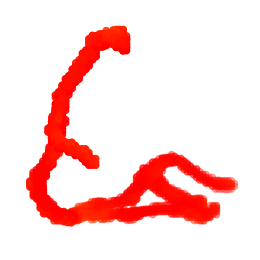}
     \end{subfigure}
     \hfill
	\begin{subfigure}[b]{0.13\linewidth}
         \centering
         \includegraphics[width=\linewidth]{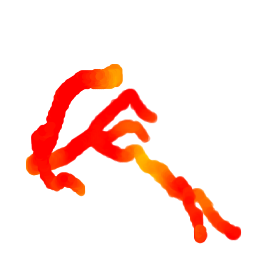}
     \end{subfigure}
     \hfill
	\begin{subfigure}[b]{0.13\linewidth}
         \centering
         \includegraphics[width=\linewidth]{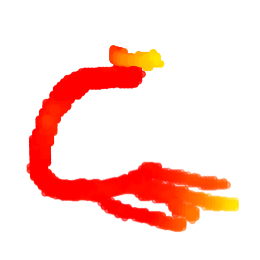}
     \end{subfigure}
     \hfill
	\begin{subfigure}[b]{0.13\linewidth}
         \centering
         \includegraphics[width=\linewidth]{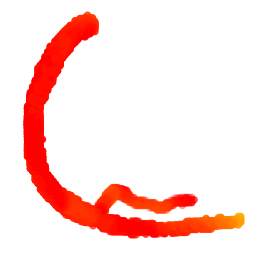}
     \end{subfigure}
     \hfill
	\begin{subfigure}[b]{0.13\linewidth}
         \centering
         \includegraphics[width=\linewidth]{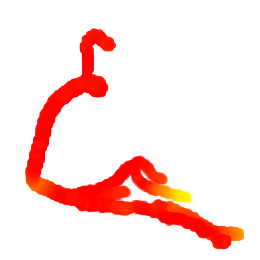}
     \end{subfigure}
     \vfill
     %%%%%% CVTG
     \begin{subfigure}[b]{0.13\linewidth}
         \centering
         \includegraphics[width=\linewidth]{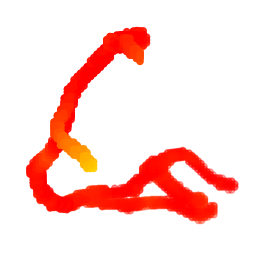}
     \end{subfigure}
     \hfill
	\begin{subfigure}[b]{0.13\linewidth}
         \centering
         \includegraphics[width=\linewidth]{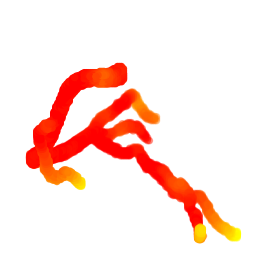}
     \end{subfigure}
     \hfill
	\begin{subfigure}[b]{0.13\linewidth}
         \centering
         \includegraphics[width=\linewidth]{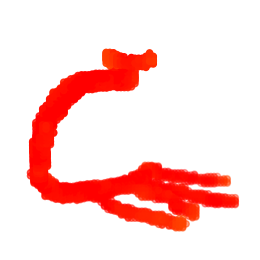}
     \end{subfigure}
     \hfill
	\begin{subfigure}[b]{0.13\linewidth}
         \centering
         \includegraphics[width=\linewidth]{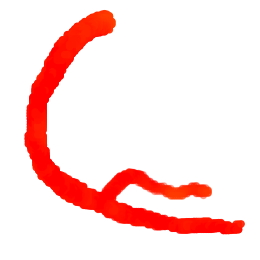}
     \end{subfigure}
     \hfill
	\begin{subfigure}[b]{0.13\linewidth}
         \centering
         \includegraphics[width=\linewidth]{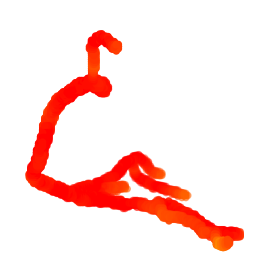}
     \end{subfigure}
     \vfill
     %%%%%% cbar
     \begin{subfigure}[b]{0.5\linewidth}
         \centering
         \includegraphics[width=\linewidth]{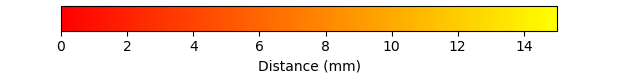}
     \end{subfigure}
     \caption{The corresponding voxel-wise prediction errors in terms of CD($\ell_2$) between the ground truth and 3D reconstruction, after rigidly registering the ground truth to the reconstructions from all the models. From left to right: 5 CCTA test data samples. From top to bottom: the prediction errors by our proposed DeepCA model, WGP, +CTLs, +DSCC, Un2+, Un3+, DSCN, and CVTG.}\label{ccta_cd}
\end{figure*}

\begin{figure*}[!h]
     \centering
     %%%%%% 3d mine
     \begin{subfigure}[b]{0.12\textwidth}
         \centering
         \includegraphics[width=\textwidth]{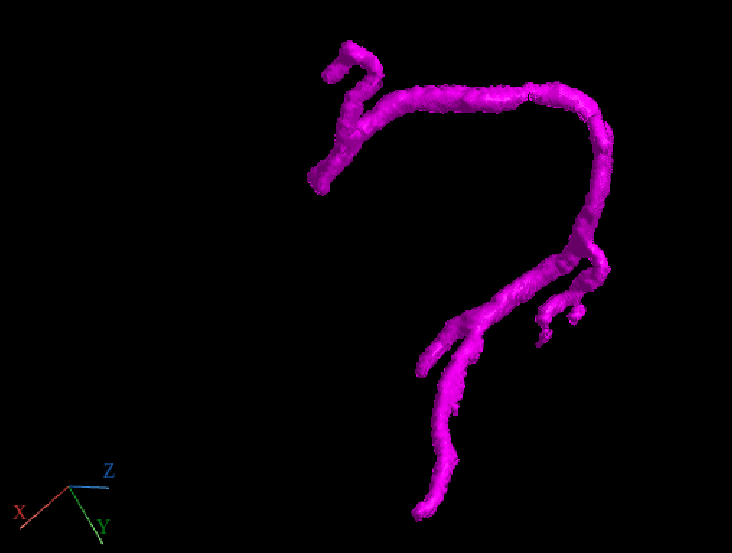}
     \end{subfigure}
     \hfill
	\begin{subfigure}[b]{0.12\textwidth}
         \centering
         \includegraphics[width=\textwidth]{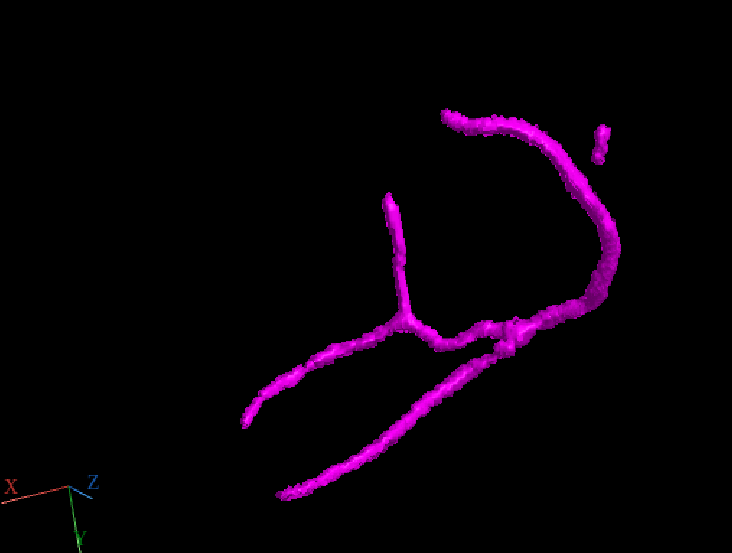}
     \end{subfigure}
     \hfill
	\begin{subfigure}[b]{0.12\textwidth}
         \centering
         \includegraphics[width=\textwidth]{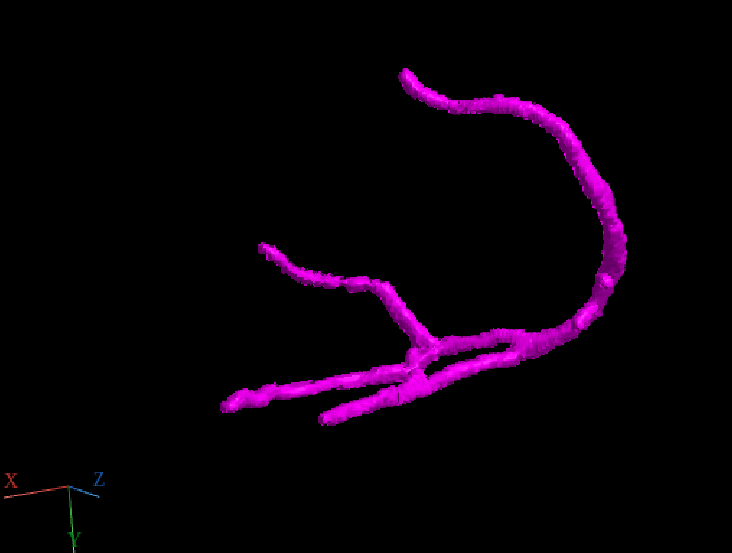}
     \end{subfigure}
     \hfill
	\begin{subfigure}[b]{0.12\textwidth}
         \centering
         \includegraphics[width=\textwidth]{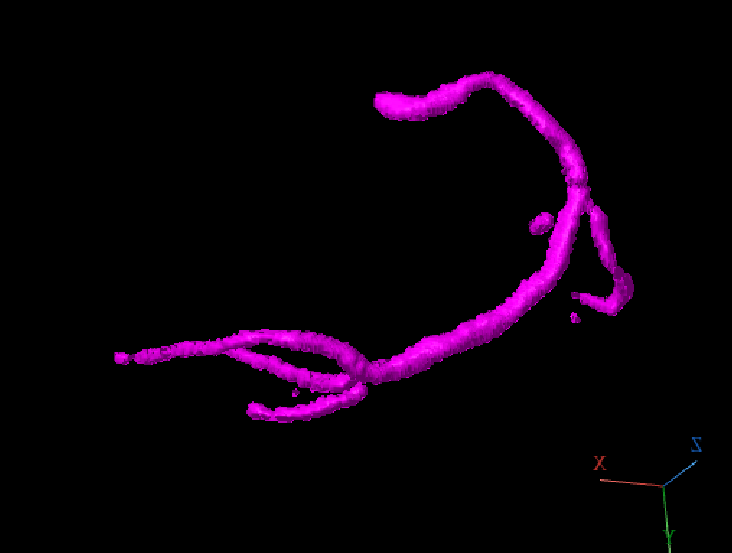}
     \end{subfigure}
     \hfill
	\begin{subfigure}[b]{0.12\textwidth}
         \centering
         \includegraphics[width=\textwidth]{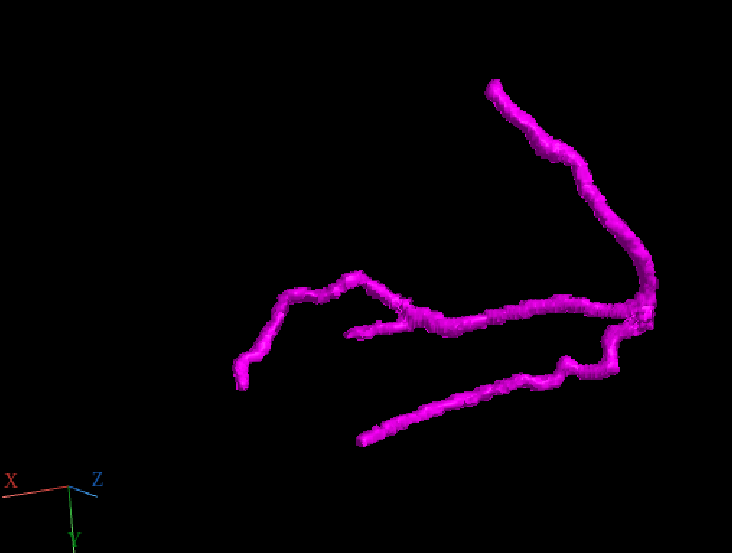}
     \end{subfigure}
     \hfill
     \begin{subfigure}[b]{0.12\textwidth}
         \centering
         \includegraphics[width=\textwidth]{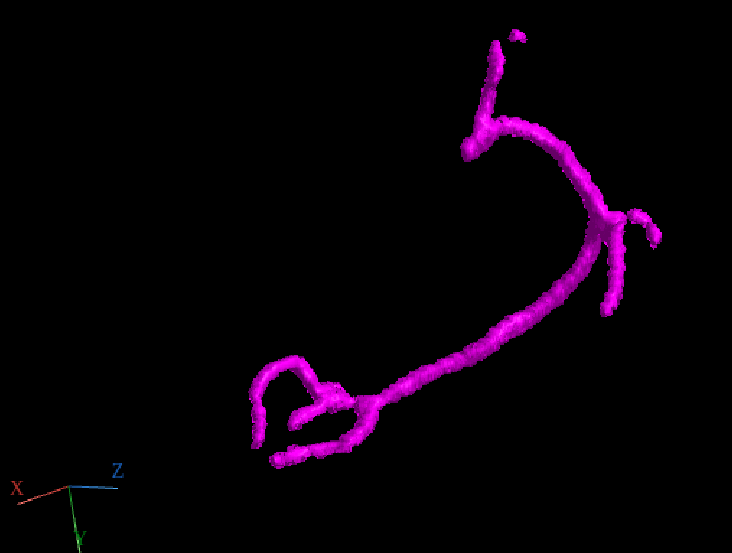}
     \end{subfigure}
     \hfill
     \begin{subfigure}[b]{0.12\textwidth}
         \centering
         \includegraphics[width=\textwidth]{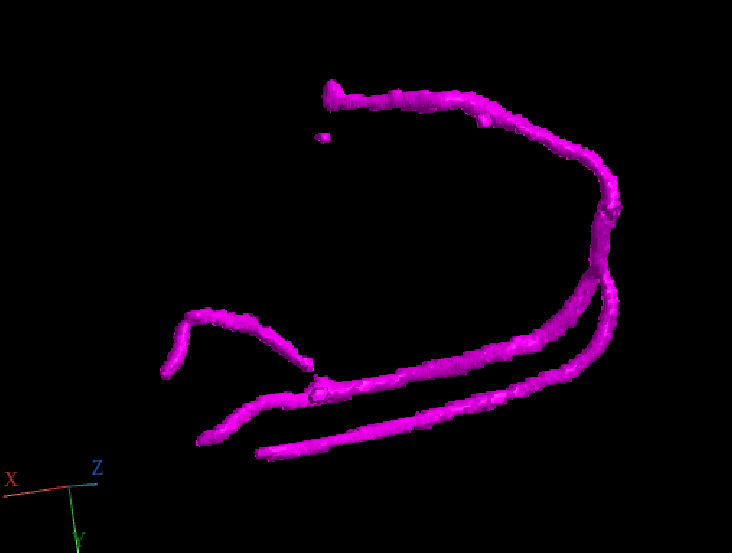}
     \end{subfigure}
     \hfill
     \begin{subfigure}[b]{0.12\textwidth}
         \centering
         \includegraphics[width=\textwidth]{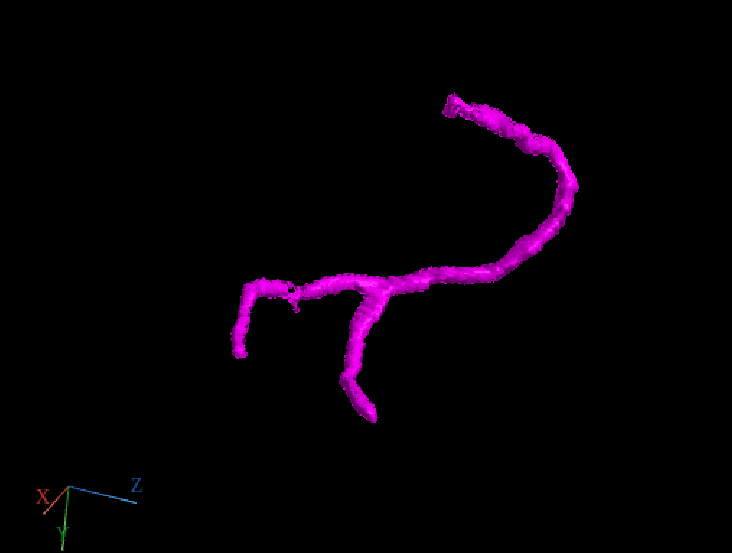}
     \end{subfigure}
     \vfill
     %%%%%% WGP
     \begin{subfigure}[b]{0.12\textwidth}
         \centering
         \includegraphics[width=\textwidth]{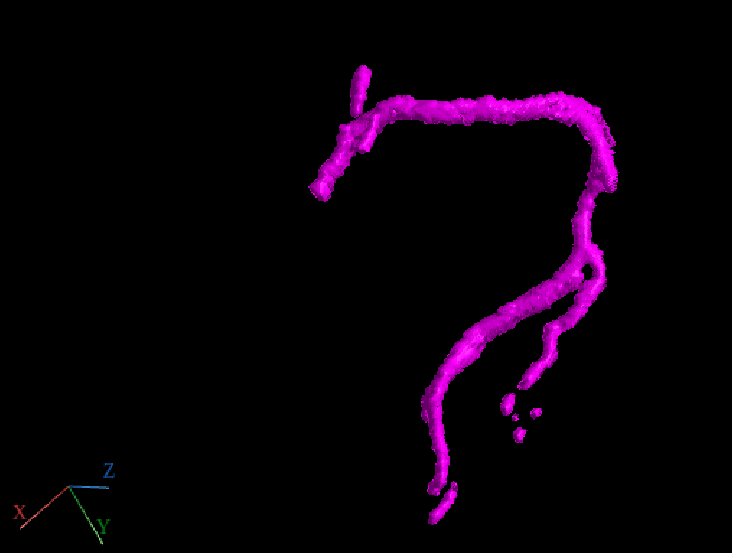}
     \end{subfigure}
     \hfill
	\begin{subfigure}[b]{0.12\textwidth}
         \centering
         \includegraphics[width=\textwidth]{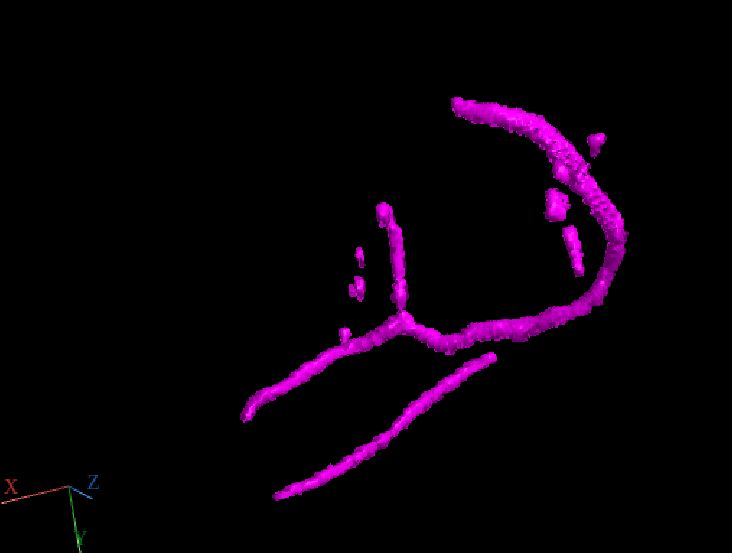}
     \end{subfigure}
     \hfill
	\begin{subfigure}[b]{0.12\textwidth}
         \centering
         \includegraphics[width=\textwidth]{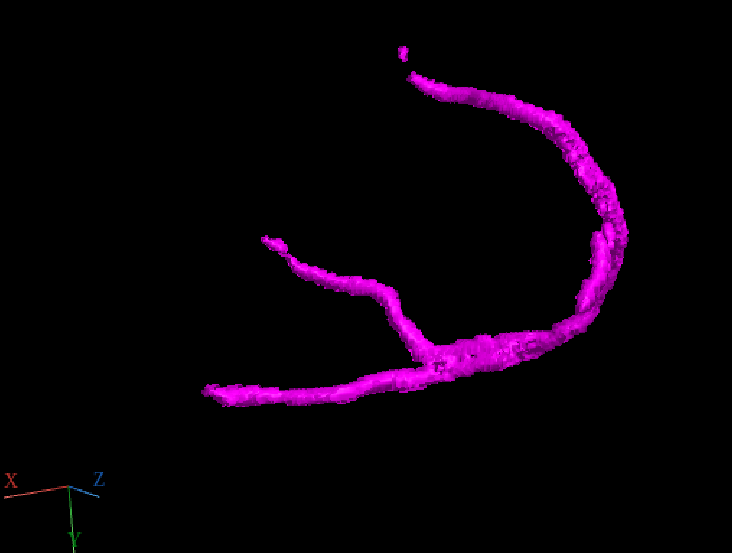}
     \end{subfigure}
     \hfill
	\begin{subfigure}[b]{0.12\textwidth}
         \centering
         \includegraphics[width=\textwidth]{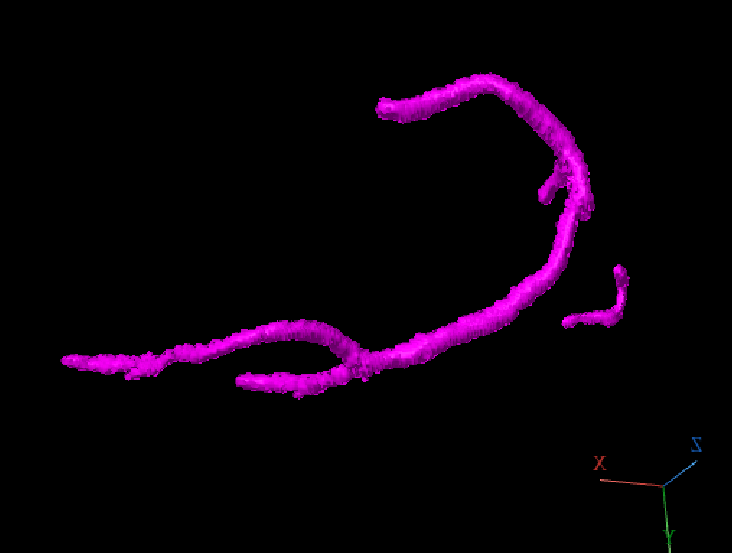}
     \end{subfigure}
     \hfill
	\begin{subfigure}[b]{0.12\textwidth}
         \centering
         \includegraphics[width=\textwidth]{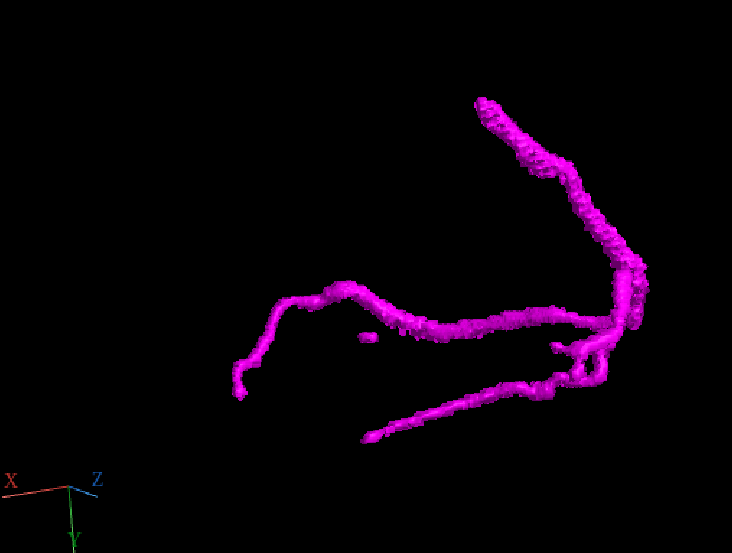}
     \end{subfigure}
     \hfill
     \begin{subfigure}[b]{0.12\textwidth}
         \centering
         \includegraphics[width=\textwidth]{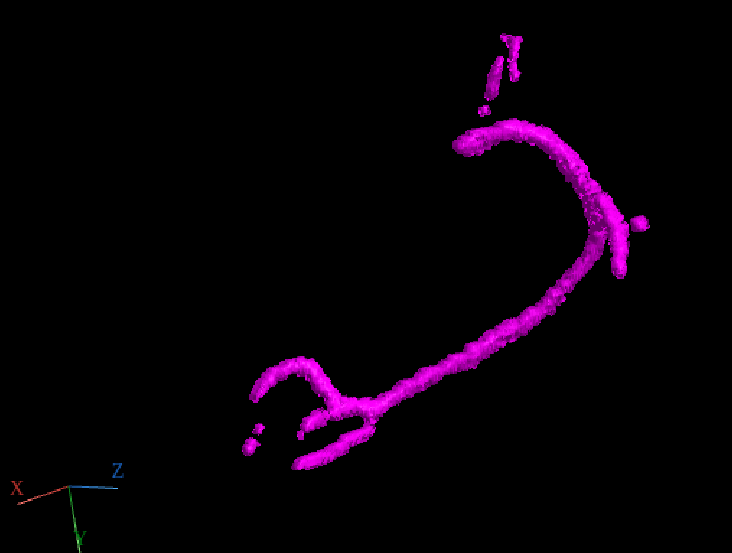}
     \end{subfigure}
     \hfill
     \begin{subfigure}[b]{0.12\textwidth}
         \centering
         \includegraphics[width=\textwidth]{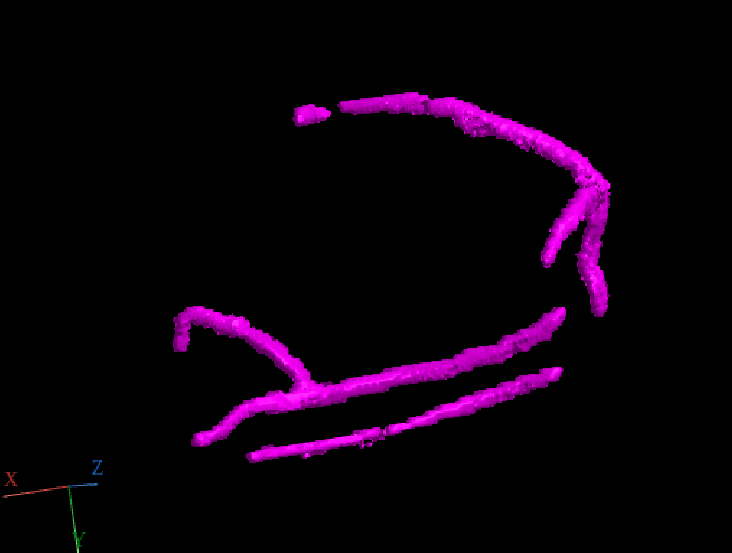}
     \end{subfigure}
     \hfill
     \begin{subfigure}[b]{0.12\textwidth}
         \centering
         \includegraphics[width=\textwidth]{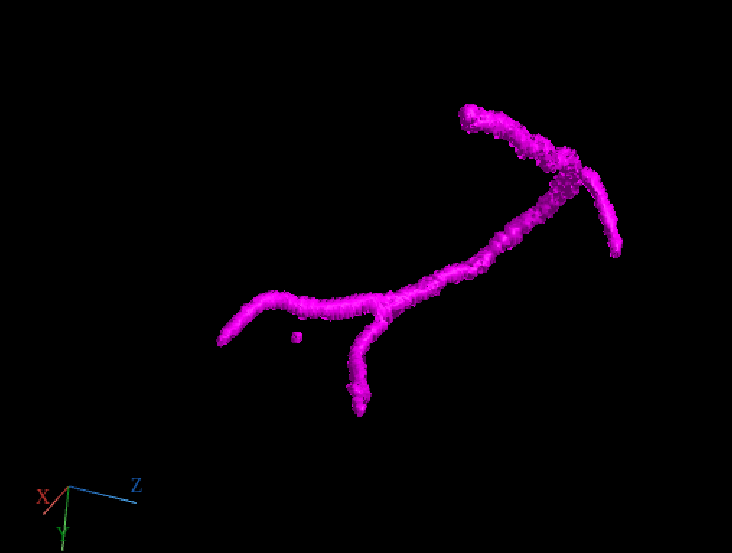}
     \end{subfigure}
     \vfill
     %%%%%% CTLs
     \begin{subfigure}[b]{0.12\textwidth}
         \centering
         \includegraphics[width=\textwidth]{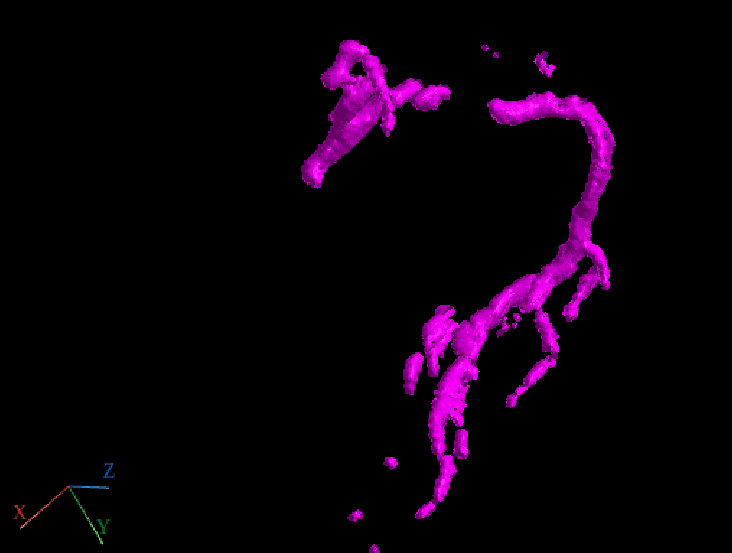}
     \end{subfigure}
     \hfill
	\begin{subfigure}[b]{0.12\textwidth}
         \centering
         \includegraphics[width=\textwidth]{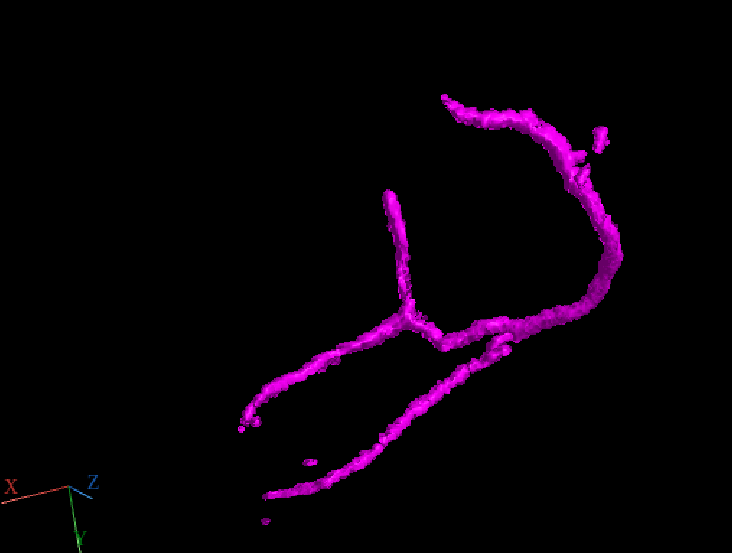}
     \end{subfigure}
     \hfill
	\begin{subfigure}[b]{0.12\textwidth}
         \centering
         \includegraphics[width=\textwidth]{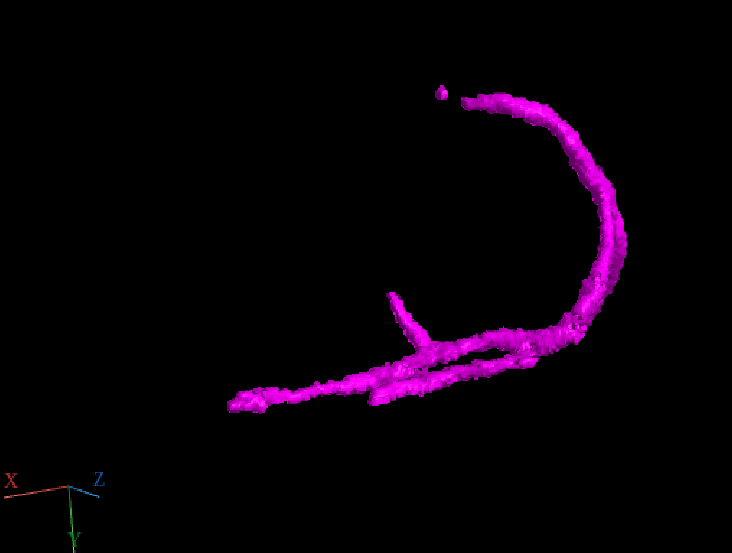}
     \end{subfigure}
     \hfill
	\begin{subfigure}[b]{0.12\textwidth}
         \centering
         \includegraphics[width=\textwidth]{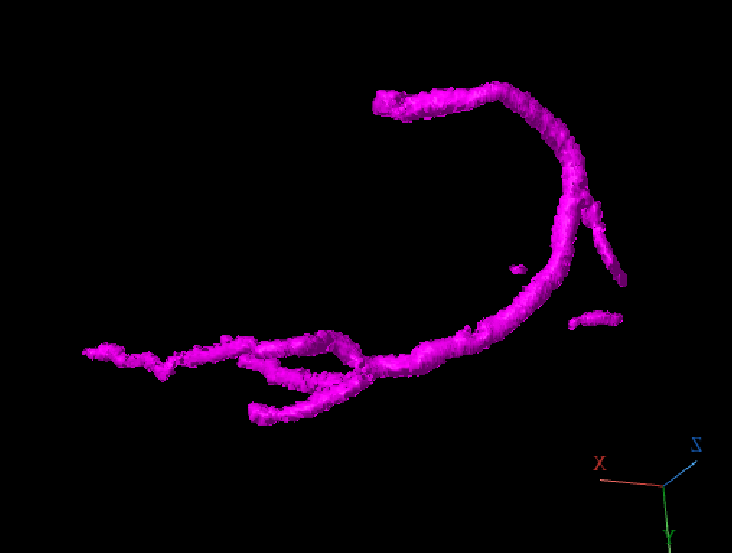}
     \end{subfigure}
     \hfill
	\begin{subfigure}[b]{0.12\textwidth}
         \centering
         \includegraphics[width=\textwidth]{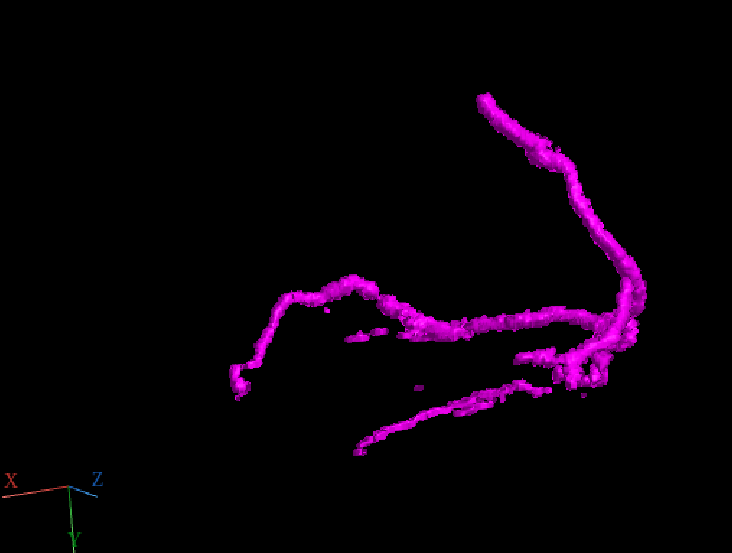}
     \end{subfigure}
     \hfill
     \begin{subfigure}[b]{0.12\textwidth}
         \centering
         \includegraphics[width=\textwidth]{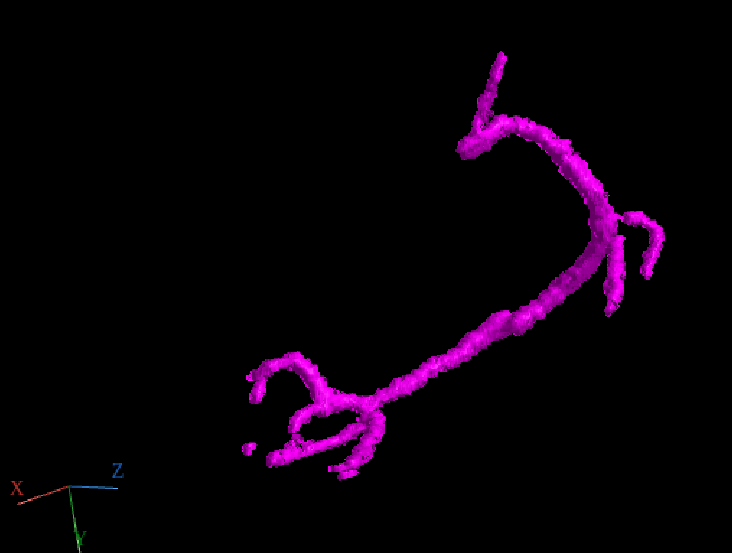}
     \end{subfigure}
     \hfill
     \begin{subfigure}[b]{0.12\textwidth}
         \centering
         \includegraphics[width=\textwidth]{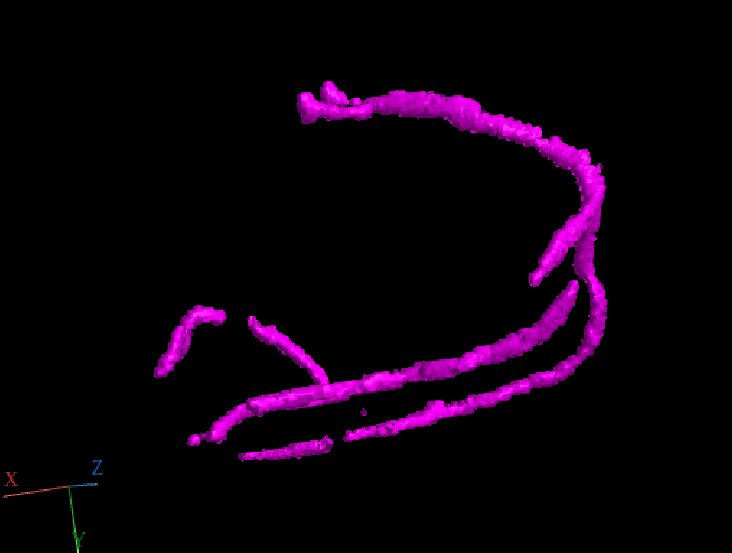}
     \end{subfigure}
     \hfill
     \begin{subfigure}[b]{0.12\textwidth}
         \centering
         \includegraphics[width=\textwidth]{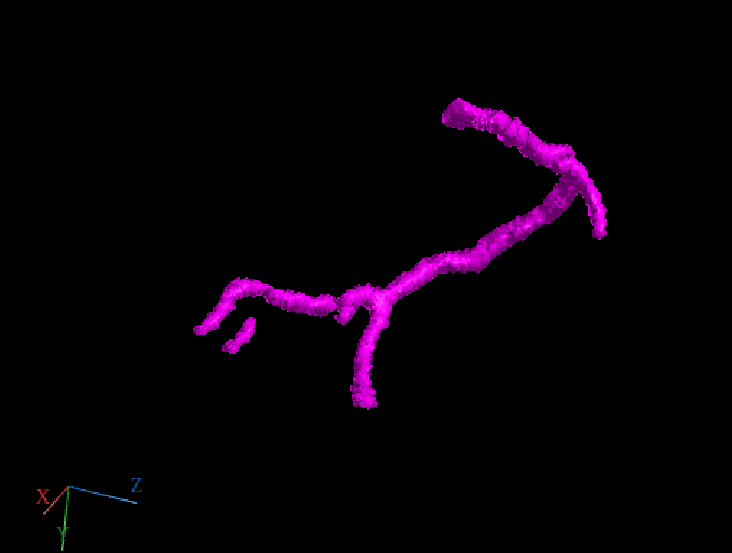}
     \end{subfigure}
     \vfill
     %%%%%% DSCC
     \begin{subfigure}[b]{0.12\textwidth}
         \centering
         \includegraphics[width=\textwidth]{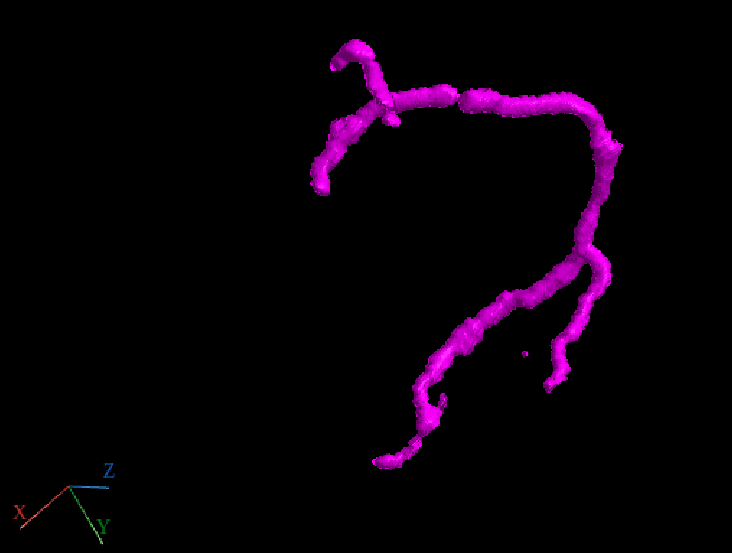}
     \end{subfigure}
     \hfill
	\begin{subfigure}[b]{0.12\textwidth}
         \centering
         \includegraphics[width=\textwidth]{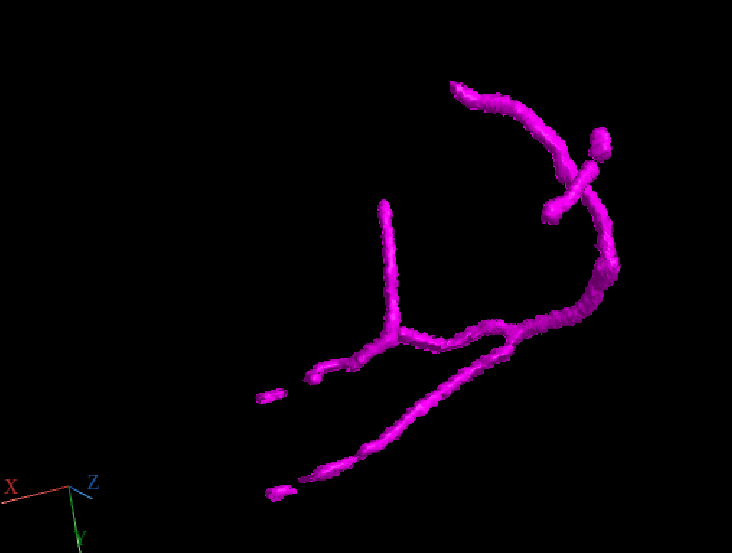}
     \end{subfigure}
     \hfill
	\begin{subfigure}[b]{0.12\textwidth}
         \centering
         \includegraphics[width=\textwidth]{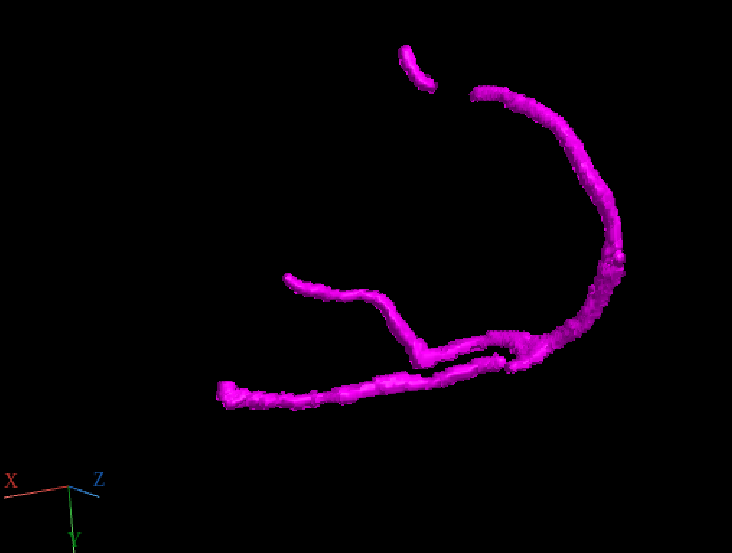}
     \end{subfigure}
     \hfill
	\begin{subfigure}[b]{0.12\textwidth}
         \centering
         \includegraphics[width=\textwidth]{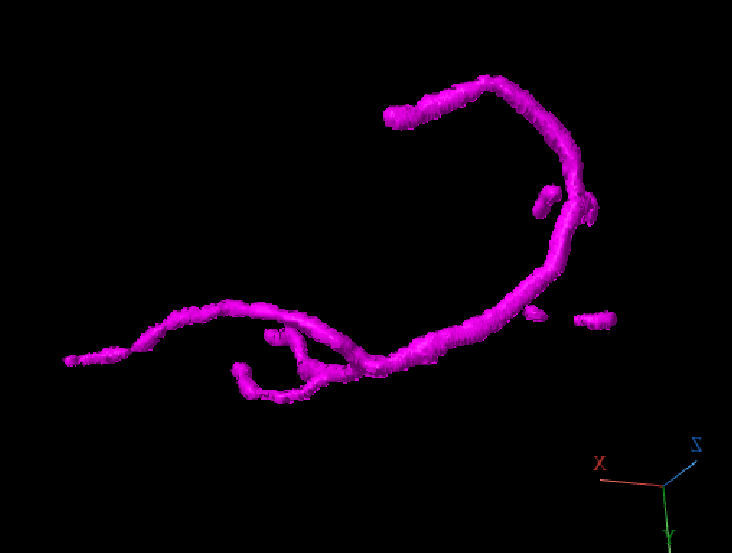}
     \end{subfigure}
     \hfill
	\begin{subfigure}[b]{0.12\textwidth}
         \centering
         \includegraphics[width=\textwidth]{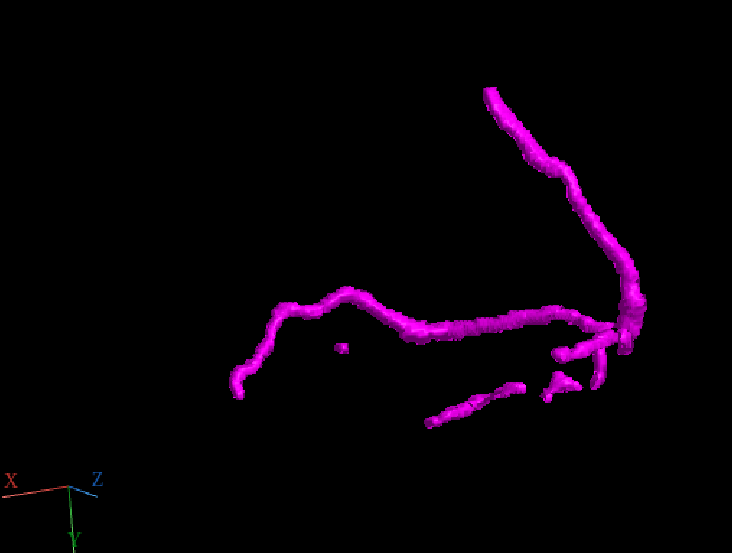}
     \end{subfigure}
     \hfill
     \begin{subfigure}[b]{0.12\textwidth}
         \centering
         \includegraphics[width=\textwidth]{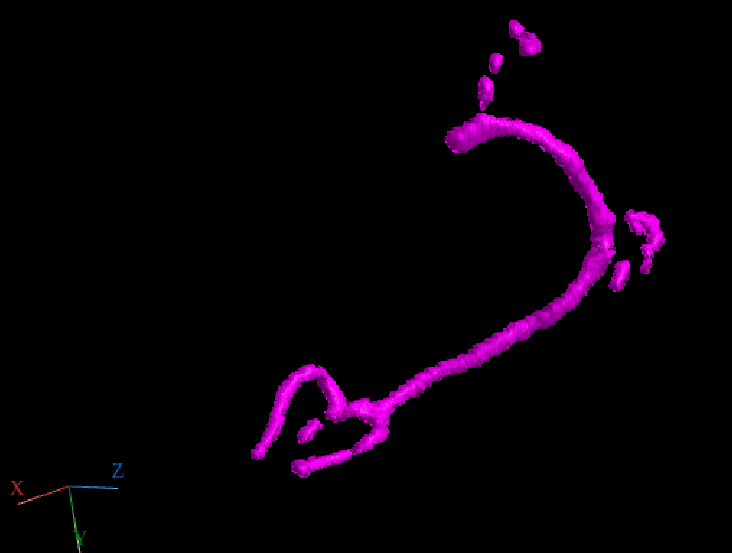}
     \end{subfigure}
     \hfill
     \begin{subfigure}[b]{0.12\textwidth}
         \centering
         \includegraphics[width=\textwidth]{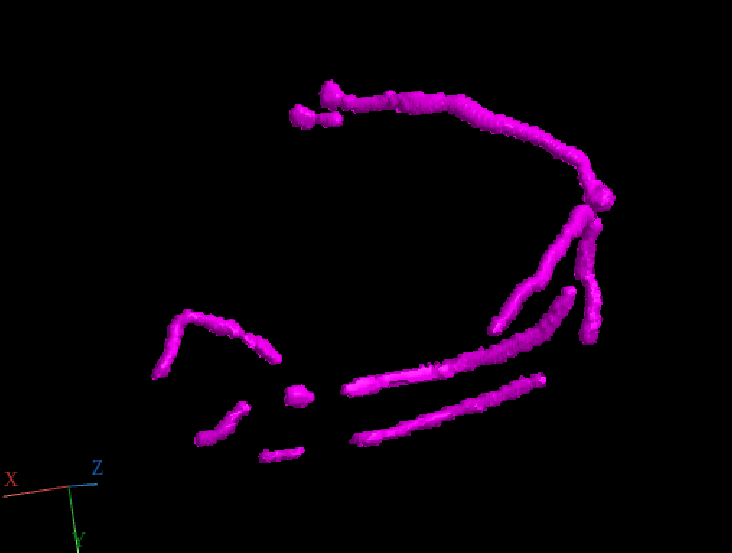}
     \end{subfigure}
     \hfill
     \begin{subfigure}[b]{0.12\textwidth}
         \centering
         \includegraphics[width=\textwidth]{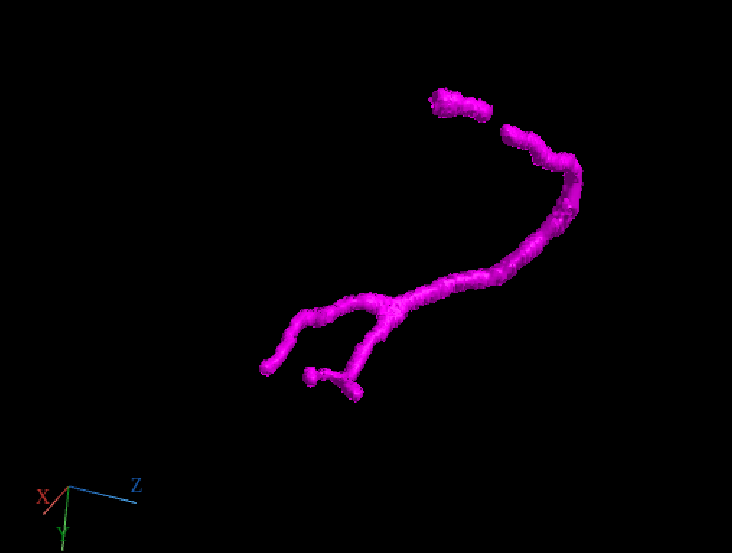}
     \end{subfigure}
     \vfill
     %%%%%% un2+ 
     \begin{subfigure}[b]{0.12\textwidth}
         \centering
         \includegraphics[width=\textwidth]{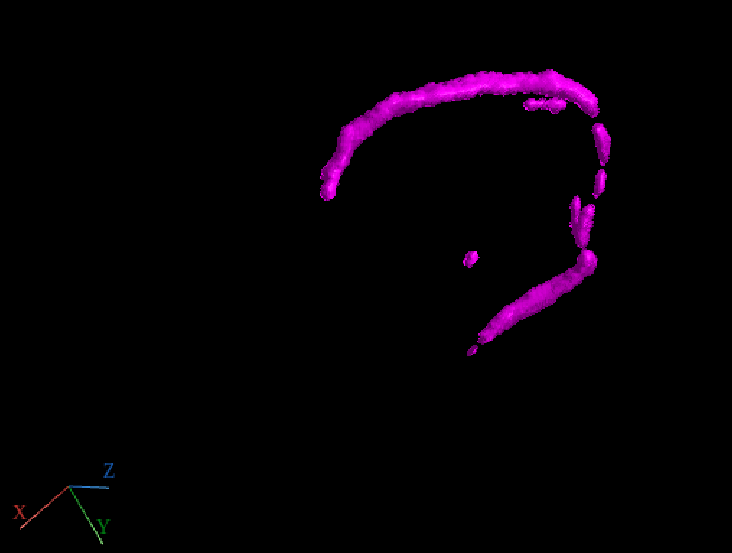}
     \end{subfigure}
     \hfill
	\begin{subfigure}[b]{0.12\textwidth}
         \centering
         \includegraphics[width=\textwidth]{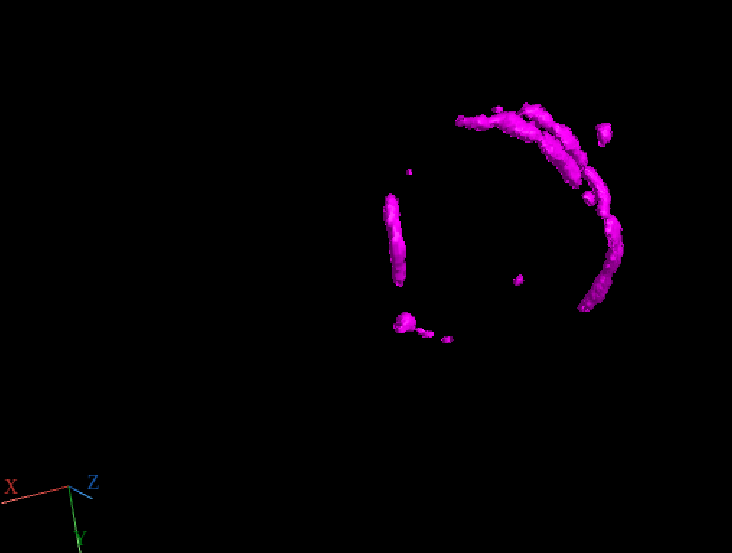}
     \end{subfigure}
     \hfill
	\begin{subfigure}[b]{0.12\textwidth}
         \centering
         \includegraphics[width=\textwidth]{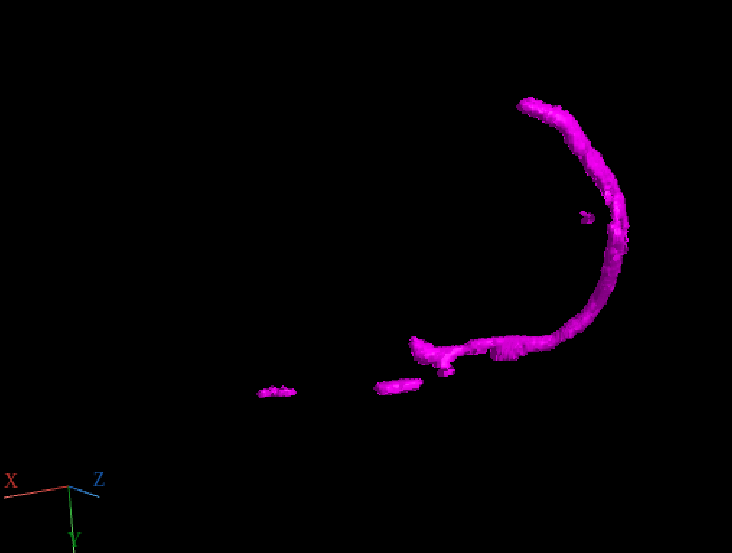}
     \end{subfigure}
     \hfill
	\begin{subfigure}[b]{0.12\textwidth}
         \centering
         \includegraphics[width=\textwidth]{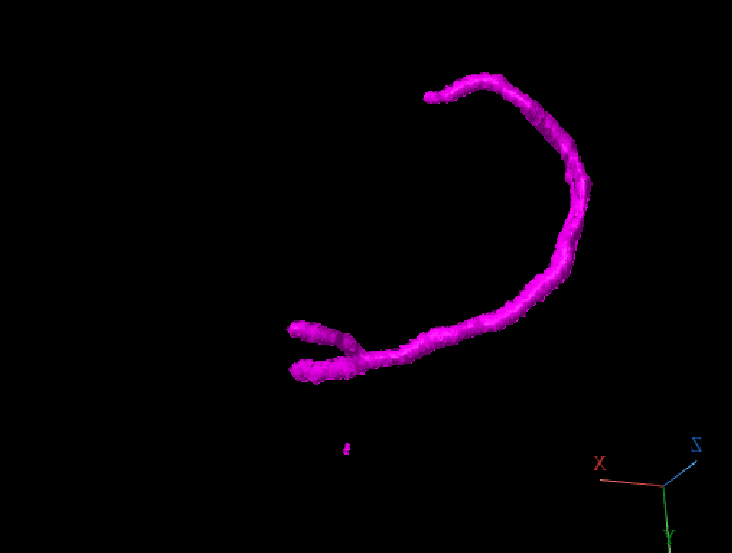}
     \end{subfigure}
     \hfill
	\begin{subfigure}[b]{0.12\textwidth}
         \centering
         \includegraphics[width=\textwidth]{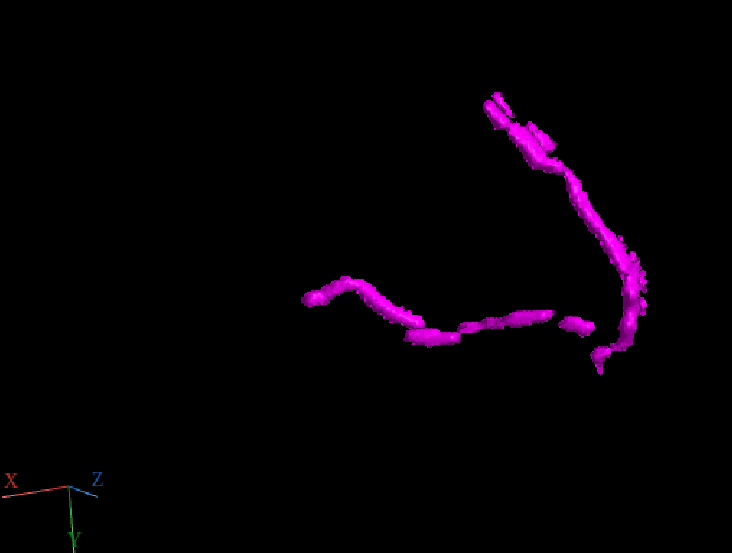}
     \end{subfigure}
     \hfill
     \begin{subfigure}[b]{0.12\textwidth}
         \centering
         \includegraphics[width=\textwidth]{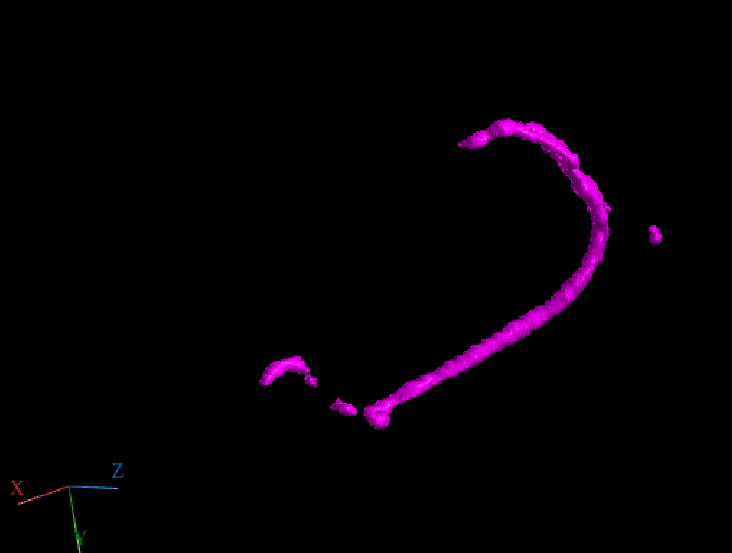}
     \end{subfigure}
     \hfill
     \begin{subfigure}[b]{0.12\textwidth}
         \centering
         \includegraphics[width=\textwidth]{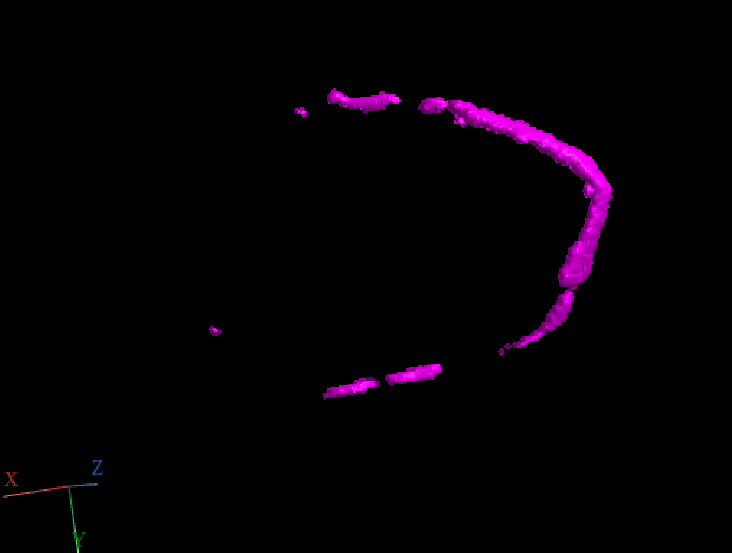}
     \end{subfigure}
     \hfill
     \begin{subfigure}[b]{0.12\textwidth}
         \centering
         \includegraphics[width=\textwidth]{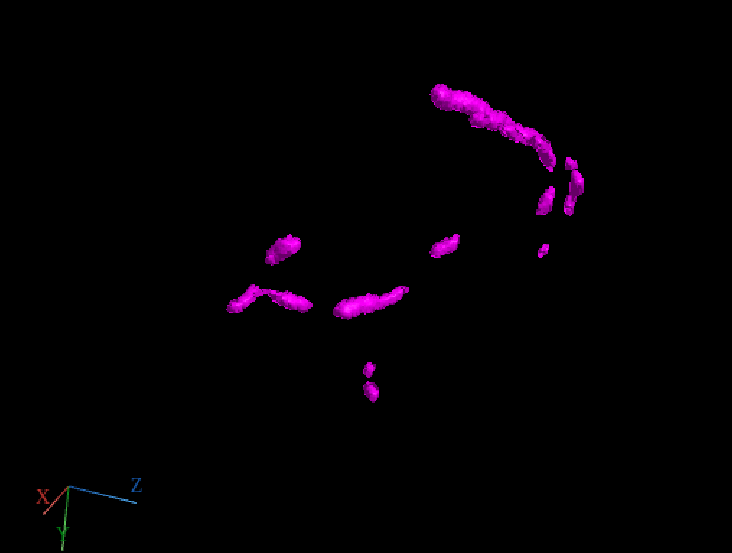}
     \end{subfigure}
     \vfill
     %%%%%% un3+
     \begin{subfigure}[b]{0.12\textwidth}
         \centering
         \includegraphics[width=\textwidth]{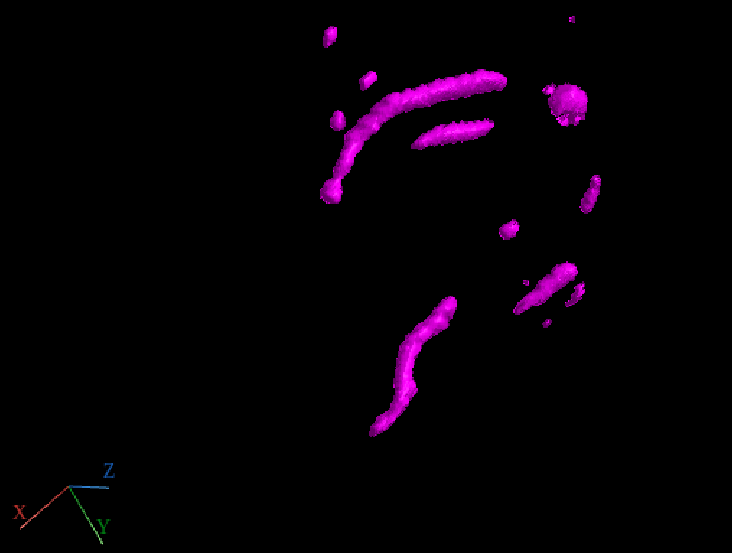}
     \end{subfigure}
     \hfill
	\begin{subfigure}[b]{0.12\textwidth}
         \centering
         \includegraphics[width=\textwidth]{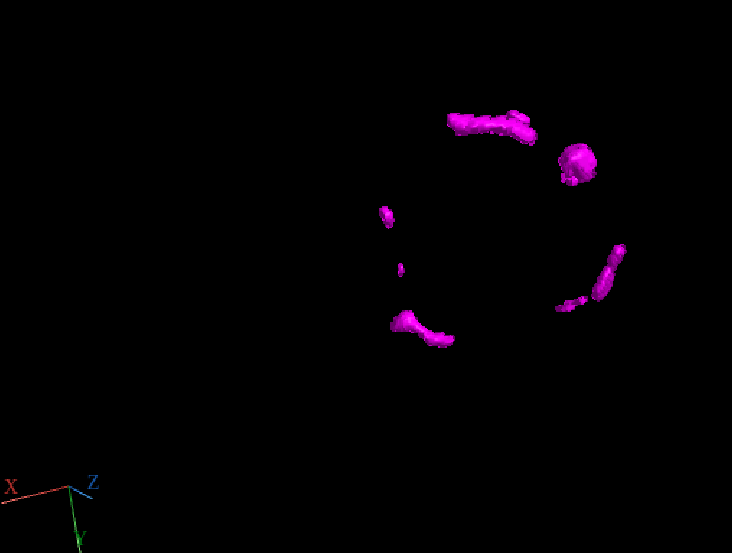}
     \end{subfigure}
     \hfill
	\begin{subfigure}[b]{0.12\textwidth}
         \centering
         \includegraphics[width=\textwidth]{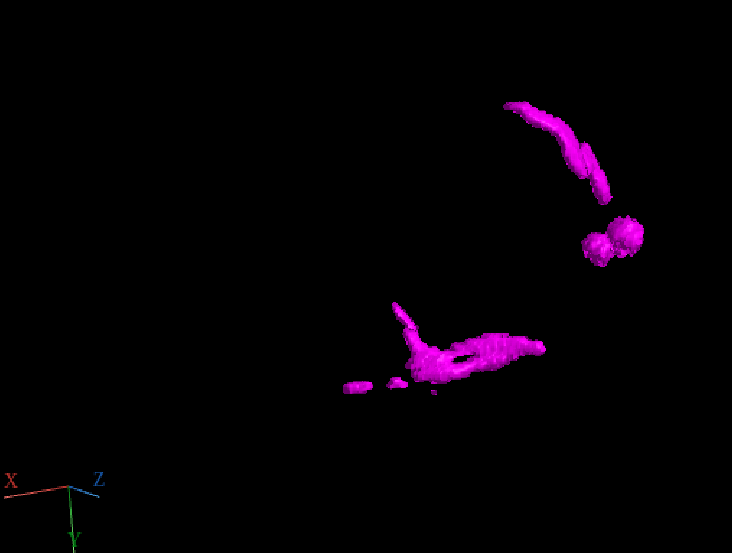}
     \end{subfigure}
     \hfill
	\begin{subfigure}[b]{0.12\textwidth}
         \centering
         \includegraphics[width=\textwidth]{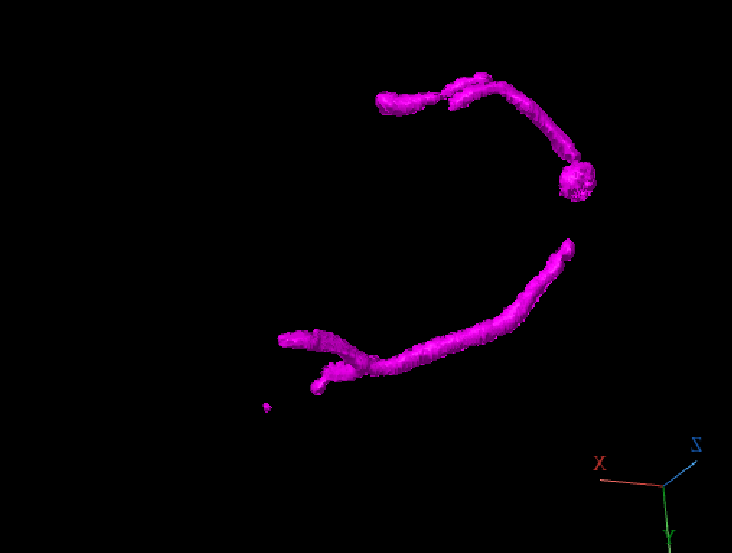}
     \end{subfigure}
     \hfill
	\begin{subfigure}[b]{0.12\textwidth}
         \centering
         \includegraphics[width=\textwidth]{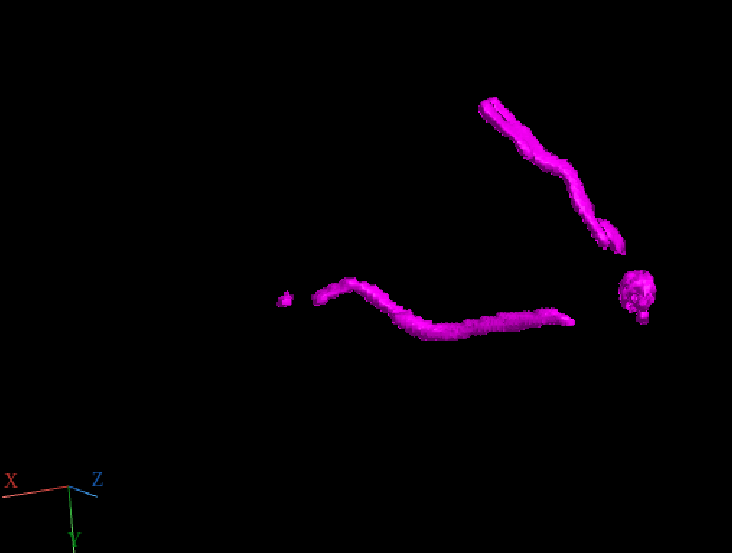}
     \end{subfigure}
     \hfill
     \begin{subfigure}[b]{0.12\textwidth}
         \centering
         \includegraphics[width=\textwidth]{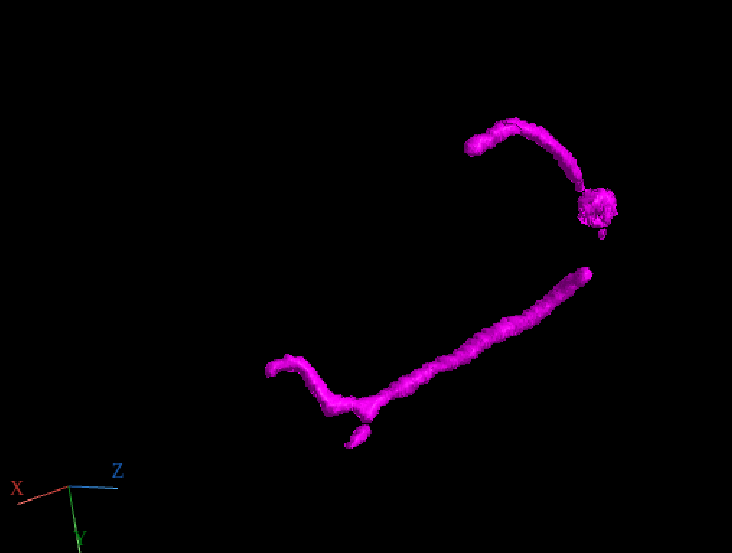}
     \end{subfigure}
     \hfill
     \begin{subfigure}[b]{0.12\textwidth}
         \centering
         \includegraphics[width=\textwidth]{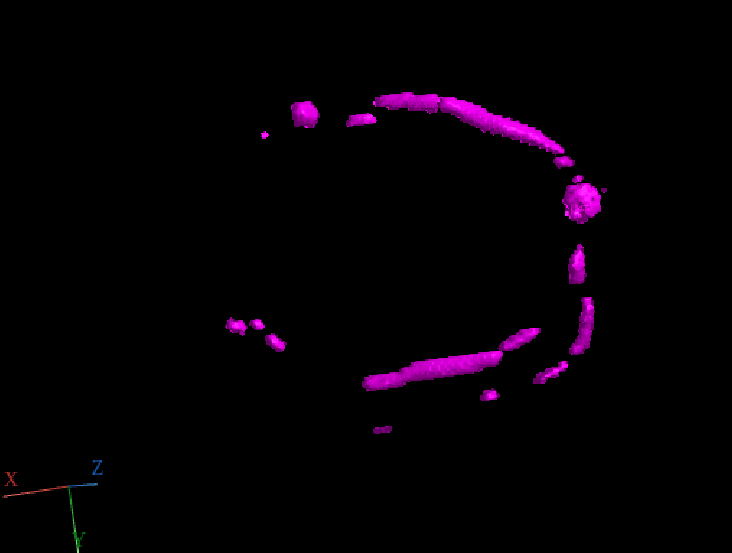}
     \end{subfigure}
     \hfill
     \begin{subfigure}[b]{0.12\textwidth}
         \centering
         \includegraphics[width=\textwidth]{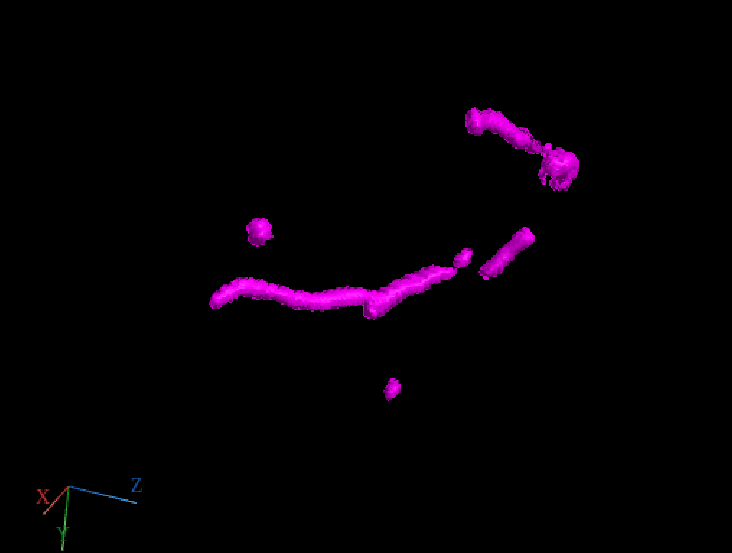}
     \end{subfigure}
     \vfill
     %%%%%% DSCN
     \begin{subfigure}[b]{0.12\textwidth}
         \centering
         \includegraphics[width=\textwidth]{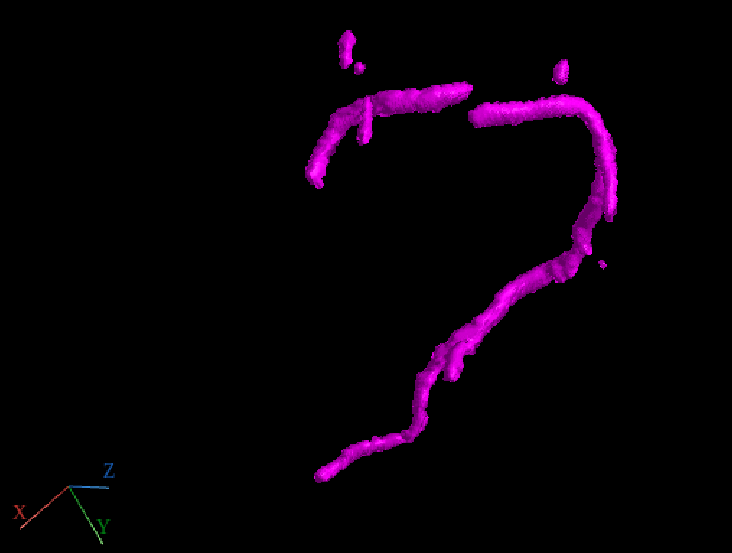}
     \end{subfigure}
     \hfill
	\begin{subfigure}[b]{0.12\textwidth}
         \centering
         \includegraphics[width=\textwidth]{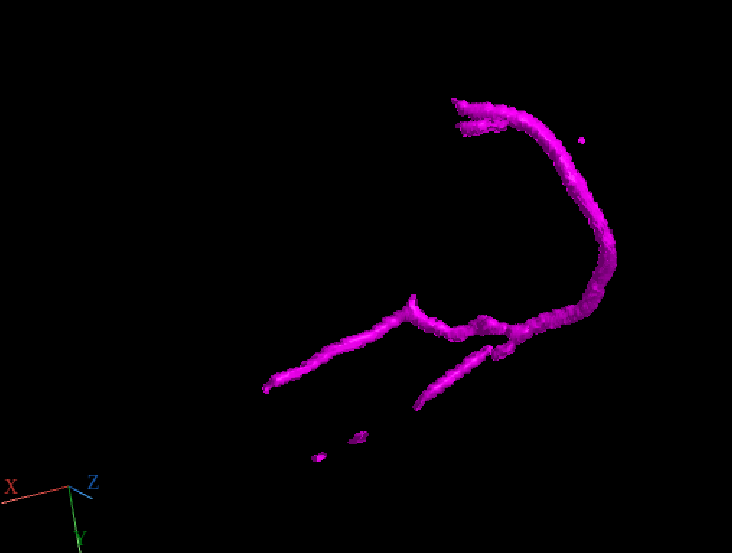}
     \end{subfigure}
     \hfill
	\begin{subfigure}[b]{0.12\textwidth}
         \centering
         \includegraphics[width=\textwidth]{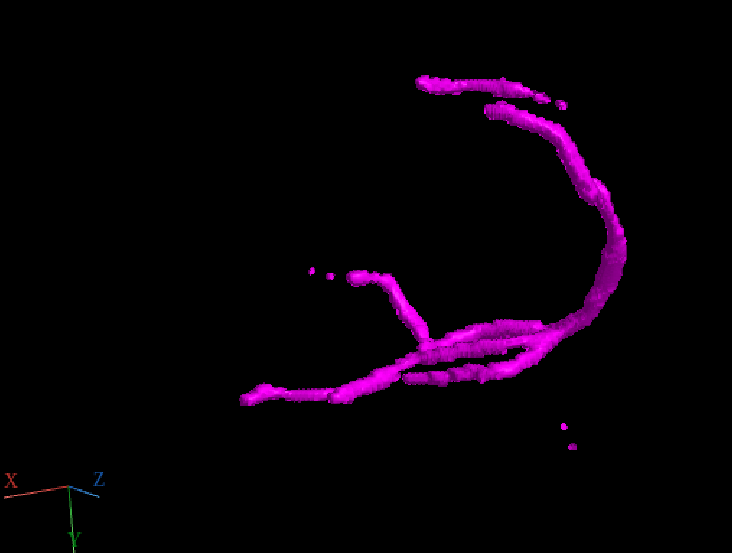}
     \end{subfigure}
     \hfill
	\begin{subfigure}[b]{0.12\textwidth}
         \centering
         \includegraphics[width=\textwidth]{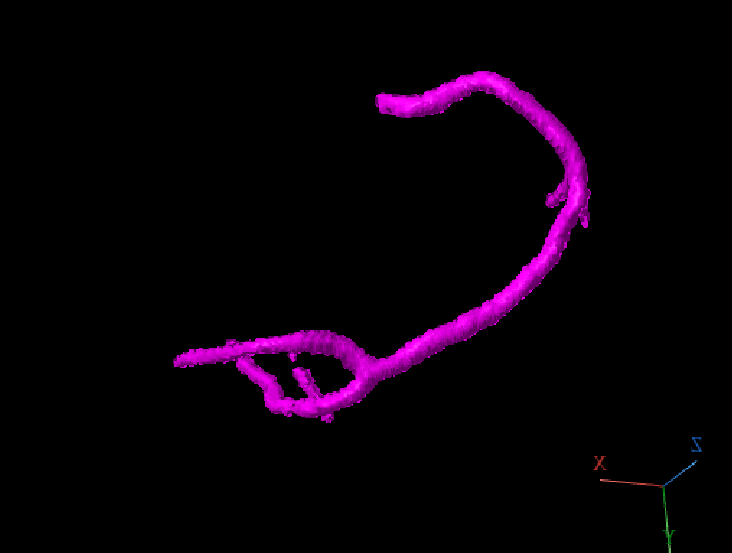}
     \end{subfigure}
     \hfill
	\begin{subfigure}[b]{0.12\textwidth}
         \centering
         \includegraphics[width=\textwidth]{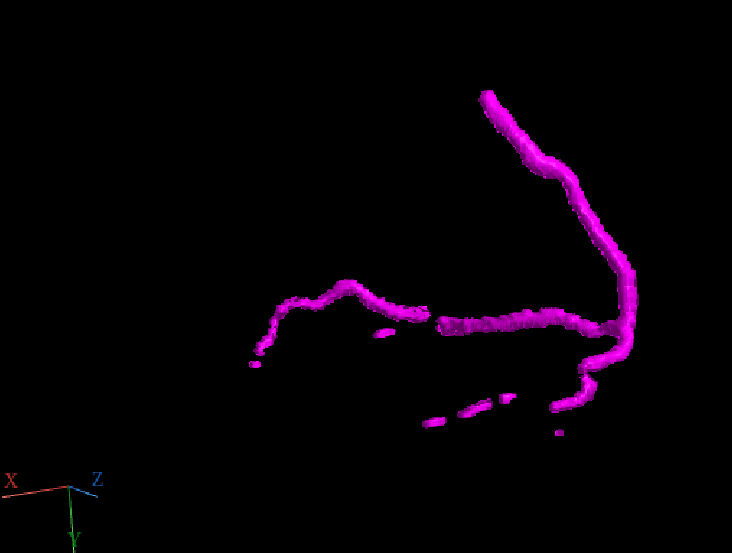}
     \end{subfigure}
     \hfill
     \begin{subfigure}[b]{0.12\textwidth}
         \centering
         \includegraphics[width=\textwidth]{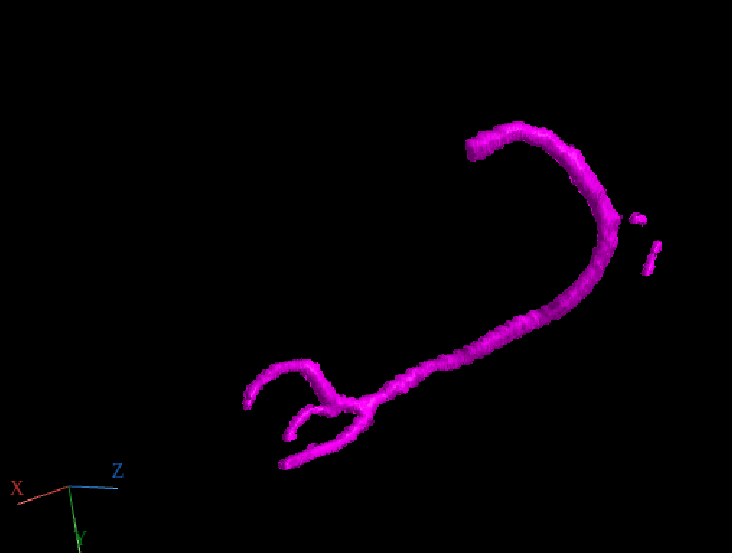}
     \end{subfigure}
     \hfill
     \begin{subfigure}[b]{0.12\textwidth}
         \centering
         \includegraphics[width=\textwidth]{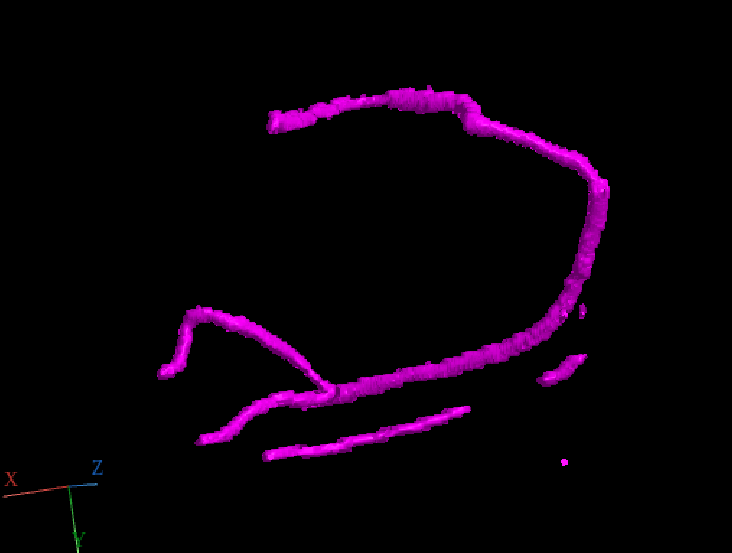}
     \end{subfigure}
     \hfill
     \begin{subfigure}[b]{0.12\textwidth}
         \centering
         \includegraphics[width=\textwidth]{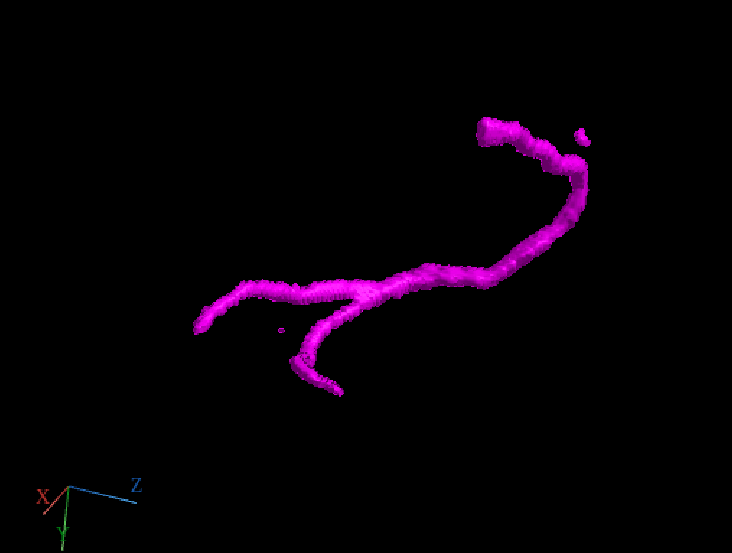}
     \end{subfigure}
     \vfill
     %%%%%% CVTG
     \begin{subfigure}[b]{0.12\textwidth}
         \centering
         \includegraphics[width=\textwidth]{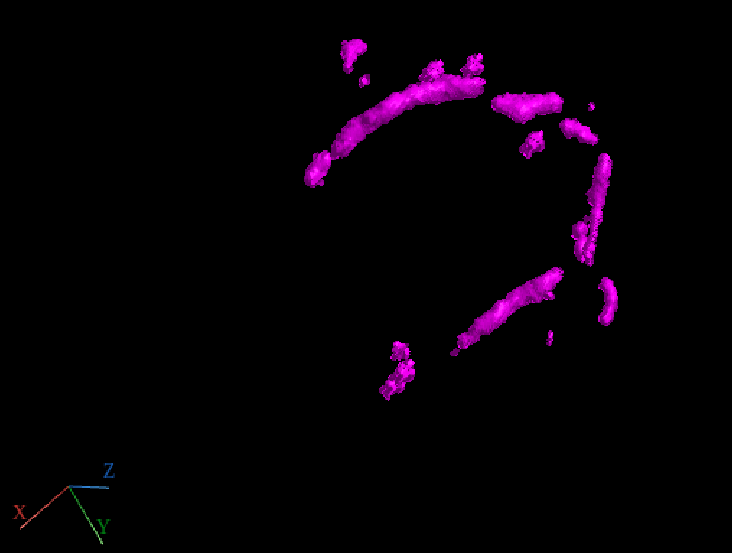}
     \end{subfigure}
     \hfill
	\begin{subfigure}[b]{0.12\textwidth}
         \centering
         \includegraphics[width=\textwidth]{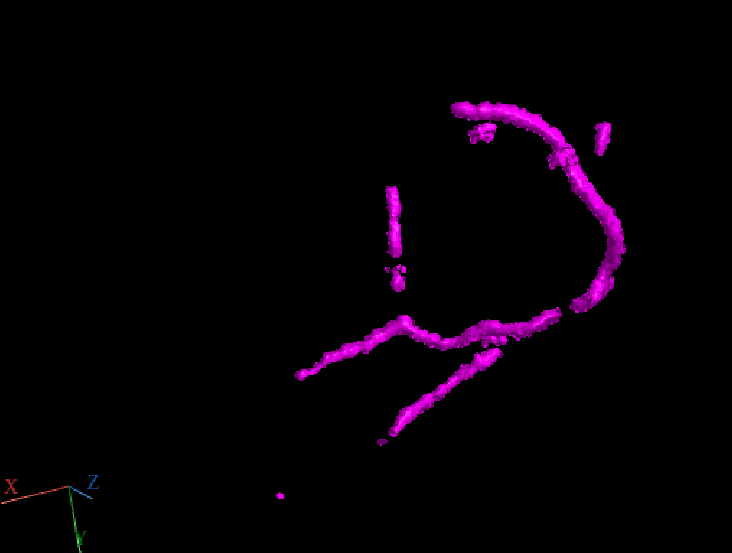}
     \end{subfigure}
     \hfill
	\begin{subfigure}[b]{0.12\textwidth}
         \centering
         \includegraphics[width=\textwidth]{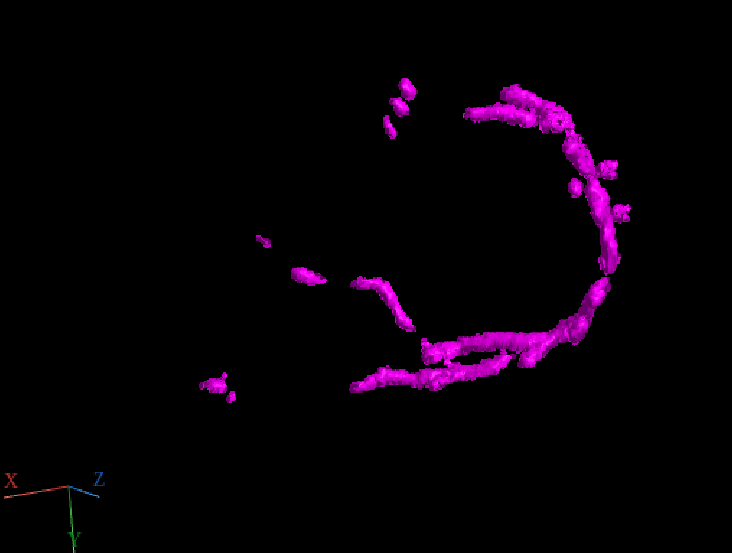}
     \end{subfigure}
     \hfill
	\begin{subfigure}[b]{0.12\textwidth}
         \centering
         \includegraphics[width=\textwidth]{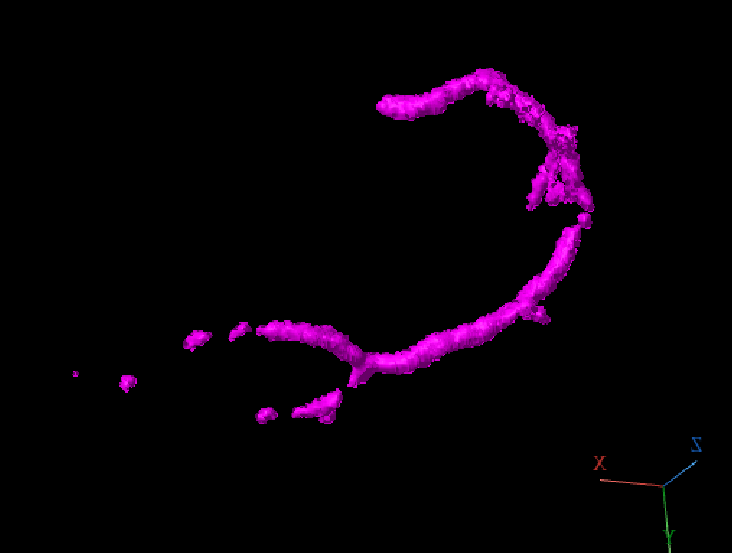}
     \end{subfigure}
     \hfill
	\begin{subfigure}[b]{0.12\textwidth}
         \centering
         \includegraphics[width=\textwidth]{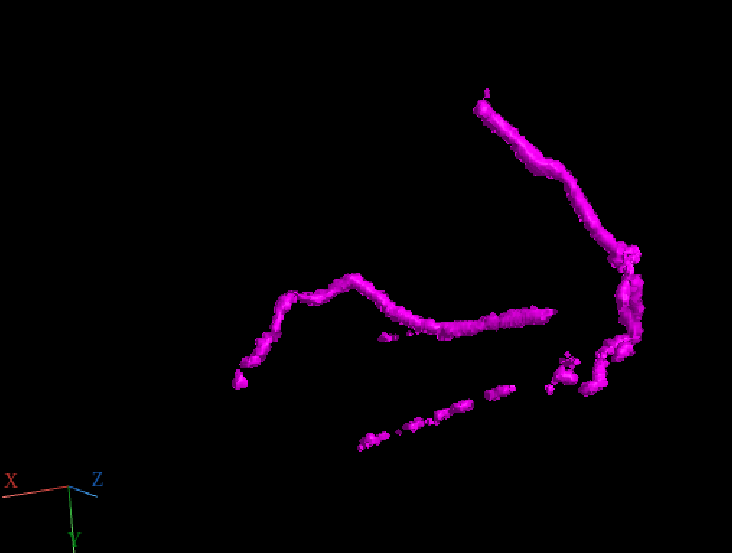}
     \end{subfigure}
     \hfill
     \begin{subfigure}[b]{0.12\textwidth}
         \centering
         \includegraphics[width=\textwidth]{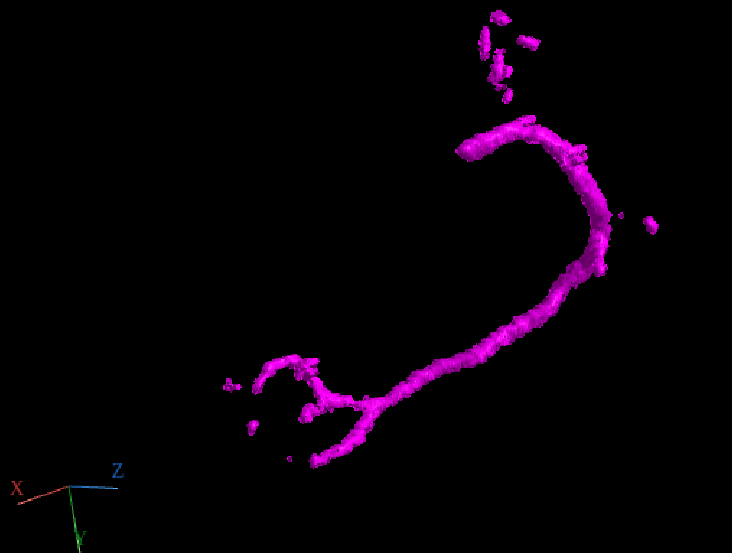}
     \end{subfigure}
     \hfill
     \begin{subfigure}[b]{0.12\textwidth}
         \centering
         \includegraphics[width=\textwidth]{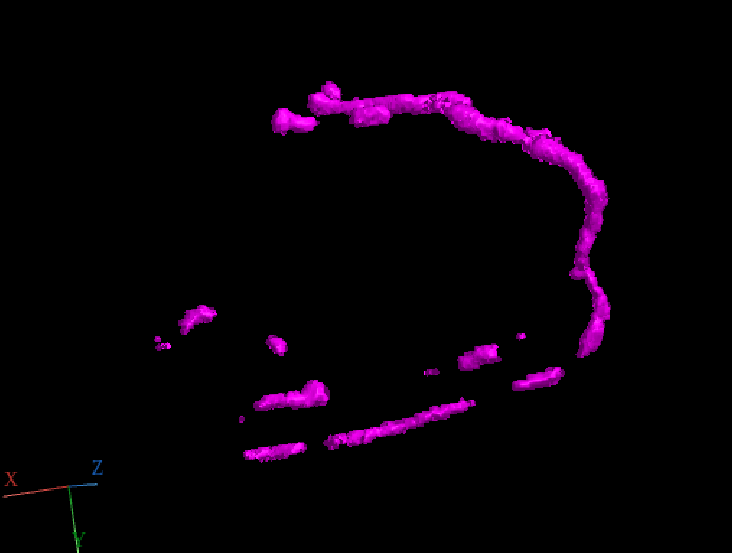}
     \end{subfigure}
     \hfill
     \begin{subfigure}[b]{0.12\textwidth}
         \centering
         \includegraphics[width=\textwidth]{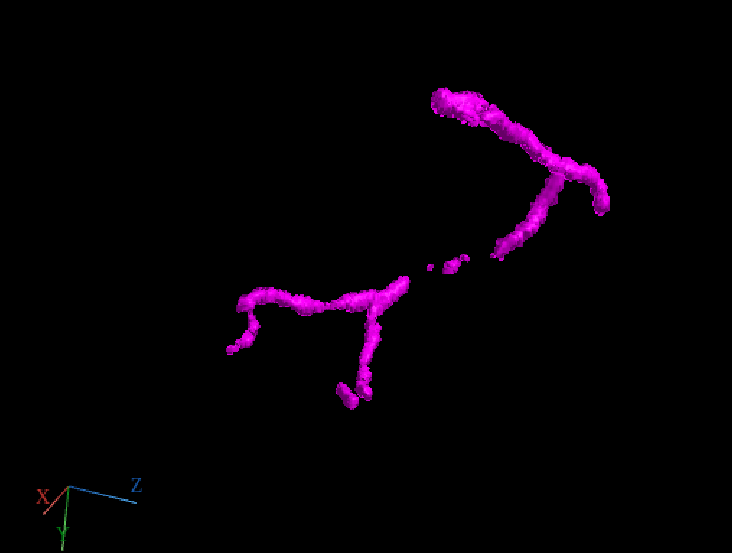}
     \end{subfigure}
     \caption{3D reconstruction results of 8 real clinical ICA data from all the models. From left to right: 8 patients. From top to bottom: 3D reconstruction results by our proposed DeepCA model, WGP, +CTLs, +DSCC, Un2+, Un3+, DSCN, and CVTG.}\label{3d_ica}
\end{figure*}

We can see from \cref{ccta_recon,ccta_cd} that our proposed DeepCA model has successfully reconstructed all the branches and maintained the vessel connectivity, while for the baseline models, there are many missing and/or broken branches visible. Although there is no corresponding 3D ground truth for the real clinical ICA data, we observe the same results in \cref{3d_ica}. In particular, some 3D reconstruction results on clinical ICA data from the baseline models miss the vascular features almost entirely, such as the reconstructions by model Un2+ and Un3+ on patients 2, 3, and 8.

\par
\begin{figure*}[!h]
     \centering
     %%%%%% first projection mine
     \vfill
     \begin{subfigure}[b]{0.11\textwidth}
         \centering
         \includegraphics[width=\textwidth]{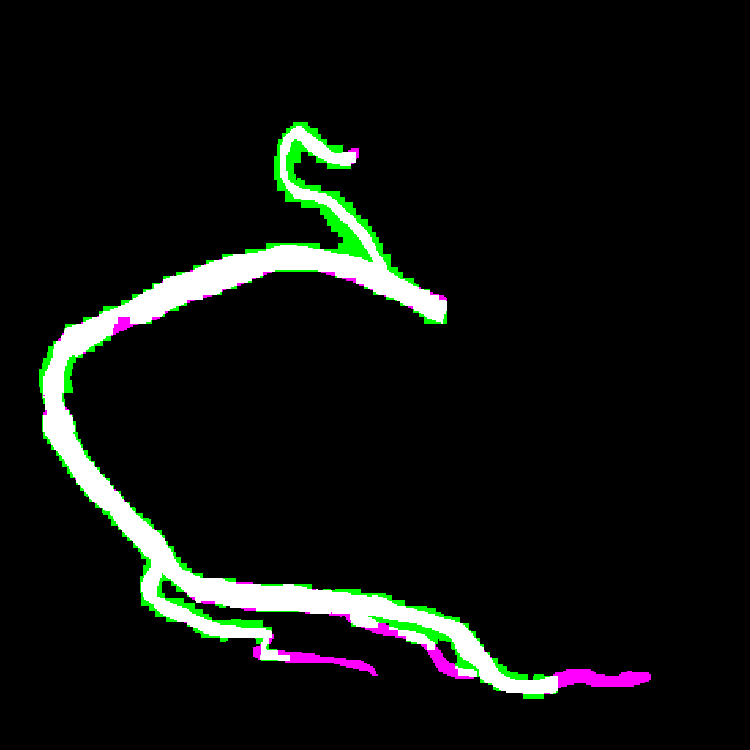}
     \end{subfigure}
     \hfill
	\begin{subfigure}[b]{0.11\textwidth}
         \centering
         \includegraphics[width=\textwidth]{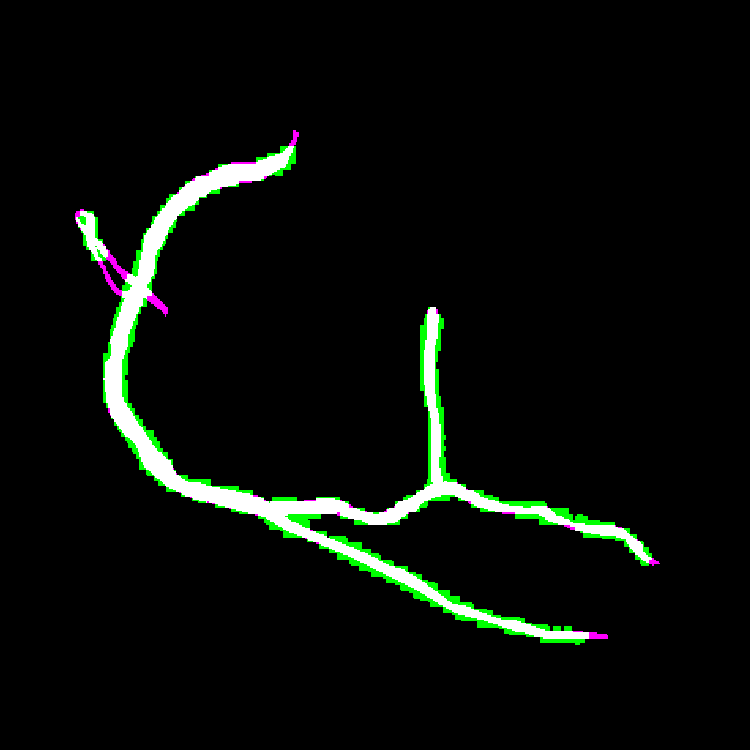}
     \end{subfigure}
     \hfill
	\begin{subfigure}[b]{0.11\textwidth}
         \centering
         \includegraphics[width=\textwidth]{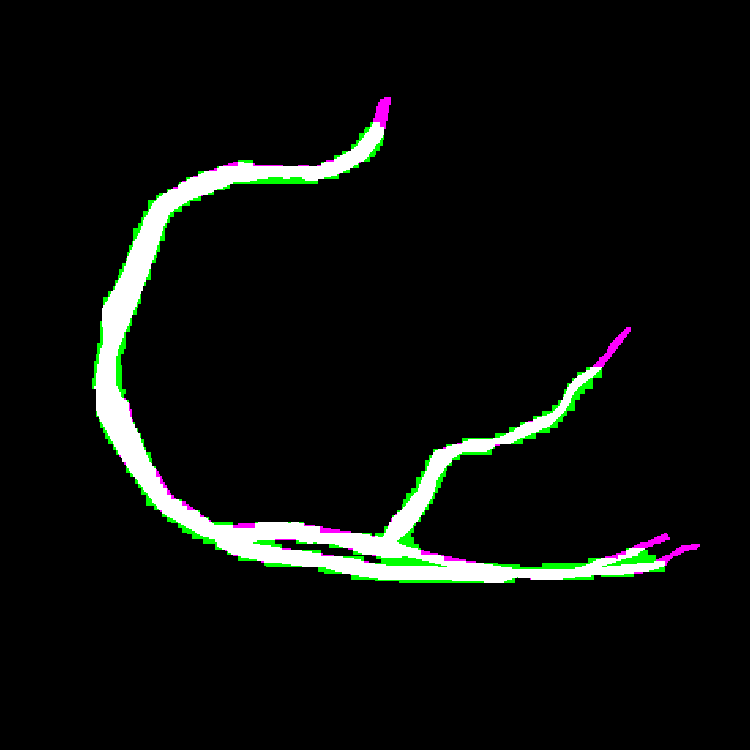}
     \end{subfigure}
     \hfill
	\begin{subfigure}[b]{0.11\textwidth}
         \centering
         \includegraphics[width=\textwidth]{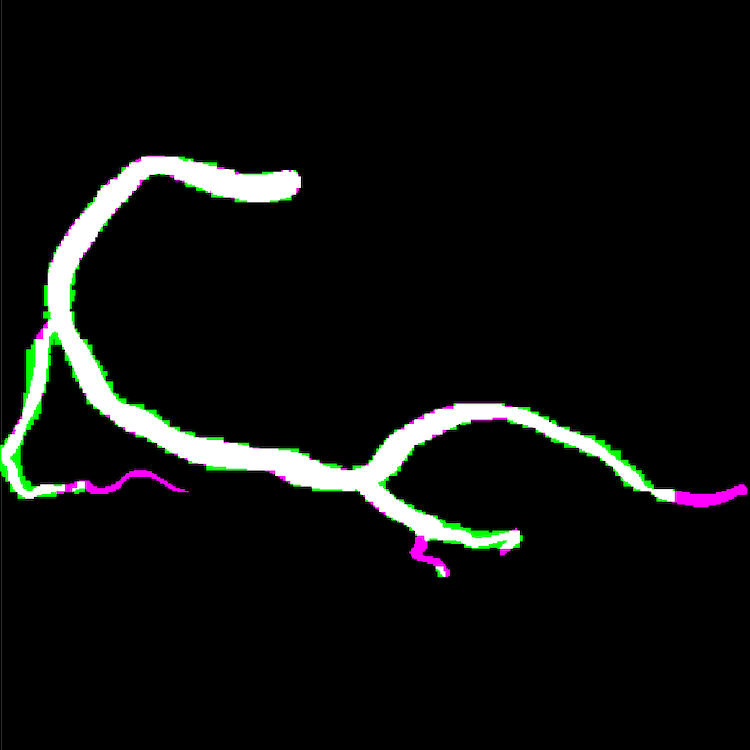}
     \end{subfigure}
     \hfill
	\begin{subfigure}[b]{0.11\textwidth}
         \centering
         \includegraphics[width=\textwidth]{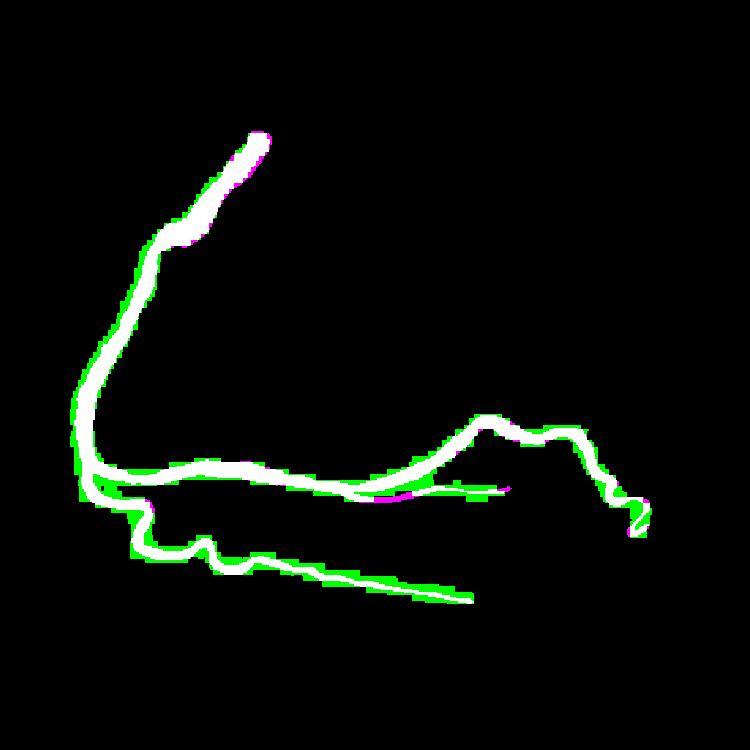}
     \end{subfigure}
     \hfill
     \begin{subfigure}[b]{0.11\textwidth}
         \centering
         \includegraphics[width=\textwidth]{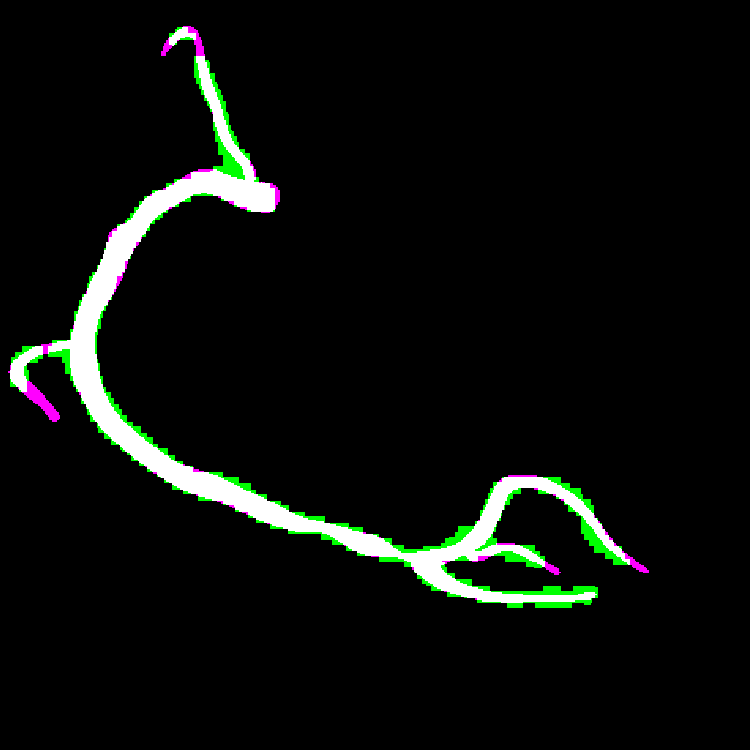}
     \end{subfigure}
     \hfill
     \begin{subfigure}[b]{0.11\textwidth}
         \centering
         \includegraphics[width=\textwidth]{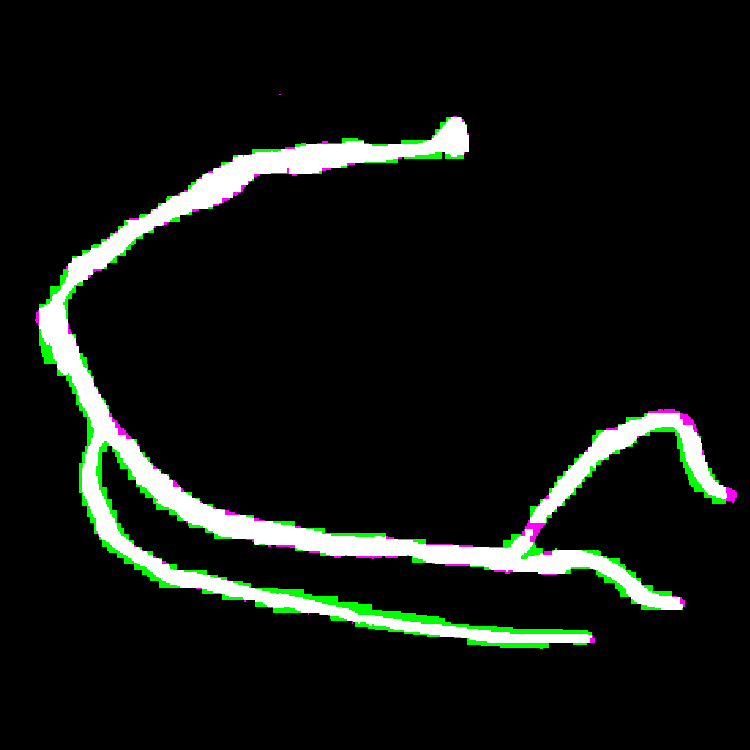}
     \end{subfigure}
     \hfill
     \begin{subfigure}[b]{0.11\textwidth}
         \centering
         \includegraphics[width=\textwidth]{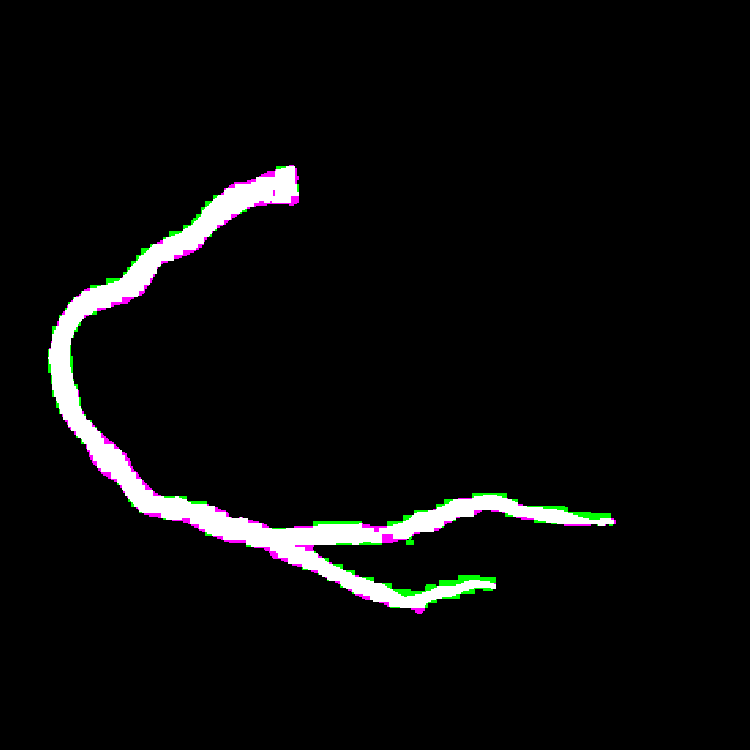}
     \end{subfigure}
     %%%%%% first projection wgp
     \vfill
     \begin{subfigure}[b]{0.11\textwidth}
         \centering
         \includegraphics[width=\textwidth]{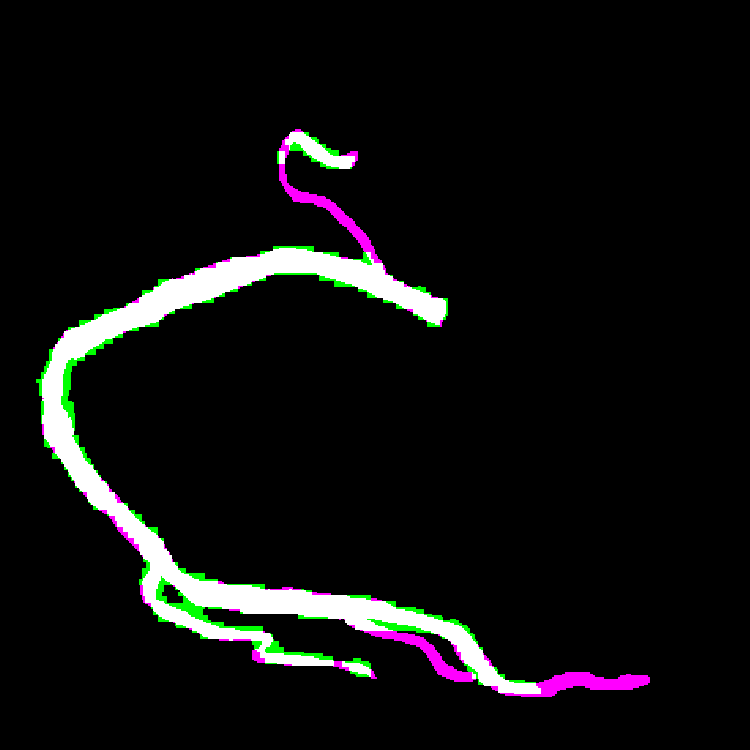}
     \end{subfigure}
     \hfill
	\begin{subfigure}[b]{0.11\textwidth}
         \centering
         \includegraphics[width=\textwidth]{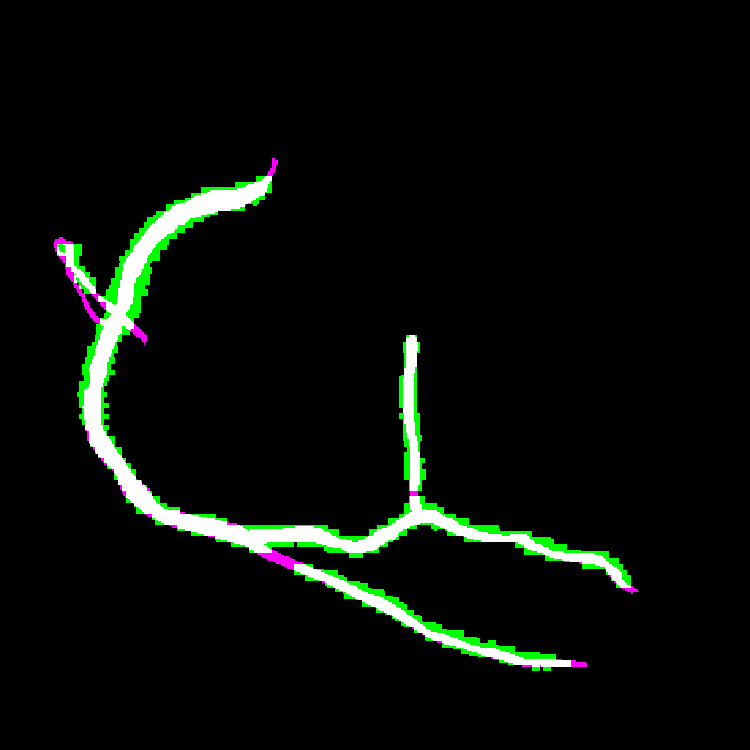}
     \end{subfigure}
     \hfill
	\begin{subfigure}[b]{0.11\textwidth}
         \centering
         \includegraphics[width=\textwidth]{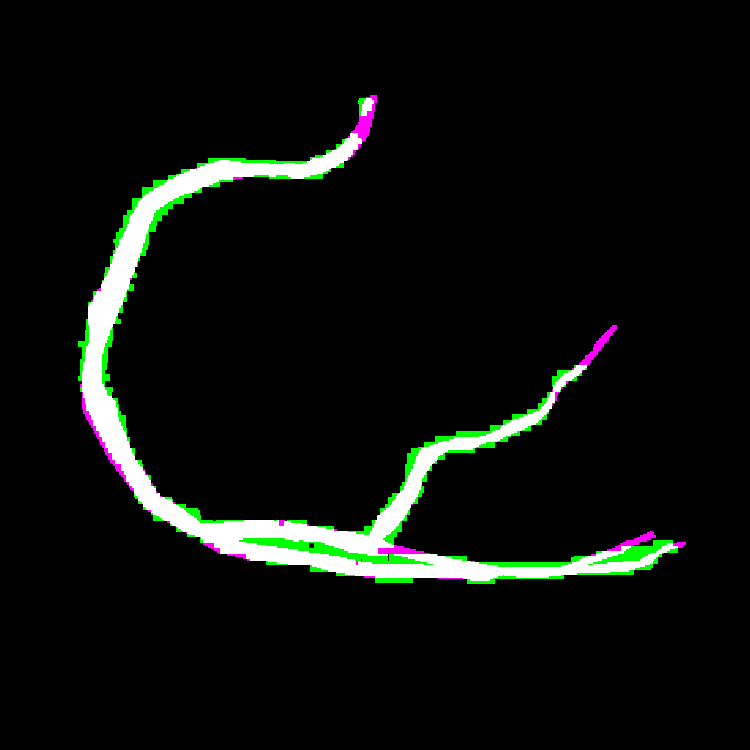}
     \end{subfigure}
     \hfill
	\begin{subfigure}[b]{0.11\textwidth}
         \centering
         \includegraphics[width=\textwidth]{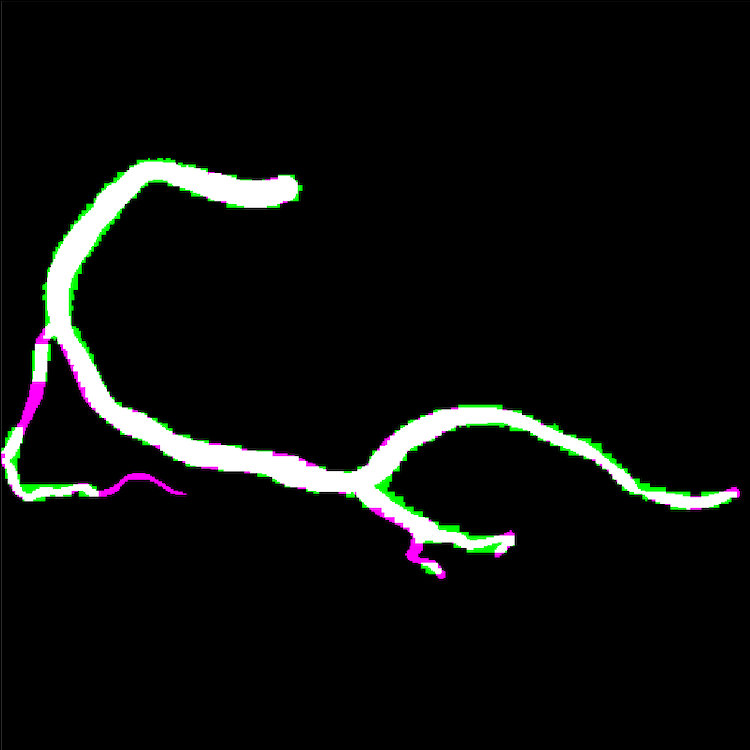}
     \end{subfigure}
     \hfill
	\begin{subfigure}[b]{0.11\textwidth}
         \centering
         \includegraphics[width=\textwidth]{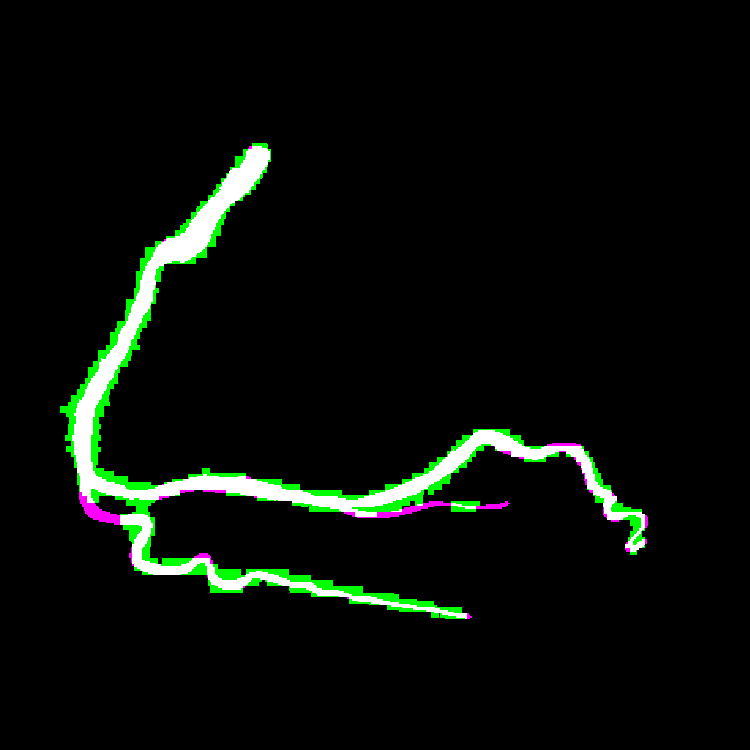}
     \end{subfigure}
     \hfill
     \begin{subfigure}[b]{0.11\textwidth}
         \centering
         \includegraphics[width=\textwidth]{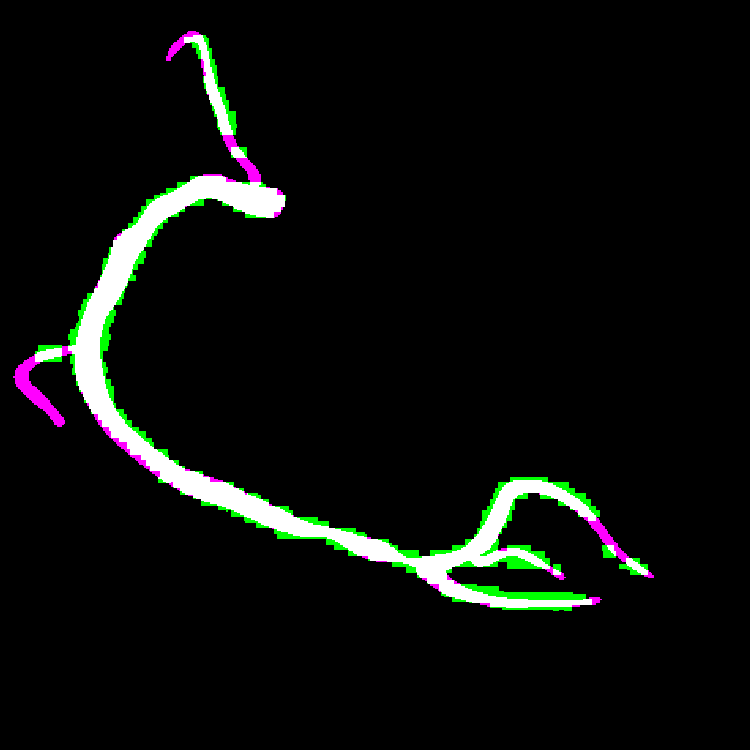}
     \end{subfigure}
     \hfill
     \begin{subfigure}[b]{0.11\textwidth}
         \centering
         \includegraphics[width=\textwidth]{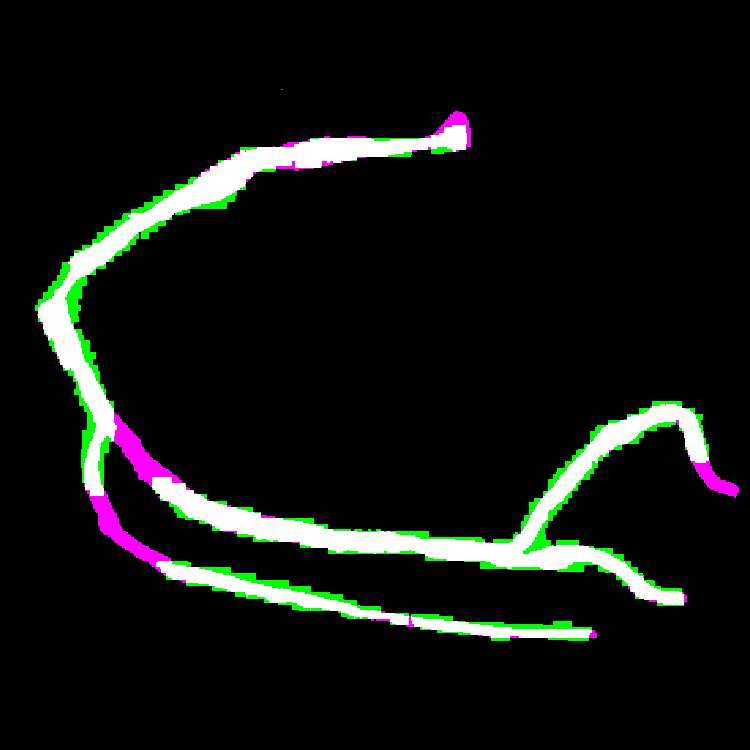}
     \end{subfigure}
     \hfill
     \begin{subfigure}[b]{0.11\textwidth}
         \centering
         \includegraphics[width=\textwidth]{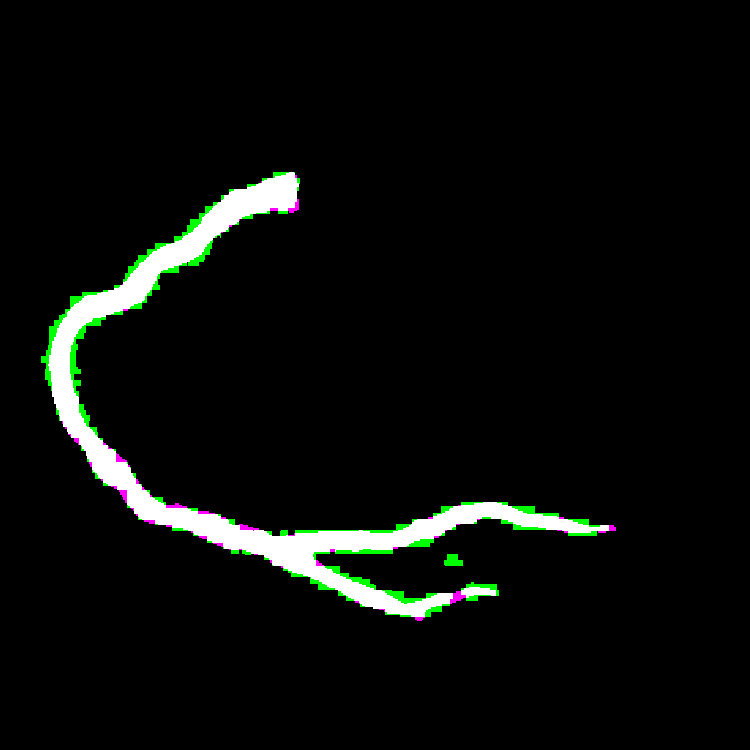}
     \end{subfigure}
     %%%%%% first projection ctls
     \vfill
     \begin{subfigure}[b]{0.11\textwidth}
         \centering
         \includegraphics[width=\textwidth]{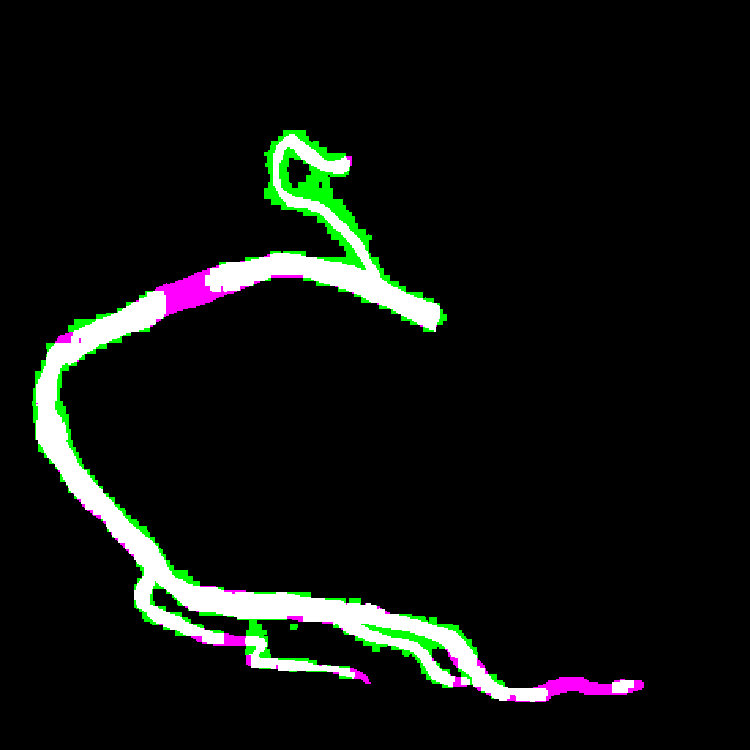}
     \end{subfigure}
     \hfill
	\begin{subfigure}[b]{0.11\textwidth}
         \centering
         \includegraphics[width=\textwidth]{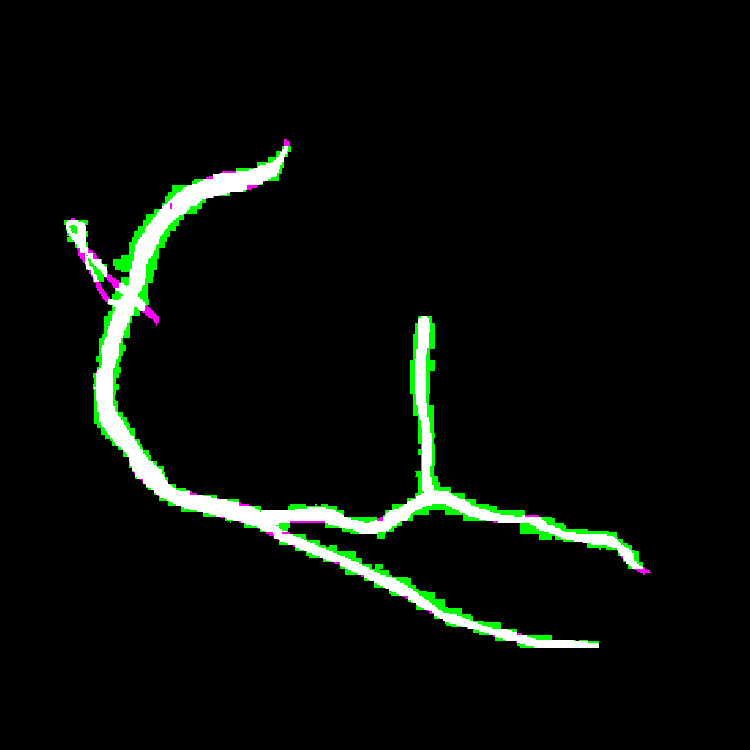}
     \end{subfigure}
     \hfill
	\begin{subfigure}[b]{0.11\textwidth}
         \centering
         \includegraphics[width=\textwidth]{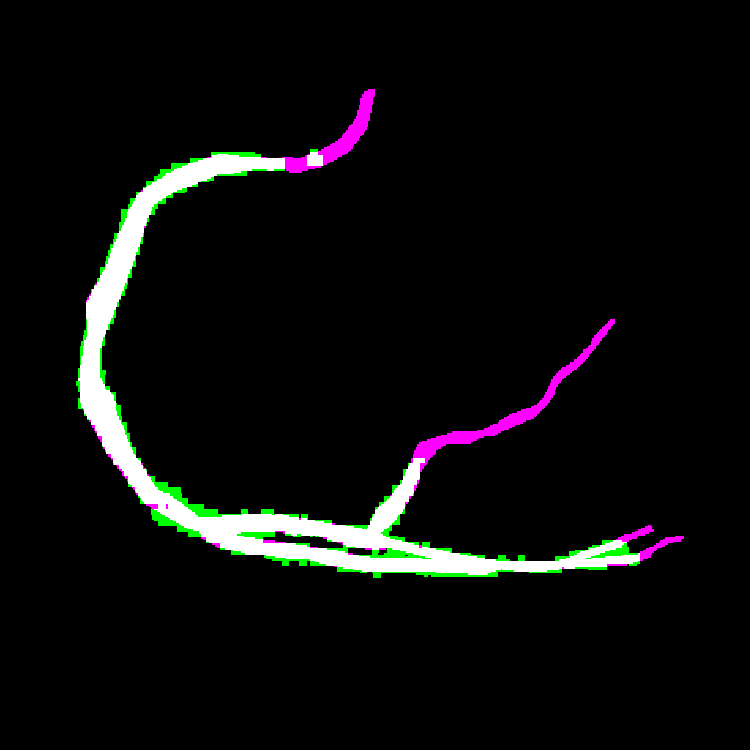}
     \end{subfigure}
     \hfill
	\begin{subfigure}[b]{0.11\textwidth}
         \centering
         \includegraphics[width=\textwidth]{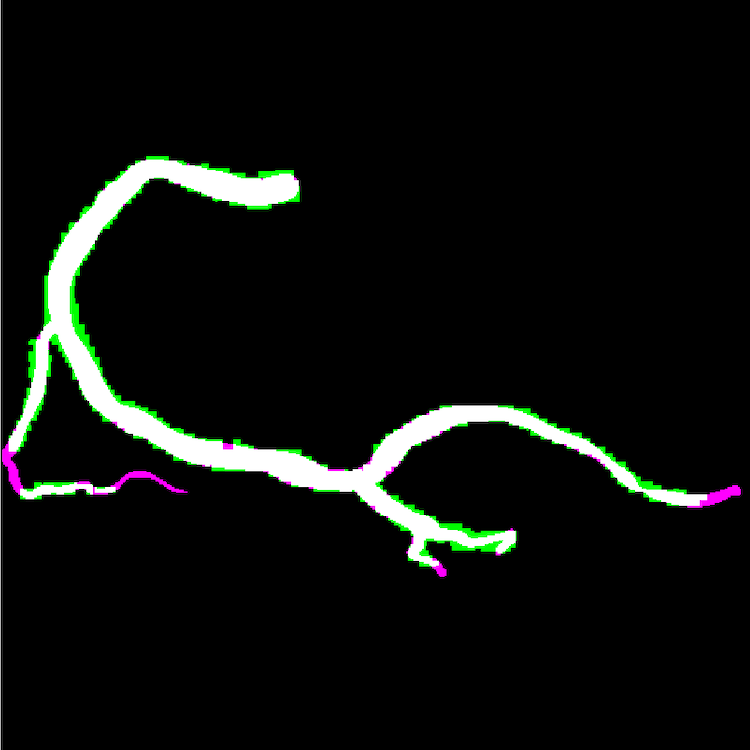}
     \end{subfigure}
     \hfill
	\begin{subfigure}[b]{0.11\textwidth}
         \centering
         \includegraphics[width=\textwidth]{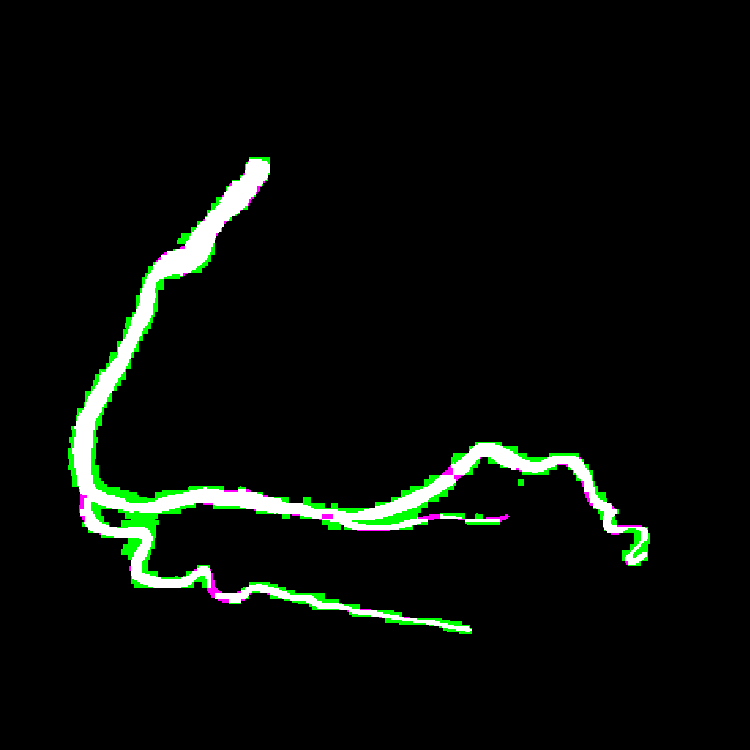}
     \end{subfigure}
     \hfill
     \begin{subfigure}[b]{0.11\textwidth}
         \centering
         \includegraphics[width=\textwidth]{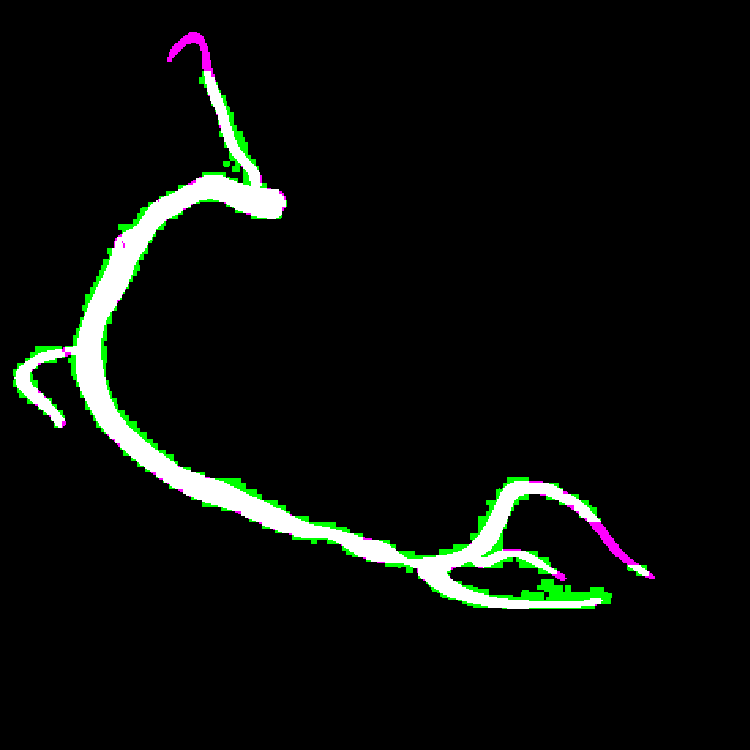}
     \end{subfigure}
     \hfill
     \begin{subfigure}[b]{0.11\textwidth}
         \centering
         \includegraphics[width=\textwidth]{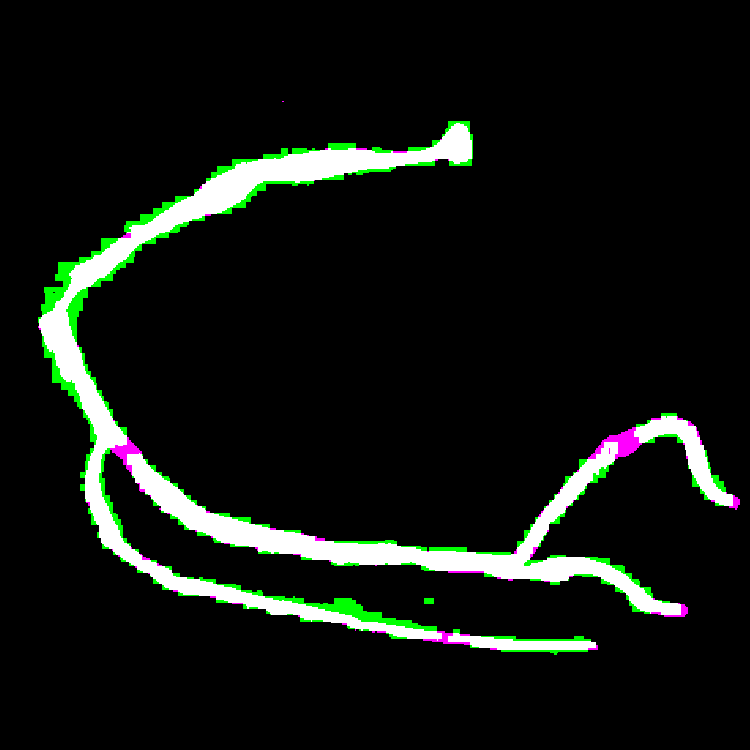}
     \end{subfigure}
     \hfill
     \begin{subfigure}[b]{0.11\textwidth}
         \centering
         \includegraphics[width=\textwidth]{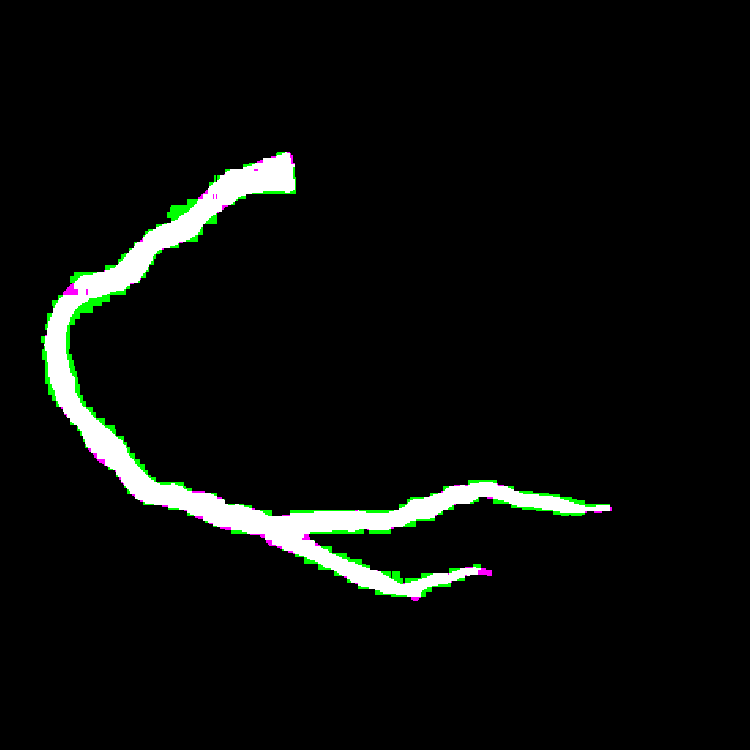}
     \end{subfigure}
     %%%%%% first projection dscc
     \vfill
     \begin{subfigure}[b]{0.11\textwidth}
         \centering
         \includegraphics[width=\textwidth]{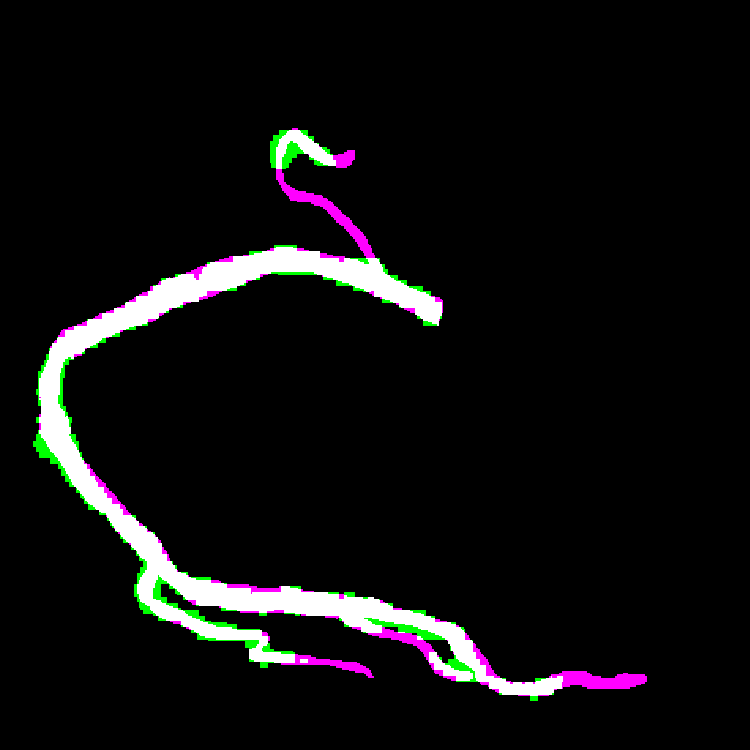}
     \end{subfigure}
     \hfill
	\begin{subfigure}[b]{0.11\textwidth}
         \centering
         \includegraphics[width=\textwidth]{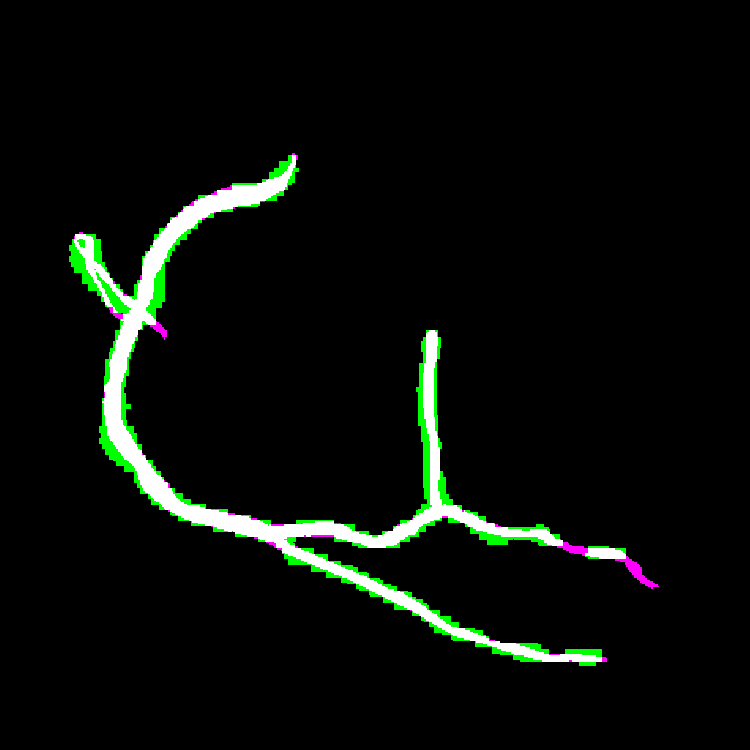}
     \end{subfigure}
     \hfill
	\begin{subfigure}[b]{0.11\textwidth}
         \centering
         \includegraphics[width=\textwidth]{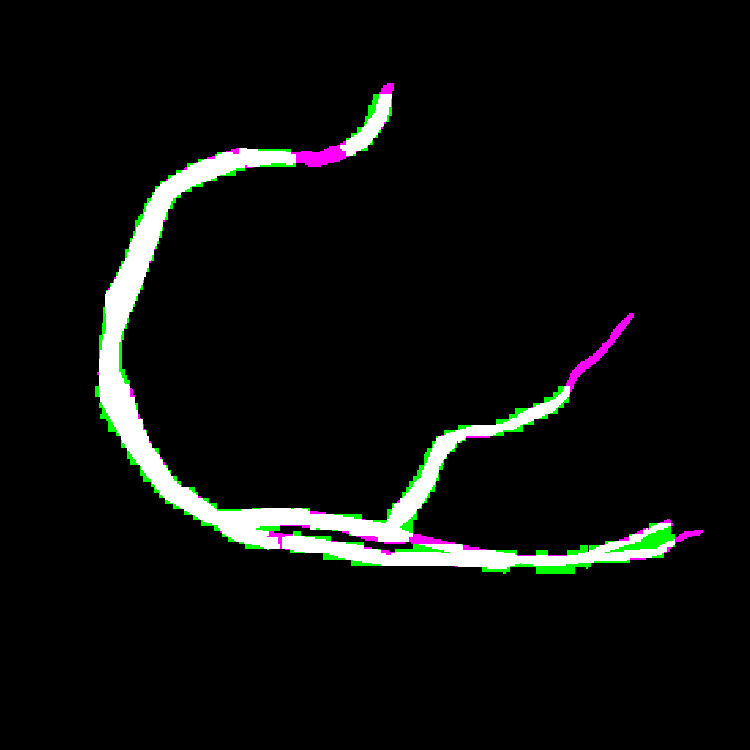}
     \end{subfigure}
     \hfill
	\begin{subfigure}[b]{0.11\textwidth}
         \centering
         \includegraphics[width=\textwidth]{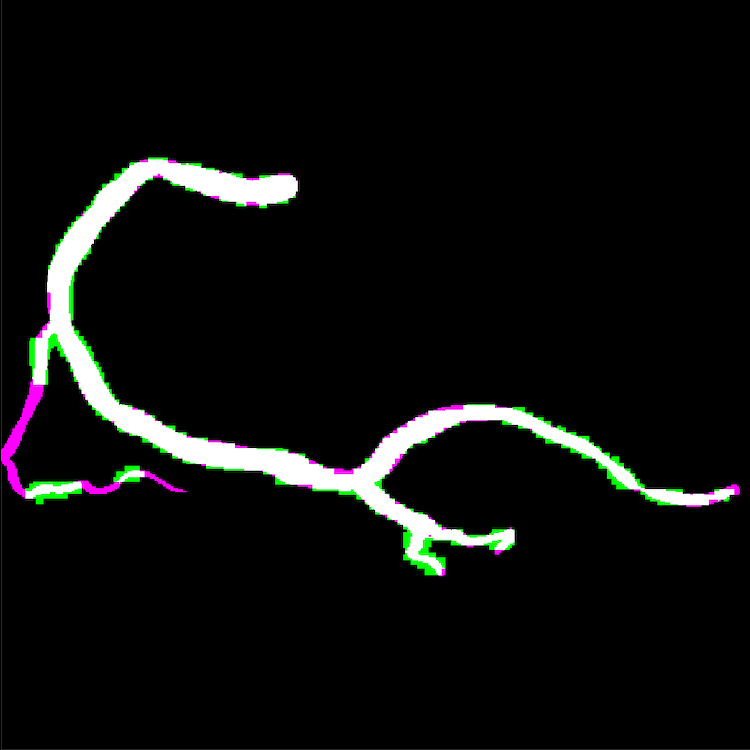}
     \end{subfigure}
     \hfill
	\begin{subfigure}[b]{0.11\textwidth}
         \centering
         \includegraphics[width=\textwidth]{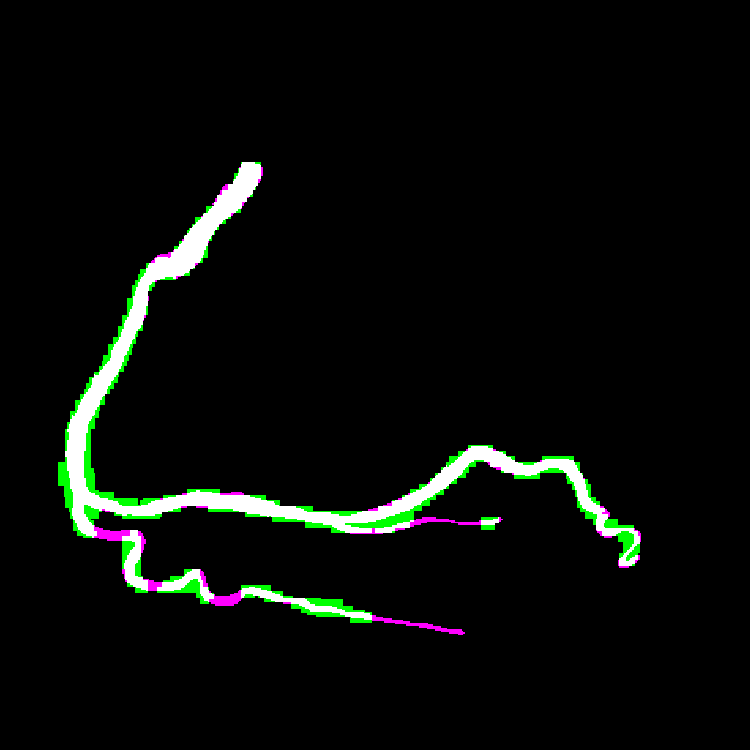}
     \end{subfigure}
     \hfill
     \begin{subfigure}[b]{0.11\textwidth}
         \centering
         \includegraphics[width=\textwidth]{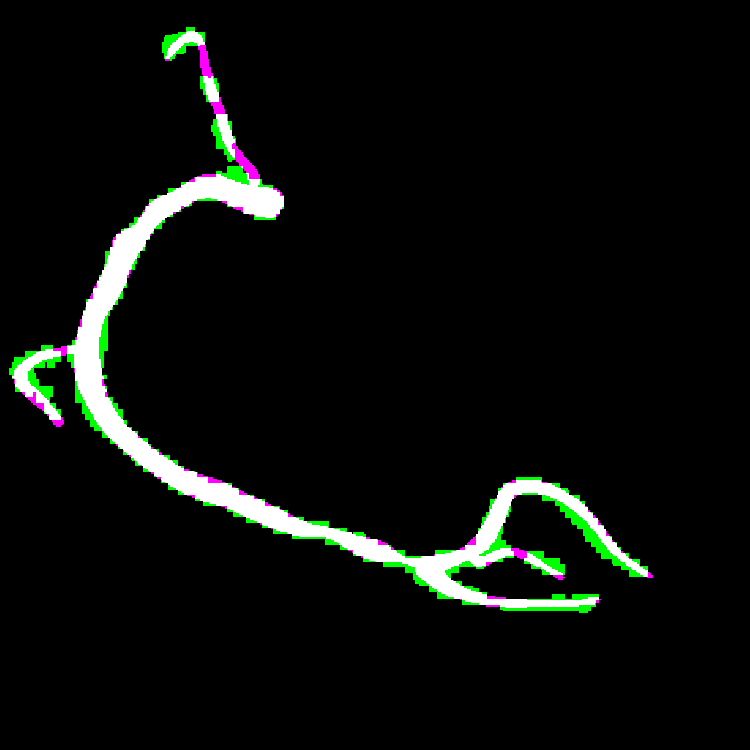}
     \end{subfigure}
     \hfill
     \begin{subfigure}[b]{0.11\textwidth}
         \centering
         \includegraphics[width=\textwidth]{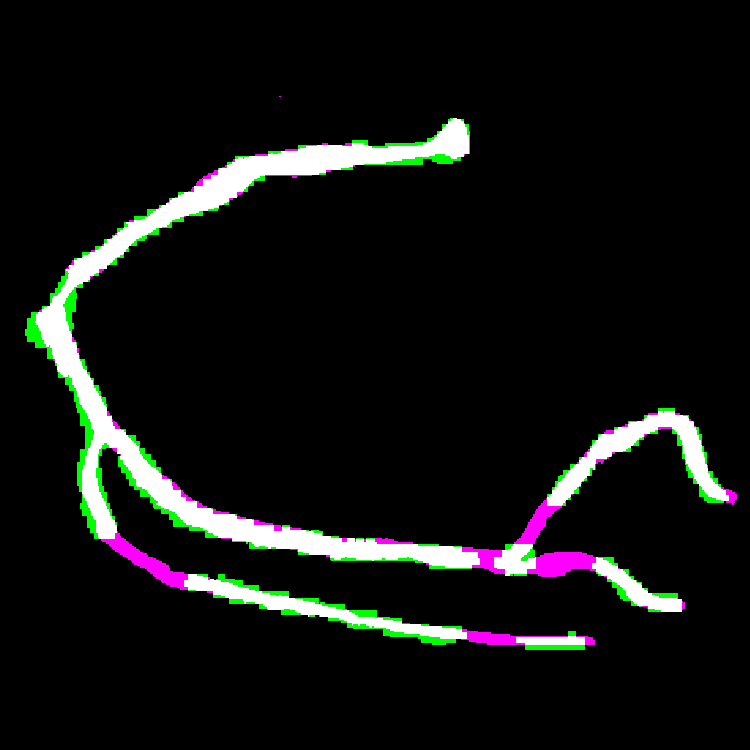}
     \end{subfigure}
     \hfill
     \begin{subfigure}[b]{0.11\textwidth}
         \centering
         \includegraphics[width=\textwidth]{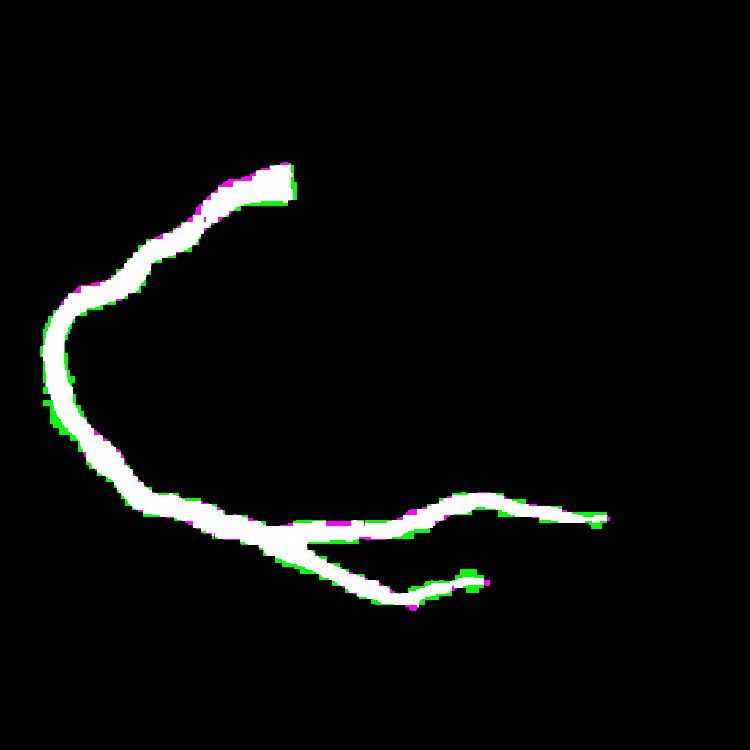}
     \end{subfigure}
     %%%%%% first projection un2+
     \vfill
     \begin{subfigure}[b]{0.11\textwidth}
         \centering
         \includegraphics[width=\textwidth]{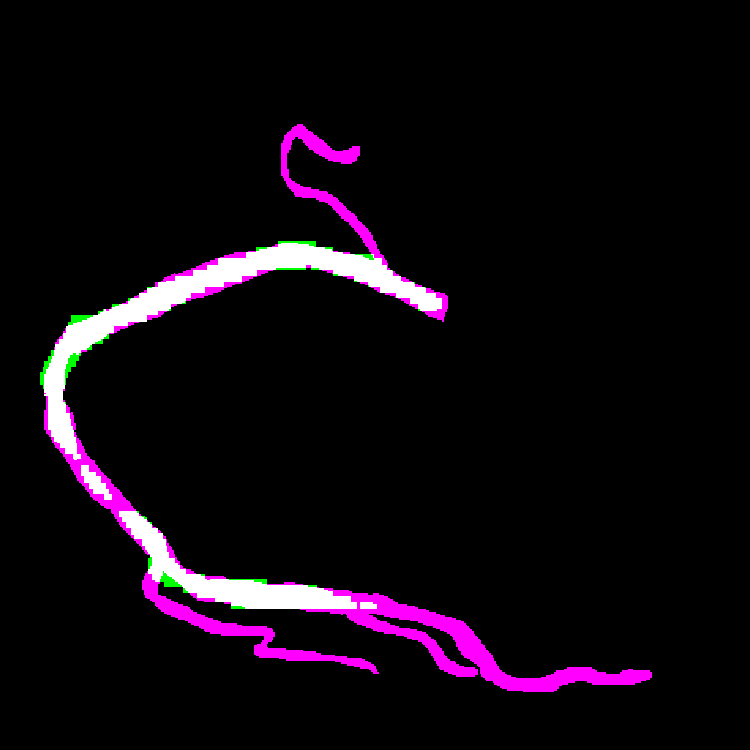}
     \end{subfigure}
     \hfill
	\begin{subfigure}[b]{0.11\textwidth}
         \centering
         \includegraphics[width=\textwidth]{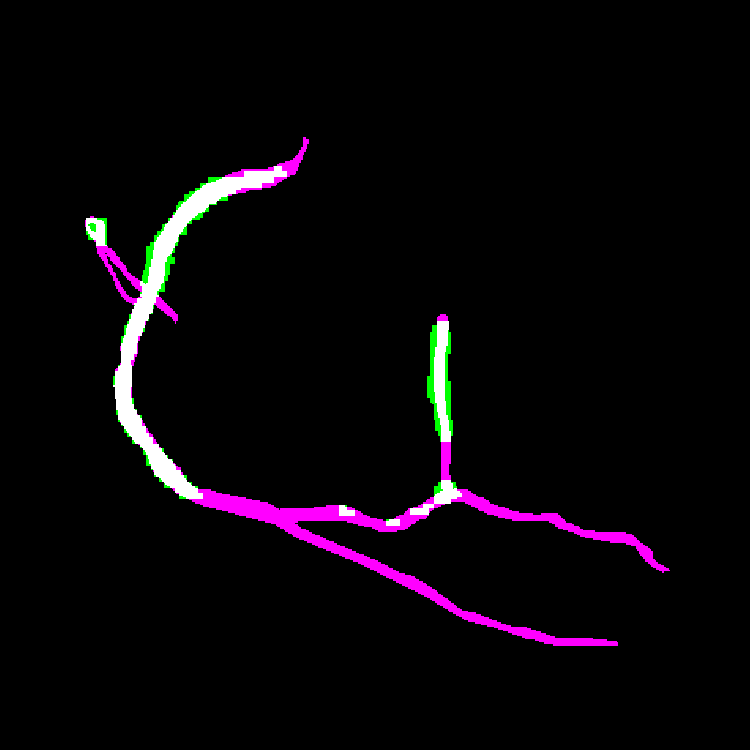}
     \end{subfigure}
     \hfill
	\begin{subfigure}[b]{0.11\textwidth}
         \centering
         \includegraphics[width=\textwidth]{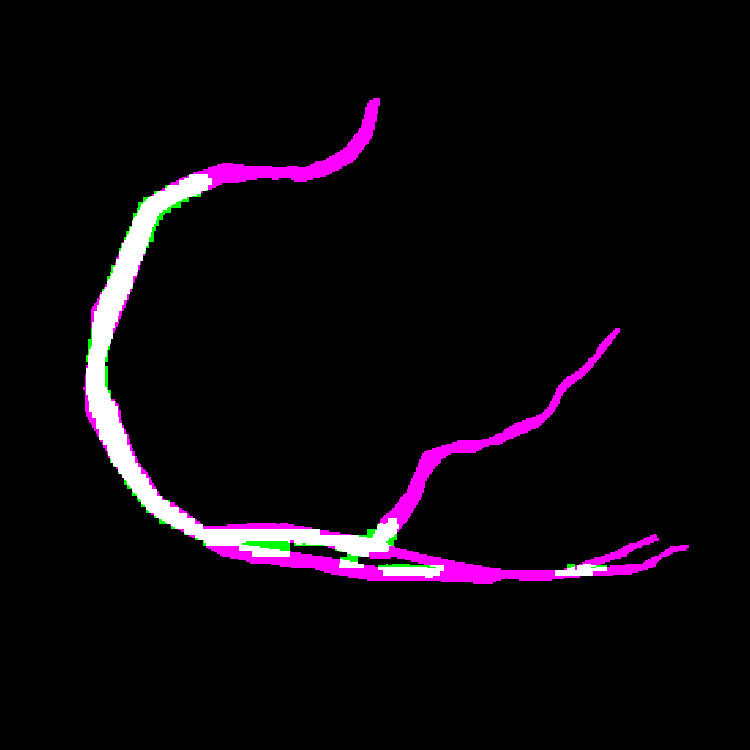}
     \end{subfigure}
     \hfill
	\begin{subfigure}[b]{0.11\textwidth}
         \centering
         \includegraphics[width=\textwidth]{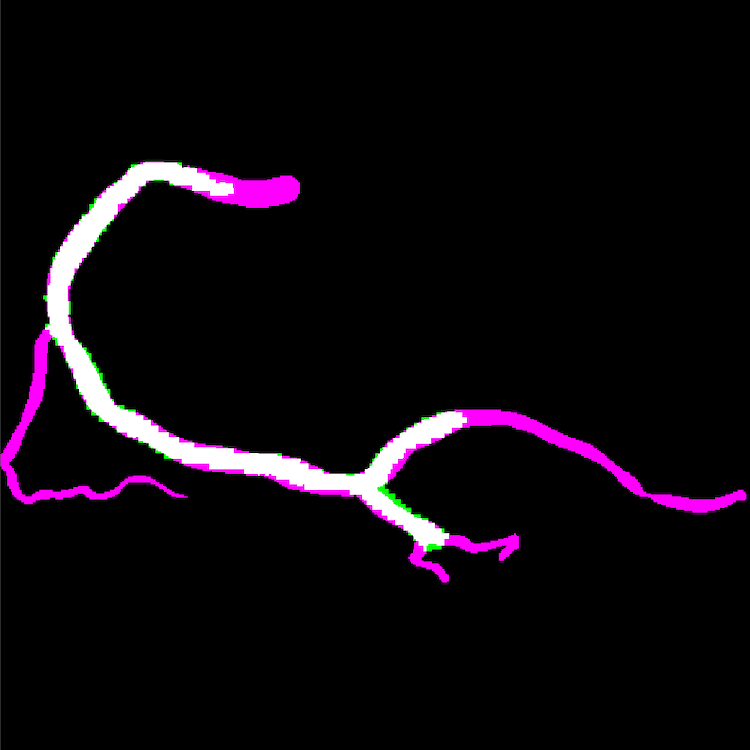}
     \end{subfigure}
     \hfill
	\begin{subfigure}[b]{0.11\textwidth}
         \centering
         \includegraphics[width=\textwidth]{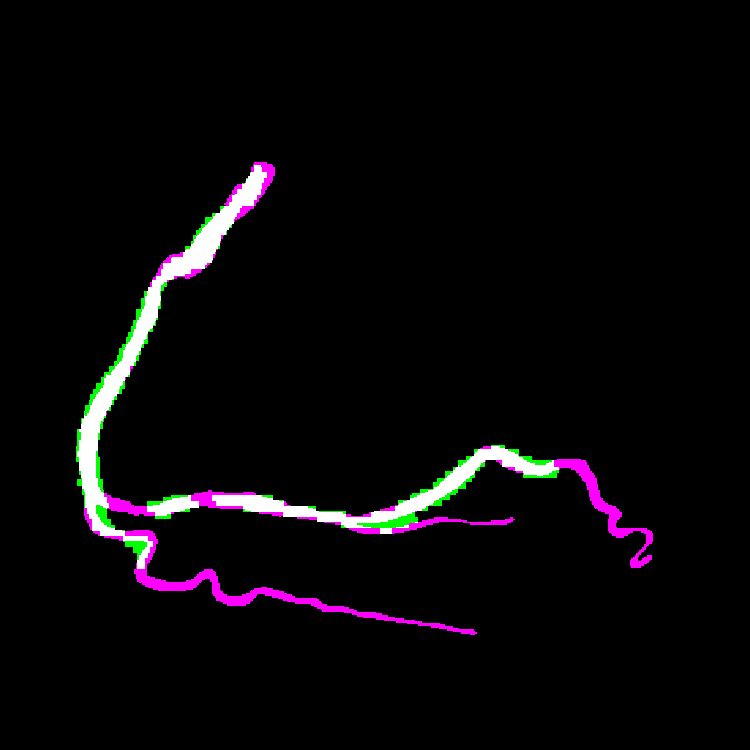}
     \end{subfigure}
     \hfill
     \begin{subfigure}[b]{0.11\textwidth}
         \centering
         \includegraphics[width=\textwidth]{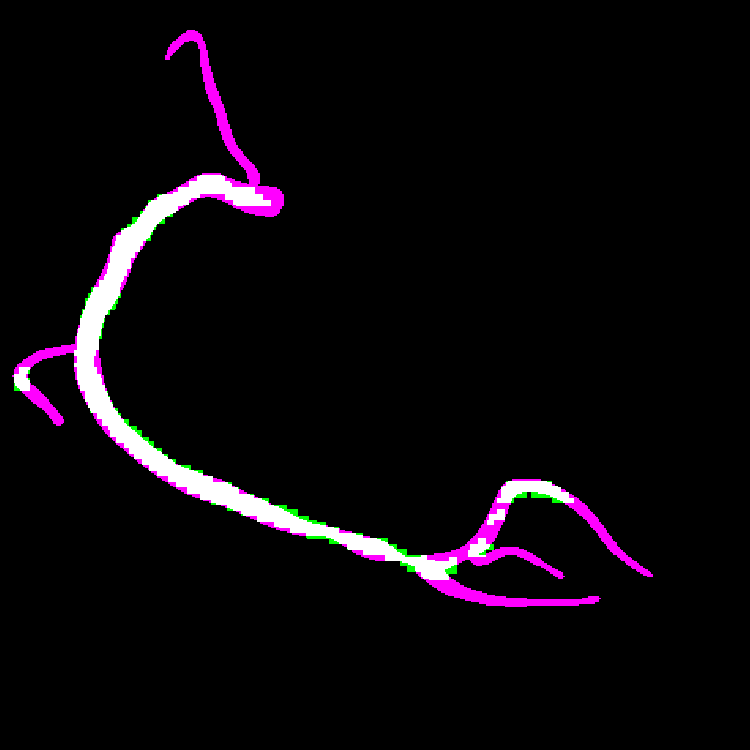}
     \end{subfigure}
     \hfill
     \begin{subfigure}[b]{0.11\textwidth}
         \centering
         \includegraphics[width=\textwidth]{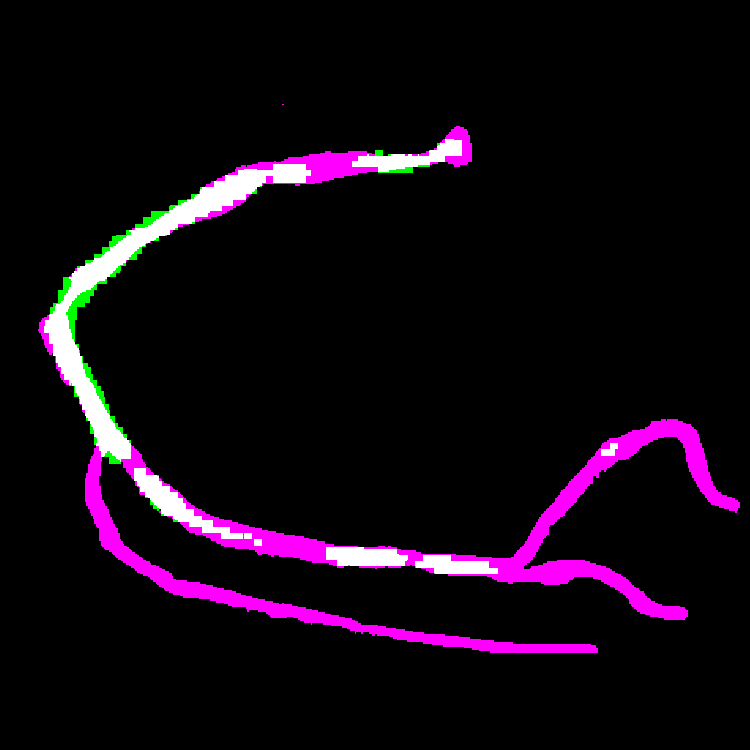}
     \end{subfigure}
     \hfill
     \begin{subfigure}[b]{0.11\textwidth}
         \centering
         \includegraphics[width=\textwidth]{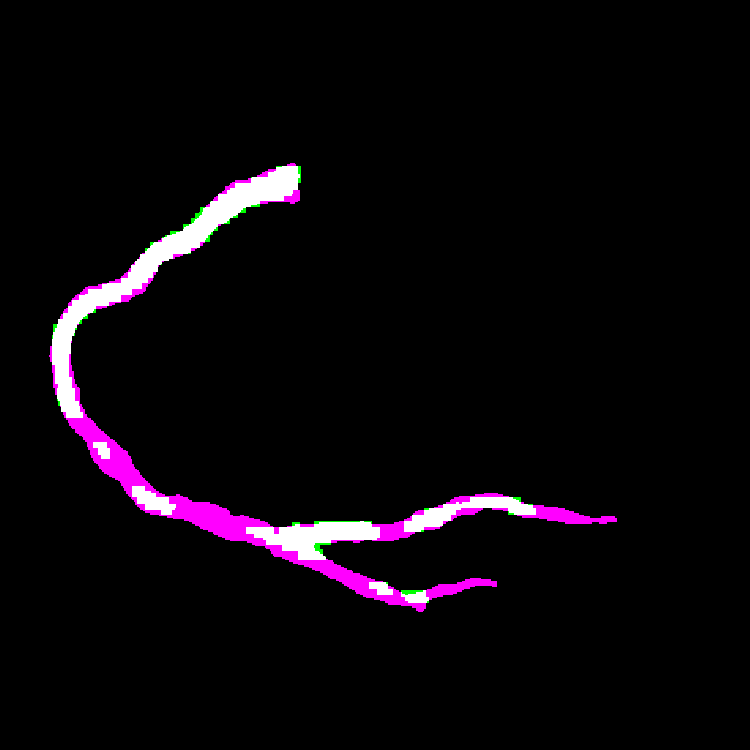}
     \end{subfigure}
	%%%%%% first projection un3+
     \vfill
     \begin{subfigure}[b]{0.11\textwidth}
         \centering
         \includegraphics[width=\textwidth]{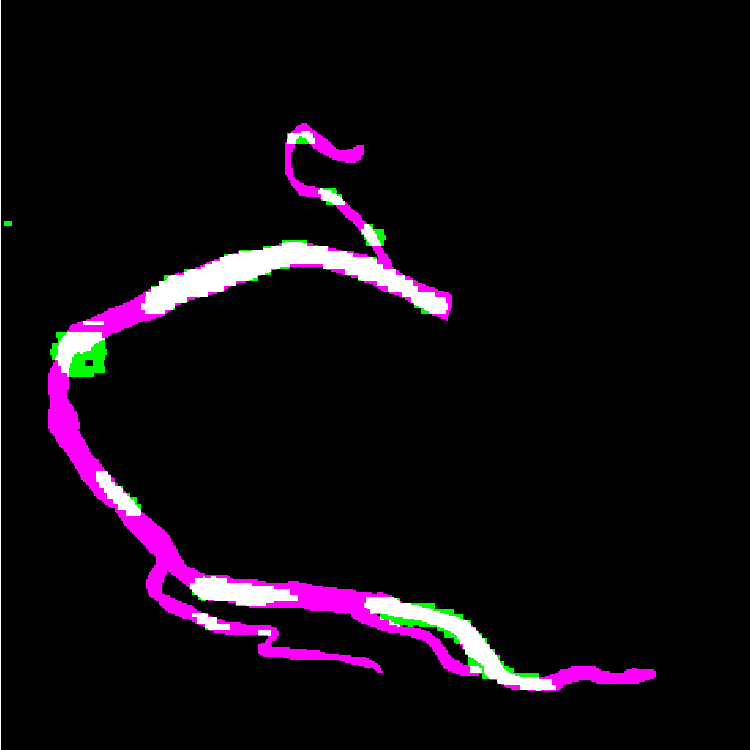}
     \end{subfigure}
     \hfill
	\begin{subfigure}[b]{0.11\textwidth}
         \centering
         \includegraphics[width=\textwidth]{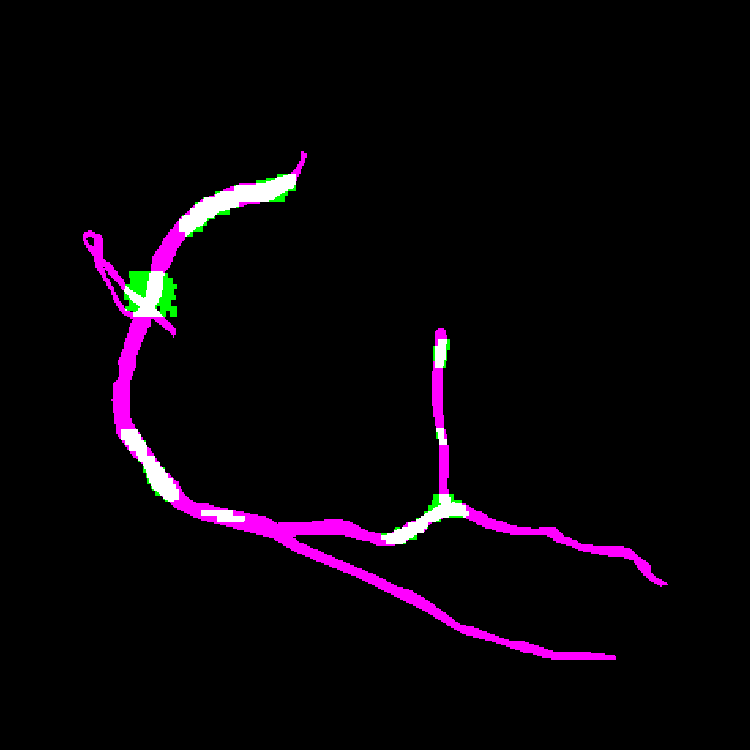}
     \end{subfigure}
     \hfill
	\begin{subfigure}[b]{0.11\textwidth}
         \centering
         \includegraphics[width=\textwidth]{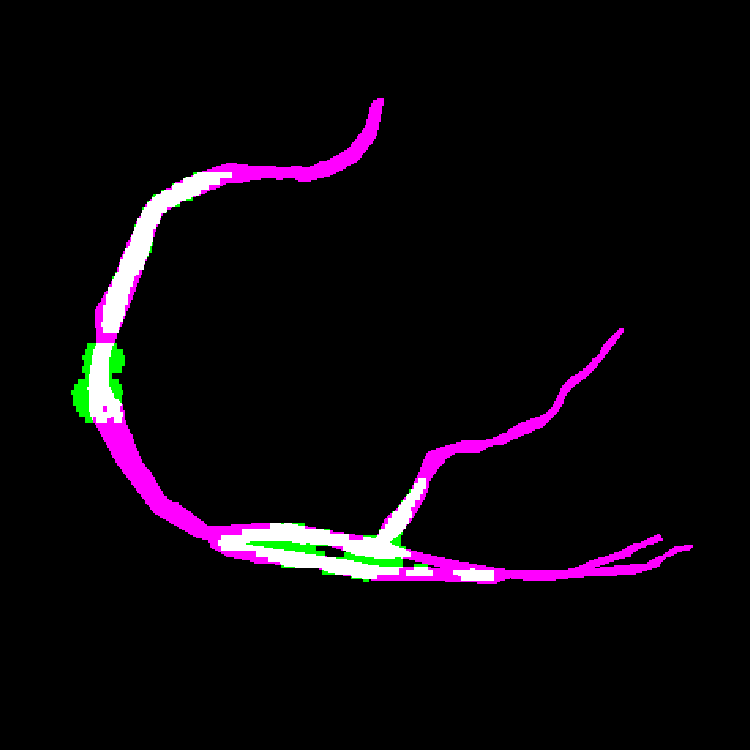}
     \end{subfigure}
     \hfill
	\begin{subfigure}[b]{0.11\textwidth}
         \centering
         \includegraphics[width=\textwidth]{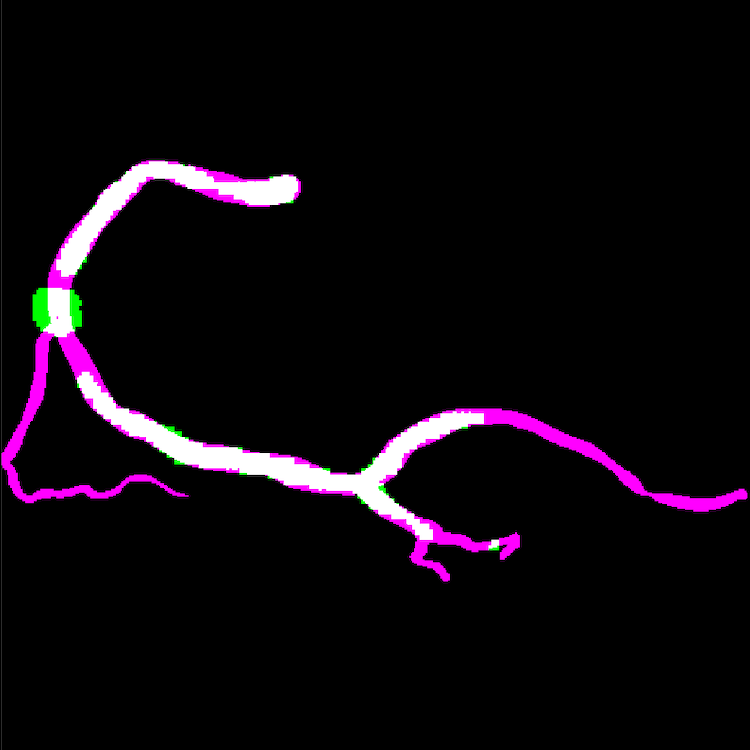}
     \end{subfigure}
     \hfill
	\begin{subfigure}[b]{0.11\textwidth}
         \centering
         \includegraphics[width=\textwidth]{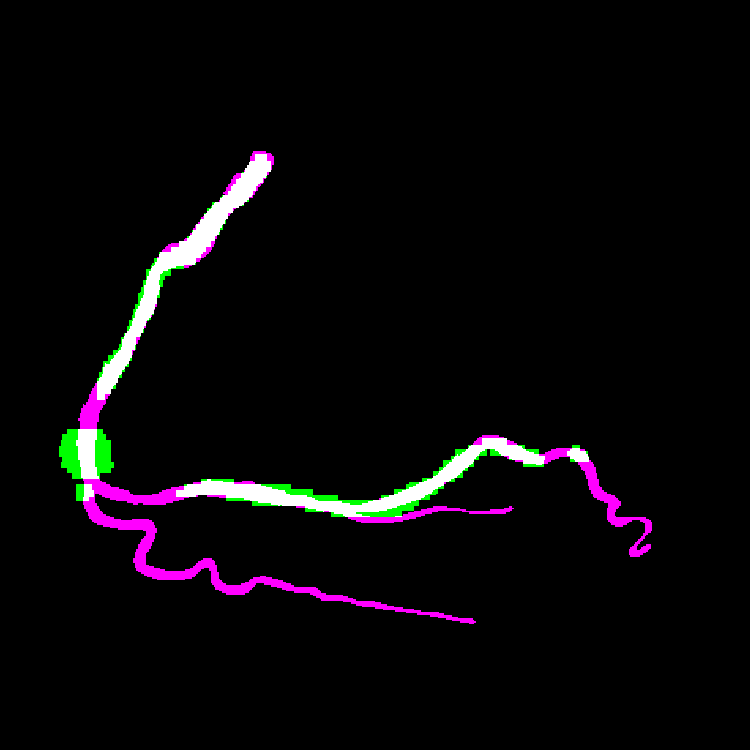}
     \end{subfigure}
     \hfill
     \begin{subfigure}[b]{0.11\textwidth}
         \centering
         \includegraphics[width=\textwidth]{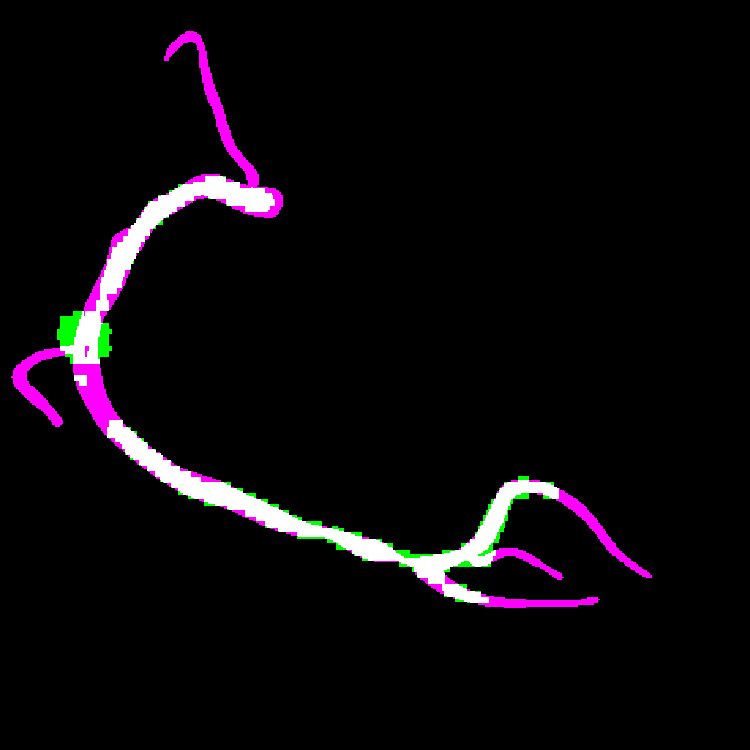}
     \end{subfigure}
     \hfill
     \begin{subfigure}[b]{0.11\textwidth}
         \centering
         \includegraphics[width=\textwidth]{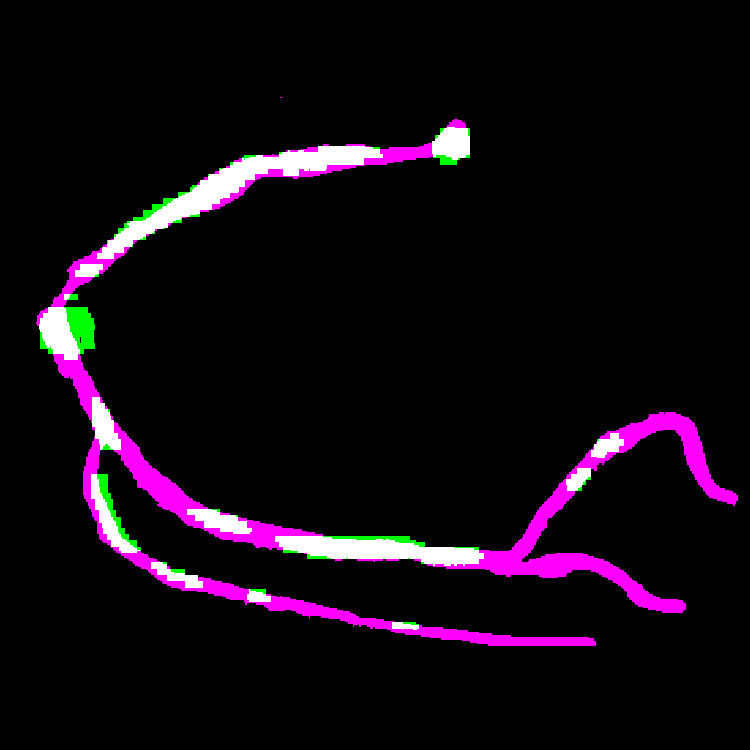}
     \end{subfigure}
     \hfill
     \begin{subfigure}[b]{0.11\textwidth}
         \centering
         \includegraphics[width=\textwidth]{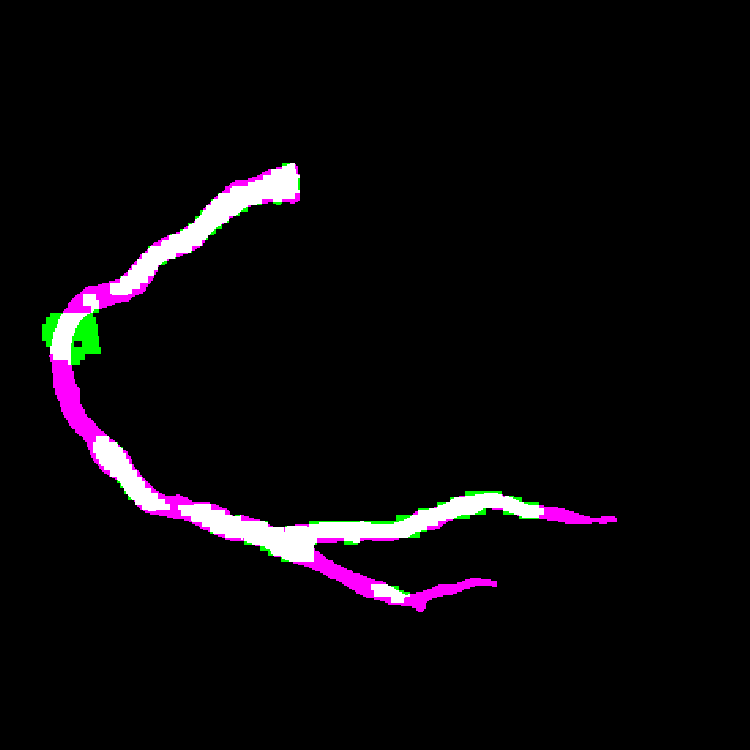}
     \end{subfigure}
     %%%%%% first projection dscn
     \vfill
     \begin{subfigure}[b]{0.11\textwidth}
         \centering
         \includegraphics[width=\textwidth]{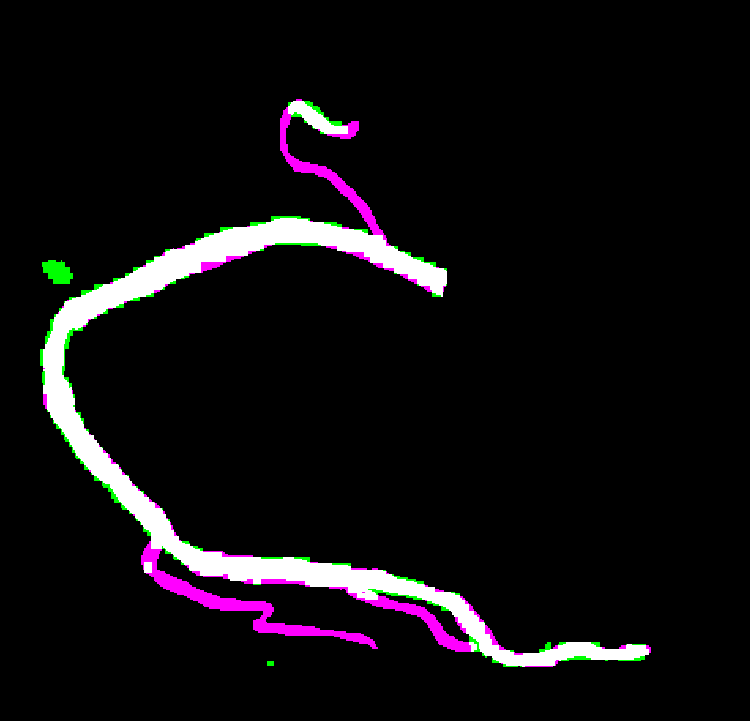}
     \end{subfigure}
     \hfill
	\begin{subfigure}[b]{0.11\textwidth}
         \centering
         \includegraphics[width=\textwidth]{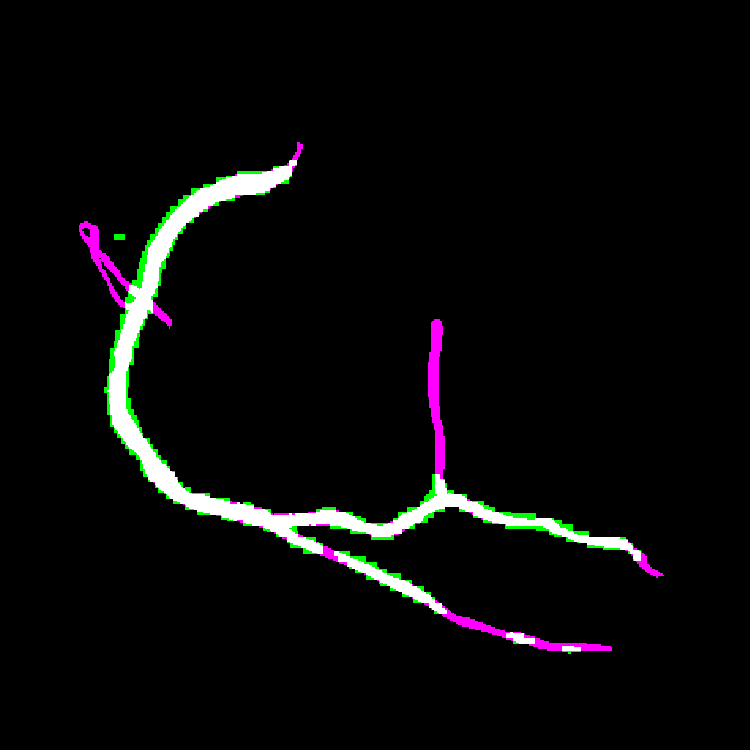}
     \end{subfigure}
     \hfill
	\begin{subfigure}[b]{0.11\textwidth}
         \centering
         \includegraphics[width=\textwidth]{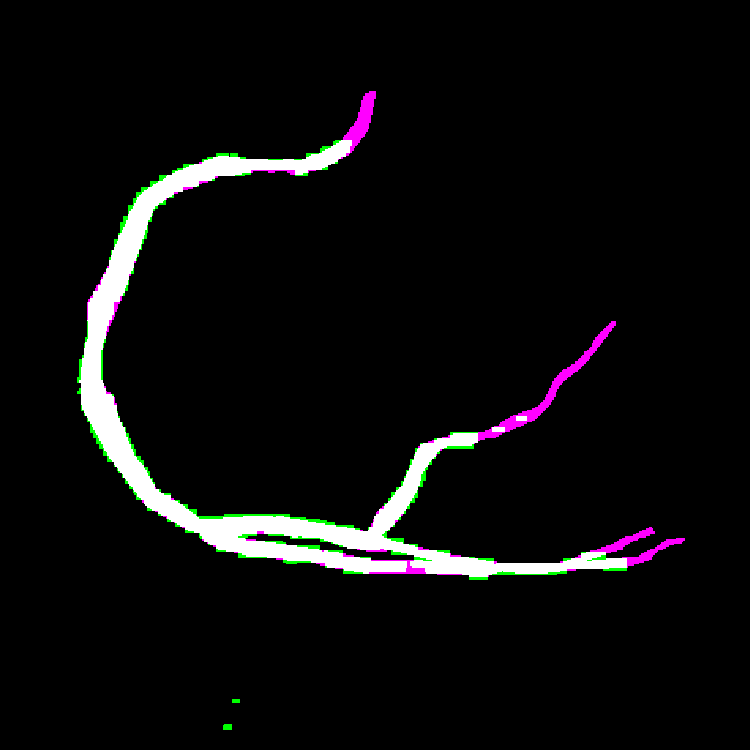}
     \end{subfigure}
     \hfill
	\begin{subfigure}[b]{0.11\textwidth}
         \centering
         \includegraphics[width=\textwidth]{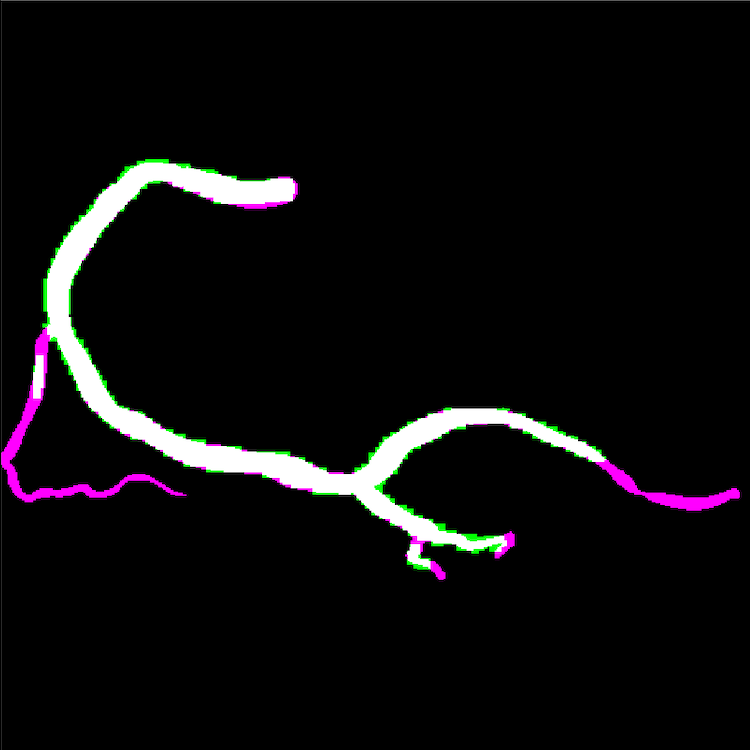}
     \end{subfigure}
     \hfill
	\begin{subfigure}[b]{0.11\textwidth}
         \centering
         \includegraphics[width=\textwidth]{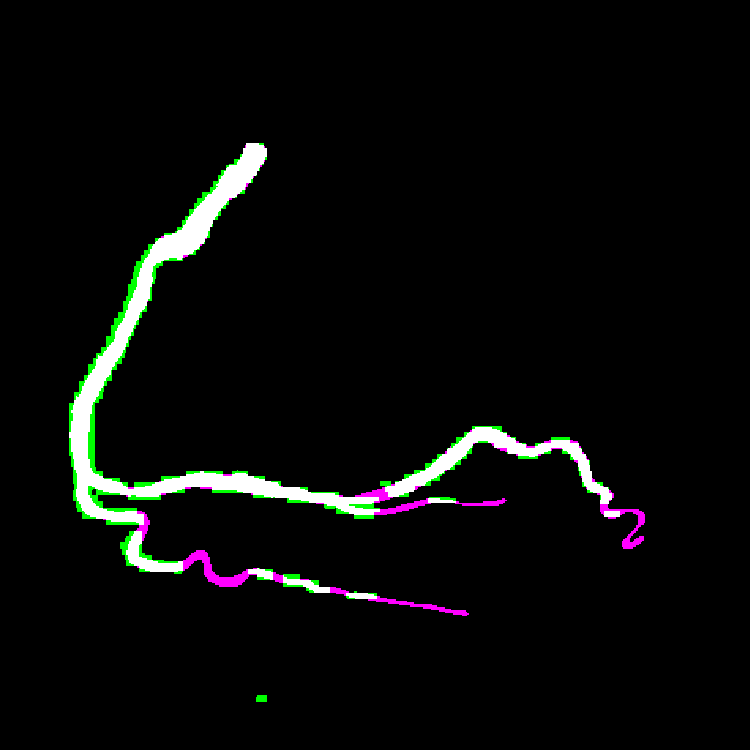}
     \end{subfigure}
     \hfill
     \begin{subfigure}[b]{0.11\textwidth}
         \centering
         \includegraphics[width=\textwidth]{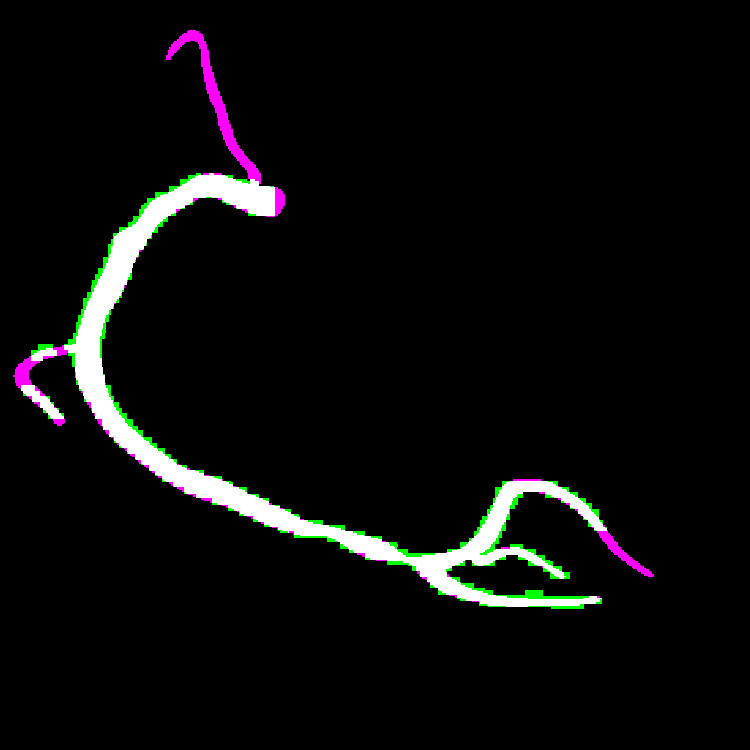}
     \end{subfigure}
     \hfill
     \begin{subfigure}[b]{0.11\textwidth}
         \centering
         \includegraphics[width=\textwidth]{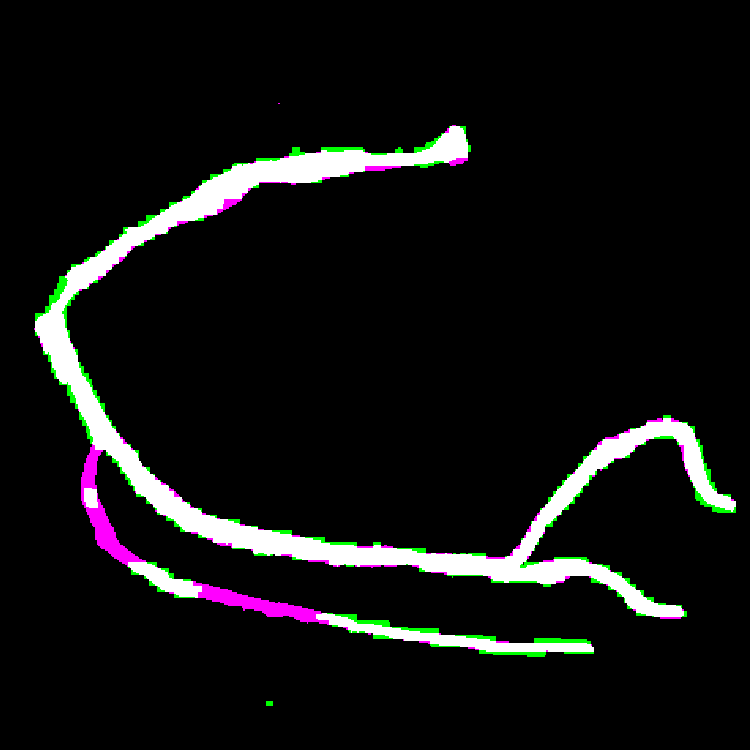}
     \end{subfigure}
     \hfill
     \begin{subfigure}[b]{0.11\textwidth}
         \centering
         \includegraphics[width=\textwidth]{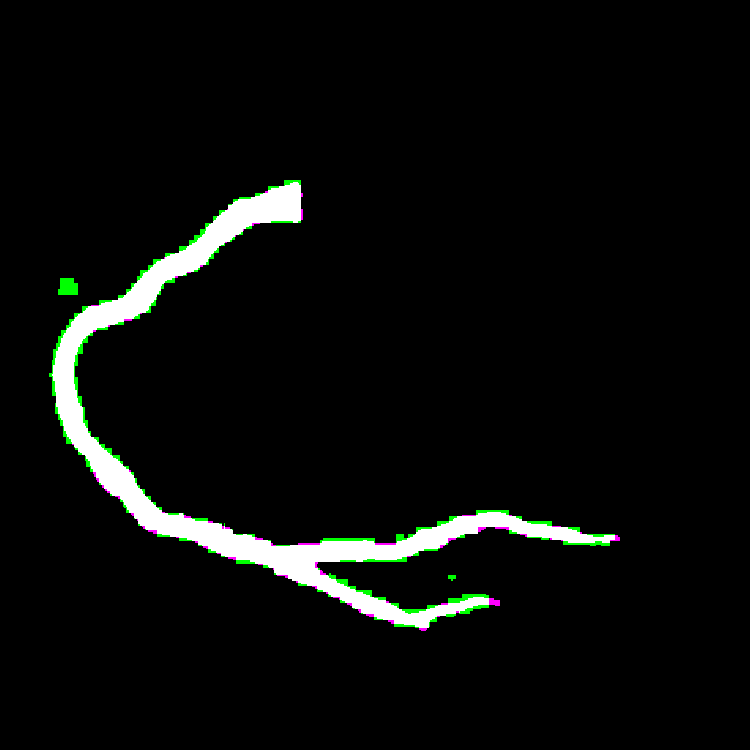}
     \end{subfigure}
     %%%%%% first projection cvtg
     \vfill
     \begin{subfigure}[b]{0.11\textwidth}
         \centering
         \includegraphics[width=\textwidth]{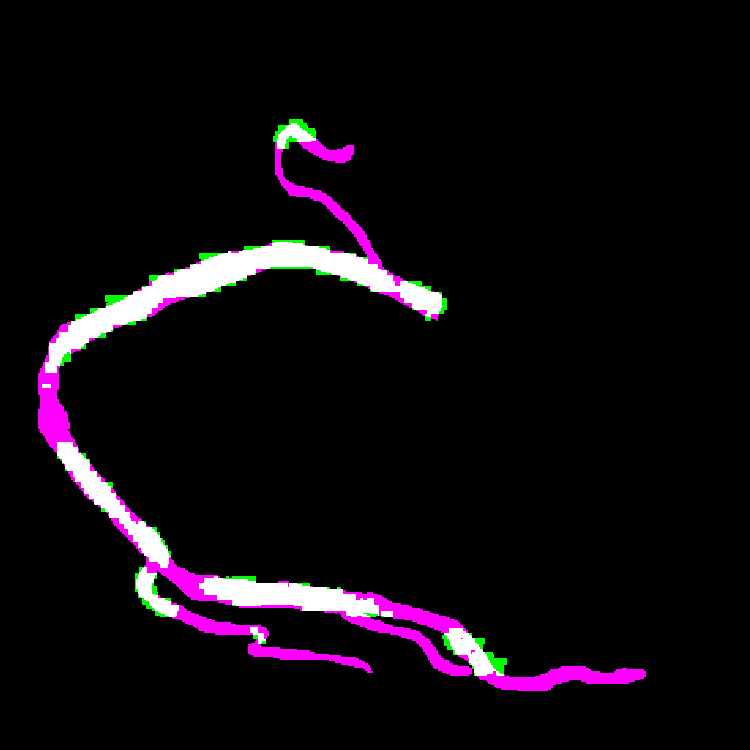}
     \end{subfigure}
     \hfill
	\begin{subfigure}[b]{0.11\textwidth}
         \centering
         \includegraphics[width=\textwidth]{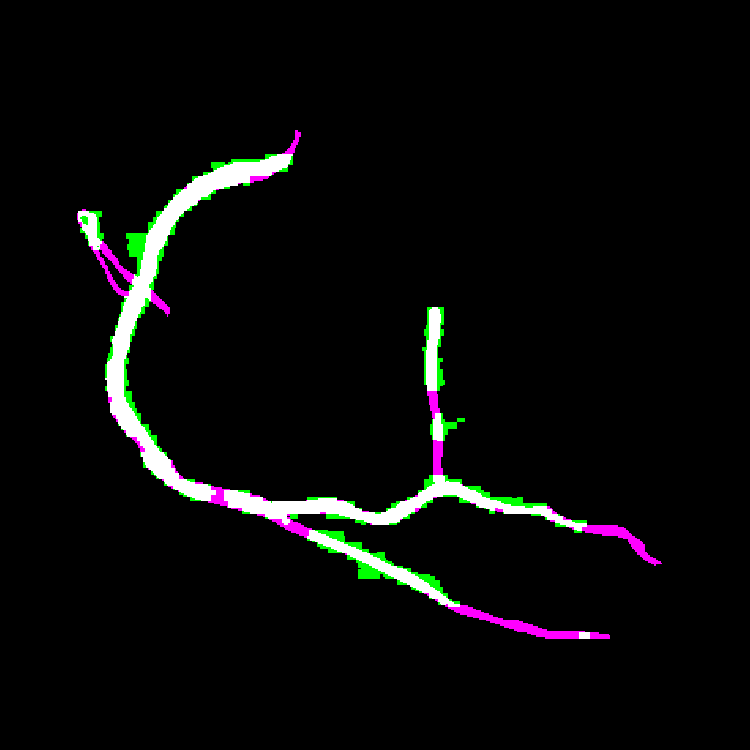}
     \end{subfigure}
     \hfill
	\begin{subfigure}[b]{0.11\textwidth}
         \centering
         \includegraphics[width=\textwidth]{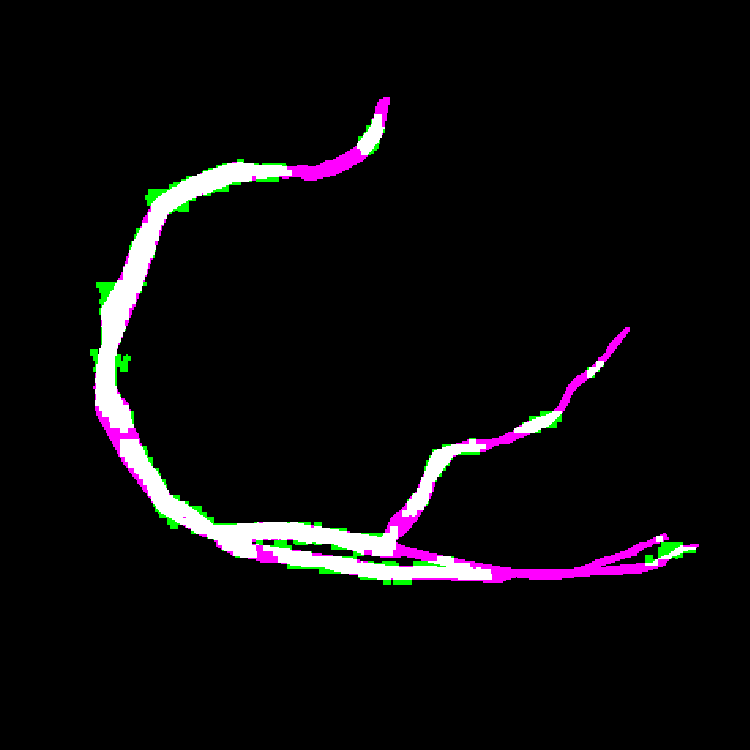}
     \end{subfigure}
     \hfill
	\begin{subfigure}[b]{0.11\textwidth}
         \centering
         \includegraphics[width=\textwidth]{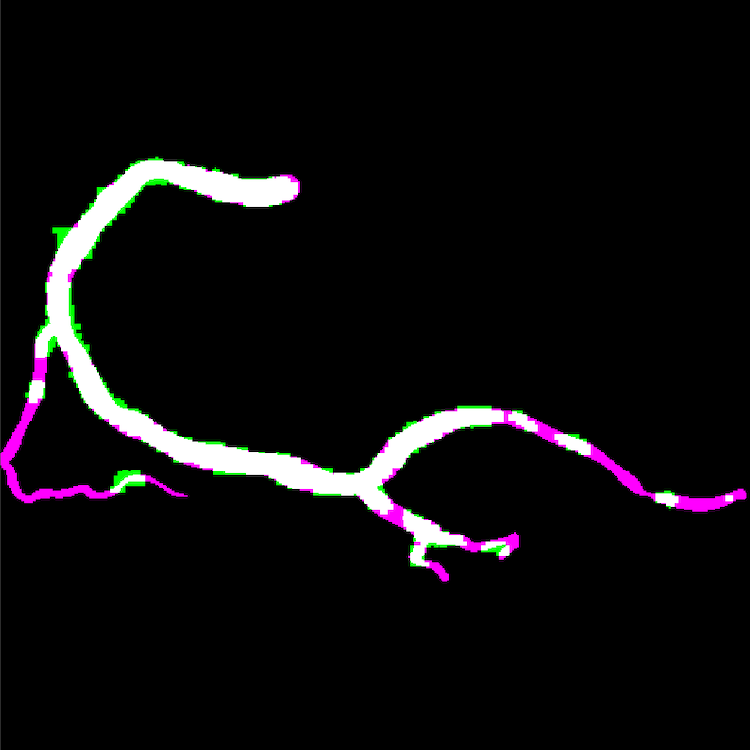}
     \end{subfigure}
     \hfill
	\begin{subfigure}[b]{0.11\textwidth}
         \centering
         \includegraphics[width=\textwidth]{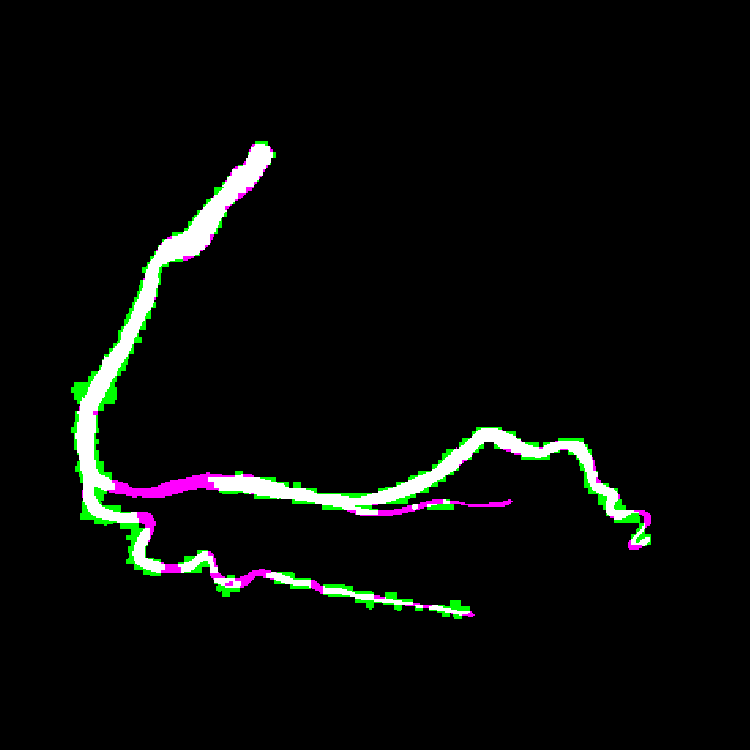}
     \end{subfigure}
     \hfill
     \begin{subfigure}[b]{0.11\textwidth}
         \centering
         \includegraphics[width=\textwidth]{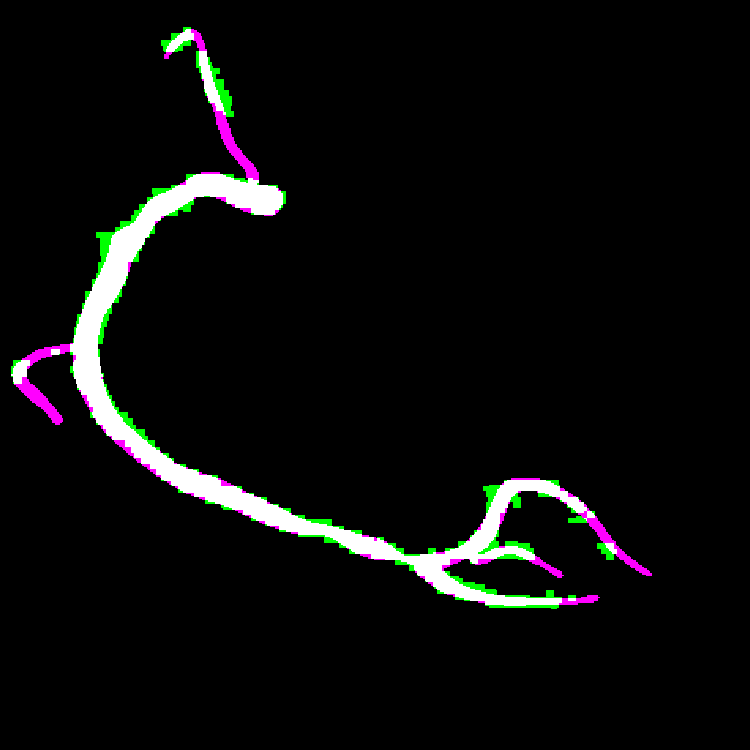}
     \end{subfigure}
     \hfill
     \begin{subfigure}[b]{0.11\textwidth}
         \centering
         \includegraphics[width=\textwidth]{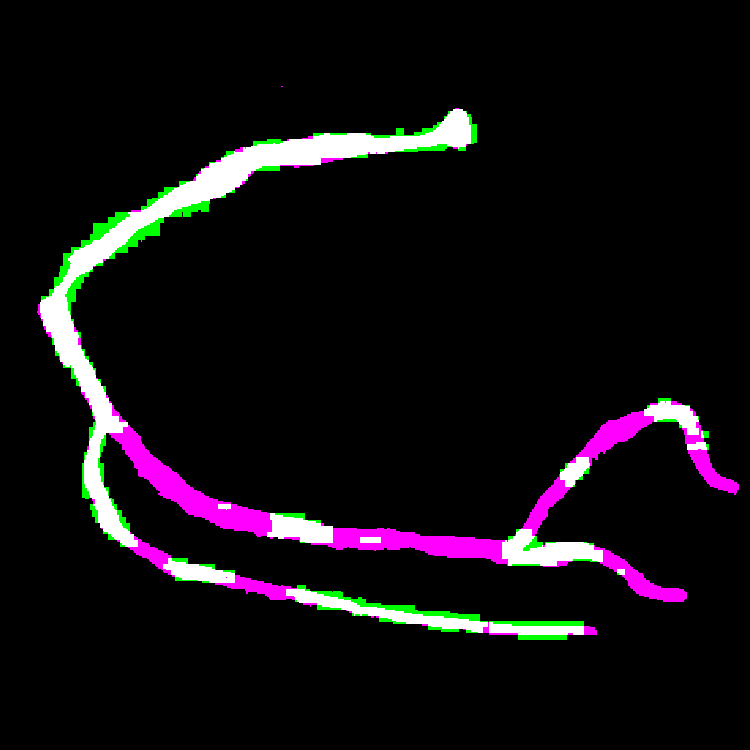}
     \end{subfigure}
     \hfill
     \begin{subfigure}[b]{0.11\textwidth}
         \centering
         \includegraphics[width=\textwidth]{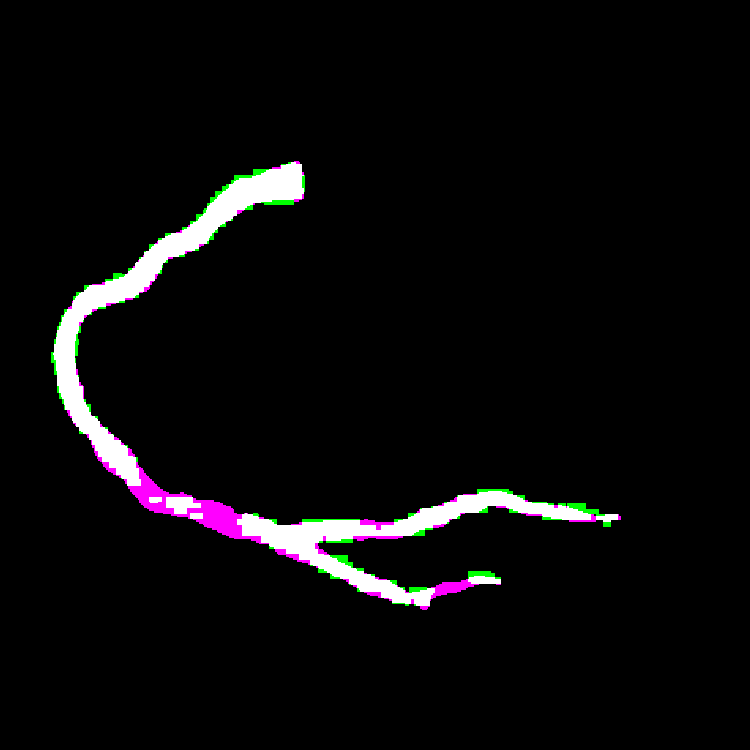}
     \end{subfigure}
     \caption{The comparisons on the first projection plane between the original clinical ICA data and the reprojections of the 3D reconstructions generated from all the models. From left to right: 8 patients. From top to bottom: comparisons between the original ICA data and the reprojections from the reconstructions by our proposed DeepCA model, WGP, +CTLs, +DSCC, Un2+, Un3+, DSCN, and CVTG. Colour purple presents original ICA data, green is reprojection, and white shows the overlap.
     }\label{ica_1}
\end{figure*}

\begin{figure*}[!h]
     \centering
     %%%%%% mine second projection
     \begin{subfigure}[b]{0.11\textwidth}
         \centering
         \includegraphics[width=\textwidth]{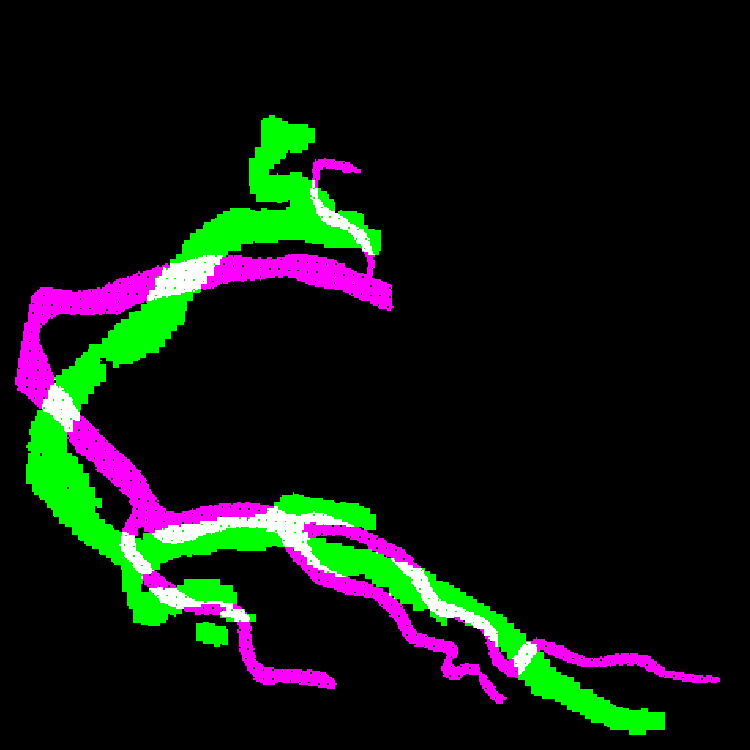}
     \end{subfigure}
     \hfill
	\begin{subfigure}[b]{0.11\textwidth}
         \centering
         \includegraphics[width=\textwidth]{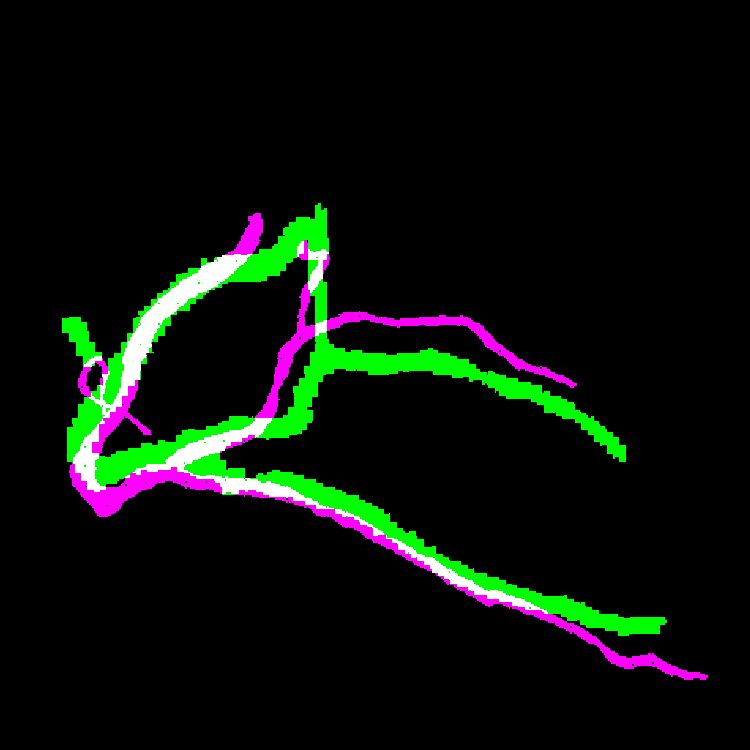}
     \end{subfigure}
     \hfill
	\begin{subfigure}[b]{0.11\textwidth}
         \centering
         \includegraphics[width=\textwidth]{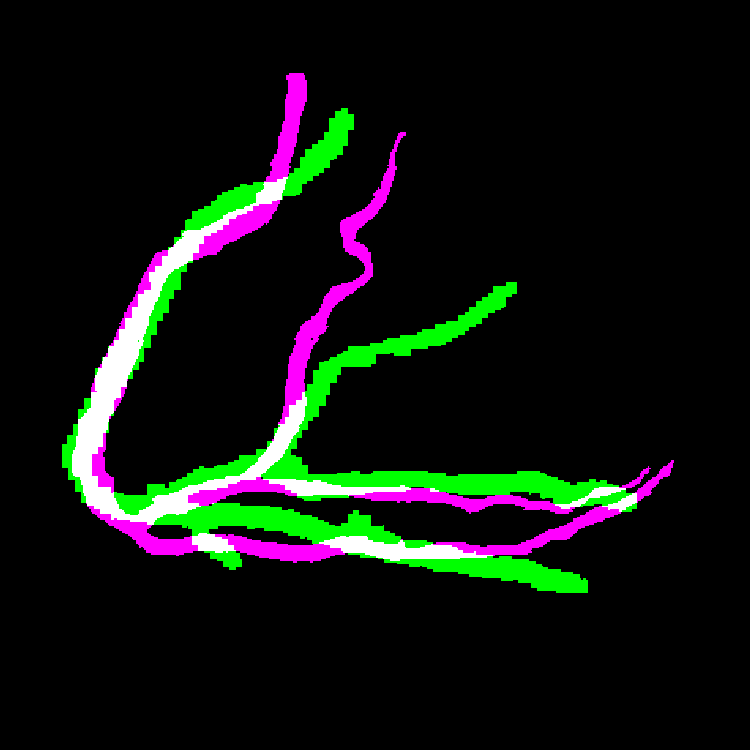}
     \end{subfigure}
     \hfill
	\begin{subfigure}[b]{0.11\textwidth}
         \centering
         \includegraphics[width=\textwidth]{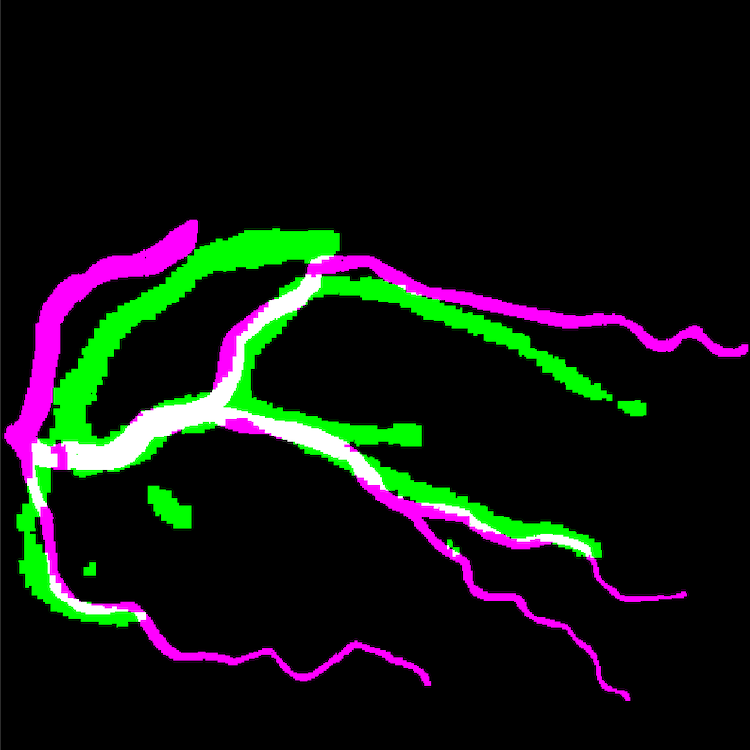}
     \end{subfigure}
     \hfill
	\begin{subfigure}[b]{0.11\textwidth}
         \centering
         \includegraphics[width=\textwidth]{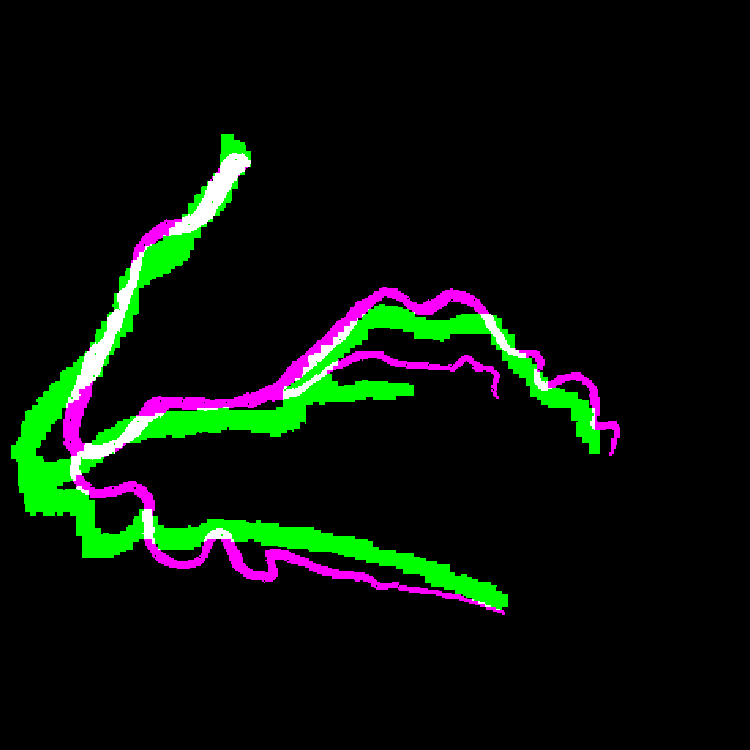}
     \end{subfigure}
     \hfill
     \begin{subfigure}[b]{0.11\textwidth}
         \centering
         \includegraphics[width=\textwidth]{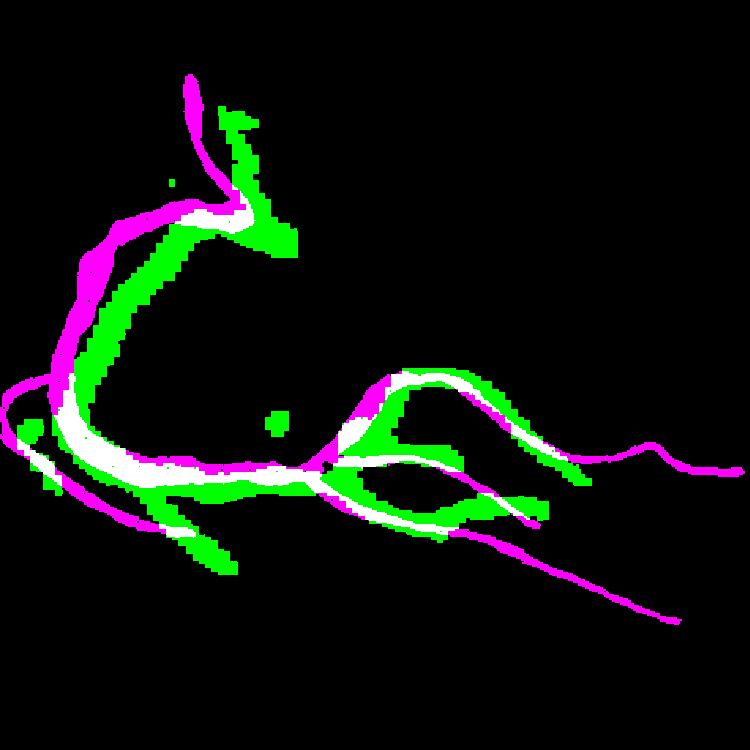}
     \end{subfigure}
     \hfill
     \begin{subfigure}[b]{0.11\textwidth}
         \centering
         \includegraphics[width=\textwidth]{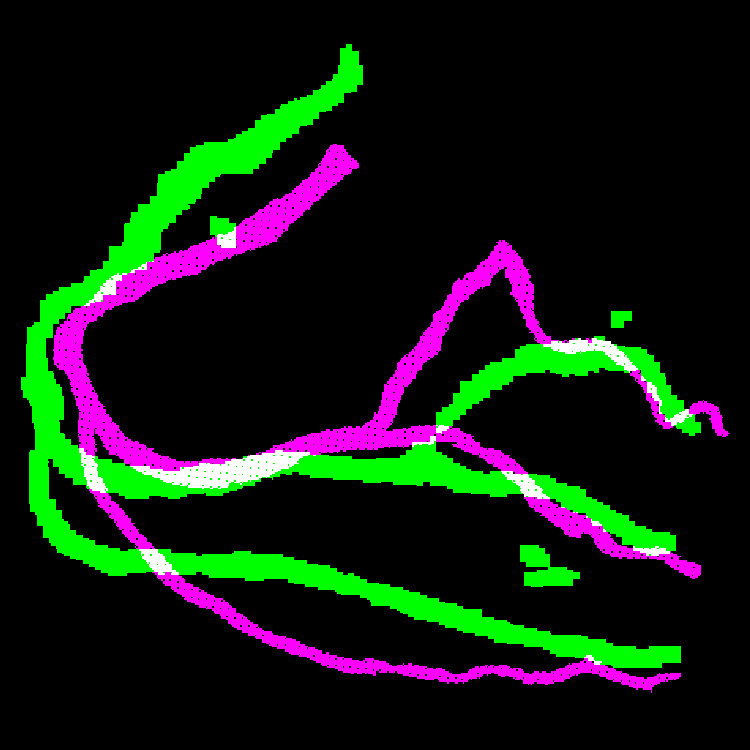}
     \end{subfigure}
     \hfill
     \begin{subfigure}[b]{0.11\textwidth}
         \centering
         \includegraphics[width=\textwidth]{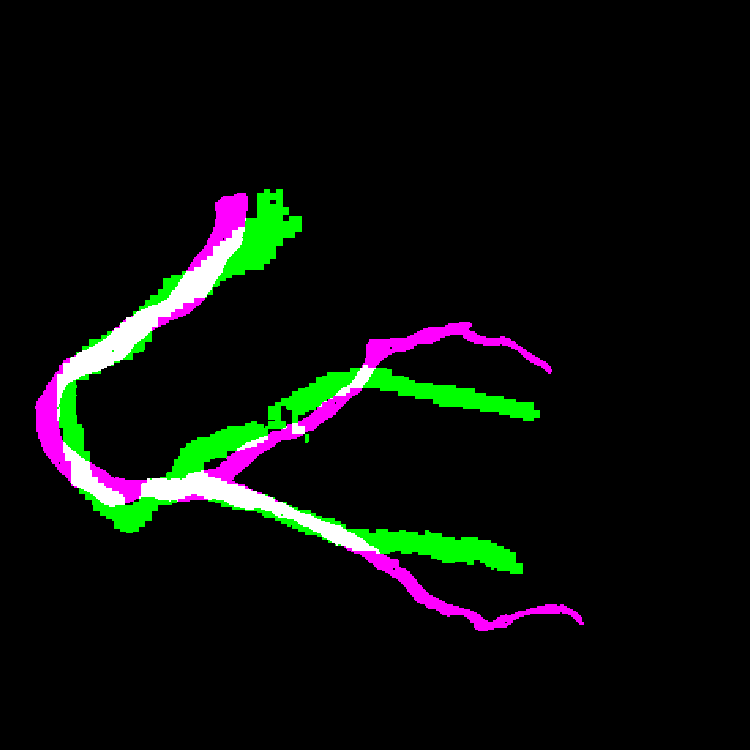}
     \end{subfigure}
     %%%%%% WGP second projection
     \vfill
     \begin{subfigure}[b]{0.11\textwidth}
         \centering
         \includegraphics[width=\textwidth]{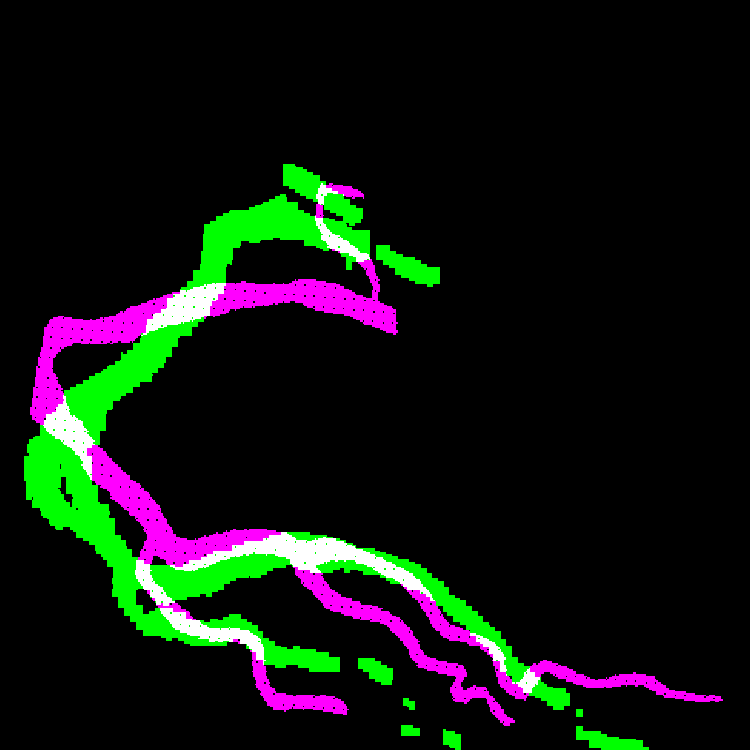}
     \end{subfigure}
     \hfill
	\begin{subfigure}[b]{0.11\textwidth}
         \centering
         \includegraphics[width=\textwidth]{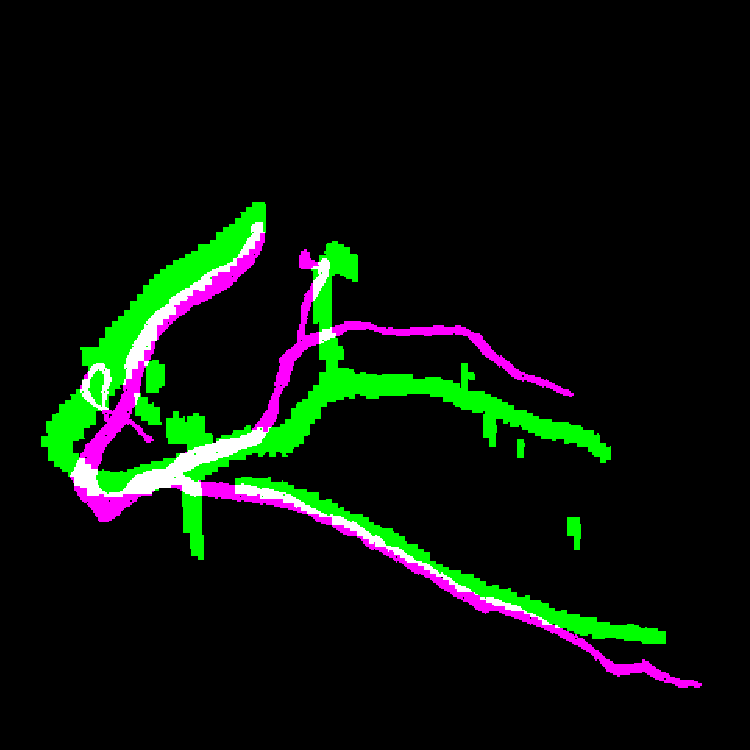}
     \end{subfigure}
     \hfill
	\begin{subfigure}[b]{0.11\textwidth}
         \centering
         \includegraphics[width=\textwidth]{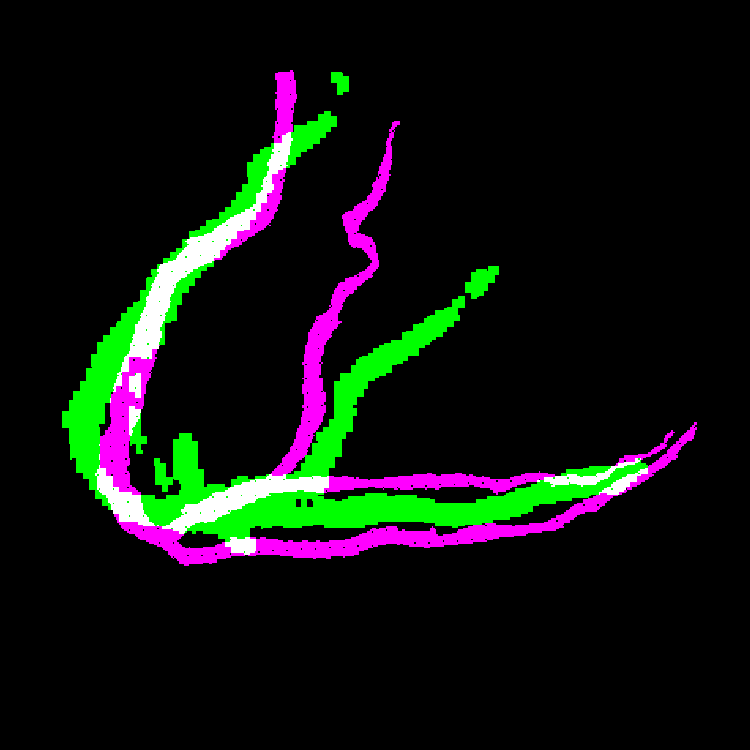}
     \end{subfigure}
     \hfill
	\begin{subfigure}[b]{0.11\textwidth}
         \centering
         \includegraphics[width=\textwidth]{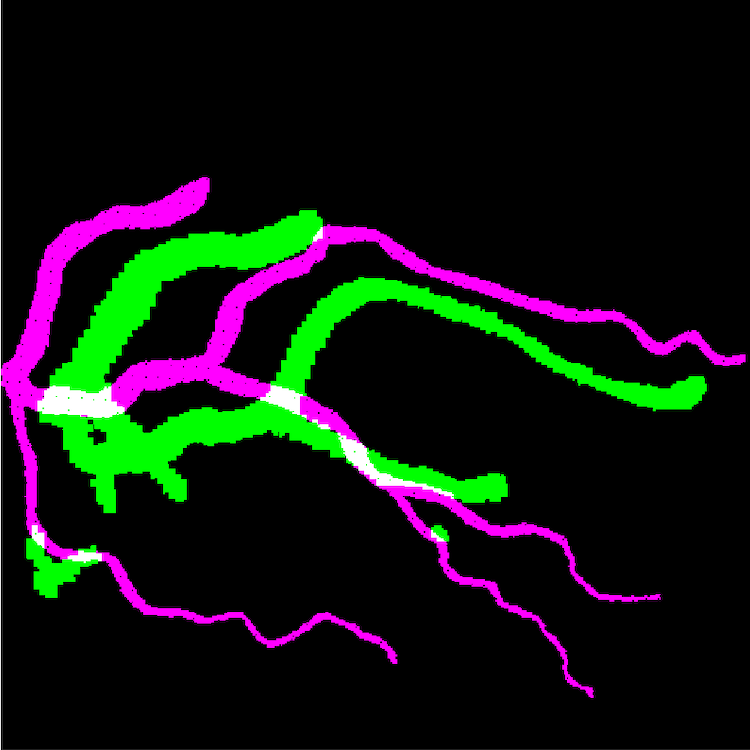}
     \end{subfigure}
     \hfill
	\begin{subfigure}[b]{0.11\textwidth}
         \centering
         \includegraphics[width=\textwidth]{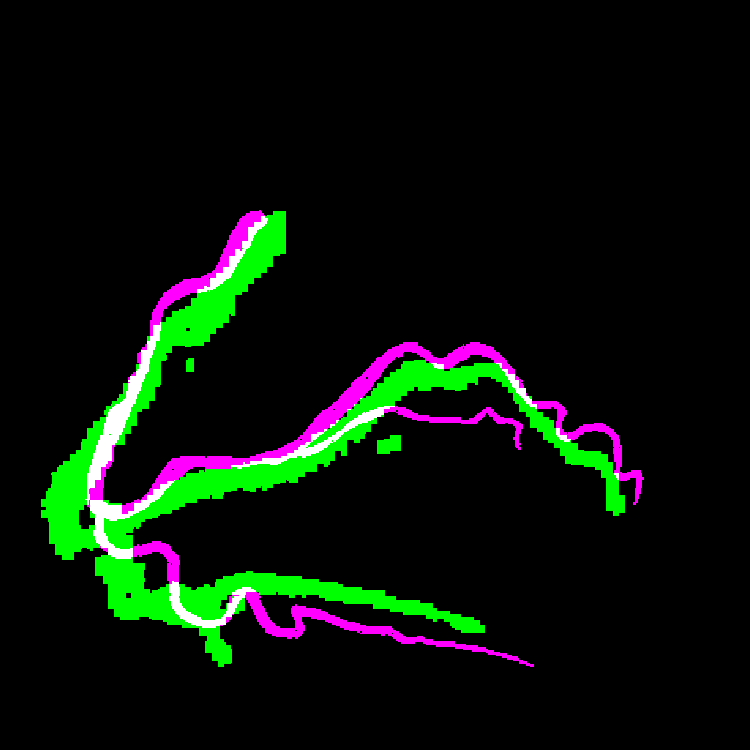}
     \end{subfigure}
     \hfill
     \begin{subfigure}[b]{0.11\textwidth}
         \centering
         \includegraphics[width=\textwidth]{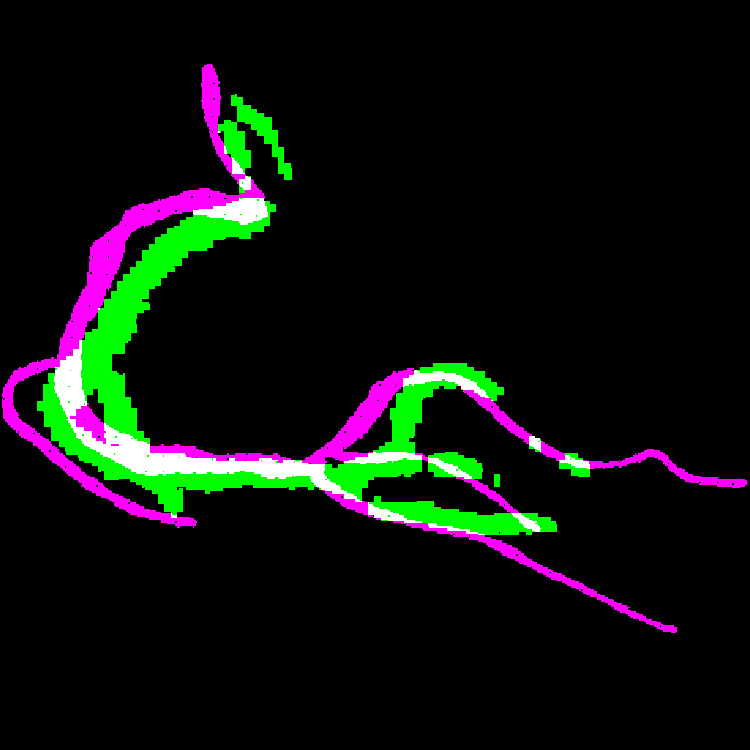}
     \end{subfigure}
     \hfill
     \begin{subfigure}[b]{0.11\textwidth}
         \centering
         \includegraphics[width=\textwidth]{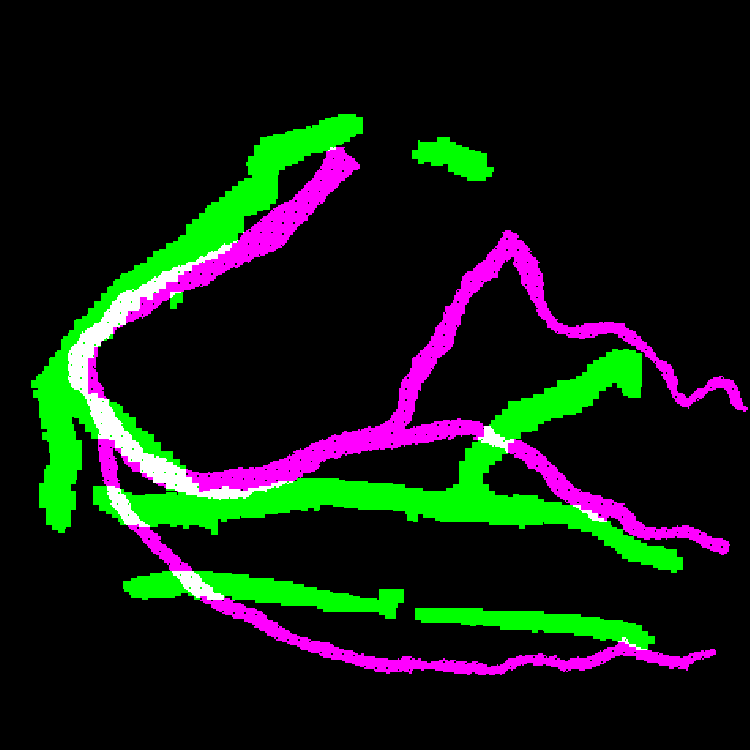}
     \end{subfigure}
     \hfill
     \begin{subfigure}[b]{0.11\textwidth}
         \centering
         \includegraphics[width=\textwidth]{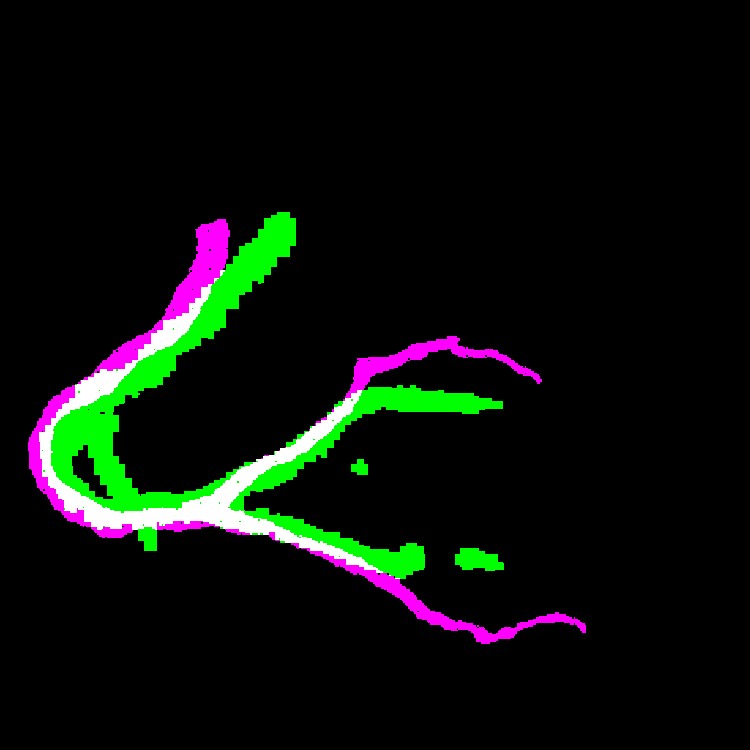}
     \end{subfigure}
     %%%%%% WGP+CTLs second projection
     \vfill
     \begin{subfigure}[b]{0.11\textwidth}
         \centering
         \includegraphics[width=\textwidth]{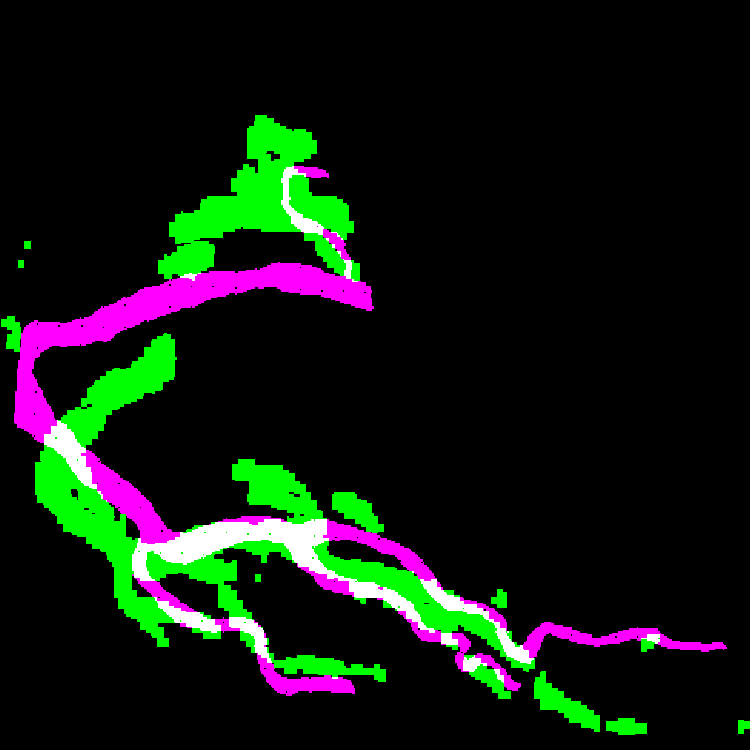}
     \end{subfigure}
     \hfill
	\begin{subfigure}[b]{0.11\textwidth}
         \centering
         \includegraphics[width=\textwidth]{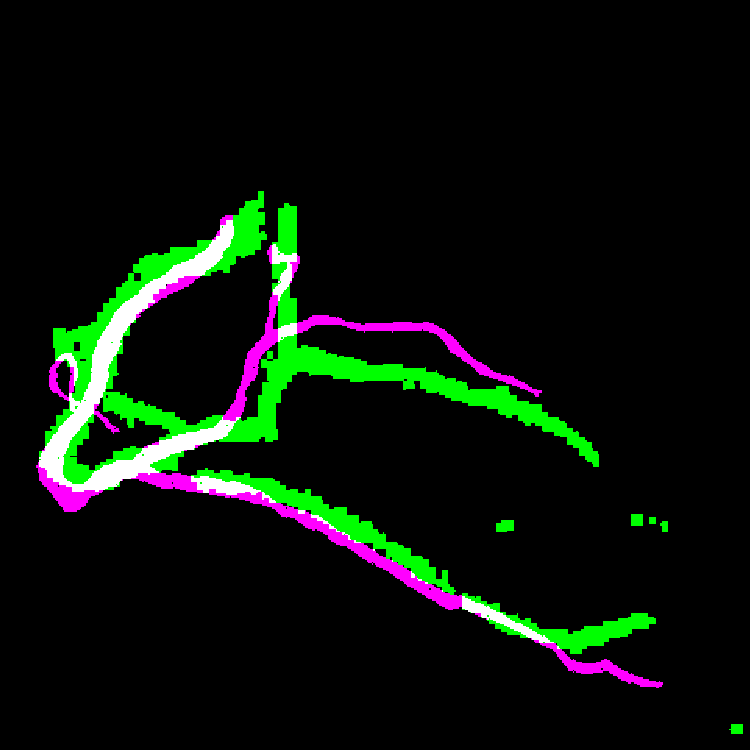}
     \end{subfigure}
     \hfill
	\begin{subfigure}[b]{0.11\textwidth}
         \centering
         \includegraphics[width=\textwidth]{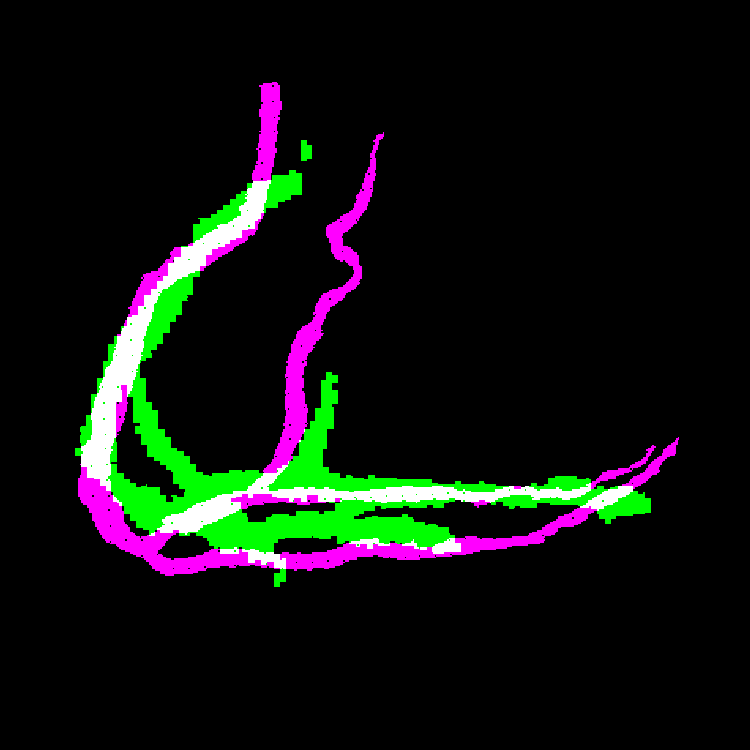}
     \end{subfigure}
     \hfill
	\begin{subfigure}[b]{0.11\textwidth}
         \centering
         \includegraphics[width=\textwidth]{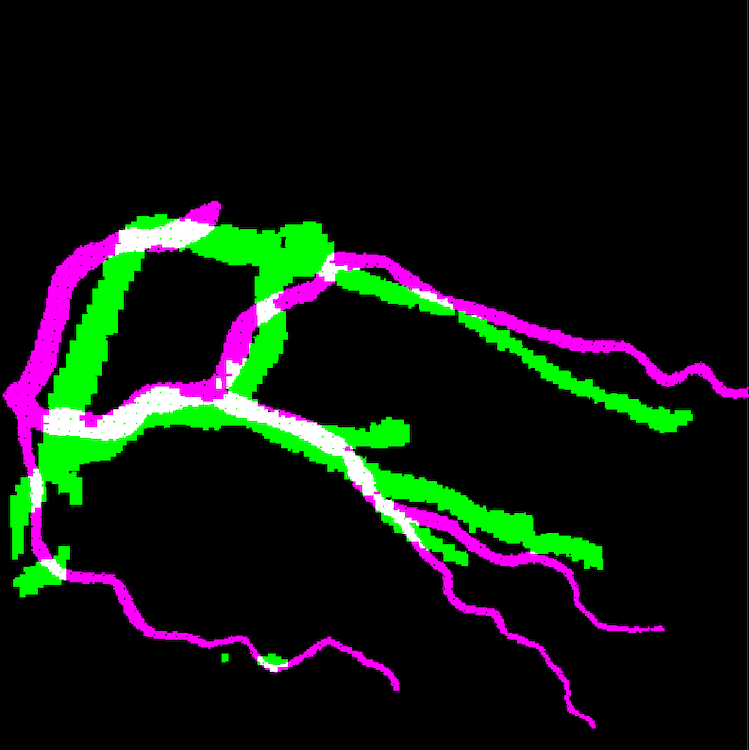}
     \end{subfigure}
     \hfill
	\begin{subfigure}[b]{0.11\textwidth}
         \centering
         \includegraphics[width=\textwidth]{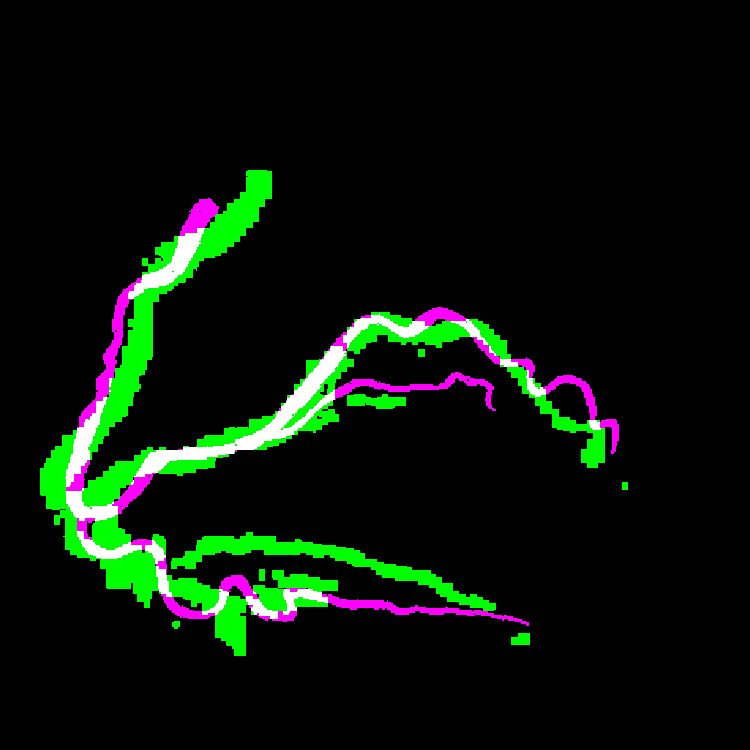}
     \end{subfigure}
     \hfill
     \begin{subfigure}[b]{0.11\textwidth}
         \centering
         \includegraphics[width=\textwidth]{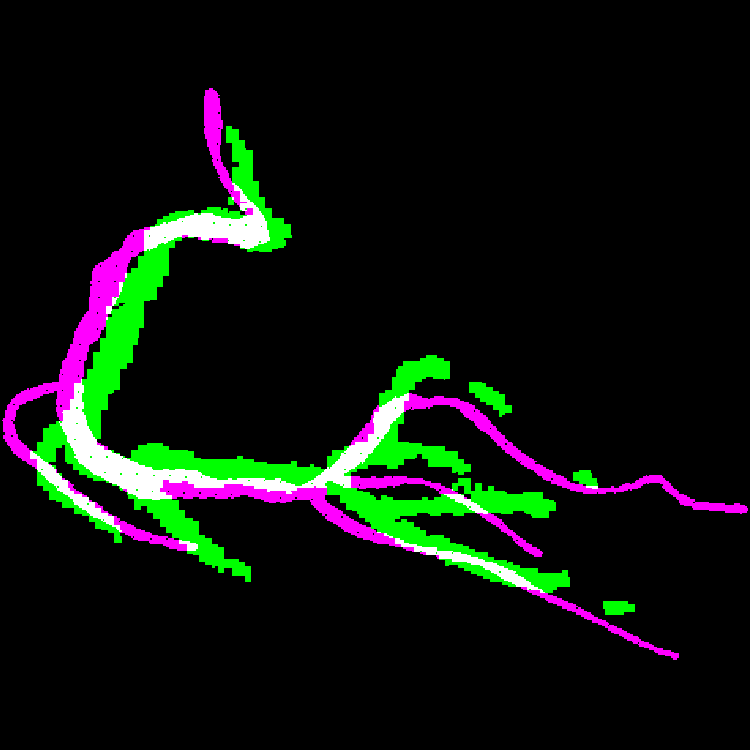}
     \end{subfigure}
     \hfill
     \begin{subfigure}[b]{0.11\textwidth}
         \centering
         \includegraphics[width=\textwidth]{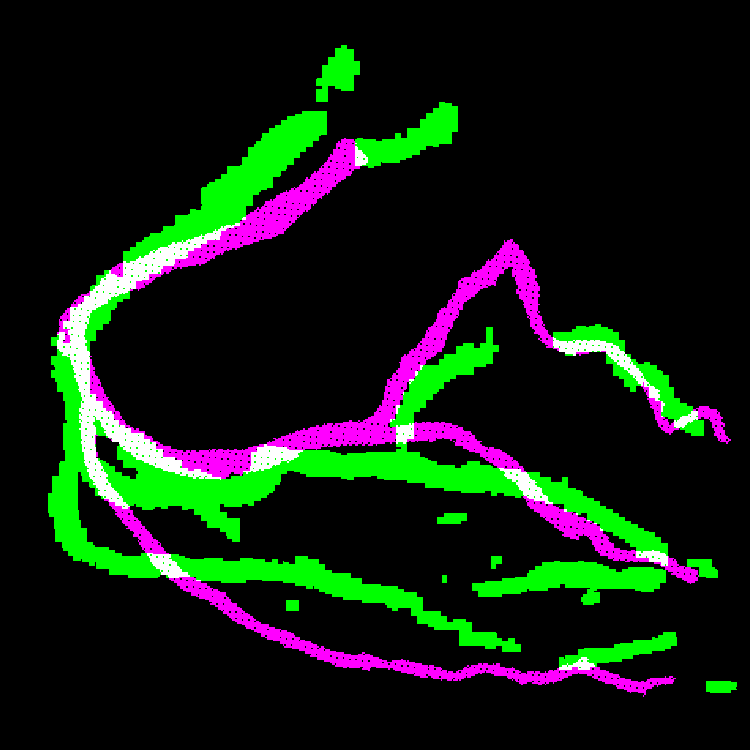}
     \end{subfigure}
     \hfill
     \begin{subfigure}[b]{0.11\textwidth}
         \centering
         \includegraphics[width=\textwidth]{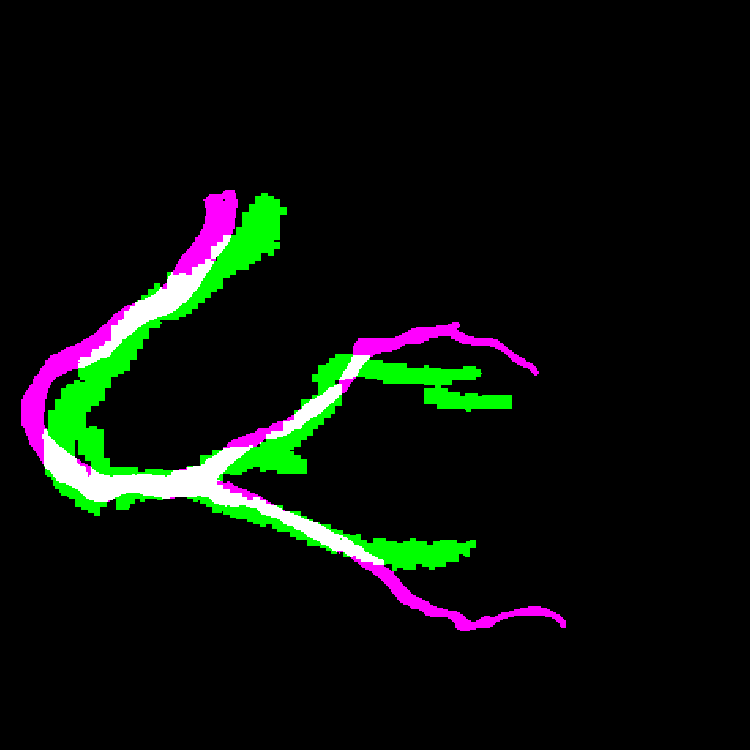}
     \end{subfigure}
     %%%%%% WGP+DSCC second projection
     \vfill
     \begin{subfigure}[b]{0.11\textwidth}
         \centering
         \includegraphics[width=\textwidth]{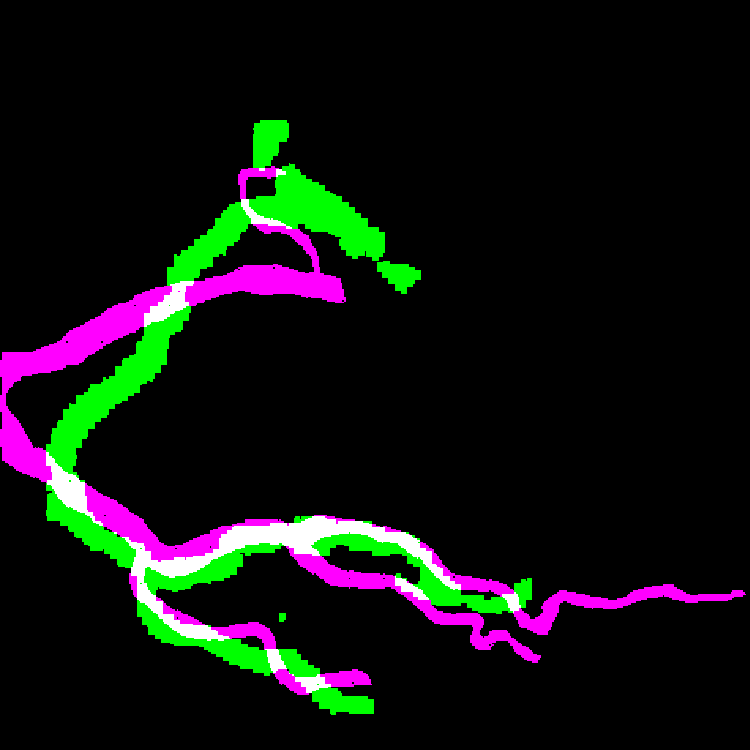}
     \end{subfigure}
     \hfill
	\begin{subfigure}[b]{0.11\textwidth}
         \centering
         \includegraphics[width=\textwidth]{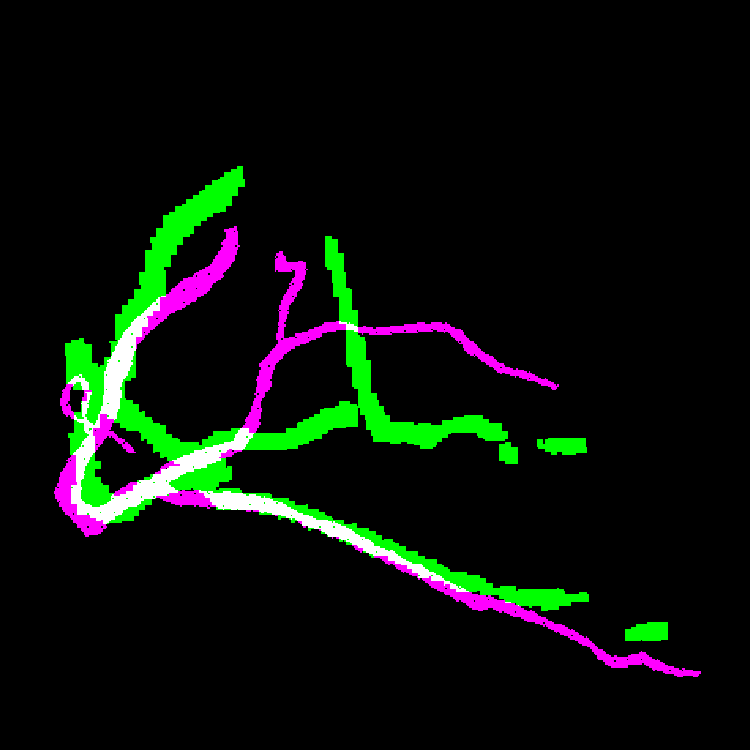}
     \end{subfigure}
     \hfill
	\begin{subfigure}[b]{0.11\textwidth}
         \centering
         \includegraphics[width=\textwidth]{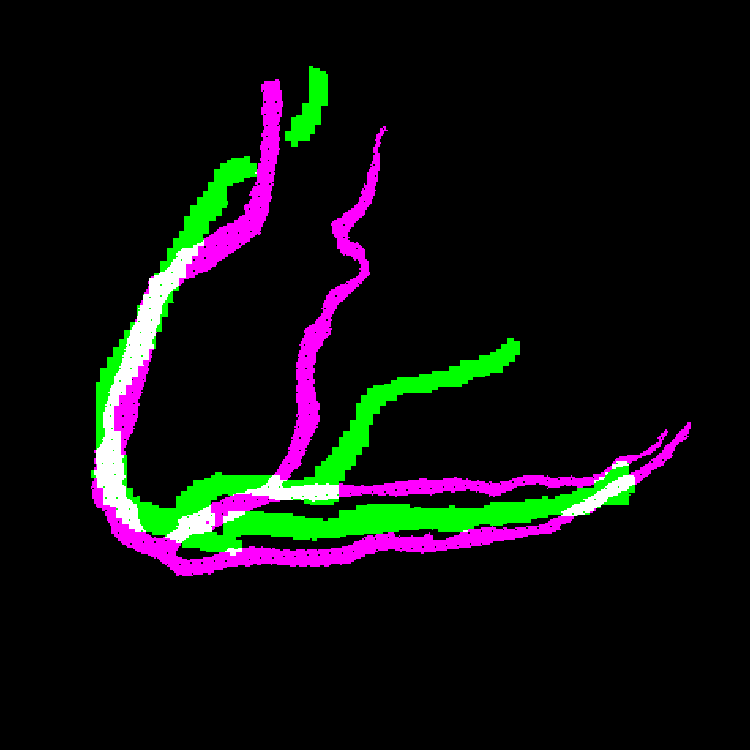}
     \end{subfigure}
     \hfill
	\begin{subfigure}[b]{0.11\textwidth}
         \centering
         \includegraphics[width=\textwidth]{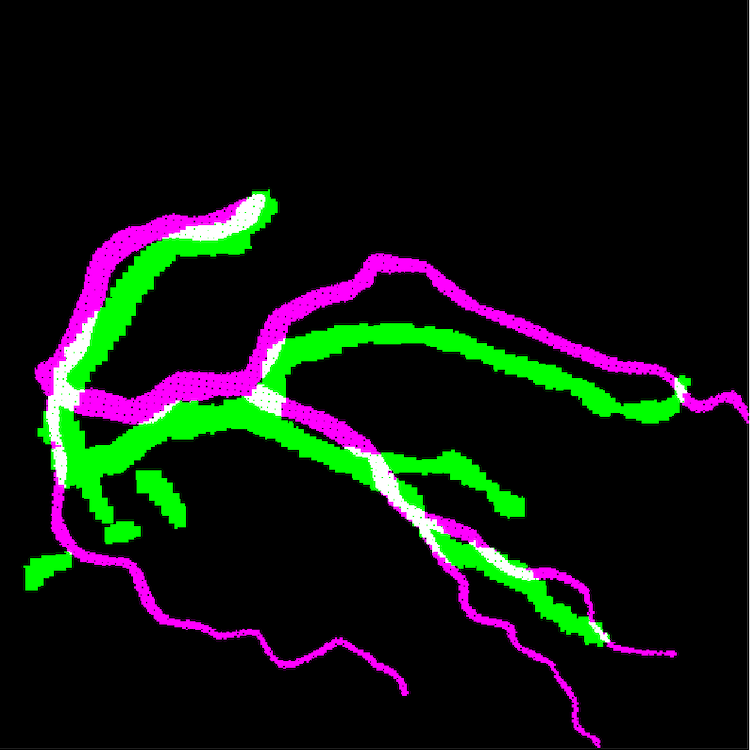}
     \end{subfigure}
     \hfill
	\begin{subfigure}[b]{0.11\textwidth}
         \centering
         \includegraphics[width=\textwidth]{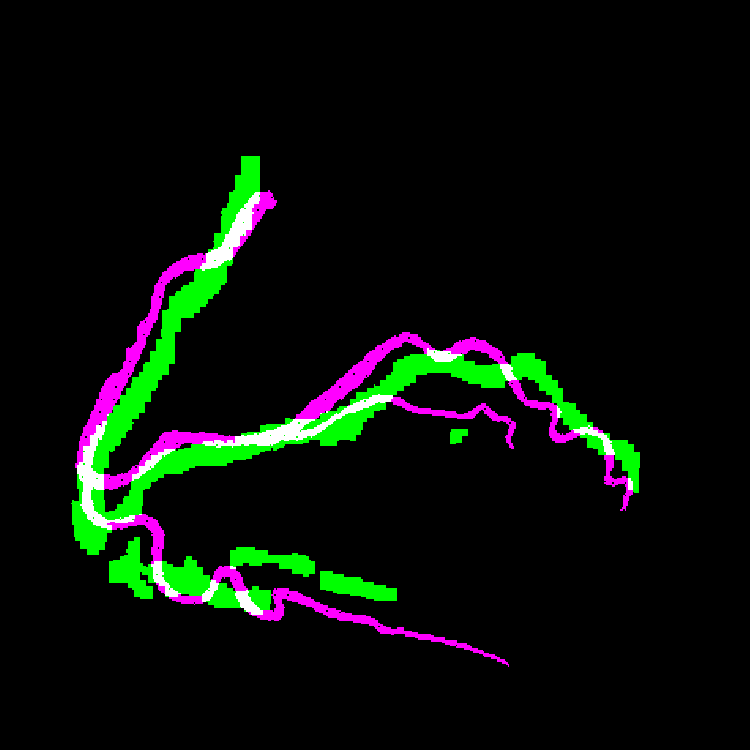}
     \end{subfigure}
     \hfill
     \begin{subfigure}[b]{0.11\textwidth}
         \centering
         \includegraphics[width=\textwidth]{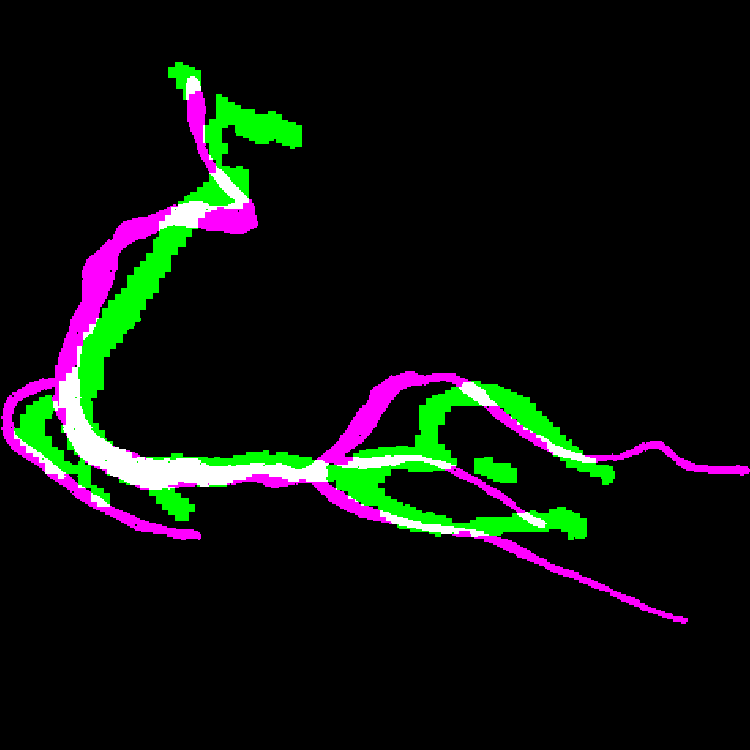}
     \end{subfigure}
     \hfill
     \begin{subfigure}[b]{0.11\textwidth}
         \centering
         \includegraphics[width=\textwidth]{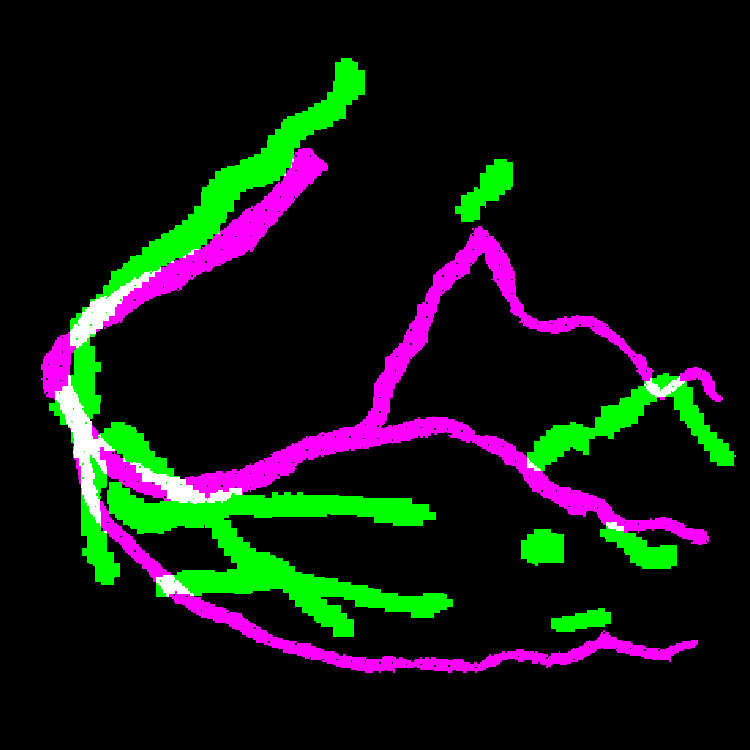}
     \end{subfigure}
     \hfill
     \begin{subfigure}[b]{0.11\textwidth}
         \centering
         \includegraphics[width=\textwidth]{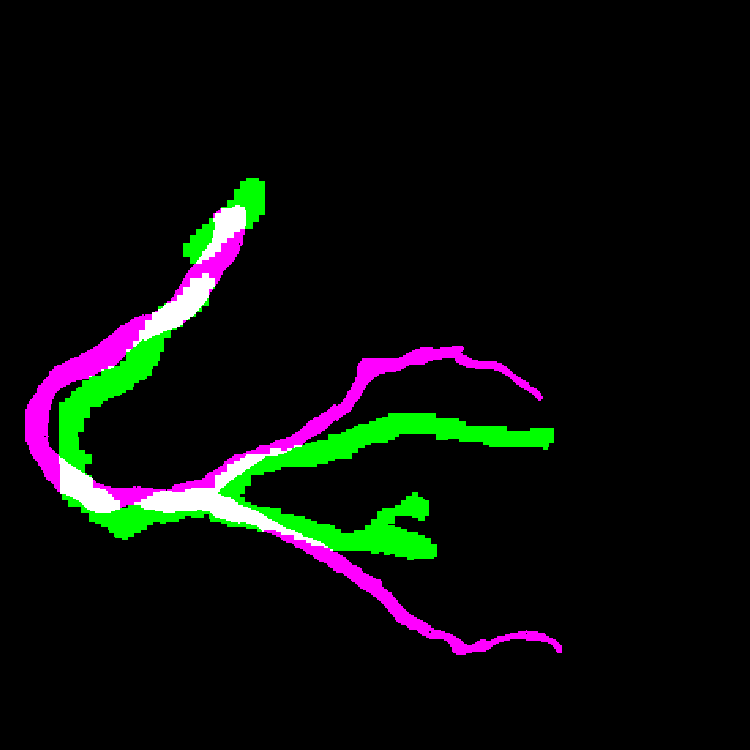}
     \end{subfigure}
     %%%%%% unet++ second projection
     \vfill
     \begin{subfigure}[b]{0.11\textwidth}
         \centering
         \includegraphics[width=\textwidth]{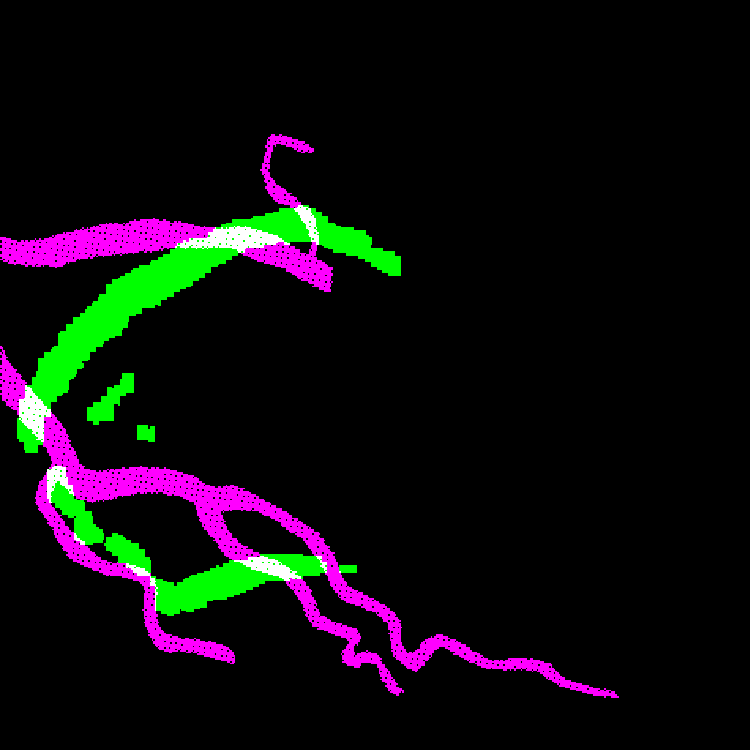}
     \end{subfigure}
     \hfill
	\begin{subfigure}[b]{0.11\textwidth}
         \centering
         \includegraphics[width=\textwidth]{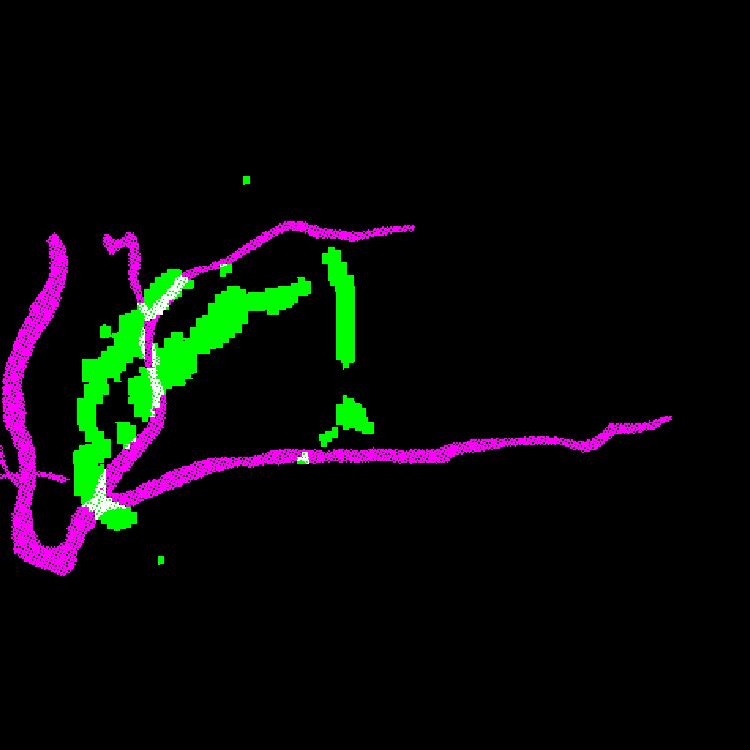}
     \end{subfigure}
     \hfill
	\begin{subfigure}[b]{0.11\textwidth}
         \centering
         \includegraphics[width=\textwidth]{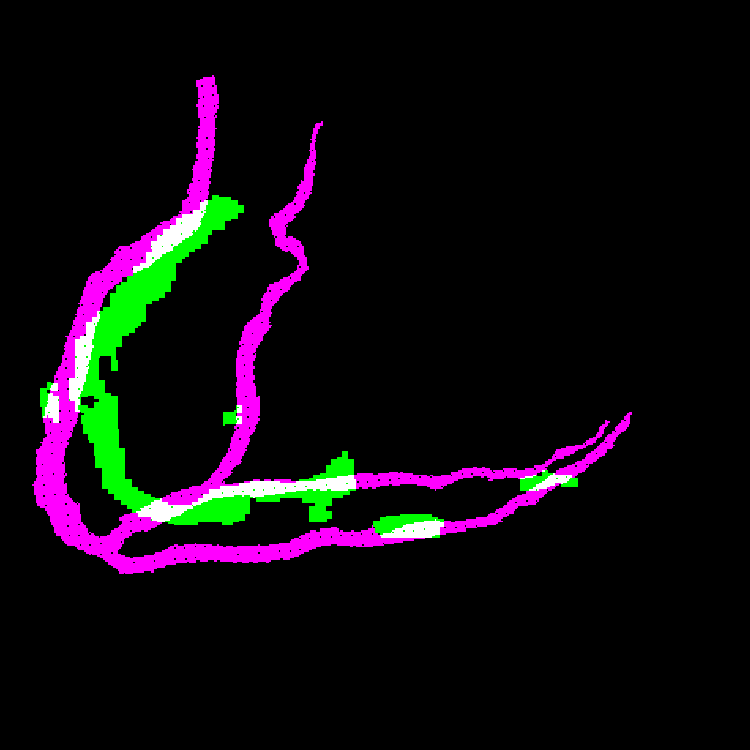}
     \end{subfigure}
     \hfill
	\begin{subfigure}[b]{0.11\textwidth}
         \centering
         \includegraphics[width=\textwidth]{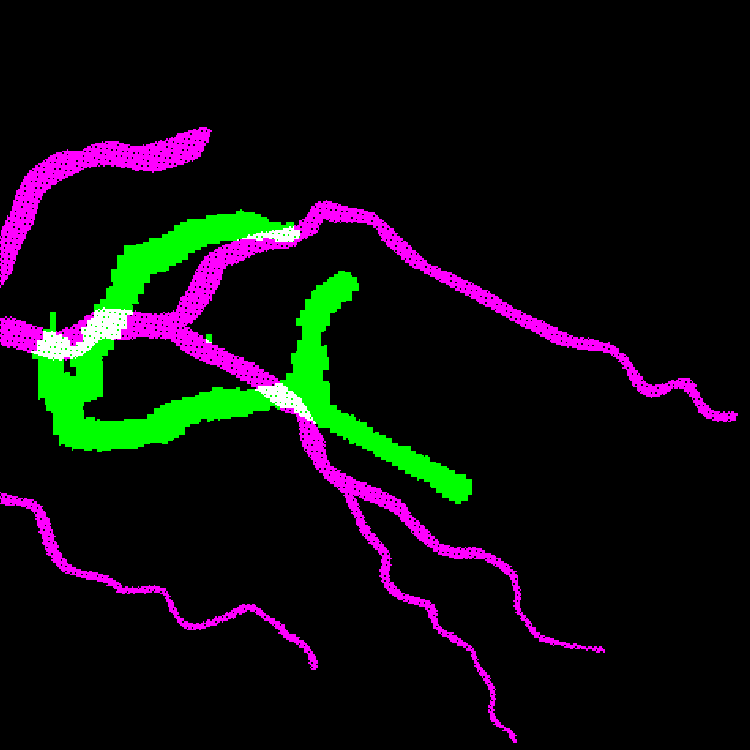}
     \end{subfigure}
     \hfill
	\begin{subfigure}[b]{0.11\textwidth}
         \centering
         \includegraphics[width=\textwidth]{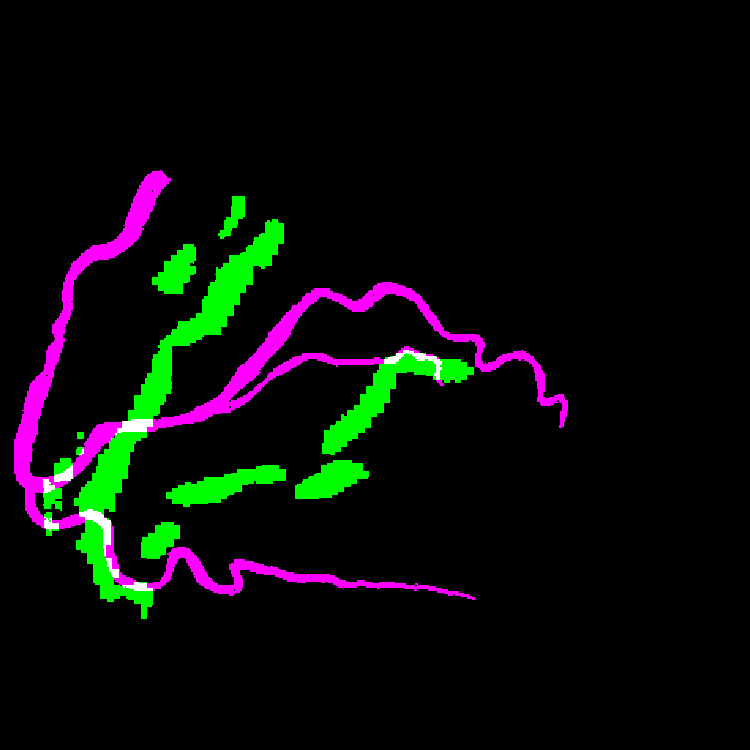}
     \end{subfigure}
     \hfill
     \begin{subfigure}[b]{0.11\textwidth}
         \centering
         \includegraphics[width=\textwidth]{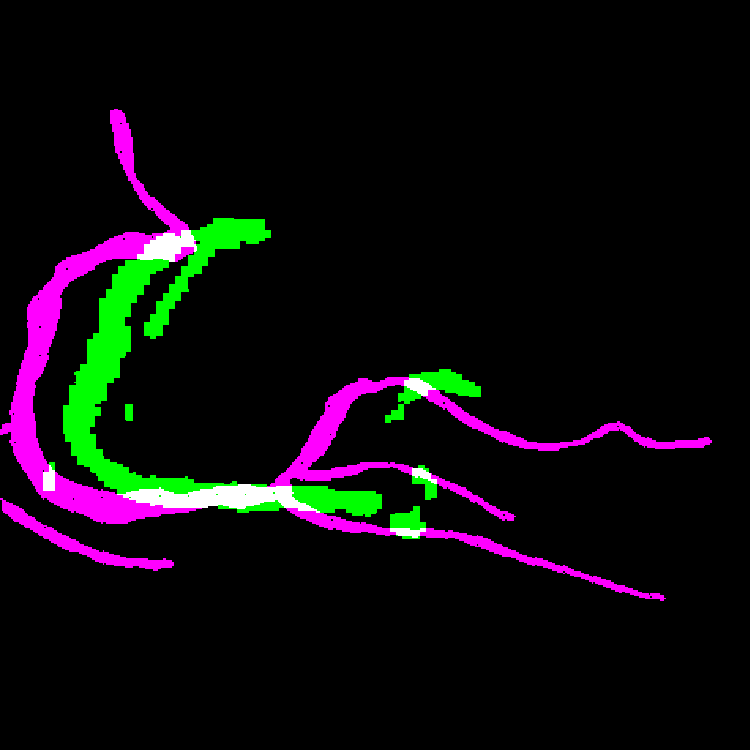}
     \end{subfigure}
     \hfill
     \begin{subfigure}[b]{0.11\textwidth}
         \centering
         \includegraphics[width=\textwidth]{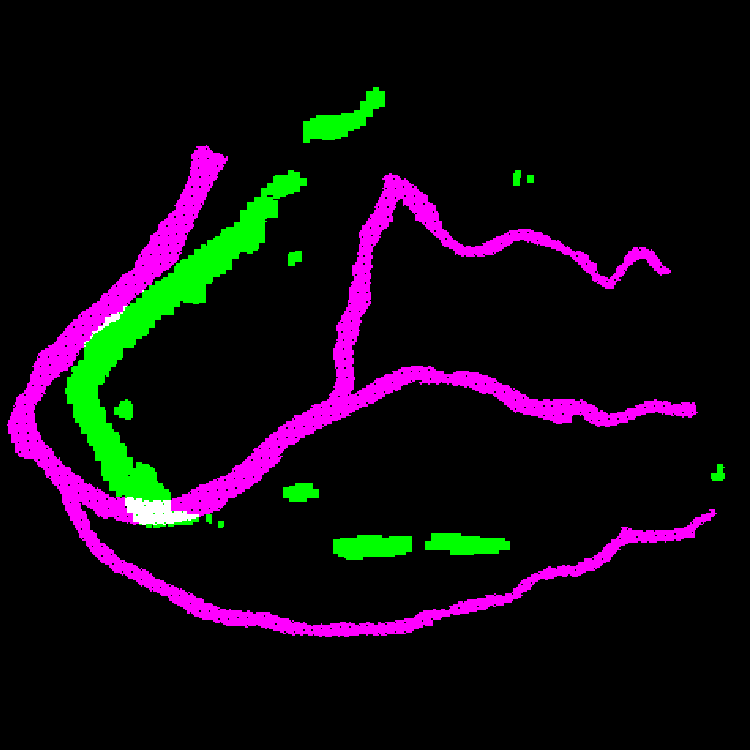}
     \end{subfigure}
     \hfill
     \begin{subfigure}[b]{0.11\textwidth}
         \centering
         \includegraphics[width=\textwidth]{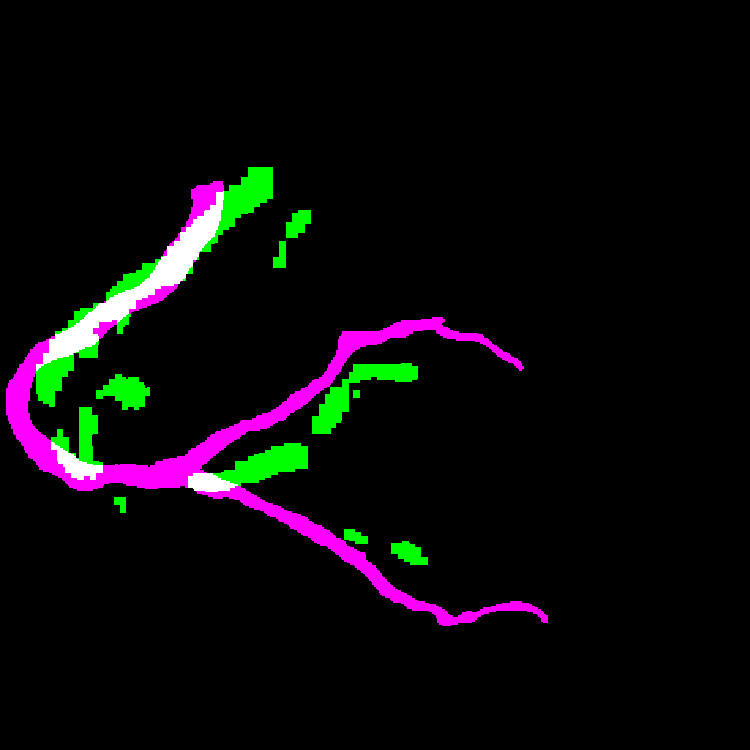}
     \end{subfigure}
     %%%%%% unet+++ second projection
     \vfill
     \begin{subfigure}[b]{0.11\textwidth}
         \centering
         \includegraphics[width=\textwidth]{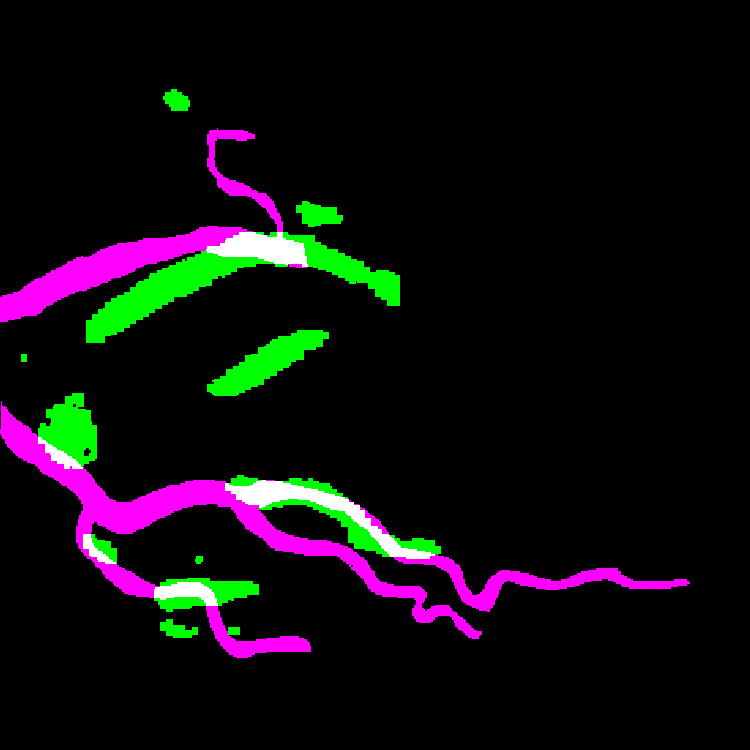}
     \end{subfigure}
     \hfill
	\begin{subfigure}[b]{0.11\textwidth}
         \centering
         \includegraphics[width=\textwidth]{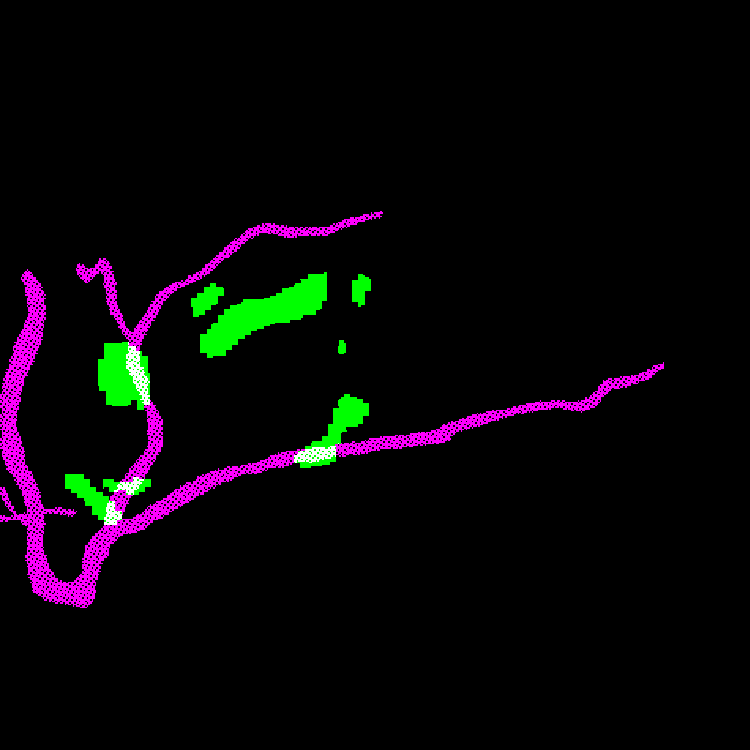}
     \end{subfigure}
     \hfill
	\begin{subfigure}[b]{0.11\textwidth}
         \centering
         \includegraphics[width=\textwidth]{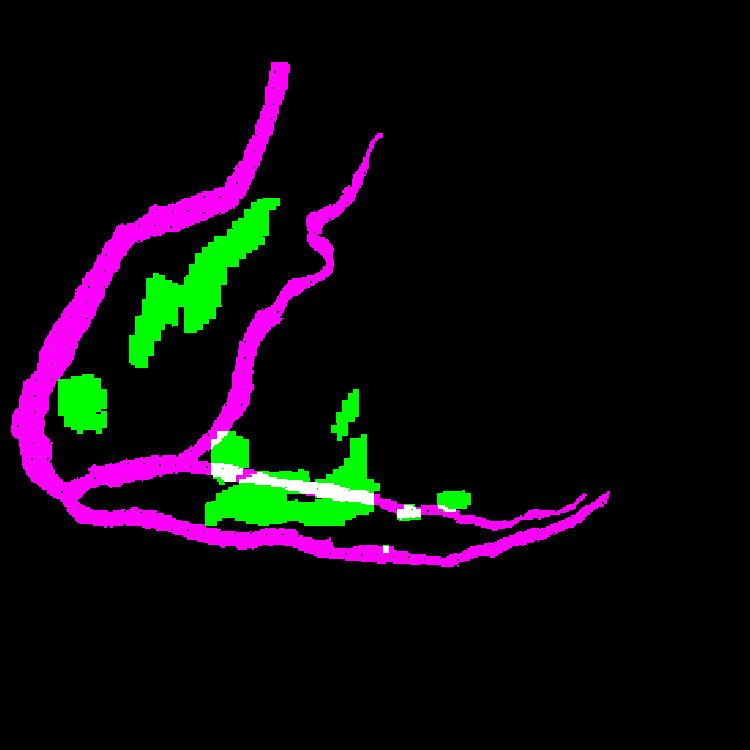}
     \end{subfigure}
     \hfill
	\begin{subfigure}[b]{0.11\textwidth}
         \centering
         \includegraphics[width=\textwidth]{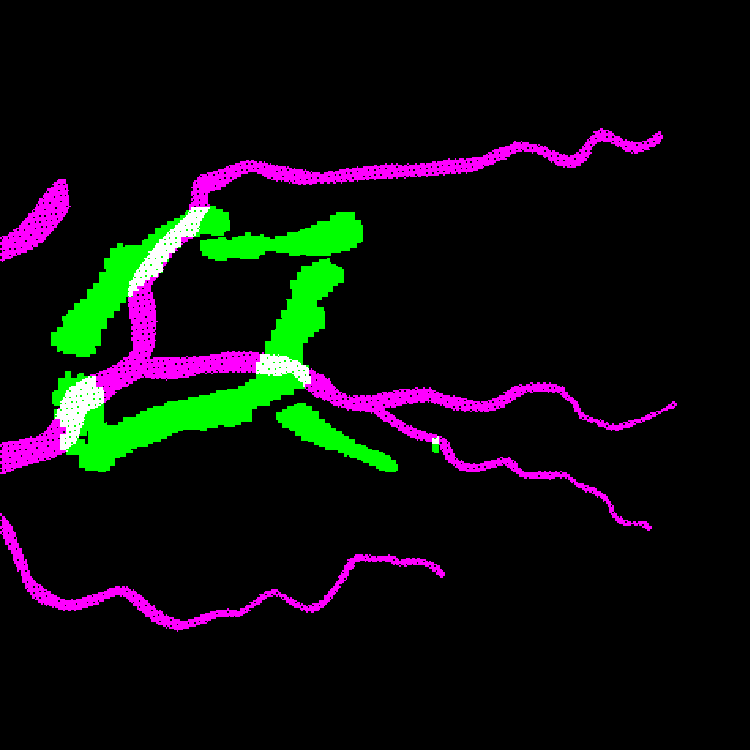}
     \end{subfigure}
     \hfill
	\begin{subfigure}[b]{0.11\textwidth}
         \centering
         \includegraphics[width=\textwidth]{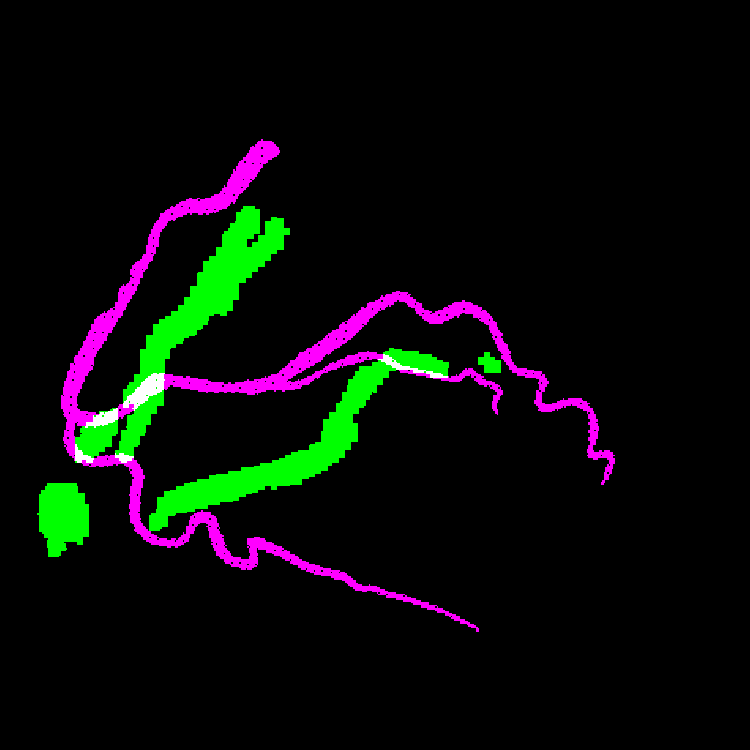}
     \end{subfigure}
     \hfill
     \begin{subfigure}[b]{0.11\textwidth}
         \centering
         \includegraphics[width=\textwidth]{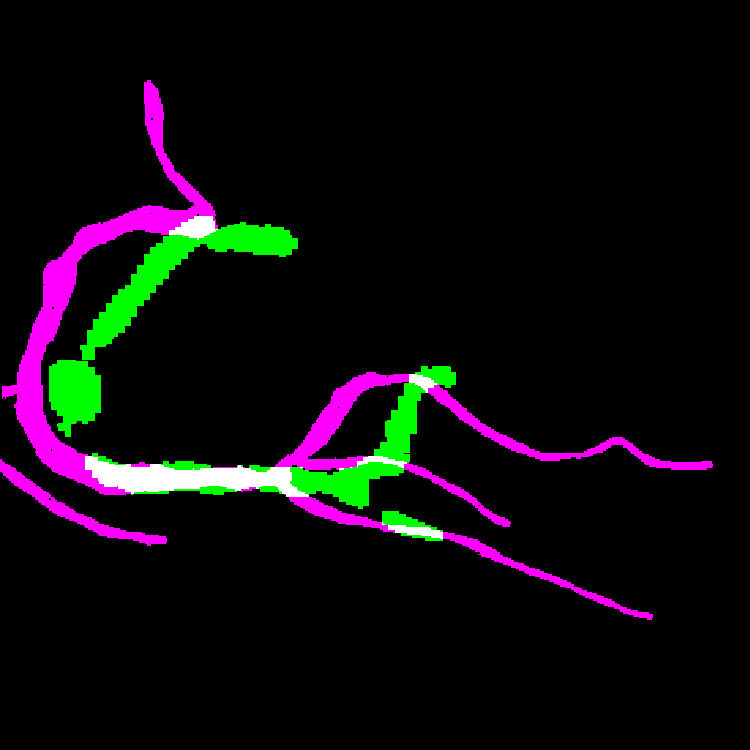}
     \end{subfigure}
     \hfill
     \begin{subfigure}[b]{0.11\textwidth}
         \centering
         \includegraphics[width=\textwidth]{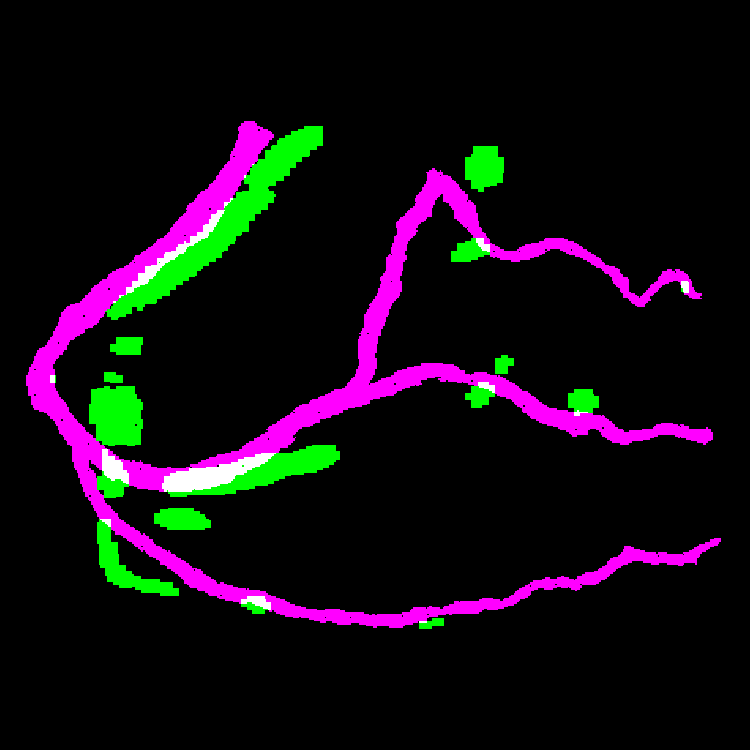}
     \end{subfigure}
     \hfill
     \begin{subfigure}[b]{0.11\textwidth}
         \centering
         \includegraphics[width=\textwidth]{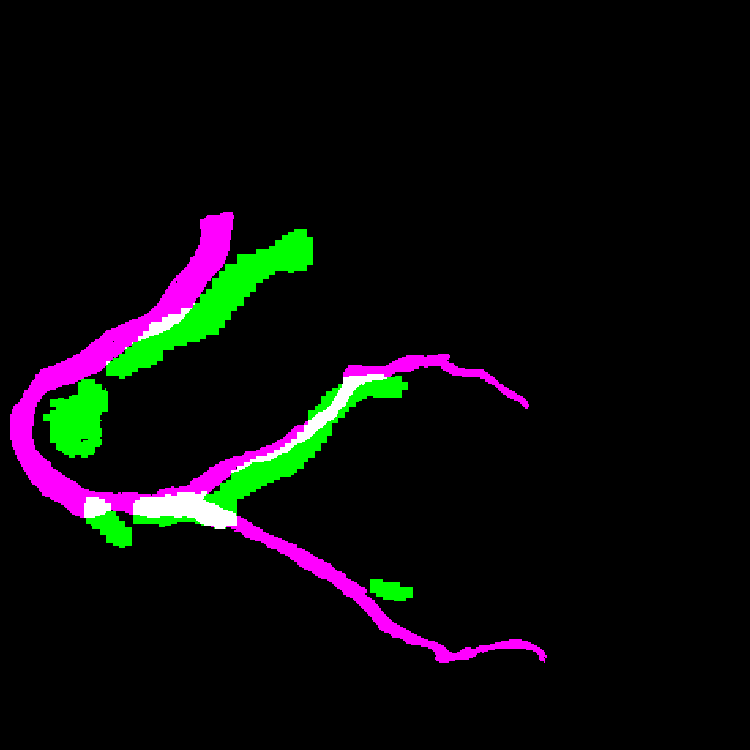}
     \end{subfigure}
     %%%%%% DSCN second projection
     \vfill
     \begin{subfigure}[b]{0.11\textwidth}
         \centering
         \includegraphics[width=\textwidth]{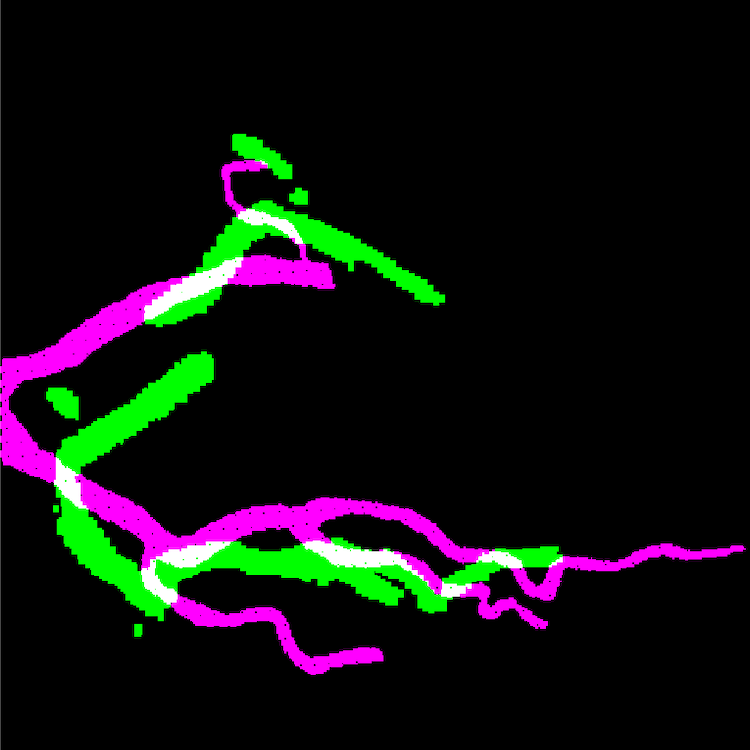}
     \end{subfigure}
     \hfill
	\begin{subfigure}[b]{0.11\textwidth}
         \centering
         \includegraphics[width=\textwidth]{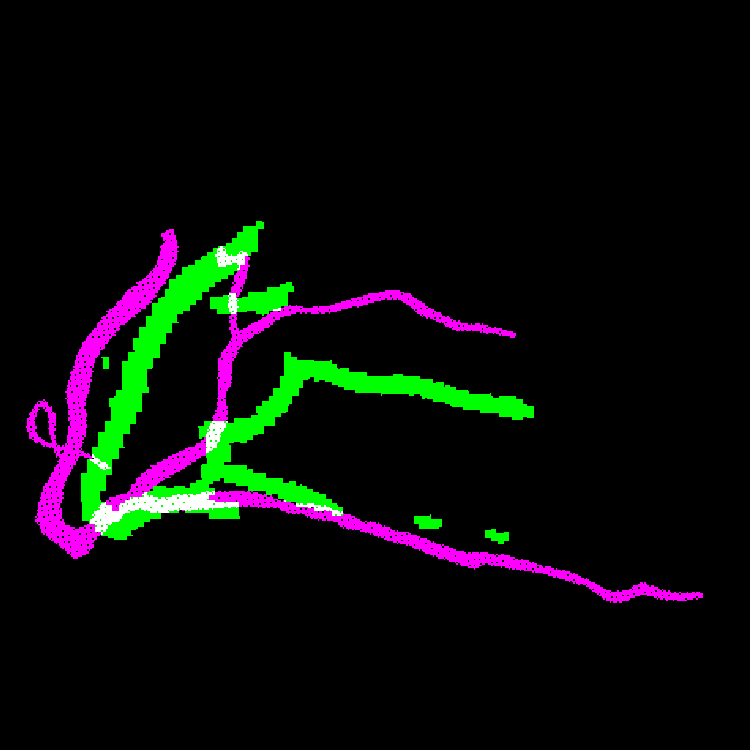}
     \end{subfigure}
     \hfill
	\begin{subfigure}[b]{0.11\textwidth}
         \centering
         \includegraphics[width=\textwidth]{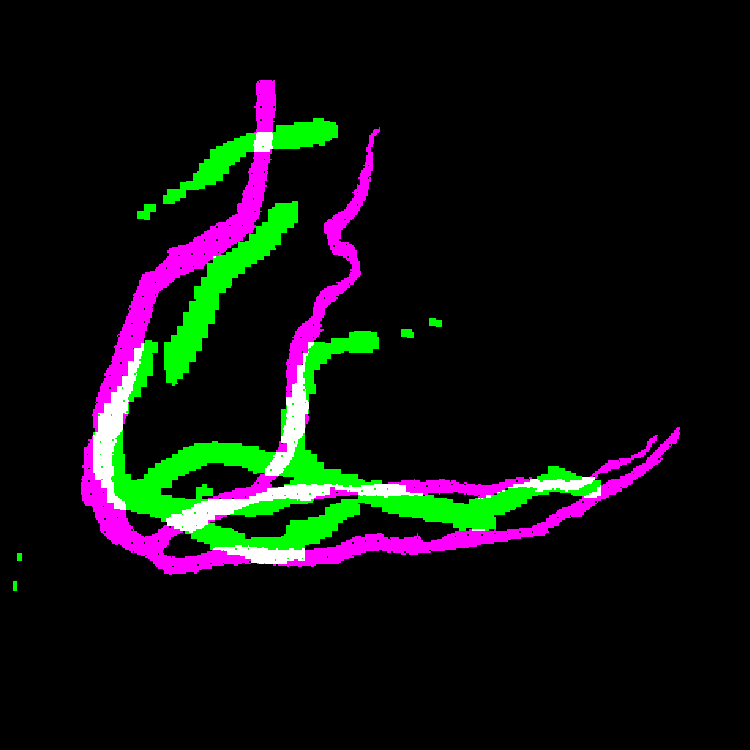}
     \end{subfigure}
     \hfill
	\begin{subfigure}[b]{0.11\textwidth}
         \centering
         \includegraphics[width=\textwidth]{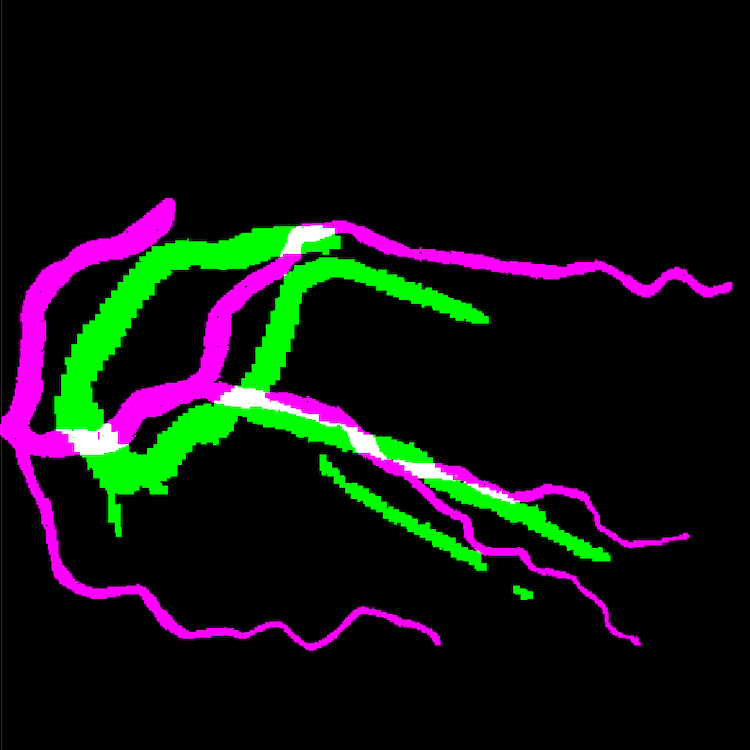}
     \end{subfigure}
     \hfill
	\begin{subfigure}[b]{0.11\textwidth}
         \centering
         \includegraphics[width=\textwidth]{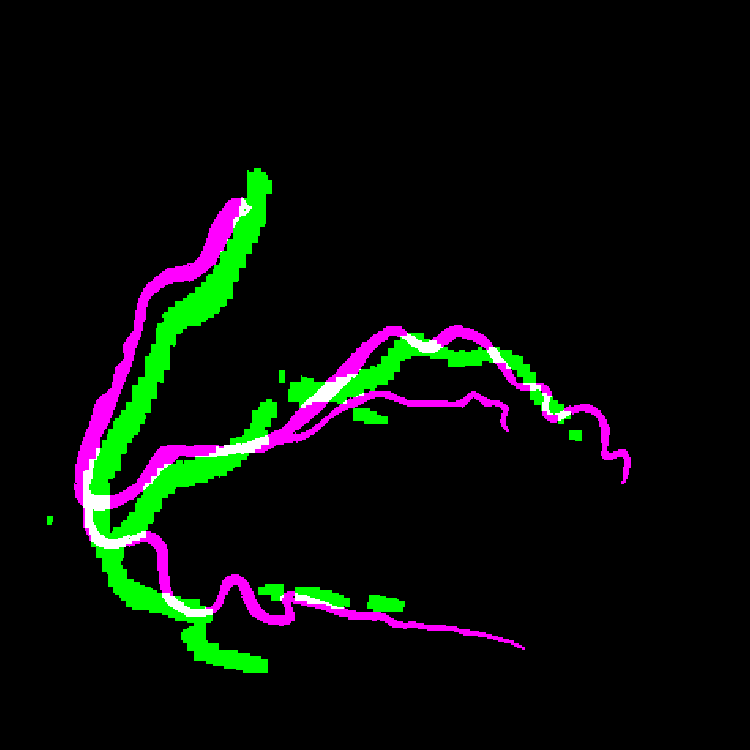}
     \end{subfigure}
     \hfill
     \begin{subfigure}[b]{0.11\textwidth}
         \centering
         \includegraphics[width=\textwidth]{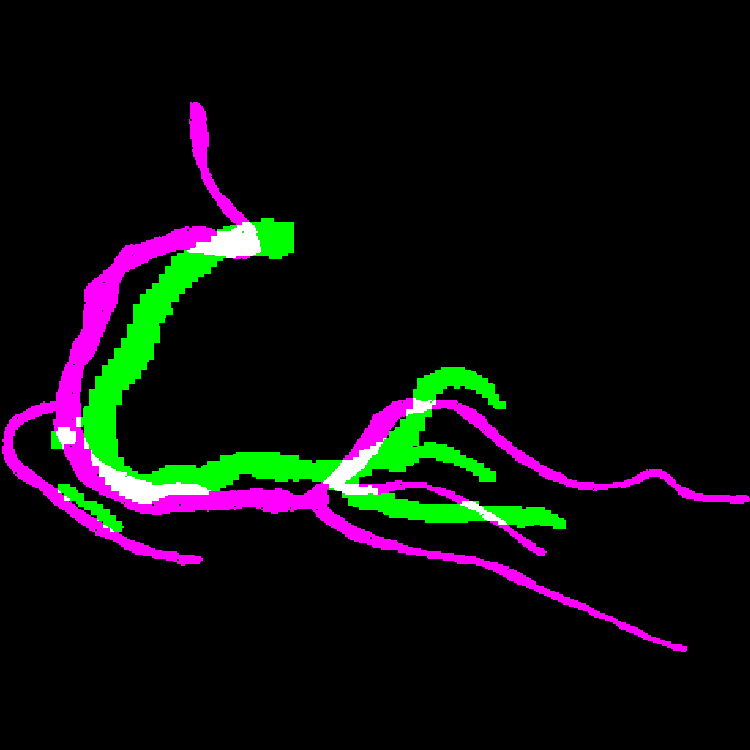}
     \end{subfigure}
     \hfill
     \begin{subfigure}[b]{0.11\textwidth}
         \centering
         \includegraphics[width=\textwidth]{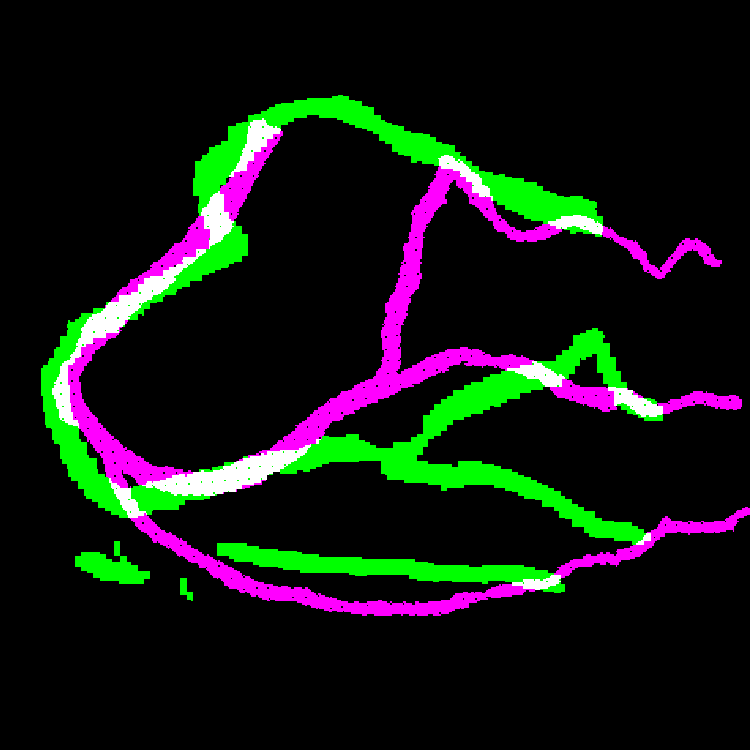}
     \end{subfigure}
     \hfill
     \begin{subfigure}[b]{0.11\textwidth}
         \centering
         \includegraphics[width=\textwidth]{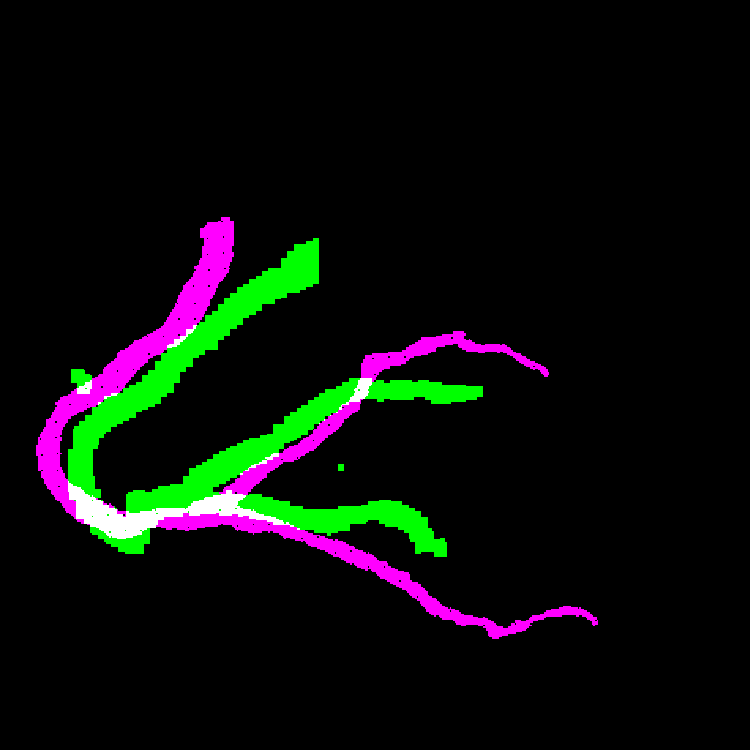}
     \end{subfigure}
     %%%%%% cvt second projection
     \vfill
     \begin{subfigure}[b]{0.11\textwidth}
         \centering
         \includegraphics[width=\textwidth]{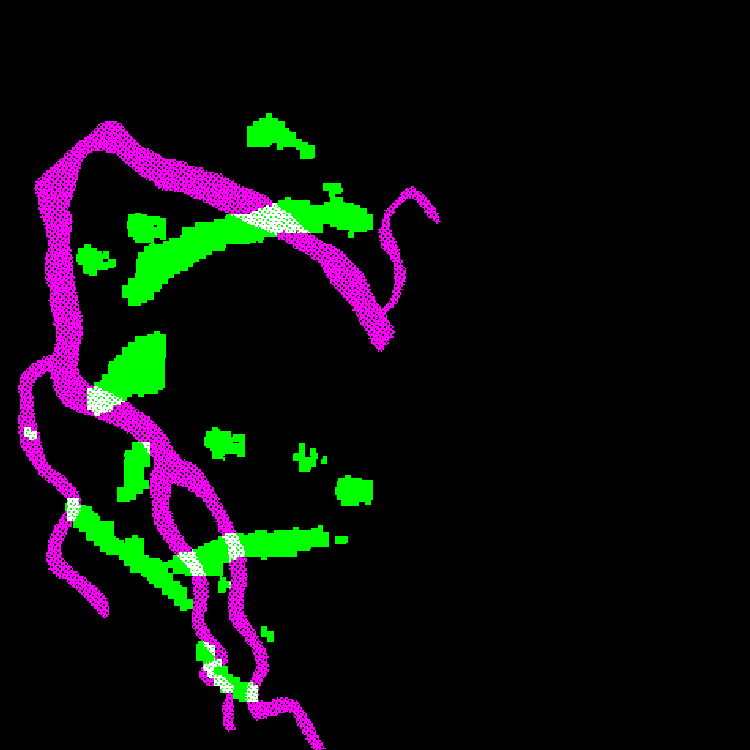}
     \end{subfigure}
     \hfill
	\begin{subfigure}[b]{0.11\textwidth}
         \centering
         \includegraphics[width=\textwidth]{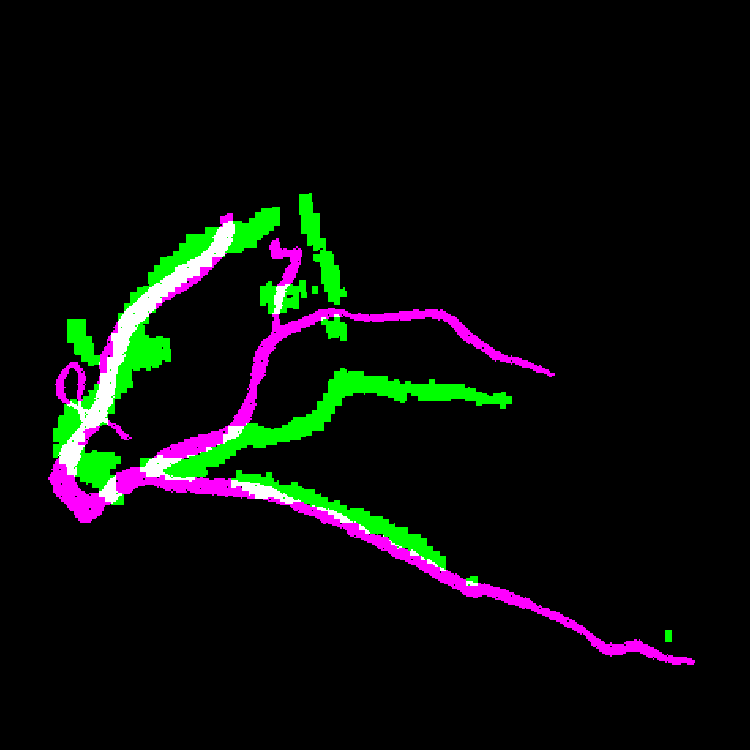}
     \end{subfigure}
     \hfill
	\begin{subfigure}[b]{0.11\textwidth}
         \centering
         \includegraphics[width=\textwidth]{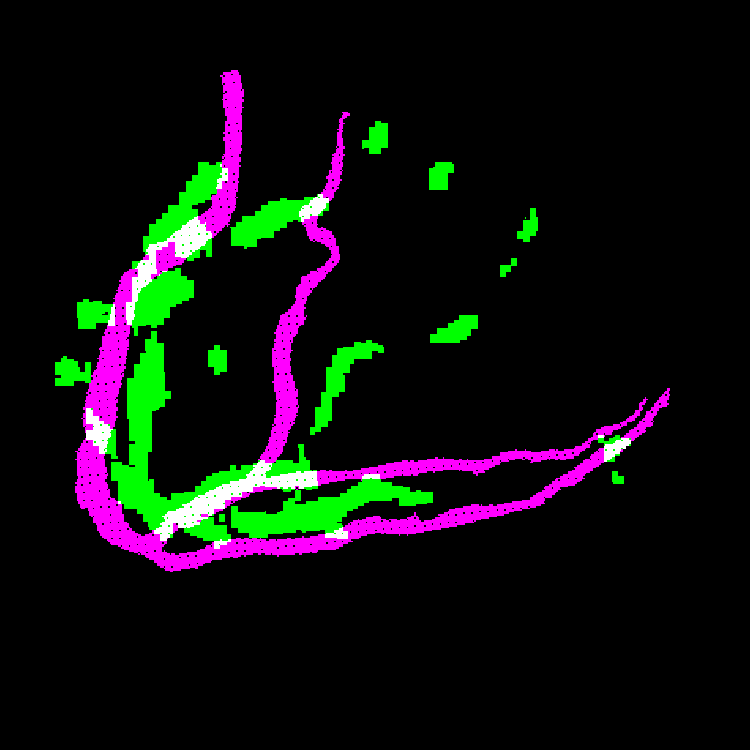}
     \end{subfigure}
     \hfill
	\begin{subfigure}[b]{0.11\textwidth}
         \centering
         \includegraphics[width=\textwidth]{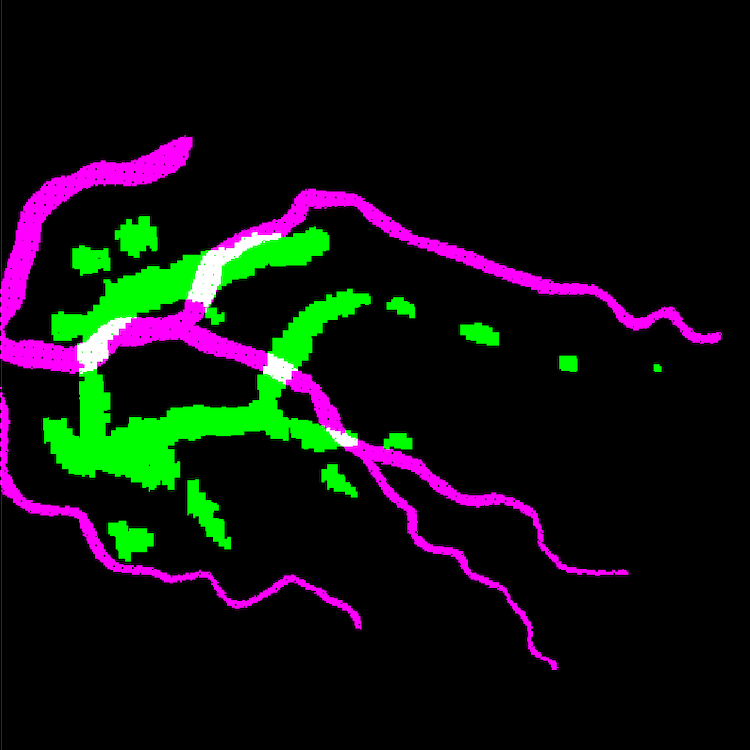}
     \end{subfigure}
     \hfill
	\begin{subfigure}[b]{0.11\textwidth}
         \centering
         \includegraphics[width=\textwidth]{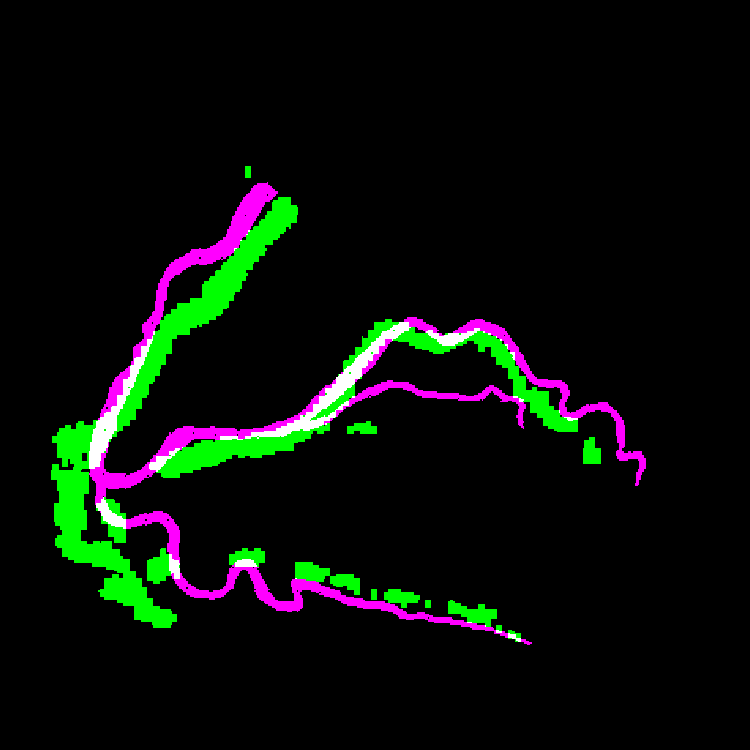}
     \end{subfigure}
     \hfill
     \begin{subfigure}[b]{0.11\textwidth}
         \centering
         \includegraphics[width=\textwidth]{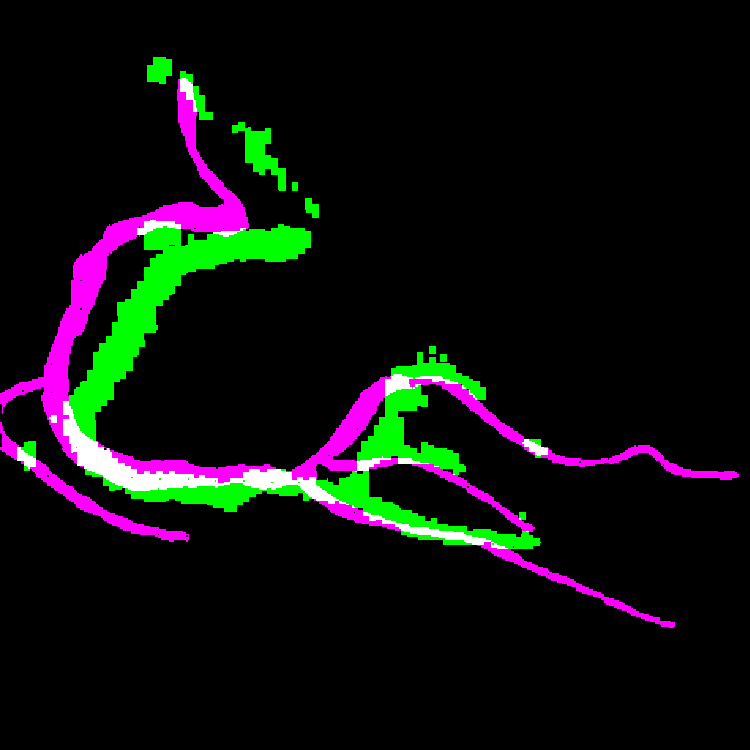}
     \end{subfigure}
     \hfill
     \begin{subfigure}[b]{0.11\textwidth}
         \centering
         \includegraphics[width=\textwidth]{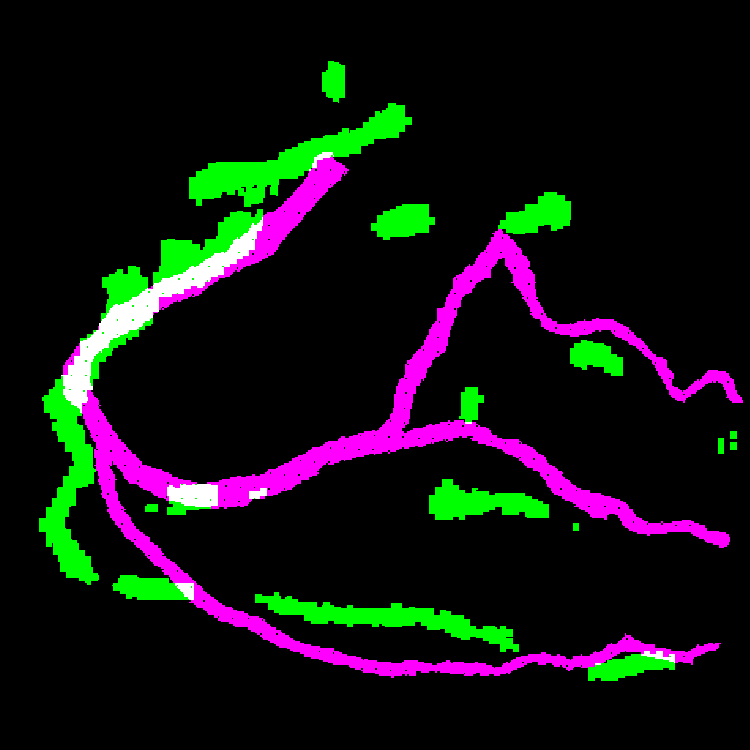}
     \end{subfigure}
     \hfill
     \begin{subfigure}[b]{0.11\textwidth}
         \centering
         \includegraphics[width=\textwidth]{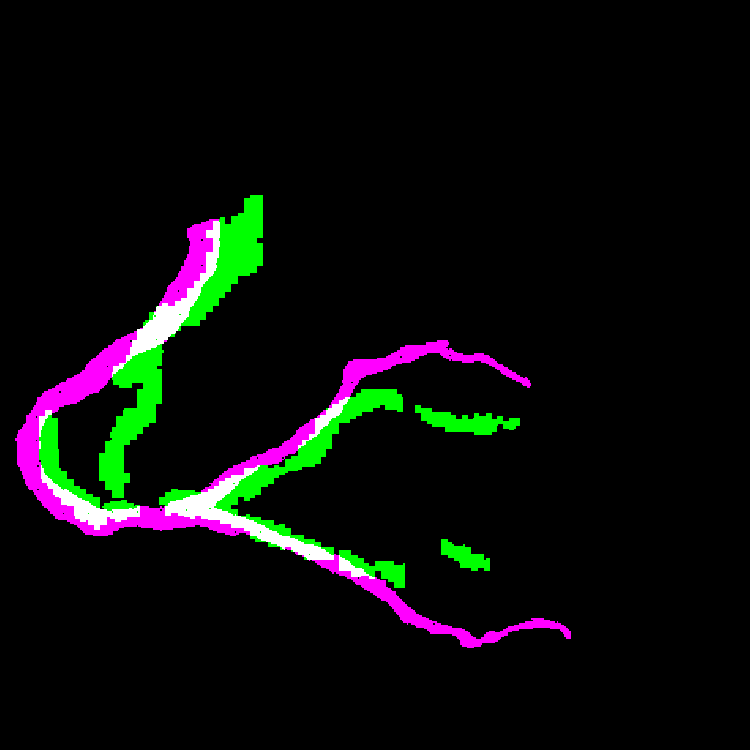}
     \end{subfigure}
     \caption{The comparisons on the second projection plane between the registered ICA data and the reprojections of the 3D reconstructions generated from all the models. The ICA data (in purple) are rigidly registered to the reprojections (in green) before comparison. From left to right: 8 patients. From top to bottom: comparisons between the registered ICA data and the reprojections from the reconstructions by our proposed DeepCA model, WGP, +CTLs, +DSCC, Un2+, Un3+, DSCN, and CVTG. Colour purple presents registered ICA data, green is reprojection, and white shows the overlap.}\label{ica_2}
\end{figure*}

\begin{figure*}[!h]
     \centering
     %%%%%% third projections 1
     \begin{subfigure}[b]{0.11\textwidth}
         \centering
         \includegraphics[width=\textwidth]{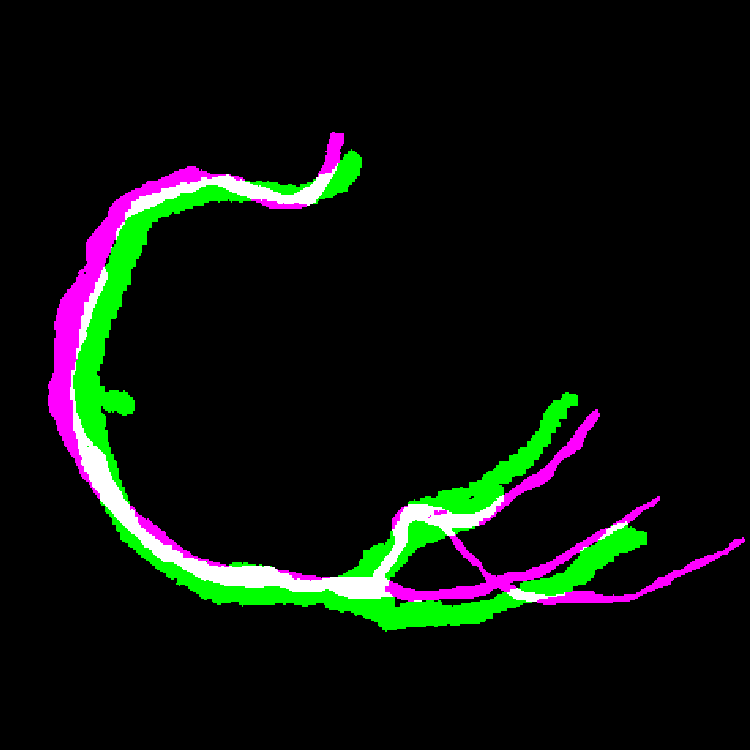}
     \end{subfigure}
     \hfill
	\begin{subfigure}[b]{0.11\textwidth}
         \centering
         \includegraphics[width=\textwidth]{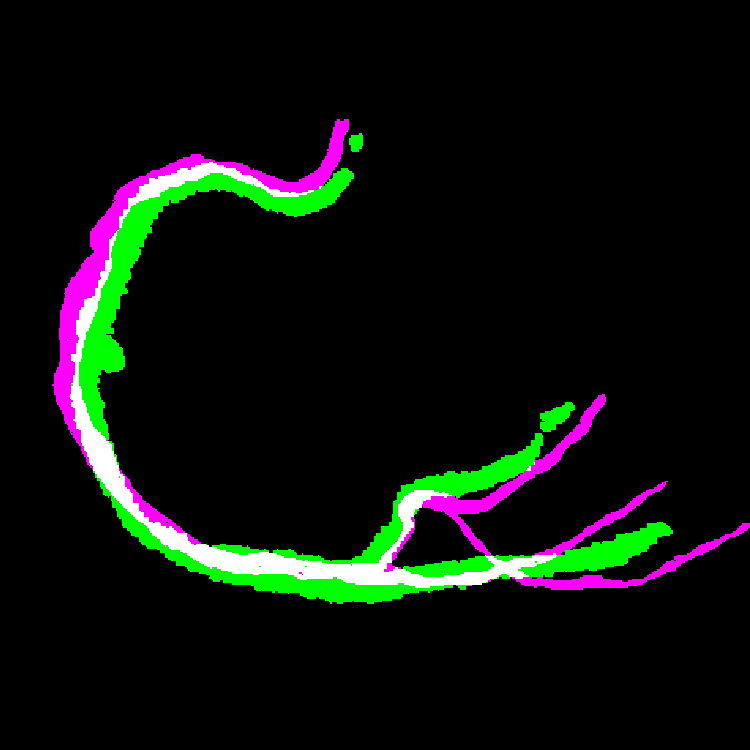}
     \end{subfigure}
     \hfill
	\begin{subfigure}[b]{0.11\textwidth}
         \centering
         \includegraphics[width=\textwidth]{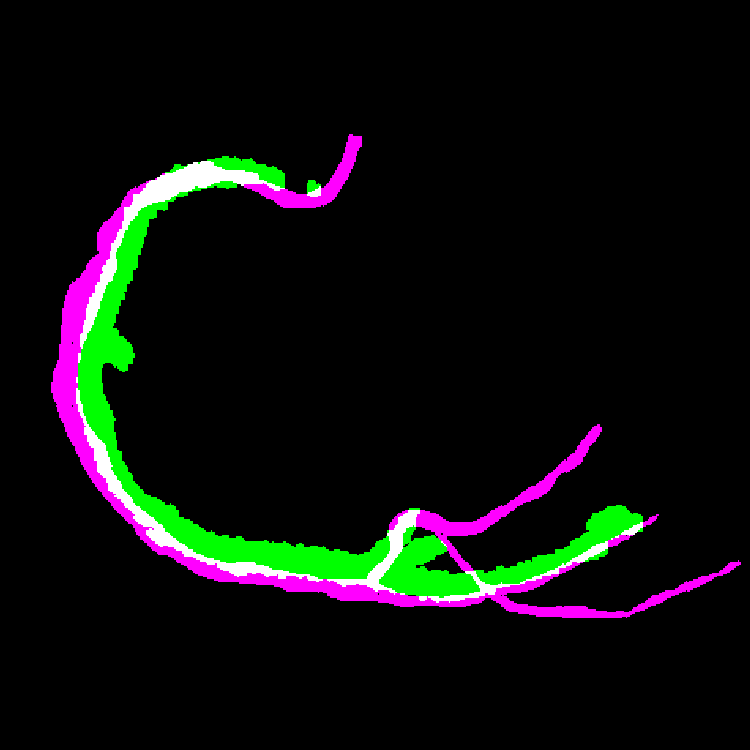}
     \end{subfigure}
     \hfill
	\begin{subfigure}[b]{0.11\textwidth}
         \centering
         \includegraphics[width=\textwidth]{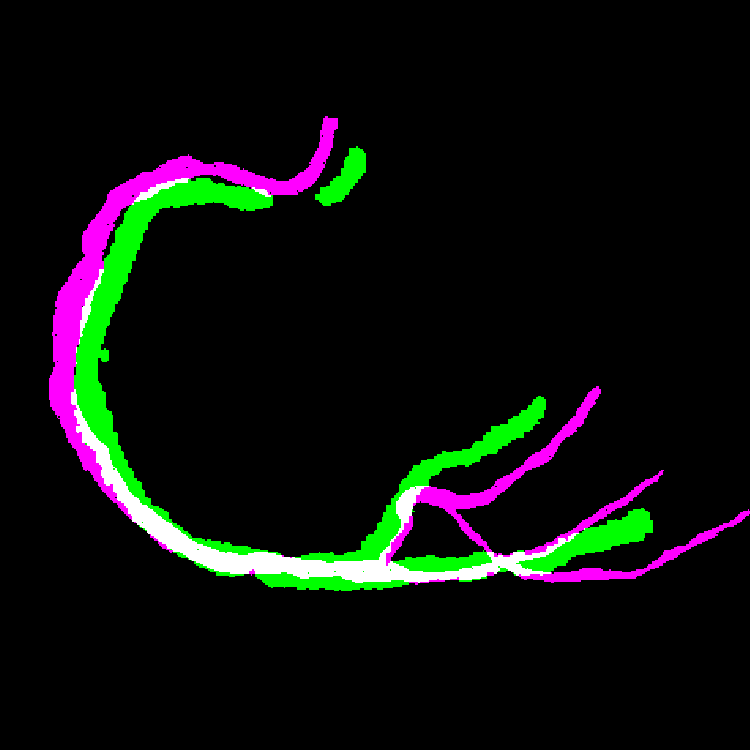}
     \end{subfigure}
     \hfill
	\begin{subfigure}[b]{0.11\textwidth}
         \centering
         \includegraphics[width=\textwidth]{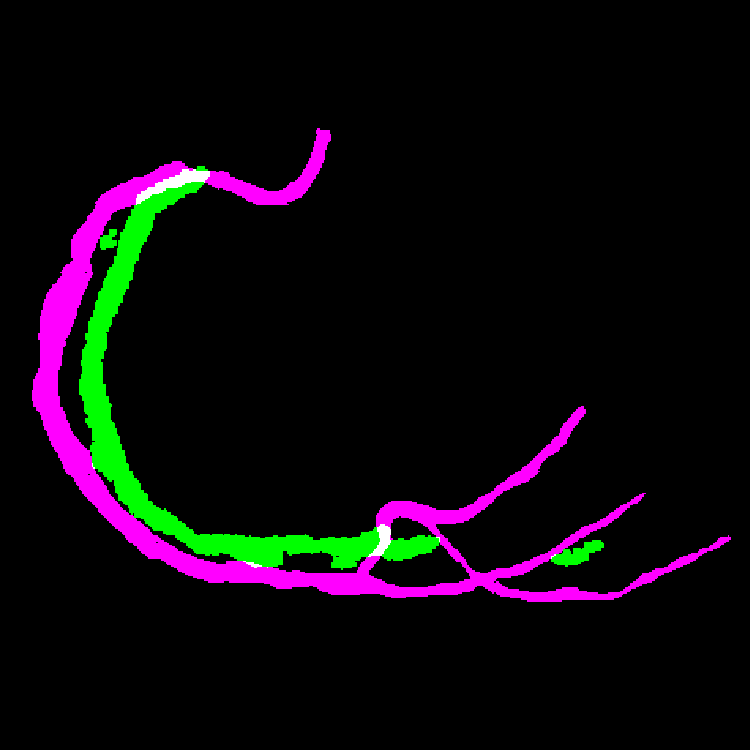}
     \end{subfigure}
     \hfill
     \begin{subfigure}[b]{0.11\textwidth}
         \centering
         \includegraphics[width=\textwidth]{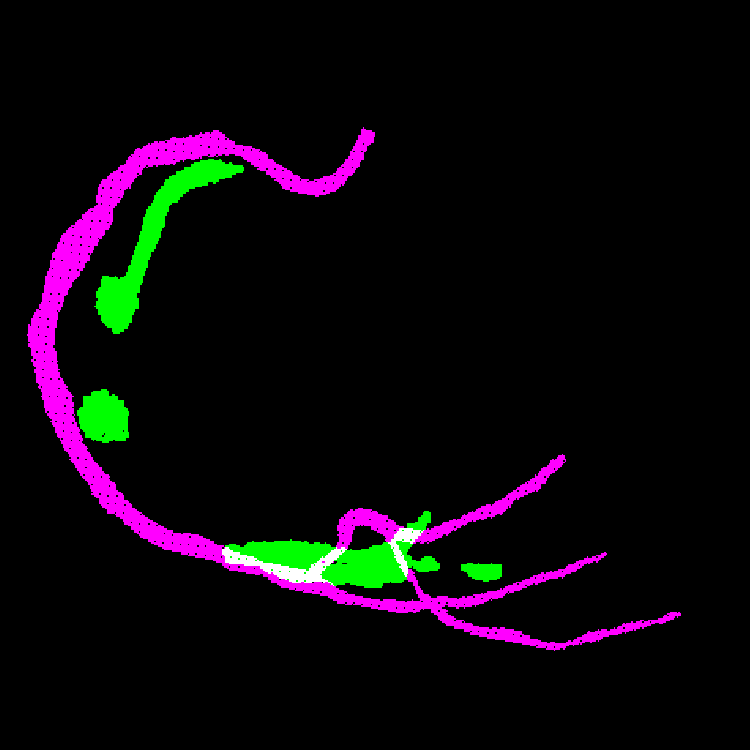}
     \end{subfigure}
     \hfill
     \begin{subfigure}[b]{0.11\textwidth}
         \centering
         \includegraphics[width=\textwidth]{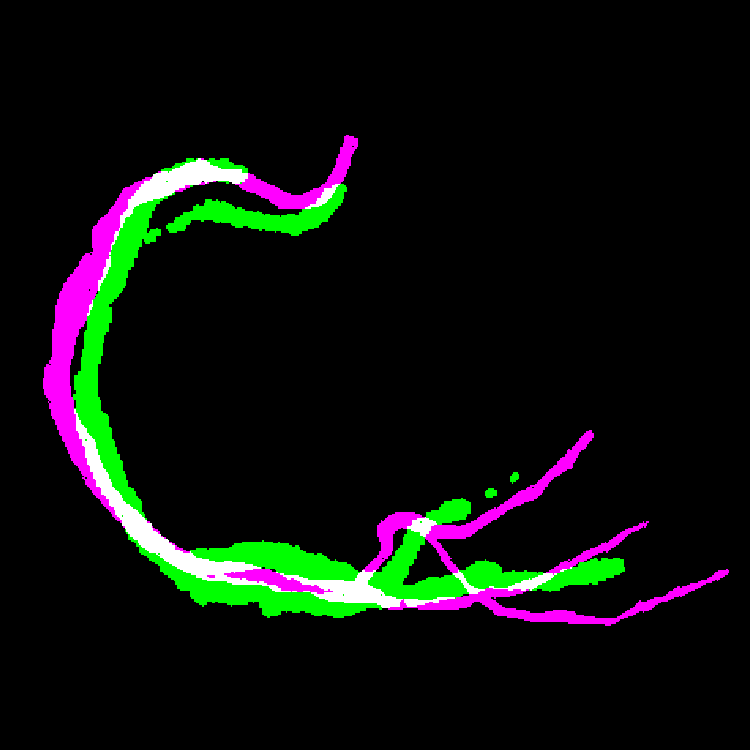}
     \end{subfigure}
     \hfill
     \begin{subfigure}[b]{0.11\textwidth}
         \centering
         \includegraphics[width=\textwidth]{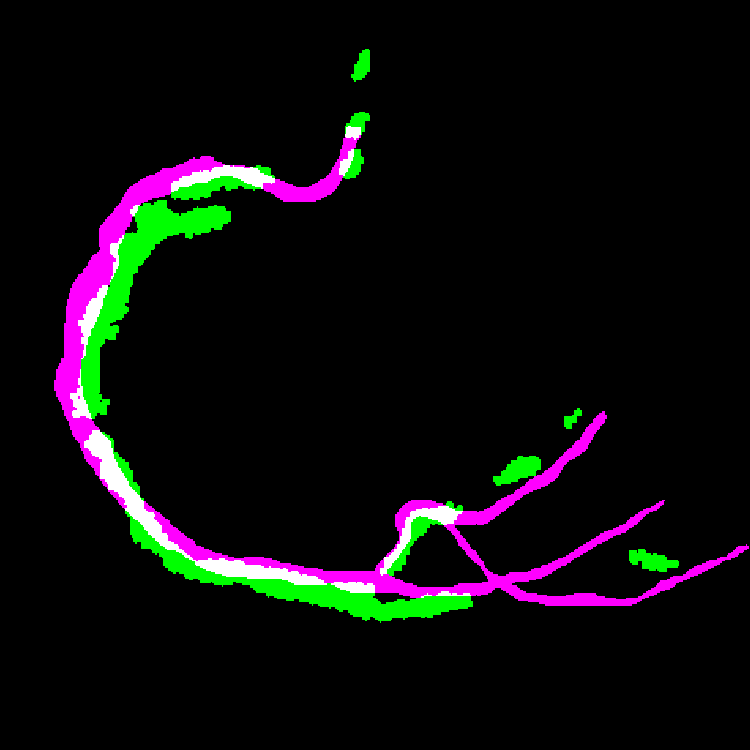}
     \end{subfigure}
     %%%%%% third projection 2
     \vfill
     \begin{subfigure}[b]{0.11\textwidth}
         \centering
         \includegraphics[width=\textwidth]{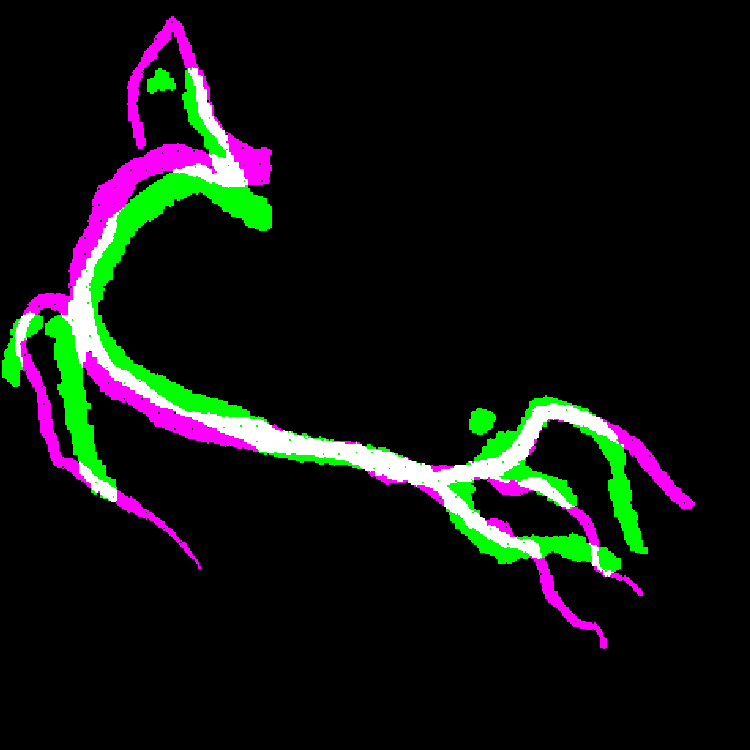}
     \end{subfigure}
     \hfill
	\begin{subfigure}[b]{0.11\textwidth}
         \centering
         \includegraphics[width=\textwidth]{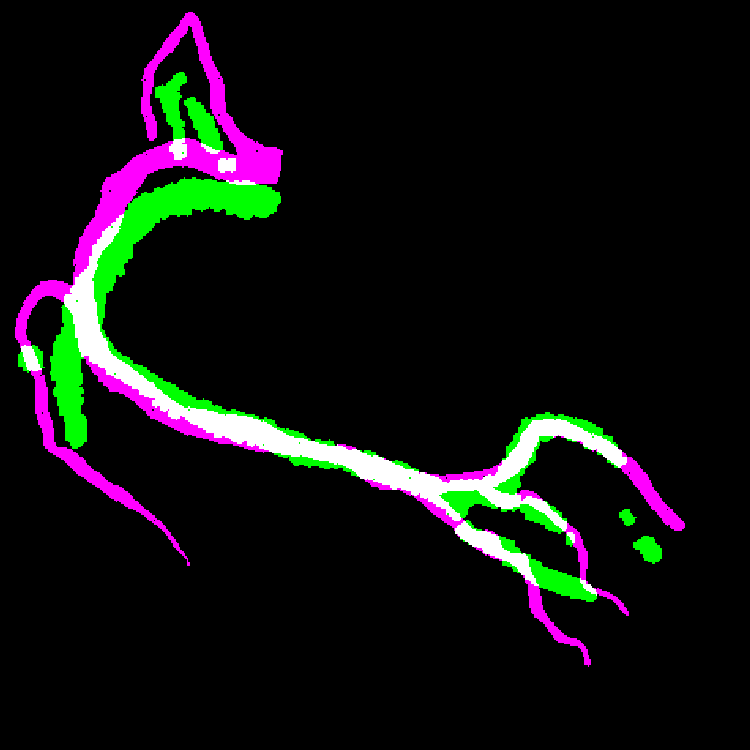}
     \end{subfigure}
     \hfill
	\begin{subfigure}[b]{0.11\textwidth}
         \centering
         \includegraphics[width=\textwidth]{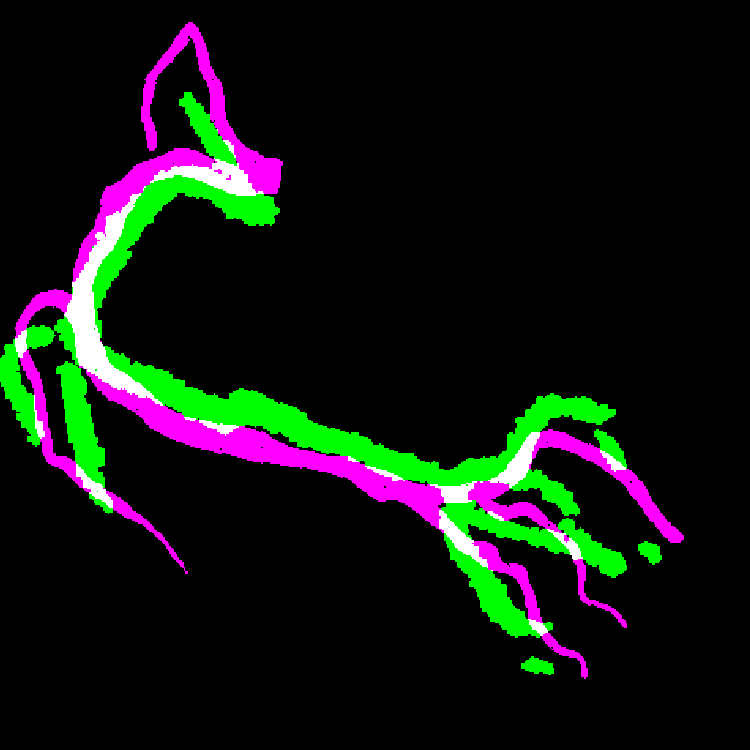}
     \end{subfigure}
     \hfill
	\begin{subfigure}[b]{0.11\textwidth}
         \centering
         \includegraphics[width=\textwidth]{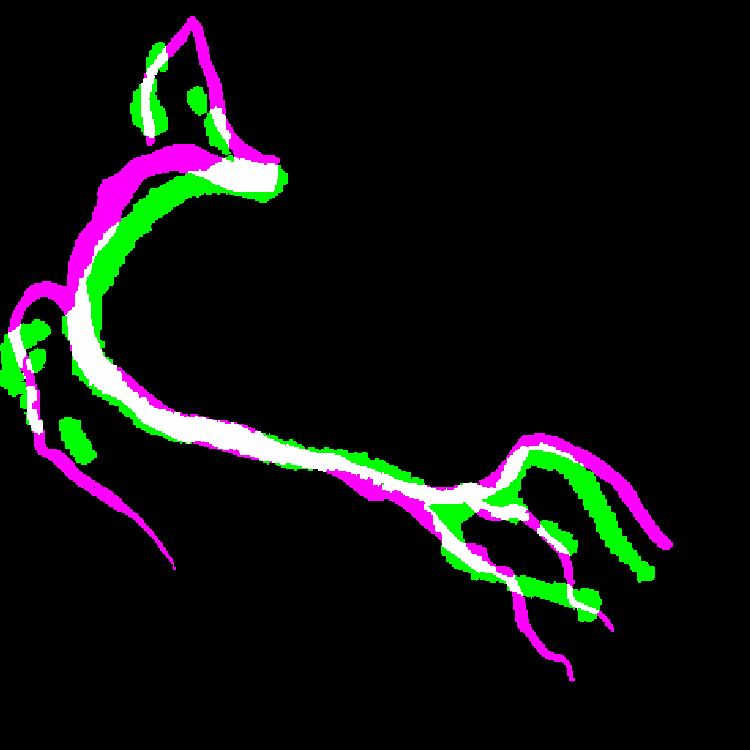}
     \end{subfigure}
     \hfill
	\begin{subfigure}[b]{0.11\textwidth}
         \centering
         \includegraphics[width=\textwidth]{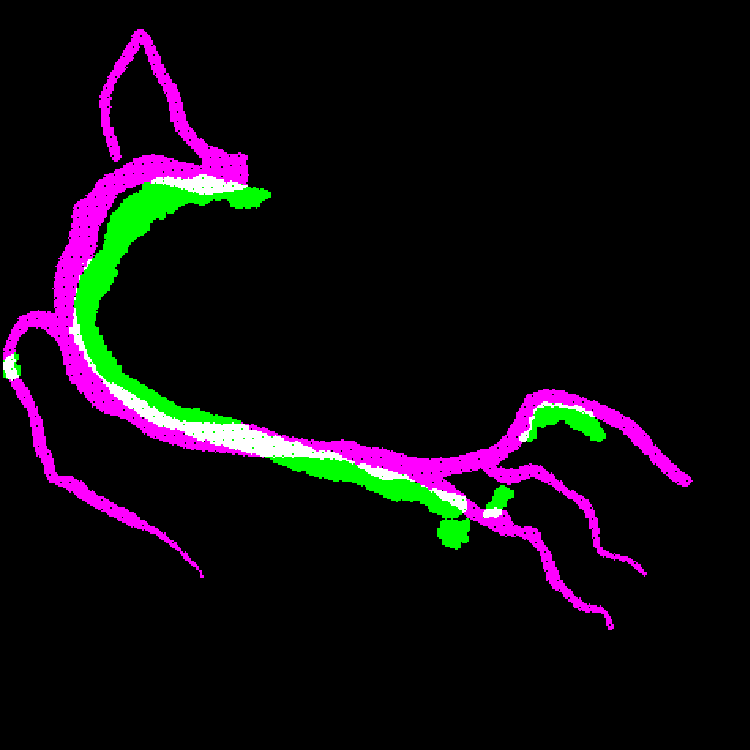}
     \end{subfigure}
     \hfill
     \begin{subfigure}[b]{0.11\textwidth}
         \centering
         \includegraphics[width=\textwidth]{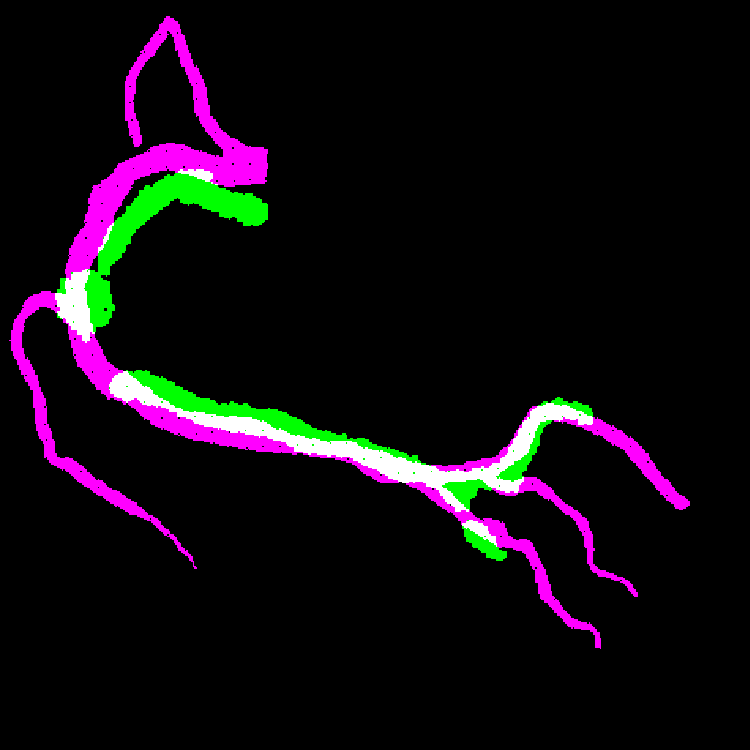}
     \end{subfigure}
     \hfill
     \begin{subfigure}[b]{0.11\textwidth}
         \centering
         \includegraphics[width=\textwidth]{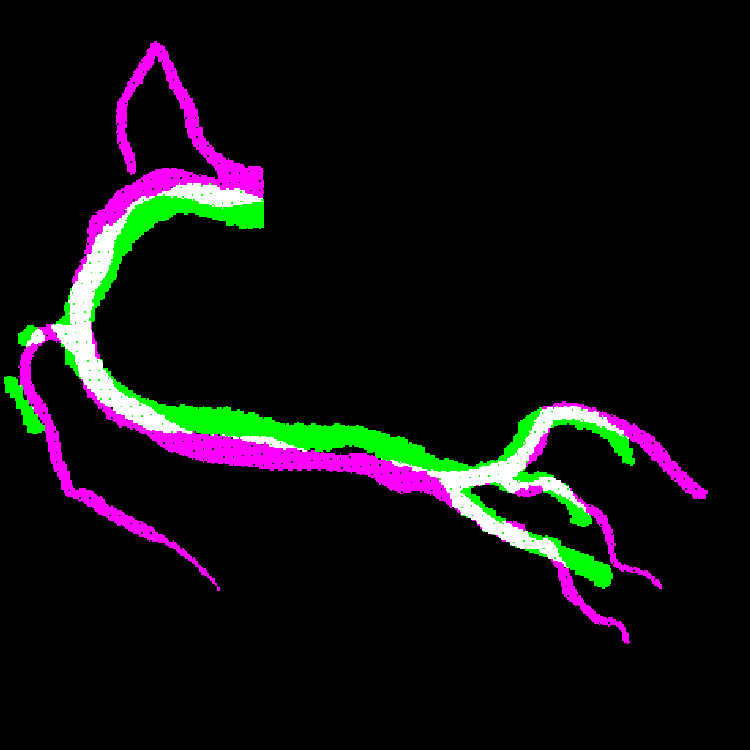}
     \end{subfigure}
     \hfill
     \begin{subfigure}[b]{0.11\textwidth}
         \centering
         \includegraphics[width=\textwidth]{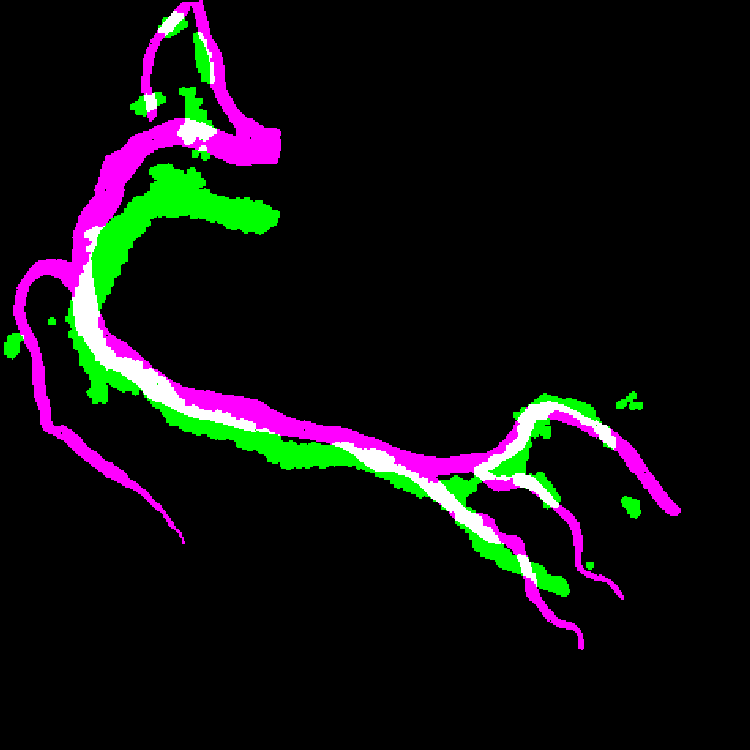}
     \end{subfigure}
     \caption{The comparisons on the additional (third) projection plane between the registered ICA data and the reprojections of the 3D reconstructions generated from all the models. The ICA data (in purple) are rigidly registered to the reprojections (in green) before comparison. From top to bottom: 2 patients who have additional ICA projections. From left to right: comparisons between the registered ICA data and the reprojections from the reconstructions by our proposed DeepCA model, WGP, +CTLs, +DSCC, Un2+, Un3+, DSCN, and CVTG. Colour purple presents registered ICA data, green is reprojection, and white shows the overlap.}\label{ica_3}
\end{figure*}

 The qualitative evaluation results on all three projection planes, as illustrated in \cref{ica_1,ica_2,ica_3}, demonstrate the superiority of our proposed DeepCA model's performance on real clinical ICA data as well. These results indicate that our proposed DeepCA model has the best performance in vessel topology preservation and recovery of missing features.

\par Moreover, in all the qualitative results from the 3 ablation models, we can find that each component of our proposed DeepCA model has contributed to the final reconstruction performance with more vascular features recovered and broken branches connected.

%\section{Projection Geometry}
\begin{table*}[!h]
\centering 
\caption{The projection geometry to simulate cone-beam forward projections on the CCTA dataset, in order to resemble the real ICA settings. The CCTA data with motion is used on the second projection plane to simulate the breathing and cardiac motions; here we rotate the CCTA data randomly for both primary and secondary angles ranging from -10$\degree$ to 10$\degree$ and add translations of -8 $mm$ to 8 $mm$ in both horizontal and vertical directions.}\label{pj_tab}
\begin{tabular}{|l|cc|}
\hline
                 & \multicolumn{1}{c|}{First Projection Plane}        & Second Projection Plane               \\ \hline
Phantom          & \multicolumn{1}{c|}{Original 3D CCTA Data}         & 3D CCTA Data with Motion                    \\ \hline
Detector Spacing & \multicolumn{2}{c|}{$0.2769 \times 0.2769$ $mm^2$ to $0.2789 \times 0.2789$ $mm^2$}        \\ \hline
Detector Size    & \multicolumn{2}{c|}{$512 \times 512$}                                                      \\ \hline
Volume Spacing   & \multicolumn{2}{c|}{$90 \times 90 \times 90$ $mm^3$ to $105 \times 105 \times 105$ $mm^3$} \\ \hline
Volume Size      & \multicolumn{2}{c|}{$128 \times 128 \times 128$}                                           \\ \hline
Distance for Source to Detector (DSD)   & \multicolumn{1}{c|}{970 $mm$ to 1010 $mm$}         & 1050 $mm$ to 1070 $mm$                \\ \hline
Distance for Source to Origin (DSO)   & \multicolumn{1}{c|}{745 $mm$ to 785 $mm$}          & $\pm$ 3 $mm$ to the First Projection  \\ \hline
Primary Angle    & \multicolumn{1}{c|}{$18 \degree$ to $42 \degree$}  & $-8 \degree$ to $8 \degree$           \\ \hline
Secondary Angle  & \multicolumn{1}{c|}{$-8 \degree$ to $8 \degree$}   & $18 \degree$ to $42 \degree$          \\ \hline
\end{tabular}
\end{table*}

{\small
\bibliographystyle{ieee_fullname}
\bibliography{refs}
}

%%%%%%%%%%%%%%
%\section{More Visual Results on 2D Clinical ICA Dataset}
%%%%%%%%%%%%%%%%%%% 3D CCTA predictions

%%%%%%%%%%%%%%%%%%% 3D CCTA CD

%%%%%%%%%%%%%%%%%%% 3D Real ICA

%%%%%%%%%%%%%%%%%%% ICA 1st

%%%%%%%%%%%%%%%%%%%% ICA 2nd

%%%%%%%%%%%%%%%%%%% ICA 3rd

%%%%%%%%% REFERENCES
% \FloatBarrier
% {\small
% \bibliographystyle{ieee_fullname}
% \bibliography{refs}
% }

\end{document}